\PassOptionsToPackage{pdfpagelabels=false}{hyperref} 
\documentclass[fleqn,usenatbib]{mnras}

\usepackage[T1]{fontenc}
\usepackage{ae,aecompl}

\usepackage{graphicx}	% Including figure files
\usepackage{amsmath}	% Advanced maths commands
\usepackage{amssymb}	% Extra maths symbols
\usepackage{color}

\def\aj{AJ}
\def\apj{ApJ}
\def\aap{A\&A}
\def\apjl{ApJL}
\def\apjs{ApJS}

\def\mnras{MNRAS}
\def\araa{ARA\&A}

\def\pasp{PASP}

\title[MAD - Star Formation Rates and Gas Metallicities on $<100pc$ Scales]
  {The MUSE Atlas of Disks (MAD):  Resolving Star Formation Rates and Gas Metallicities on $<100pc$ Scales\thanks{Based on observations obtained at the Very Large Telescope (VLT) of the European Southern Observatory, Paranal, Chile (ESO Programme IDs 095.B-0532, 096.B-0309, 097.B-0165, 098.B-0551 and 099.B-0242.)}}

\author[S. Erroz-Ferrer et al.]
{Santiago Erroz-Ferrer,$^{1}$\thanks{Email: serroz86@gmail.com} C. Marcella Carollo,$^{1}$ Mark den Brok,$^{1,2}$ Masato Onodera,$^{1,3}$     
\newauthor  Jarle Brinchmann,$^{4,5}$ Raffaella A. Marino,$^{1}$ Ana Monreal-Ibero,$^{6,7}$ Joop Schaye,$^{4}$
\newauthor   Joanna Woo,$^{1,8}$ Anna Cibinel,$^{9}$ Victor P. Debattista,$^{10}$ Hanae Inami,$^{11}$
\newauthor   Michael Maseda,$^{4}$ Johan Richard,$^{11}$ Sandro Tacchella$^{1,12}$ and Lutz Wisotzki$^{2}$\\ 
$^{1}${Department of Physics, ETH Zurich, CH-8093 Zurich, Switzerland}\\
$^{2}${Leibniz-Institut f\"ur Astrophysik Potsdam (AIP), An der Sternwarte 16, 14482 Potsdam, Germany}\\
$^{3}${Subaru Telescope, National Astronomical Observatory of Japan, HI 96720 Hilo, USA}\\
$^{4}${Leiden Observatory, Leiden University, PO Box 9513, NL-2300 RA Leiden, the Netherlands}\\
$^{5}${Instituto de Astrof\'isica e Ci\^encias do Espa\c{c}o, Universidade do Porto, CAUP, Rua das Estrelas, PT4150-762 Porto, Portugal}\\
$^{6}${Instituto de Astrof\'isica de Canarias, V\'ia L\'actea s/n 38205 La Laguna, Spain}\\
$^{7}${Departamento de Astrof\'isica, Universidad de La Laguna, 38206 La Laguna, Spain}\\
$^{8}${Dept. of Physics \& Astronomy, University of Victoria, PO Box 1700 STN CSC Victoria, Canada}\\
$^{9}${Astronomy Centre, Department of Physics and Astronomy, University of Sussex, Brighton, BN1 9QH, UK}\\
$^{10}${Jeremiah Horrocks Institute, University of Central Lancashire, Preston PR1 2HE, UK}\\
$^{11}${Univ Lyon, Univ Lyon1, Ens de Lyon, CNRS, Centre de Recherche Astrophysique de Lyon UMR5574, F-69230, Saint-Genis-Laval, France}\\
$^{12}${Harvard-Smithsonian Center for Astrophysics, 60 Garden St, Cambridge, MA 02138, USA}}
\date{Accepted 2019 January 9. Received 2019 January 9; in original form 2017 December 22.}

\pubyear{2019}

\begin{document}
\label{firstpage}
\pagerange{\pageref{firstpage}--\pageref{lastpage}}
\maketitle

\begin{abstract}
We study the physical properties of the ionized gas in local disks using the sample of 38 nearby $\sim10^{8.5-11.2}$M$_\odot$ Star-Forming Main Sequence (SFMS) galaxies observed so far as part of the MUSE Atlas of Disks (MAD). Specifically, we use all strong emission lines in the MUSE wavelength range 4650-9300\,\AA\ to investigate the resolved ionized gas properties on $\sim$\,100 pc scales. This spatial resolution enables us to disentangle H{\sc ii} regions from the Diffuse Ionized Gas (DIG) in the computation of gas metallicities and Star Formation Rates (SFRs) of star forming regions.

The gas metallicities generally decrease with radius. The metallicity of the H{\sc ii} regions is on average $\sim$0.1 dex higher than that of the DIG, but the metallicity radial gradient in both components is similar. The mean metallicities within the inner galaxy cores correlate with the total stellar mass of the galaxies. On our <\,100 pc scales, we find two correlations previously reported at kpc scales: a spatially resolved Mass-Metallicity Relation (RMZR) and a spatially resolved SFMS (RSFMS). We find no secondary dependency of the RMZR with the SFR density. We find that both resolved relations have a local origin, as they do not depend on the total stellar mass. The observational results of this paper are consistent with the inside-out scenario for the growth of galactic disks.
\end{abstract}


%%%%%%%%%%%%%%%%%%%%%%%%%%%%%%%%%%%%%%%%%%%%%%%%%%

%%%%%%%%%%%%%%%%% BODY OF PAPER %%%%%%%%%%%%%%%%%%

\section{Introduction}

The emission line spectrum emitted by ionized gas is the observational key to the properties of not only the ionized gas in galaxies, but also of the massive young stellar population that has recently formed inside galactic disks, as the line fluxes are a direct tracer of the properties of the H{\sc ii} regions that are ionized by such population. Also, far from and in-between H{\sc ii} regions,  Diffuse Ionized Gas (DIG; also known as warm ionized medium, WIM) adds further information about  the  global  gas-stellar  cycle that is globally sustained within galactic disks. The DIG  is warm ($\sim$10$^4$K), ionized, low density (0.1 cm$ ^{-3} $) gas with low ionization parameter that can extend 1 kpc or more above the disk plane (see  \citealt{Mathis2000} and \citealt{Haffner2009} for reviews). Most of the ionized gas in a galaxy is in the form of DIG \citep{Walterbos1998}. This gas component is understood to be ionized by stars (O, B and hot evolved stars) in the disk whose Lyman continuum photons travel large path lengths (non-local sources of ionization), and its emission is superimposed on that coming from the local ionization inside H{\sc ii} regions (\citealt{Mathis1986,Mathis2000}; \citealt{Domgorgen1994}; \citealt{Sembach2000}; \citealt{Wood2004}; \citealt{Wood2010}). This component  can be pushed above the galactic disk by superbubbles created by supernova events (e.g., \citealt{Wood2010}). \citet{Weilbacher2017} recently found  that the UV photons leaked by the H{\sc ii} regions in the Antennae can explain the ionization of the DIG there.

The metallicity of the gas phase (referred to as $ Z_{\rm gas} \equiv 12+{\rm log(O/H)} $, oxygen abundance or metallicity throughout this paper) is  a product of the whole formation and evolutionary history of a galaxy.  Oxygen is synthesized in high-mass stars and can be released into the interstellar medium (ISM) via supernova explosions or into the circumgalactic medium through galactic outflows. In addition, pristine metal-poor gas may accrete during the evolution of the galaxy (see e.g. \citealt{Sanchez-Almeida2014} and references therein), changing the metallicity of a galaxy but also triggering star formation (SF) which would enrich the galactic disk.  On global galactic scales, the relationship between total galaxy stellar mass and total gas metallicity (i.e., the mass-metallicity relation, MZR), is well established over a wide range of galaxy masses (e.g., \citealt{Lequeux1979}; \citealt{Tremonti2004}; \citealt{Lee2006}; \citealt{Kewley2008}; \citealt{Zhao2010}) . The MZR has been shown to evolve with redshift (e.g., \citealt{Savaglio2005}; \citealt{Maier2006}; \citealt{Erb2006}; \citealt{Maiolino2008}; \citealt{Onodera2016}).  

Although still controversial (\citealt{Barrera-Ballesteros2017}; \citealt{Sanchez2017,Sanchez2018}), it has also been argued that the total star-formation rate (SFR) is a second parameter in the MZR (with SFR being anti-correlated with gas metallicity),  and that possibly the three-parameter relation of mass, gas metallicity and SFR is universal at all epochs  (\citealt{Ellison2008}; \citealt{Mannucci2010}; \citealt{Lara-Lopez2010}).  This so-called ``Fundamental Metallicity Relation'' (FMR) is well explained by self-regulated 'bathtub' galaxy evolution models (e.g., \citealt{Bouche2010}; \citealt{Lilly2013}). \cite{Lilly2013} showed that the FMR emerges naturally in their 'regulator' model, due to the fact that the evolution of the specific star formation rate, sSFR, closely follows the specific accretion rate of dark matter halos. However, other  'regulator' models identify different parameters as  the key driver of the internal galactic (semi)equilibrium  (e.g., \citealt{Finlator2008}; \citealt{Dave2012}; \citealt{Forbes2014}). Studying the relationship between $Z_{\rm gas}$, $M_{\star}$ and SFR is important to understand the relative importance of different physical parameters in establishing the physical balance within galactic disks  (\citealt{Lilly2013}; \citealt{Ma2016}).  

Important details are coming from the novel Integral Field Unit spectrographs (IFUs), which are enabling the study of the existence and properties of MZR- and FMR-equivalent relationships on resolved local scales within individual galaxies. The resolved spectra enable us not only to disentangle H{\sc ii} regions from DIG emission, and gas with SF excitation properties from gas that is excited by shocks and/or AGNs, but also to understand whether the global MZR, FMR and SFMS relations emerge from averaging the local contributions to the total mass, metallicity and SFR. Several IFU surveys such as the Calar Alto Legacy Integral Field Area (CALIFA, \citealt{Sanchez2012}), Mapping Nearby Galaxies at Apache Point Observatory (MaNGA, \citealt{Bundy2015}), SAMI Galaxy Survey \citep{Croom2012} and others have investigated  the relation between  stellar surface mass density ($\Sigma_{\star}$), SFR surface density and gas metallicity resolved on kpc scales within galaxies (e.g., \citealt{Rosales-Ortega2012}; \citealt{Barrera-Ballesteros2016}; \citealt{Barrera-Ballesteros2017}; \citealt{Sanchez2017,Sanchez2018}). While  \citealt{Rosales-Ortega2012} find a local FMR, \citealt{Sanchez2013}; \citealt{Barrera-Ballesteros2017} and \citealt{Sanchez2017,Sanchez2018} find a lack of a secondary relation between the MZR and the SFR at one effective radius ($R_{e}$). Using IFU data, \citet{Cano-Diaz2016}; \citet{Hsieh2017} and \citet{Medling2018} among others have studied the relationship between $\Sigma_{\star}$ and $\Sigma _{{\rm H}\alpha}$ (or $\Sigma_{\rm SFR}$) on kpc scales, finding a resolved SFMS (RSFMS) relation at these kpc scales.

The availability of spatially-resolved information for the ionized gas component of galactic disks also enables more detailed  determination of the  radial profiles of oxygen abundances within galaxies, i.e., of gas metallicity gradients within the disks. Their  shape, together with the azimuthal variations around the radial gradients, add crucial  information on the past chemical enrichment histories of the whole galaxies and their main structural components such as  bulges, bars  and disks.  The above-mentioned IFU surveys, which disentangle the ionized gas within galaxies on kpc-sized scales, have shown  that negative gas metallicity gradients, i.e., outward-decreasing metallicity profiles, are quite common in the local Universe (e.g., \citealt{Sanchez2012b,Sanchez2014}; \citealt{Sanchez-Menguiano2017}). Importantly, however, the precise shape of these gradients depends on the specific metallicity calibrator that is adopted to convert line fluxes into oxygen abundances (e.g., \citealt{Sanchez-Menguiano2016,Sanchez-Menguiano2017}; \citealt{Vogt2017}).  

The overall result of widespread  negative gas metallicity gradients is not very surprising,  given that inward gas flows over the lifetime of a galaxy would almost inevitably cause radial metallicity gradients like those observed (e.g.; \citealt{Lacey1985}).   Such negative gradients are also however consistent with an inside-out formation scenario of the disks,  which is reasonable to expect  for  cosmological reasons (e.g.; \citealt{White1991}; \citealt{Mo1998}).   Other processes  such as  secular evolution, radial migration and merging can also occur over cosmic time; in general, their global effect  seems to be in the direction of flattening these profiles within a large radial extent  (e.g., \citet{Vilacostas1992};  \citealt{Zaritsky1994}; \citealt{Martin1994,Martin1995}; \citealt{Friedli1994,Friedli1995};  \citealt{Vilchez1996}; \citealt{Roy1997}; \citealt{Portinari2000}; \citealt{Cavichia2014}; \citealt{Marino2012,Marino2016}).

The Multi Unit Spectroscopic Explorer (MUSE, \citealt{Bacon2010}) spectrograph on the Very Large Telescope (VLT) of the European Southern Observatory (ESO)  is opening a new avenue of studies of   far and near galactic populations.  In particular, MUSE, 
within its field-of-view (FoV) of 1 arcmin$^2$, delivers higher combination of spatial and spectral resolutions than any other current instrument. Targeting a sample of  $\sim$50 disk galaxies that are close enough to be dissected by MUSE at an average spatial resolution of $\sim100$ pc or less, our MUSE Atlas of Disks (MAD) survey (Carollo et al., in prep.) provides a new view of the local disk population on the small physical scales that are relevant for probing the physical conditions on which gas and stars  directly affect each other and thus the physical origin of the global state of galactic disks. The MAD sample covers the galaxy mass range between $10^{8.5}$ M$_\odot$ and  $10^{11}$ M$_\odot$ and traces, within this range, the so-called  'Star Forming Main Sequence' relation between total stellar mass and total SFR (SFMS; \citealt{Noeske2007}; \citealt{Daddi2007}; \citealt{Elbaz2007}; \citealt{Peng2010};  \citealt{Tomczak2016}).

In  this paper we focus our attention on the physical properties of the ionized gas inside the 38 MAD galaxies observed and analysed so far, with emphasis on the gas metallicity and SFR of both H{\sc ii} and DIG components. Companion articles will present other aspects of the MAD survey, including its specifications and goals (MAD1; Carollo et al. in prep.), the main properties of  stellar and gas kinematics across the SFMS (MAD3; den Brok et al. in prep) and the analysis of the stellar continuum and line absorption spectra for stellar ages,  abundances and abundance ratios determinations (MAD4; Onodera et al. in prep). More specifically, exploiting the unprecedented $<100$pc resolution of the MAD data, in this paper we separate the emission of H{\sc ii} and DIG regions and give an overview of the SFRs and metallicities of the star-forming regions at our sub-kpc scales. We also study the relations between surface stellar mass density, gas metallicity and surface SFR density separately for the H{\sc ii} regions and for the surrounding DIG to understand how the chemical enrichment happens at these local scales and across the SFMS.

This paper is organized as follows. Sect.~\ref{sectiondata} describes the MAD sample that we use in this study and the basic data reduction;  Sect.~\ref{sectionanalysis} describes the analysis performed on MUSE data in order to obtain the line flux diagnostics on which we base this study, including correction for dust effects. Sect.~\ref{sectionresults} presents the basic diagnostics. The results are presented in Sect.~\ref{section6} and discussed in Sect.~\ref{sectiondiscussion}. We summarize in Sect.~\ref{section7}. In Appendix \ref{App:AppendixA} we provide additional information on the individual galaxies and show their flux and derived-diagnostic maps; the latter are available for download together with the reduced MUSE cubes\footnote{https://www.mad.astro.ethz.ch/data-products}. Throughout this paper, we adopt a $H_0=73\pm 5~\mathrm{km~s^{-1}~Mpc^{-1}}$ and  $1\arcsec$ corresponds to $\approx 50~\mathrm{pc}$ at a distance of 10 Mpc. 

%%%%%%%%%%%%%%%%%%%%%THE DATA%%%%%%%%%%%%%%%%%%%%%%%%%%%%%

\section{MAD data}
\label{sectiondata}

\subsection{The sample}

We analyse 38 galaxies, out of which 36 were observed during the MAD Guaranteed Time Observations (GTO) runs conducted between April 2015 and September 2017, plus NGC~337 from the MUSE Commissioning Programme 60.A-9100(C) and NGC~1097 from 097.B-0640(A). These 36 constitute all the MAD observations before the AO was mounted on the VLT.  A detailed table with the properties of the MAD galaxies and description of their observations is presented in MAD1. In Table~\ref{tableprops} we present some of the global properties of the 38 galaxies studied in this paper. Briefly, these galaxies are nearby ($z<0.013$), spiral galaxies with inclination < 70$\degr$ and stellar masses between 10$^{8.5}$ and 10$^{11.2}$ $M_{\sun}$, which lie in the $z=0$ SFMS. They show a variety of structural components (bars, star forming rings, AGN, bulges and pseudo-bulges).

\begin{table*}
\caption{Global properties of the 38 MAD galaxies studied in this paper, ordered by total stellar mass. \textit{Column~I)} Galaxy name. \textit{Column~II)} Morphological classification from The Third Reference Catalogue of Bright Galaxies (RC3; \citealt{RC3}). \textit{Columns~III) and IV)} Galaxy redshift and adopted values of the distances, from the NASA/IPAC Extragalactic Database (NED). \textit{Column~V)} Effective radius in arcsec, obtained from 2-D decomposition to photometric images (procedure explained in MAD1). \textit{Columns~VI) and VII)} Total stellar mass and star formation rate (SFR), computed from Spectral Energy Distribution (SED) fitting (details given in MAD1). \textit{Column~VIII)} Average gas metallicity inside 0.5 $R_{\rm e}$. Columns IX) and X) Metallicity gradient for all the regions computed when normalizing the galactocentric distance by $R_{\rm e}$ and by physical distance, respectively.}
 \label{tableprops}
\begin{tabular}{l|c|c|c|c|c|c|c|c|cc|c|c|c|c}
\hline
Galaxy name  & Morphology&\textit{z}      &  $  D$  & $R_{e}$ & log(M$_{\star}$) & SFR & $\left<Z_{\rm gas}\right>_{0.5Re}$&$ \triangledown Z_{\rm gas}$ & $\triangledown Z_{\rm gas}$\\
&   && (Mpc)   &(")   & log(M$_{ \sun} $) & ($M_{ \sun} $ yr$^{-1}$) & (dex)  & (dex/R$_{e}$) & (dex/kpc) \\
\hline
%name		&     d    &  Re    & mass    &med	std	&med noAGN	std noAGN	&med HII	std HII&	med DIG	std DIG  &gradients	sigma   &   gradients	sigmas		\\
NGC~4030    & SA(s)bc       & 0.004887 &  29.9 &  31.8 & 11.18  & 11.08  &9.00 $\pm$	0.07	&  -0.13	$\pm$ 0.01	  & -0.03	$\pm$ 0.01		\\
NGC~3521    & SAB(rs)bc     & 0.002672 &  14.2 &  61.7 & 11.15  & 2.98   &8.85 $\pm$	0.06	&  -0.09	$\pm$ 0.02	  & -0.02	$\pm$ 0.01		\\
NGC~3256    & pec           & 0.009354 &  38.4 &  26.6 & 11.14  & 3.10   &8.77 $\pm$	0.09	&  -0.04    $\pm$ 0.04	  & -0.01	$\pm$ 0.01		\\
NGC~4603    & SA(s)c?       & 0.008647 &  32.8 &  44.7 & 11.10  & 0.65   &8.82 $\pm$	0.10	&  -0.17	$\pm$ 0.04	  & -0.02	$\pm$ 0.01		\\
NGC~3393    & (R')SB(rs)a?  & 0.012509 &  55.2 &  21.1 & 11.09  & 7.06   &8.86 $\pm$	0.02	&  0.04     $\pm$ 0.02	  & 0.01	$\pm$ 0.01		\\
NGC~1097    & SB(s)b        & 0.004240 &  16.0 &  55.1 & 11.07  & 4.66   &8.99 $\pm$	0.08	&  -0.07	$\pm$ 0.08	  & -0.03	$\pm$ 0.01		\\
NGC~289     & SB(rs)bc      & 0.005434 &  24.8 &  27.0 & 11.00  & 3.58   &8.93 $\pm$	0.11	&  -0.09	$\pm$ 0.03	  & -0.03	$\pm$ 0.01		\\
NGC~4593    & (R)SB(rs)b    & 0.009000 &  25.6 &  63.3 & 10.95  & 4.10   &8.94 $\pm$	0.14	&  -0.38	$\pm$ 0.01	  & -0.01	$\pm$ 0.02		\\
IC~2560     & (R')SB(r)b?   & 0.009757 &  32.2 &  37.8 & 10.89  & 3.76   &8.81 $\pm$	0.11	&  -0.05	$\pm$ 0.04	  & -0.01	$\pm$ 0.01		\\
NGC~5643    & SAB(rs)c      & 0.003999 &  17.4 &  60.7 & 10.84  & 1.46   &8.82 $\pm$	0.11	&  -0.25	$\pm$ 0.12	  & -0.06	$\pm$ 0.03		\\
NGC~3081    & (R)SAB0/a(r)  & 0.007976 &  33.4 &  18.9 & 10.83  & 1.47   &8.97 $\pm$	0.08	&  -0.08	$\pm$ 0.01	  & -0.02	$\pm$ 0.01		\\
NGC~4941    & (R)SAB(r)ab?  & 0.003696 &  15.2 &  64.7 & 10.80  & 3.01   &8.93 $\pm$	0.13	&  -0.22	$\pm$ 0.02	  & -0.06	$\pm$ 0.02		\\
NGC~5806    & SAB(s)b       & 0.004533 &  26.8 &  27.2 & 10.70  & 3.61   &8.84 $\pm$	0.07	&  -0.04	$\pm$ 0.02	  & 0.00	$\pm$ 0.01		\\
NGC~3783    & (R')SB(r)ab   & 0.009730 &  40.0 &  27.7 & 10.61  & 6.93   &8.82 $\pm$	0.06	&  -0.04	$\pm$ 0.01	  & -0.01	$\pm$ 0.01		\\
NGC~5334    & SB(rs)c?      & 0.004623 &  32.2 &  51.2 & 10.55  & 2.45   &8.64 $\pm$	0.11	&  -0.30	$\pm$ 0.03	  & -0.05	$\pm$ 0.01		\\
NGC~7162    & SA(s)c        & 0.007720 &  38.5 &  18.0 & 10.42  & 1.73   &8.76 $\pm$	0.06	&  -0.11	$\pm$ 0.01	  & -0.03	$\pm$ 0.01		\\
NGC~1084    & SA(s)c        & 0.004693 &  20.9 &  23.8 & 10.40  & 3.69   &8.69 $\pm$	0.05	&  -0.11	$\pm$ 0.01	  & -0.05	$\pm$ 0.01		\\
NGC~1309    & SA(s)bc?      & 0.007125 &  31.2 &  20.3 & 10.37  & 2.41   &8.57 $\pm$	0.04	&  -0.11	$\pm$ 0.01	  & -0.04	$\pm$ 0.01		\\
NGC~5584    & SAB(rs)cd     & 0.005464 &  22.5 &  63.5 & 10.34  & 1.29   &8.49 $\pm$	0.10	&  -0.31	$\pm$ 0.03	  & -0.05	$\pm$ 0.01		\\
NGC~4900    & SB(rs)c       & 0.003201 &  21.6 &  35.4 & 10.24  & 1.00   &8.67 $\pm$	0.08	&  -0.15	$\pm$ 0.02	  & -0.04	$\pm$ 0.01		\\
NGC~7496    & SB(s)b        & 0.005365 &  11.9 &  66.6 & 10.19  & 1.80   &8.71 $\pm$	0.10	&  -0.17	$\pm$ 0.04	  & -0.05	$\pm$ 0.01		\\
NGC~7552    & (R')SB(s)ab   & 0.005500 &  14.8 &  26.0 & 10.19  & 0.59   &8.93 $\pm$	0.13	&  -0.12	$\pm$ 0.03	  & -0.06	$\pm$ 0.02		\\
NGC~1512    & SB(r)a        & 0.002995 &  12.0 &  63.3 & 10.18  & 1.67   &8.79 $\pm$	0.06	&   -   				  & -0.03	$\pm$ 0.01		\\
NGC~7421    & SB(rs)bc      & 0.005979 &  25.4 &  29.6 & 10.09  & 2.03   &8.81 $\pm$	0.12	&  -0.18	$\pm$ 0.02	  & -0.05	$\pm$ 0.01		\\
ESO~498-G5  & SAB(s)bc pec  & 0.008049 &  32.8 &  19.8 & 10.02  & 0.56   &8.78 $\pm$	0.09	&  -0.02	$\pm$ 0.02	  & 0.00	$\pm$ 0.01		\\
NGC~1042    & SAB(rs)cd     & 0.004573 &  15.0 &  63.7 & 9.83   & 2.41   &8.76 $\pm$	0.12	&  -0.14	$\pm$ 0.01	  & -0.03	$\pm$ 0.02		\\
IC~5273     & SB(rs)cd?     & 0.004312 &  15.6 &  33.8 & 9.82   & 0.83   &8.46 $\pm$	0.08	&  -0.13	$\pm$ 0.01	  & -0.04	$\pm$ 0.01		\\
NGC~1483    & SB(s)bc?      & 0.003833 &  24.4 &  19.0 & 9.81   & 0.43   &8.27 $\pm$	0.09	&  -0.10	$\pm$ 0.01	  & -0.05	$\pm$ 0.01		\\
NGC~2835    & SB(rs)c       & 0.002955 &  8.8  &  57.4 & 9.80   & 0.38   &8.64 $\pm$	0.10	&  -0.27	$\pm$ 0.03	  & -0.08	$\pm$ 0.01		\\
PGC~3853    & SAB(rs)d      & 0.003652 &  11.3 &  73.1 & 9.78   & 0.35   &8.57 $\pm$	0.10	&  -0.31	$\pm$ 0.01	  & -0.06	$\pm$ 0.02		\\
NGC~337     & SB(s)d        & 0.005490 &  18.9 &  24.6 & 9.77   & 0.57   &8.36 $\pm$	0.08	&  -0.12	$\pm$ 0.01	  & -0.05	$\pm$ 0.01		\\
NGC~4592    & SA(s)dm?      & 0.003566 &  11.7 &  37.9 & 9.68   & 0.31   &8.15 $\pm$	0.08	&  -0.06	$\pm$ 0.01	  & -0.03	$\pm$ 0.01		\\
NGC~4790    & SB(rs)c?      & 0.004483 &  16.9 &  17.7 & 9.60   & 0.39   &8.47 $\pm$	0.08	&  -0.09	$\pm$ 0.01	  & -0.04	$\pm$ 0.01		\\
NGC~3513    & SB(rs)c       & 0.003983 &  7.8  &  55.4 & 9.37   & 0.21   &8.46 $\pm$	0.09	&  -0.29	$\pm$ 0.06	  & -0.13	$\pm$ 0.01		\\
NGC~2104    & SB(s)m pec    & 0.003873 &  18.0 &  16.5 & 9.21   & 0.24   &8.25 $\pm$	0.07	&  -0.08	$\pm$ 0.01	  & -0.05	$\pm$ 0.01		\\
NGC~4980    & SAB(rs)a pec? & 0.004767 &  16.8 &  13.0 & 9.00   & 0.18   &8.15 $\pm$	0.10	&  -0.06	$\pm$ 0.01	  & -0.05	$\pm$ 0.01		\\
NGC~4517A   & SB(rs)dm?     & 0.005087 &  8.7  &  46.8 & 8.50   & 0.10   &8.27 $\pm$	0.12	&  -0.15	$\pm$ 0.10	  & -0.13	$\pm$ 0.01		\\
ESO~499-G37 & SAB(s)d?      & 0.003186 &  18.3 &  18.3 & 8.47   & 0.14   &8.00 $\pm$	0.14	&  -0.01	$\pm$ 0.02	  & 0.00	$\pm$ 0.01		\\
               \hline
\end{tabular}
\end{table*}

%%%%%%%%%%%%%%%%%%%%%%%DATA ANALYSIS%%%%%%%%%%%%%%%%%%%%%%%%%%%

\subsection{Observations}

The datacubes studied in this paper have the MUSE spatial sampling of 0\farcs2 and spectral sampling of 1.25\,\AA. These MAD galaxies were observed with the Wide Field (i.e., FoV of 1 arcmin$^2$) and nominal modes (i.e. wavelength range from 4650 to 9300\,\AA) for one hour on target, with seeing values between 0\farcs4 and 0\farcs9. Offset sky observations were taken before or after the target observations in order to subtract the sky. With one pointing per galaxy (targeting the central 1 arcmin$^2$), the spatial coverage varies from 0.3 to 4 \,$R_e$ (from 3 to 15 kpc on physical scale).

\subsection{Data reduction}

The details of the basic data reduction are fully described in MAD1. Briefly, the data were reduced using the MUSE pipeline \citep{Weilbacher2012}. This initial basic reduction step includes bias and dark subtraction, flat fielding, wavelength calibration and drizzling of the different IFU slices into one final datacube for each individual exposure. Each of the individual datacubes were then aligned, sky-subtracted with the {\it Zurich Atmosphere Purge} ({\sc zap}, \citealt{Soto2016}) algorithm, median-combined using a 10$\sigma$ clipping algorithm to remove cosmic rays until finally obtaining one single MUSE datacube per galaxy.

\section{Fits to emission lines: Methodology}
\label{sectionanalysis}

The MUSE spectral coverage includes several strong emission lines including H$\beta$, [O{\sc iii}]$\lambda$4959,  [O{\sc iii}]$\lambda$5007, [N{\sc ii}]$\lambda$6548,  H$\alpha$,  [N{\sc ii}]$\lambda$6583,  [S{\sc ii}]$\lambda$6717 and  [S{\sc ii}]$\lambda$6731. The MUSE spectra of a number of galaxies (in some spatial regions) also show weaker emission lines. The analysis of these additional lines will be reported  in forthcoming papers.

We  correct each spectrum for  Milky Way extinction using the $ E(B-V) $ values from NED. These values are obtained from the  \citet{Schlafly2011} recalibration of the \citet{Schlegel1998} dust map. This recalibration assumes a \citet{Fitzpatrick1999} reddening law with R$ _{V} $ = 3.1.

\subsection{Subtraction of the stellar continuum}

We removed the contribution of the stellar components from the spectra in order to obtain pure emission-line spectra (assuming that contamination to the continuum from  nebular emission can be neglected, see e.g., \citealt{Byler2016}; MAD4). The continuum-fitting procedure is explained in detail in MAD4. Briefly, the stellar continuum was computed by performing full spectral fitting to each spectrum using the \texttt{pPXF} package \citep{cappellari2004} with the ELODIE stellar libraries \citep{Leborgne2004} between Z=0.004 and Z=0.1 and ages between 1 Myr and 13 Gyr.  To increase the accuracy of the continuum fits, the 2-D spectra were tessellated using the Voronoi adaptive binning package of  \citet{Cappellari2003} so as to achieve  a signal-to-noise ratio (S/N) of 50 in each Voronoi cell in the wavelength range 5650-5750~\AA\,(i.e., a region without emission lines). 

The best-fit stellar continuum determined for each cell was subtracted, after rescaling in flux, from the total spectrum in each spaxel encompassed within that cell; resulting in a pure emission line spectrum for each spaxel. This step assumes that the stellar properties are identical for each spaxel within a Voronoi cell, which is not necessarily true; it is however a good compromise that avoids fitting the stellar continua to poor S/N data, which would lead to unreliable results.

\subsection{Determination of the Stellar Mass Surface Density}
We use in this paper the stellar mass surface density maps, $\Sigma_{\star}$, derived by fitting stellar population models to the stellar continuum spectra of our galaxies; details about these fits are provided in MAD1 and MAD4. Briefly, a second continuum fitting was performed on the stellar Voronoi tessellation (S/N=50 on the continuum) in order to get ages, metallicities and $\Sigma_{\star}$. This second {\sc pPXF} run was done using the stellar templates from {\sc miles} \citep{Sanchez-Blazquez2006}. A discussion of the robustness of {\sc pPXF} when obtaining these stellar properties can be found in \citet{Ge2018} and MAD4.

Then, we transform the resulting $\Sigma_{\star}$ map to a spaxel-by-spaxel map, assuming that the continuum is the same at each stellar Voronoi bin.
\subsection{Emission line fitting}
\label{ELfitting}

The Voronoi tessellation performed on the stellar continuum is not ideal for constructing 2-D maps of the emission line signal. First, the regions where the stellar continuum is brightest may not coincide with the H{\sc ii} regions or generally with regions of high emission line flux. Second, the Voronoi binning based on the stellar continuum may lead to dilution of the emission line flux within a cell;  important but weak emission lines  may disappear within a stellar Voronoi bin.

We therefore performed the study of the emission line features in a spaxel by spaxel basis, masking those where the S/N $<$ 3 in all the studied lines. Taking into account the data from all the galaxies in this paper, the total number of spaxels are $\sim$1330000, with a median physical scale of 20 pc. In each spaxel, the emission  spectrum may have contributions from several physical sources of emission such as narrow  and/or broad AGN emission, inflows and or outflows that give rise to blue-/red-shifted lines relative to the bulk disk emission, and so on.  We see, at some physical locations, clear double components in at least four of our 38 galaxies, namely NGC~3256 (a merger relic, \citealt{deVaucouleurs1956}), NGC~5643 (a Seyfert-2 galaxy, \citealt{Condon1998}), NGC~7496 and NGC~7552 (LIRG, \citealt{Sanders2003}). We nevertheless assumed in the current analysis that each line is well described by a single Gaussian profile. The use of multiple-component decompositions of the emission line spectra is postponed to future papers. 

The MUSE Line Spread Function (LSF, i.e., the shape of a line in the spectral domain)  varies with wavelength. Although  ESO provides  a tabulated non-parametric profile of the MUSE LSF,  we  measured the variations of the LSF in our own lamp observations as well as in our sky frames in order to achieve a more accurate description of the LSF for our data. Our LSF analysis is presented in MAD3; here we take into account this correction when computing the emission line fluxes for our analysis. Specifically, we fit our emission lines with a Gaussian:

\begin{equation}
G(x) = \frac{F_{\mathrm{\lambda}}}{\sqrt{2\pi} \sigma_\lambda}\exp\left(-\frac{(x-\lambda_c)^2}{2\sigma^2_\lambda}\right),
\end{equation}
where $F_{\mathrm{\lambda}}$ is the emission line flux, $\lambda_{c}$ is the position of the peak of the line, $x$ is the central wavelength and $\sigma_{\lambda}$ is:
\begin{equation}
\sigma_{\lambda}=\sqrt{\sigma_{target}^{2}+\sigma_{LSF}^{2}},
\end{equation}
where $ \sigma_{LSF} $ is the contribution to the instrumental broadening per wavelength due to the LSF.

To obtain the pure emission line fluxes, we simultaneously fitted  two independent groups of emission lines. Specifically, we fit H$\beta$ and H$\alpha$ together and, in the other group, we performed a simultaneous fit to the [O{\sc iii}]$\lambda$4959, [O{\sc iii}]$\lambda$5007, He{\sc i}$\lambda$5876, [O{\sc i}]$\lambda$6300, [N{\sc ii}]$\lambda$6548, [N{\sc ii}]$\lambda$6583, [S{\sc ii}]$\lambda$6717, [S{\sc ii}]$\lambda$6731 lines. The fits were carried out by assuming identical velocities and velocity dispersions for each line inside each of the groups.  In the following  we focus on the analysis of  the emission line diagnostics based on line fluxes and flux ratios, in order to investigate the SF, ionization and metallicity properties of the ionized gas.

\subsection{Uncertainties on emission line fluxes}

In order to assess the uncertainty in the emission line fluxes, we estimate the errors in the fitting code plus residuals due to continuum subtraction imperfections.

First, we use 100,000 simulated spectra of a Gaussian emission line to test our fits to the observed spectra. Specifically, we create 1000 simulated spectra for each value of S/N between 1 and 100 (in S/N steps of 1), and fit the resulting spectra using both the same software and approach used for our observed spectra. The errors on the uncertainties of the emission line flux reach about 10\% in the low S/N$\lesssim$3 regime, decreasing to about 2\% for higher  S/N values. 

Second, systematic errors due to uncertainties in the continuum subtraction must be added to the error budget. Pessimistic errors for the H$\alpha$ flux estimates can be inferred by considering the maximum stellar absorption Equivalent Width, which are $\sim50$\AA~for the youngest regions (higher than the corrections proposed for the {\sc granada} models; \citealt{Gonzalez-Delgado2005}). To place empirical limits on the incorrect stellar continuum subtraction on our emission line flux estimates, we perform a simple test on our observed spectra. Specifically we extract four ``extreme" stellar absorption spectra from the four corners of the stellar age versus stellar metallicity plane that includes all stellar spaxels for the whole galaxy sample (5 and 95\% of the cumulative distribution function of both age and metallicity distributions; see MAD4 for details). We then represent   those ``extreme" spectra covered by our sample through Single Stellar Population models with age and metallicity values close to those spectra; the four templates combine ages of 1.6 and 12.00 Gyr with metallicities of Z=0.0004 and Z=0.1, respectively. 

 We then subtract from each total spectrum in each of the four age-metallicity corners not only its optimal stellar continuum spectrum but also  the other three age-metallicity combinations, i.e., also the ``maximally inaccurate" stellar continua. We calculate  the emission line fluxes from the four resulting spectra, and compare the distribution of line fluxes produced by the four continuum subtractions. From this test we estimate that stellar continuum subtraction  errors produce uncertainties on the  H$\alpha$ and H$\beta$  fluxes  of order 4\% for S/N=10, decreasing to 2\% for S/N=20 and 1\% for S/N=40. We take these conservative errors as fiducial systematic errors arising from a possibly incorrect continuum subtraction from our spectra.

\subsection{Dust correction and dust-corrected emission line flux maps}
\label{sectdust}

We computed  our 2-D fiducial dust  corrections with the colour excess $E(B-V)$ using the determined values of  the Balmer decrement assuming  case B recombination. The theoretical ratio $(\mathrm{H}\alpha/\mathrm{H}\beta)_\mathrm{theo}$ depends on the temperature and density; here we adopt a fiducial  $T=10^4$~K and $n=10^2$~cm$^{-3}$, which gives $(\mathrm{H}\alpha/\mathrm{H}\beta)_\mathrm{theo}=2.86$ \citep{Osterbrock2006}. The colour-excess $E(B-V)$ is then given by:
\begin{equation}
E(B-V) = \frac{2.5}{k(\mathrm{H}\beta) - k(\mathrm{H}\alpha)}\log\left\{\frac{(\mathrm{H}\alpha/\mathrm{H}\beta)_\mathrm{obs}}{(\mathrm{H}\alpha/\mathrm{H}\beta)_\mathrm{theo}}\right\},
\end{equation}
where $k(\lambda)$ is the value from the Milky Way extinction curve by \citet{Cardelli1989} and \citet{ODonnell1994} with $R_V=3.1$. Spaxels with unphysical ratios of H$ \alpha $/H$ \beta <2$ were assumed to have zero dust attenuation (i.e., their Balmer decrement was set to zero, e.g. \citealt{Groves2012}).

We then corrected all emission line flux maps using these dust-correction maps as follows:
\begin{equation}\label{eqfluxcorr}
F_{\lambda,corr}=F_{\lambda,obs} 10^{0.4 A_{\lambda}}=F_{\lambda,obs} 10^{0.4E(B-V) k(\lambda)}.
\end{equation}

These dust-corrected emission line flux maps were then used to compute the gas diagnostic maps analysed in this paper.  The dust-corrected line emission flux maps and the dust correction maps of each galaxy are presented in Appendix~\ref{App:AppendixA} .

\section{Diagnostics of ionized gas at MAD resolution}
\label{sectionresults}

The high spatial resolution of the MUSE data enables us to handle two key issues that are of importance in the determination of the SFR densities and gas metallicities inside disk galaxies. Specifically, we  are able to $(i)$  compute SFRs and gas metallicities largely unaffected by contamination from regions of non-thermal emission;  and $(ii)$  establish the relative contributions to the SFR of compact H{\sc ii} regions and the diffuse component, as well as the separate contributions of these components to gas metallicities. Previous 2-D spectroscopic surveys have looked into these issues as well, e.g., CALIFA, MaNGA and SAMI \citep{Croom2012}; but the $\sim 10$ times higher spatial resolution of the MAD data boosts the ability to disentangle the different components, in addition to providing information on scales that are still largely unexplored in terms of a systematic study as a function of the location on the SFMS.

\subsection{Resolved BPT diagrams}
\label{secBPT}

\begin{figure}
\begin{center}
 \includegraphics[width=90mm]{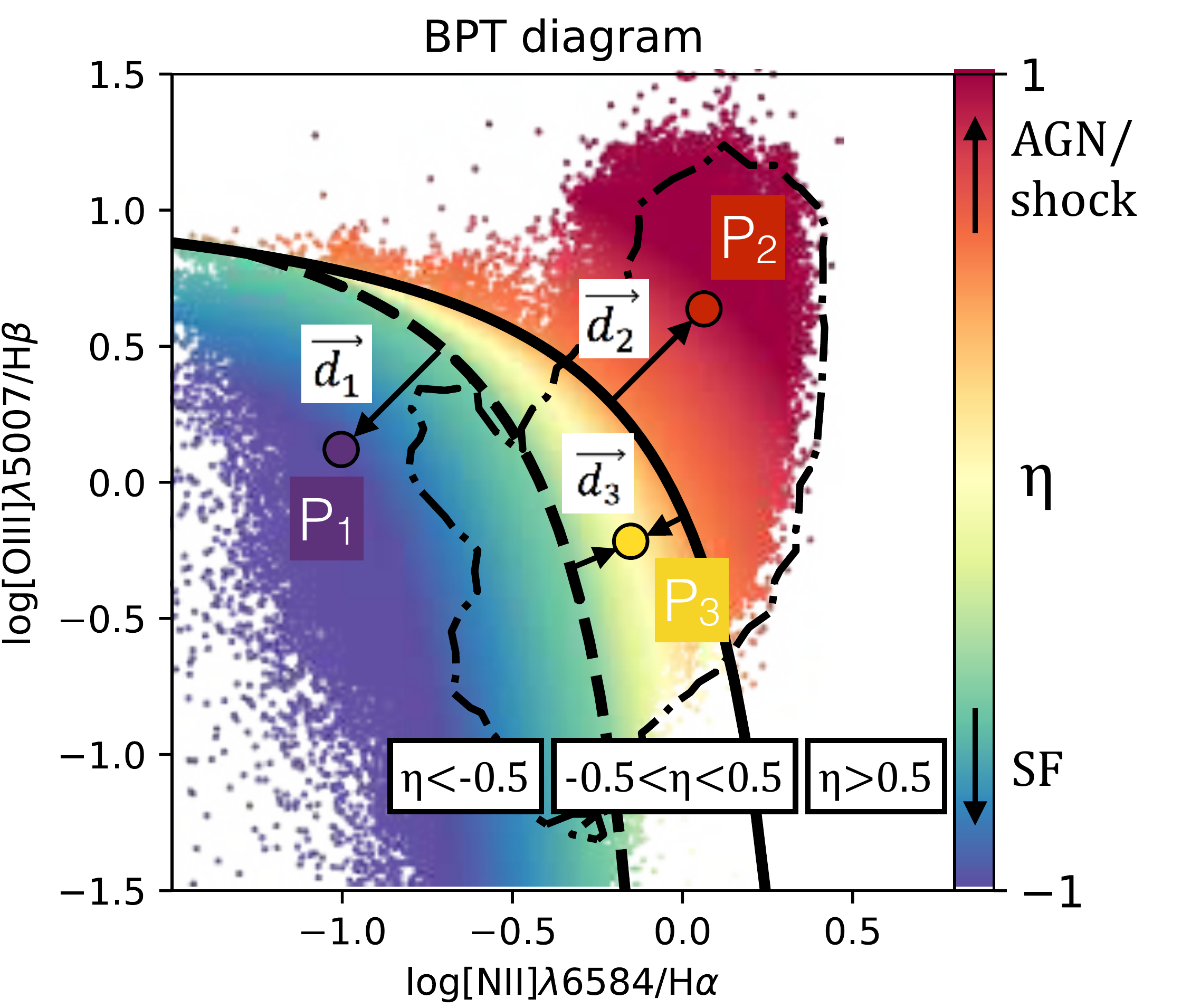}
\caption{BPT diagram of all the spaxels from all the galaxies in the sample. The colour scheme has been selected as a continuous variation from purple/blue (star formation) to red (AGN/shock dominated), using the $\eta$ parameter. To explain how $\eta$ is defined, we compute the minimum distances ($d_1$, $d_2$ and $d_3$) of three general points (P$_1$, P$_2$ and P$_3$ respectively, each one in a different region of the BPT diagram) to the two lines that define the three BPT regions. The spaxels with EW(H$\alpha$)<6 \AA ~lie inside the dashdot area (see text).}
\label{eta}
\end{center}
\end{figure}

We use the well-known [O{\sc iii}]/H$ \beta $ vs [N{\sc ii}]/H$ \alpha $ Baldwin, Phillips \& Terlevich (hereafter BPT; \citealt{Baldwin1981}) diagram to identify  regions within the galaxies that are ionized by UV radiation from newly born stars (regions of star-formation -- hereafter, the SF gas) and regions that are ionized by AGN-emitted radiation, shocks or AGB stars. In these diagrams (e.g., Fig~\ref{eta}), the upper, solid line is the photoionization model of \citet{Kewley2001} which defines  the upper envelope of the region of parameter space occupied by extreme starbursts;  this line is a conservative boundary between SF gas and AGN/shocks. The lower, dashed line shows an equivalent but empirical demarcation derived by \citet{Kauffmann2003b}. The area of BPT parameter space that lies between these two lines is  regarded as identifying regions of gaseous emission with an intermediate ionization spectrum.
 
It is also well known that the equivalent width (EW) is an extra proxy to constrain the possible ionization source of the ionized gas, restricting the SF regions to those with EW(H$\alpha$) > 6 \AA~(e.g., \citealt{Cid-Fernandes2011}; \citealt{Sanchez2014}; \citealt{Barrera-Ballesteros2016}; \citealt{Belfiore2017}). The EW(H$\alpha$) is measured by dividing the H$\alpha$ intensity by the flux density of the underlying continuum of the emission-line-free spectra, computed as the mean flux density of two windows of 30 \AA~wide and centered at $\pm$60 \AA~from the observed H$\alpha$ wavelength. The spaxels with EW(H$\alpha$) < 6 \AA~are presented in Fig.~\ref{eta} as those inside the dash-dotted area. The ratio between all the spaxels with EW(H$\alpha$) < 6 \AA ~ and all the spaxels under the empirical SF demarcation of the BPT is 4\%. Although this ratio is lower than  2\% for 24 out of 38 galaxies, it is significant for some galaxies such as NGC~3521 (23\%), IC5273 (11\%) or NGC~4941 (8\%).
 
To obtain a single-valued continuous parameter that conveys the  location of each spaxel on its galaxy's BPT diagram, we define the variable $\eta$ as either $(i)$ the inward-pointing orthogonal distance from the dashed line, identifying SF gas; or $(ii)$ the outward-pointing orthogonal distance from the solid line, identifying AGN/shocked-ionized gas; or $(iii)$ the inward- or outward-pointing orthogonal distance from the $\eta=0$ line, taken to be the line which splits vertically the intermediate region of the BPT diagram in half. $\eta$ is normalized such that $\eta$ for SF gas is $\eta\leq-0.5$, for AGN/shock-ionized gas is  $\eta\geq+0.5$ and the intermediate region has $-0.5 < \eta < +0.5$. A detailed explanation of how the parameter $\eta$ is computed can be found in Appendix~\ref{App:AppendixB}.

The variable $\eta$ is presented throughout the paper with a continuous colour map starting from blue in the SF regions, going to yellow for the intermediate regions and finishing in red for the regions ionized by AGN/shocks/AGB stars, as shown in Fig.~\ref{eta}. The individual BPT diagrams for each of the galaxies are shown in panel e) of the figures in  Appendix~\ref{App:AppendixA}.  The 2D maps showing the EW and resolved BPT properties of the individual galaxies, using the local values of  $\eta$, are presented in panel f) of the figures in Appendix~\ref{App:AppendixA}.

The dominant ionization mechanism in most spaxels is star formation, except for some galaxies with clear AGN/shock emission in the central (e.g., NGC~4941 or NGC~5643) or bar regions (e.g., NGC~289, NGC~1042, NGC~1512, NGC~4603, 5806 or NGC~7421), where intermediate  ionization spectra in between the SF and AGN/shock regions are found. Furthermore, there is a relationship between the gas ionization mechanism and the substructure, e.g, star-forming rings in NGC~1097, NGC~1512, NGC~3081, NGC~5806 and NGC~7552, or the intermediate ionization in the interarm regions of NGC~289 and NGC~4030. 

Here we use the knowledge of the resolved EW and BPT properties of the galaxies to exclude the  non-thermal  gas spaxels when estimating gas metallicities and SFR; the calibrations may break down when using gas ionized by hard radiation fields such as from AGNs or shocks (see Sect.~\ref{oh12section}).

\subsection{Distributions of star formation rates}

In regions of SF, the H$ \alpha $ line traces emission from massive young stars and thus the very recent SF occurring on timescales $ < $\,20 Myr. H$ \alpha $ is less sensitive to dust attenuation than the UV (although dust effects on H$ \alpha $ are not entirely negligible, e.g., \citealt{Cardelli1989}, \citealt{Erroz-Ferrer2013}). 

To measure the local SFR density in each spaxel, we first corrected the measured H$\alpha$ flux for dust extinction using the methodology explained in Sect.~\ref{sectdust}. Second, we converted this flux into an H$\alpha$ luminosity using the distance $D$ (Table \ref{tableprops}) and the equation $L(\mathrm{H}\alpha_{\rm corr}) = 4 \pi D^2 F(\mathrm{H}\alpha_{\rm corr})$. The H$\alpha$ luminosity is then converted into a SFR using the calibration from \citet{Hao2011}:

\begin{equation}
\mathrm{SFR}[M_{\sun}\mathrm{yr}^{-1}] = 10^{-41.27} \times L(\mathrm{H}\alpha_{\mathrm{corr}})[erg~s^{-1}]
\end{equation}

This equation assumes a \citet{Kroupa2001} stellar IMF with a mass range of 0.1-100 $M_{\odot} $,  an electron temperature of $ T_{\rm e} =10^{4} $ K and electron density $ n_{\rm e} =100$ cm$ ^{-3} $. Variations in $ T_{\rm e} $ from 5000 to 20000 K would result in a variation of the calibration coefficient ($10^{-41.27}$) of $\approx$15\%. Variations of $ n_{\rm e}  = 100-10^{6} $ cm$ ^{-3} $ would result in variations in the calibration coefficient below 1\% \citep{Osterbrock2006}. This calibration also assumes that over timescales  $>$\,6 Myr, star formation remains constant, and no information about the previous star formation history is given (see \citealt{Kennicutt2012} and references therein). Some of these assumptions may break down when studying resolved SFR (i.e., $ \Sigma$SFR) maps  for a number of reasons: (i)  an incomplete sampling of the IMF, especially at regions of low ($ \lesssim $0.01 $ M_{\sun} $ yr$ ^{-1} $) SFRs; (ii) the assumption that the star formation remains constant may not be true when the spatial resolution encloses single young clusters; (iii) the spatial resolution may be smaller than the  Str\"omgren diameter of the H{\sc ii} regions. We refer to \citet{Weilbacher2001} for a thorough modelling showing the effects which have an impact on the ratio between L(H$\alpha$) and SFR. The main consequence of an incomplete sampling of the IMF would typically be a suppressed H$\alpha$ flux (e.g. \citealt{Lee2009}; \citealt{Fumagalli2011}) and thus an underestimate of the SFR. A variable star formation history can lead to both a lower and higher conversion factor between H$\alpha$ luminosity and SFR which is likely to add some scatter in the relations below, especially for the DIG. As discussed above, there are regions in our galaxies which are not ionized by young stars, but from nuclear activity, shocks or post-AGB stars. We therefore restrict our SFR calculations to the SF regions (obtained from the BPT diagram).

Keeping these caveats in mind, we present the resolved H$ \alpha $-based SFR maps in panel (d) of the figures in Appendix~\ref{App:AppendixA}. The strong patchiness of the SFR maps reflects the highly inhomogeneous distribution of the H$\alpha$ emission, which is highly concentrated in the H{\sc ii} regions across the disks. The maps show nuclear SF rings in NGC~1097, NGC~1512, NGC~3081, NGC~5806 and NGC~7552; inner SF rings in NGC~3783, NGC~4941 and IC~2560; and outer SF rings in  NGC~3081 and NGC~5806, most of those previously identified in \citet{Comeron2010,Comeron2014}. As also discussed in e.g. \cite{Erroz-Ferrer2015} (and references therein), some bars in our sample show enhanced star-formation while  others do not; it is unclear whether this is entirely due to the presence/absence of gas or also to different gas conditions in similarly gas-rich galaxies.

\subsection{Identification of the Diffuse Ionized Gas}
\label{DIG_identification}

There are two ways to identify the DIG emission in our data: (i) using an H$ \alpha $ flux threshold to isolate the H{\sc ii} regions, and identifying the remaining gas as the DIG (as done e.g., by \citealt{Marino2013} for the CALIFA sample,  using the {\sc HIIexplorer} package presented in \citealt{Sanchez2012b} that  identifies the H{\sc ii} regions); or (ii) following the method developed in \citet{Blanc2009} to compute the fraction of flux coming from DIG and from H{\sc ii} regions.  The idea behind this method is that the observed H$ \alpha $ flux ($F({\rm H}\alpha) $)  includes the emission from both H{\sc ii} regions and the surrounding DIG. Here we follow the method by \citet{Blanc2009}. 

Quantitatively, we compute the fraction of $F({\rm H}\alpha) $ coming from H{\sc ii} regions ($ C_{\rm HII} $) and from DIG ($ C_{\rm DIG} $) following:
\begin{equation}
\begin{split}
F({\rm H}\alpha) & = F({\rm H}\alpha)_{\rm HII} + F({\rm H}\alpha)_{\rm DIG} \\
& = C_{\rm HII}F({\rm H}\alpha) + C_{\rm DIG}F({\rm H}\alpha),
\end{split}
\end{equation}
where $C_{\rm HII}= 1 - C_{\rm DIG}$. We refer the reader to \citet{Blanc2009} and  \citet{Kaplan2016} for a full description of the method. Briefly,  we use $F({\rm H}\alpha)$ (distinguishing between the bright H{\sc ii} regions and the lower, diffuse emission) and the ratio [S{\sc ii}]/H$ \alpha $, which is observed to be different in the H{\sc ii} regions vs DIG  of the Milky Way \citep{Madsen2006}. This method yields: (i) for each galaxy, a threshold value $F({\rm H}\alpha)_0$ below which all the $F({\rm H}\alpha)$ comes from DIG; (ii) for each region, $ C_{\rm HII} $ and $ C_{\rm DIG} $. The DIG is then defined as those regions with $F({\rm H}\alpha)<F({\rm H}\alpha)_0$ and significant contribution from the DIG (i.e., $ C_{\rm HII} < 0.6$). The forthcoming results in this paper do not change when a value of  $ C_{\rm HII} < 0.5$ or $ C_{\rm HII} < 0.7$ is chosen. In order to compare the properties of the DIG and H{\sc ii} regions, we restrict our study to the DIG that is located in the SF part of the BPT diagram. Panel (h) of the figures in Appendix~\ref{App:AppendixA} shows in red the spatial location of the DIG, coinciding with the low-flux,  diffuse ${\rm H}\alpha$ emission.

\subsection{Resolved gas-phase metallicities}
\label{oh12section}

The gas-phase metallicity  $Z_{\rm gas}$ is an important diagnostic for constraining the past star-formation and assembly histories of galaxies, and the origin of their gas components.  Several calibrations -either empirical, theoretical or hybrid- have been proposed over the years to derive gas metallicities from emission line fluxes.  We refer to the recent studies by \citet{Barrera-Ballesteros2017} and \citet{Sanchez2017,Sanchez2018} for a detailed comparison between metallicity calibrators.

In our work we explored a number of calibrations, to understand their impact on our results. Specifically we applied to our data:
 (a) the \citet{Marino2013} (M13 hereafter) fully empirical calibration, based on the O3N2 indicator, $\mathrm{O3N2}\equiv  \log\frac{[{\rm OIII}]\lambda5007}{\mathrm{H}\beta}\frac{\mathrm{H}\alpha}{[{\rm NII}]\lambda6584}$; (b) the calibration of \citealt{Pettini2004}, which uses the O3N2 indicator and is based on a hybrid combination of  oxygen measurements in galaxies and photoionization models; and (c) the calibration of \citet{Dopita2016} (DOP16 hereafter). This calibration was inferred from photo-ionization models and requires only the H$\alpha$, [N{\sc ii}] and [S{\sc ii}] emission lines to infer the oxygen abundance. Since both calibrators have been derived for regions where the ionizing mechanism is SF, we exclusively consider SF-only spaxels  (i.e., excluding spaxels with an AGN or shock spectrum).

\begin{figure*}
\begin{center}
 \includegraphics[width=165mm]{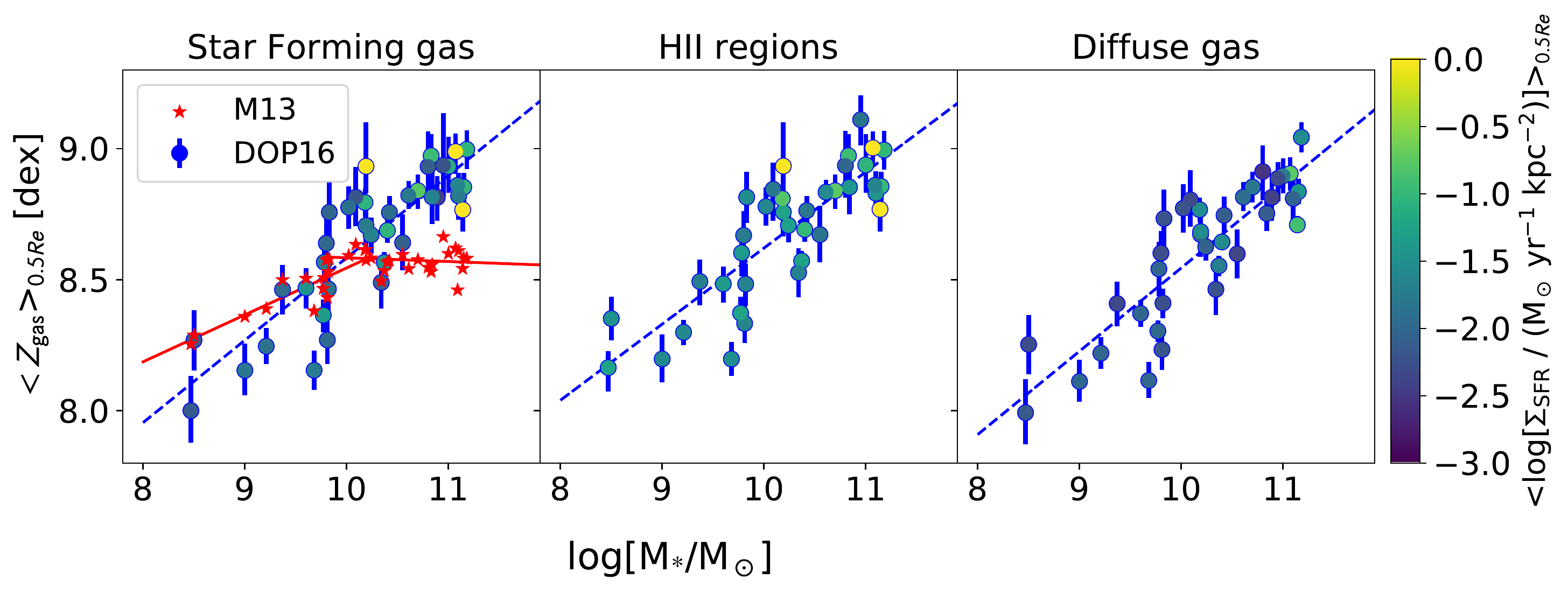}
\caption{Mean gas metallicity inside 0.5 $R_{e}$  as a function of the total stellar mass of the galaxy, colour-coded by the mean SFR density inside 0.5 $R_{e}$. From left to right,  mean metallicity  inside 0.5 $R_{e}$ derived from all SF spaxels  according to the BPT classification, only spaxels belonging to H{\sc ii} regions, and only spaxels belonging to the DIG. The left panel, displaying the metallicity of the gas with SF BPT properties, shows the comparison between  our fiducial  DOP16  calibration (circles colour-coded according to the bar to the right) and the M13 calibration  (red stars).}
\label{oh12stats}
\end{center}
\end{figure*}

There are indeed differences between the calibrations. The absolute metallicity values may differ: for the same O3N2 ratio, PP04  delivers systematically higher metallicities than M13 due to the photoionization models used in PP04. The M13 calibration covers a smaller dynamical range than DOP16.  Consequently, as expected, the standard deviations using M13 calibration are lower than those in DOP16 due to  the smaller dynamical range of the former.  We present some results of these tests in Fig.~\ref{oh12stats} where we show the average metallicity inside 0.5 $R_{e}$, using the DOP16 and M13 calibrations, plotted  as a function of the total stellar mass and colour-coded by the average SFR density inside 0.5$R_{e}$. With both calibrations, there is a positive correlation between  the average  $Z_{\rm gas}$ within  0.5$R_{e}$ and the total stellar mass $ M_{\star} $ (i.e., the MZR holds when the average metallicity  inside 0.5$R_{e}$ is used instead of the metallicity of the entire galaxy in our reduced sample galaxies, see e.g., \citealt{Sanchez2017} for a study of the MZR using a large set of metallicity calibrators for a set of 734 galaxies). The slopes of the relation are respectively $\alpha=0.31 \pm 0.03$ dex/log$M_{\sun}$ using the DOP16 calibration, and $\alpha=0.18 \pm 0.03$ dex/log$M_{\sun}$ up to $10^{10} M_{\sun}$ using M13. However, we note that M13 reaches a plateau metallicity for $ M_{\star}\gtrsim10^{10} M_{\sun}$. This is not the case for the DOP16 calibration, which covers a larger dynamic range than the M13 calibration. 
 
Although the DOP16 calibration depends on the N/O ratio, we adopt it as our fiducial calibration because it reproduces super-solar oxygen abundances. Fig.~\ref{oh12stats} shows that the mass-metallicity relation inside 0.5 $R_{e}$ holds not only when the total SF gas is taken into account, but also when either only the H{\sc ii} regions or only the DIG components are considered. Note that we see no clear trend of the average SFR density with either total stellar mass or gas metallicity inside 0.5 $R_{e}$.

\section{Results}
\label{section6}

\subsection{Gas Metallicities: Comparison between Diffuse Gas and H{\sc ii} regions}

Exploiting the high spatial resolution of the MAD data, we study the distinct contributions from the H{\sc ii} regions and DIG to the properties of the SF gas in SFMS disks. We discuss the differences in chemical enrichment  that we detect in these two components of the ISM.

\subsubsection{Emission line ratios in H{\sc ii} regions and Diffuse Ionized Gas}
\label{DIGsection}

The typical size of an H{\sc ii} region is between few tenths of pc to $\sim$200 pc (\citealt{Kennicutt1984}; \citealt{Garay1999}; \citealt{Kim2001}; \citealt{Hunt2009}); the resolution of the MAD data therefore makes it possible to identify and isolate a substantial fraction of H{\sc ii} regions from the surrounding DIG, although few H{\sc ii} regions may remain unresolved. As noted in the previous section, it is important to use only the star-forming gas when applying certain calibrations. \citet{Kewley2002}; \citet{Kewley2008} and \citet{Yuan2012} excluded contaminated  (non-SF) regions when lacking the spatial resolution required to avoid computing incorrect gas metallicities with some calibrations. Hence we study here the emission line ratios of H{\sc ii} regions and DIG of the SF regions only, distinguished using the methodology explained in Sect.~\ref{DIG_identification}.

\begin{figure}
\begin{center}
 \includegraphics[width=90mm]{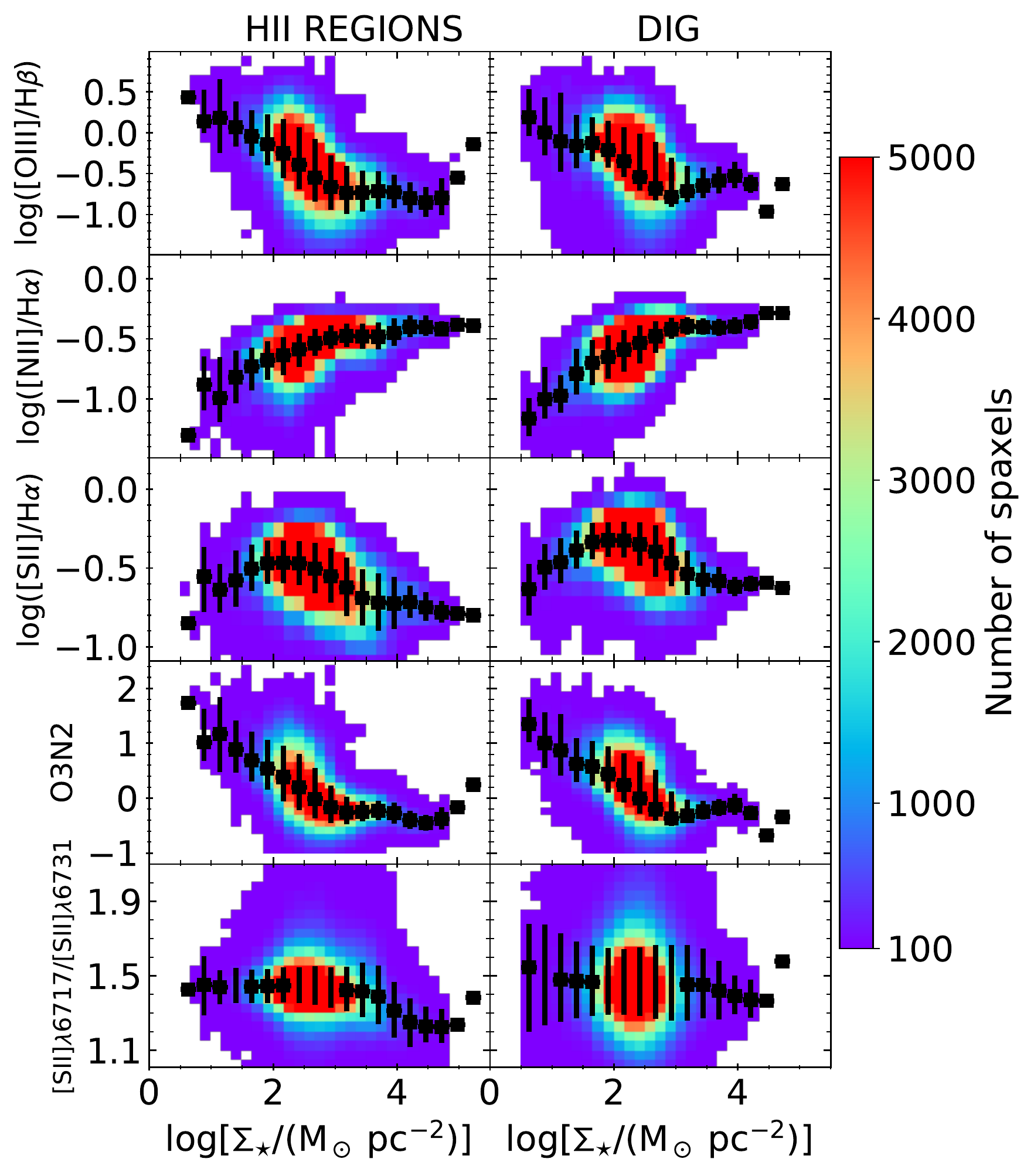}
\caption{Emission line ratios corrected for dust extinction for SF regions with dominant H{\sc ii} emission (\textit{left}) and for the DIG (\textit{right}), plotted as a function of $\Sigma_{\star}$ and colour-coded by the number of spaxels at each x-y bin. In each panel, the squares denote the median y-value at a fixed $\Sigma_{\star}$ and the error bars are the 1-$ \sigma $ error of the distribution. \textit{Top and second rows}: [O{\sc iii}]/H$ \beta $ and [N{\sc ii}]/H$\alpha$ ratios respectively. $ Third$ $row $: The [S{\sc ii}]/H$ \alpha $ ratio, used to identify regions of DIG emission. $Fourth$ $row$: The O3N2 ratio, an indicator of $Z_{\rm gas} $. $Last$ $row$: The [S{\sc ii}]6717/[S{\sc ii}]6731 ratio, which traces the electron density.}
\label{DIGratios}
\end{center}
\end{figure}

The ionization source for DIG and H{\sc ii} regions may be different (\citealt{Mathis2000}; \citealt{Haffner2009}). If so, we may expect the gas properties to differ in these regions. In order to explore  this issue, we present in Fig.~\ref{DIGratios} the ratios of the dust-corrected emission line fluxes of [O{\sc iii}]/H$ \beta $, [N{\sc ii}]/H$ \alpha $, [S{\sc ii}]/H$ \alpha $, O3N2 and [S{\sc ii}]$ \lambda $6717/[S{\sc ii}]$ \lambda $6731 for both H{\sc ii} regions and DIG. Thanks to the great spatial resolution of our data, it is possible to measure the median emission line fluxes at very low and high values of $\Sigma_{\star}$ ($10^{0.5}$ and  $10^{5}$ M$_\odot$~pc$^{-2}$ respectively). The first two rows show the difference in BPT ionization between H{\sc ii} regions and DIG. Since we study the H{\sc ii} regions and DIG that are in the SF area of the BPT diagram and show EW(H$ \alpha $)>6 \AA, we expect that both the [O{\sc iii}]/H$ \beta $ and [N{\sc ii}]/H$ \alpha $ ratios are similar for both components. The third row in the figure shows  the [S{\sc ii}]/H$ \alpha $ ratio, which is almost constant with $\Sigma_{\star}$ for both H{\sc ii} regions and DIG. The median value for this ratio in H{\sc ii} regions is, as expected, systematically lower than in the DIG (given the way the DIG is identified). The fourth row of the figure shows the O3N2 ratio. We observe a correlation between O3N2 and  $ \Sigma_{\star} $, i.e., a resolved MZR on the local scales of MAD.  Note that the ratio [S{\sc ii}]$ \lambda $6717/[S{\sc ii}]$ \lambda $6731, a tracer of the electron density, is similar for the H{\sc ii} regions and DIG, although with larger variation for the latter.

Overall  the emission line ratios in the DIG are similar than those found in H{\sc ii} regions, i.e., the range of BPT, $ Z_{\rm gas} $ and $ n_{e} $ properties. This is in part inherited from the restriction to study SF-only regions and from the constraints on the [S{\sc ii}]/H$ \alpha $   ratio that are applied to define and identify the DIG.

\subsubsection{Metallicity Radial Gradients}
\label{sectionmetalgradients}

The prediction from an inside-out formation scenario (e.g.; \citealt{White1991}; \citealt{Mo1998}), where the stars in the central parts are formed before the stars in the outer parts, is that the metallicity decreases with galactocentric radius (i.e., a negative gradient). Non-interacting galaxies show, in general, negative gas metallicity gradients \citep{Searle1971}, and, at least in some studies, the slope of this gradient has been found not to depend strongly on galaxy properties (e.g., \citealt{Zaritsky1994}; \citealt{Sanchez2014}; \citealt{Ho2015}). Other studies (e.g., \citealt{Belfiore2017}) however find that the metallicity gradient steepens with stellar mass. \citet{Carton2015} found a correlation between the metallicity gradients and the gas content of  50 isolated, late-type galaxies. Interacting galaxies seem to have flatter metallicity gradients than isolated galaxies, likely because of  gas flows induced by the mergers (\citealt{Vilacostas1992}; \citealt{Krabbe2008,Krabbe2011}, \citealt{Rupke2010}, \citealt{Kewley2010}, \citealt{Rosa2014} and \citealt{Torres-Flores2014}). Recently, using IFU data, \citet{Sanchez2012b,Sanchez2014}, \citet{Kaplan2016} and \citet{Sanchez-Menguiano2017} have  found that barred and unbarred spirals  show similar metallicity gradients, in contrast with some previous studies based on long-slit data (e.g., \citealt{Vila-Costas1992}; \citealt{Martin1994}; \citealt{Zaritsky1994}; \citealt{Dutil1999}; \citealt{Henry1999}). 

\begin{figure*}
\begin{center}
 \includegraphics[width=165mm]{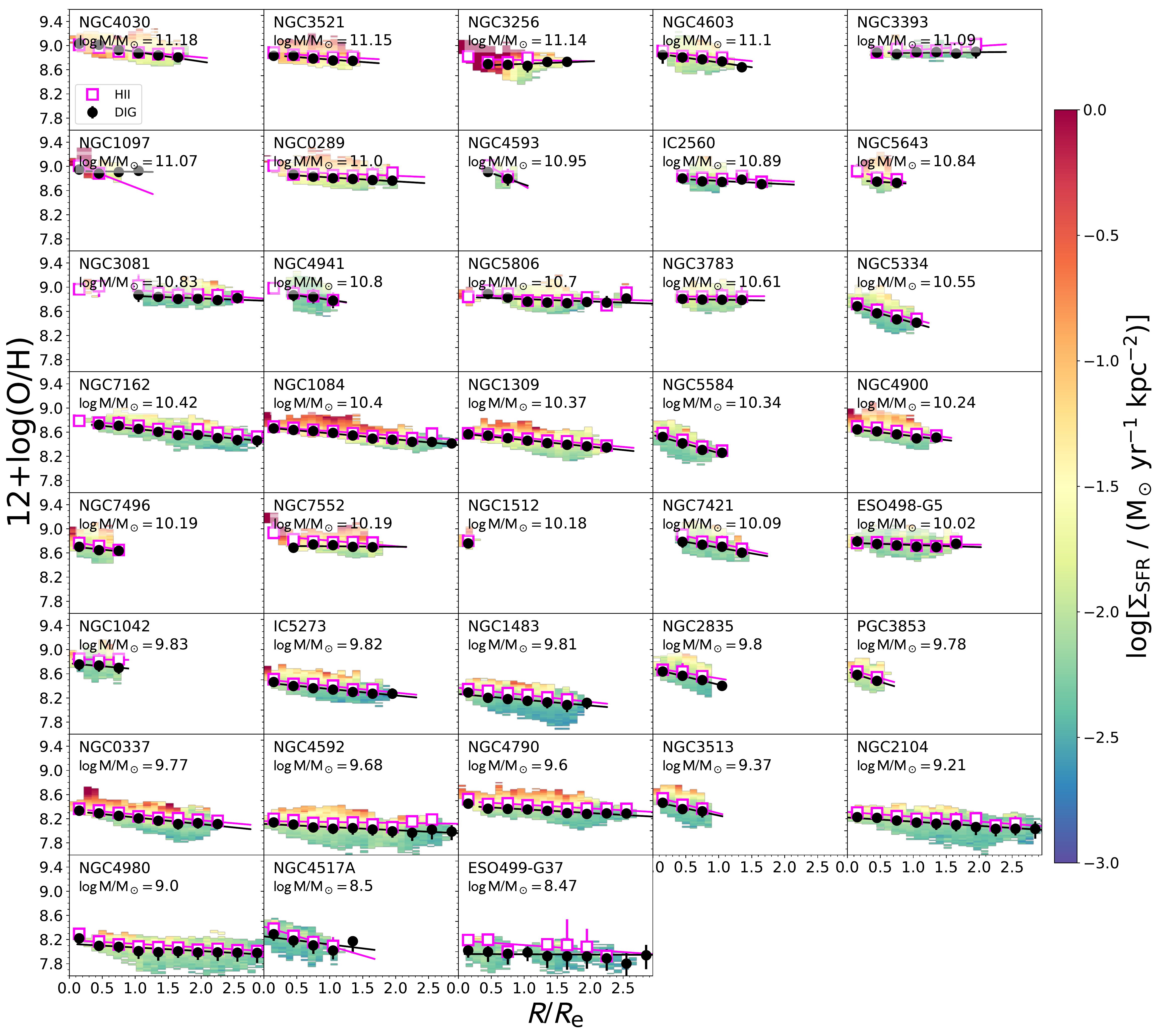}
\caption{Gas-phase metallicity of the SF regions computed using the DOP16 calibration, as a function of galactocentric distance normalized by the effective radius ($R/R_e$), colour-coded by SFR density. We compute the median metallicities in bins of 0.3 $R/R_e$ for the H{\sc ii} regions (pink squares) and DIG (black circles), respectively. The gradient $ \alpha $ for both the H{\sc ii} regions and DIG regions has been measured as the slope of the fitted line, and coloured in pink and black for the H{\sc ii} regions and DIG respectively. The galaxies are ordered according to their total stellar mass -- starting with the highest stellar mass at  the top-left, down to the lowest stellar mass in the bottom-right panel. Note that no gradient is measured for NGC~1512, as its radial extent does not reach 0.6 $R/R_e$.}
\label{oh12scatter_DIG}
\end{center}
\end{figure*}

To fairly explore the metallicity radial profiles in galaxies with different radial coverage and of different sizes/masses,  the galactocentric distances  are normalized to the half-light radius (listed in Table~\ref{tableprops}). The galactocentric distance has been binned in units of 0.3~$R/R_{e}$, and the median metallicity for each bin is shown in Fig.~\ref{oh12scatter_DIG}. Note that no gradient is measured for NGC~1512, as its radial extent does not reach 0.6~$R/R_e$. We computed the gradient $\alpha$ to be the slope of the fitted line to the median points up to the probed radius.  We note  that the qualitative trends that we report do not change when using  the galactocentric distance in physical units (kpc). 

MAD spatial resolution enables us to study the gas metallicity profiles of H{\sc ii} regions and DIG separately. To this aim we compare, for every galaxy, the metallicity gradient that results when using H{\sc ii}-regions only spaxels (pink open squares  in Fig.~\ref{oh12scatter_DIG}) to the gradient that results when using DIG regions (black circles in the figure). We only consider (and compute a median value for) radial bins in which the number of spaxels covered by SF regions is at least 10\% of all spaxels in that radial bin. 

The metallicity gradients that we measure in the inner SF regions of disks are mostly negative (29 out 37 cases) or flat (8 out 37 cases). Outward-decreasing metallicities, i.e., negative gradients,  are consistent with a disk evolution following the classical inside-out scenario. Flat metallicity gradients, also reported by other authors (e.g., \citealt{Marino2012}; \citealt{Marino2016}), point towards scenarios that include the presence of bars, changes in the SF histories or coincidence with the corotation radius.  

%When comparing the gas metallicity gradients of HII regions and DIG separately (Fig. 5), we find that their metallicity gradients are similar.  However the median metallcities of HII regions are on average about 0.1 dex higher than those of the DIG, consistent with our findings in section 5.1.  
When comparing the gas metallicity profiles of H{\sc ii} regions and DIG separately (Fig.~\ref{oh12scatter_DIG}), we find that their metallicity gradients are similar. However, the median values of the gas metallicity in each radial bin are higher for the H{\sc ii} regions  than for the DIG regions by $\sim$0.1 dex on average. Fig.~\ref{oh12scatter_DIG} is colour-coded by the $\Sigma_{\rm SFR}$:   we note that, at any given radius, regions of lower metallicity have lower median SFR density.

\begin{figure}
\begin{center}
 \includegraphics[width=70mm]{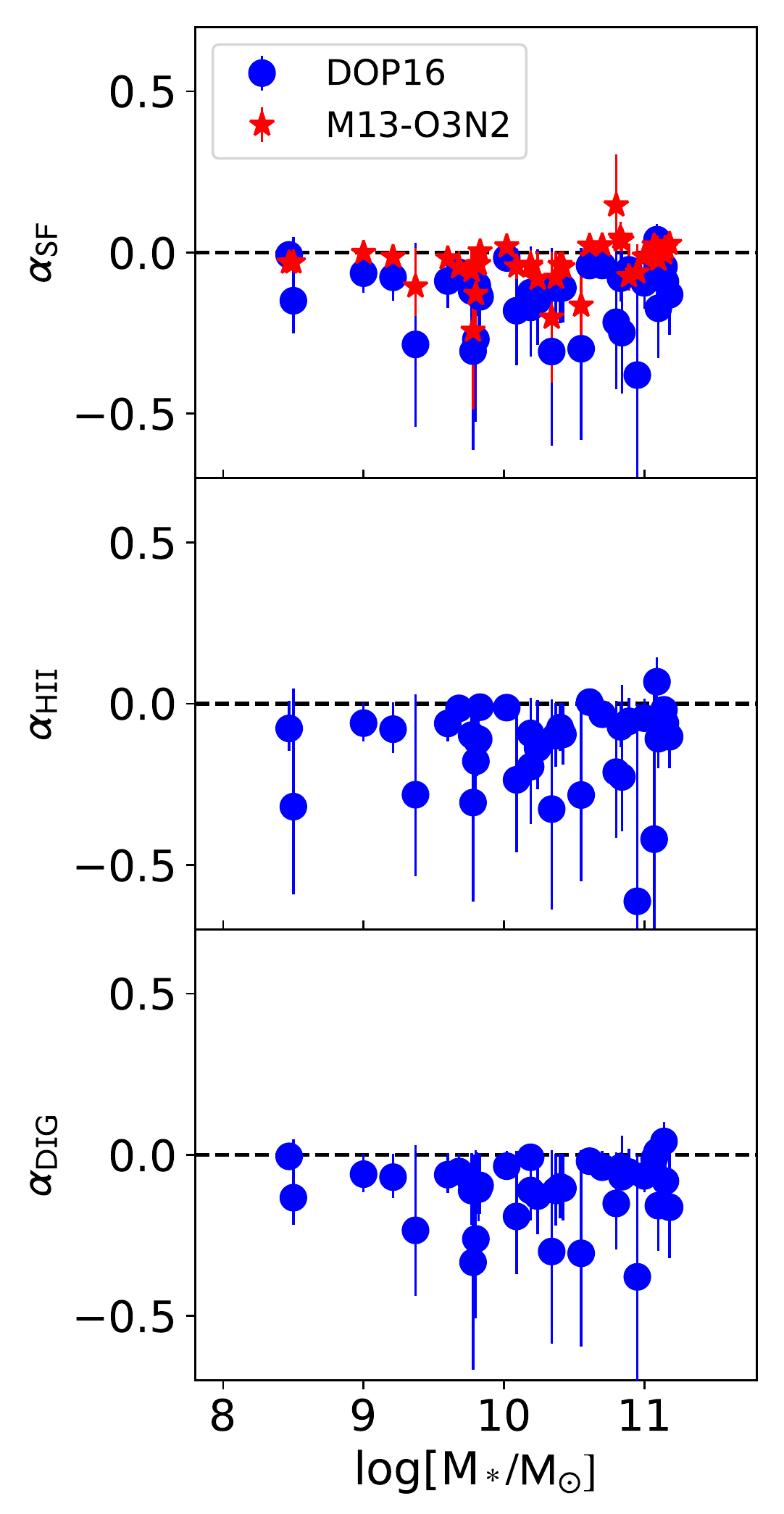}
\caption{Gas metallicity gradients $ \alpha $ of (\textit{top panel}) SF regions using two different metallicity calibrators: DOP16 (blue dots) and the M13 calibration based on the O3N2 ratio (red stars); \textit{(second panel}) H{\sc ii} regions and \textit{(third panel)} DIG regions. The dashed line at $ \alpha=0 $ is shown to guide the eyes in separating outward-decreasing (negative) from outward-increasing (positive) gradients.}
\label{oh12gradients}
\end{center}
\end{figure}

%the large impact of the adopted metallicity calibration on the measured gradients.
Fig.~\ref{oh12gradients} shows the metallicity gradients as a function of the total stellar mass for our three analyses: SF (no AGN/shock ionization) regions, H{\sc ii} regions and DIG. We find that the gradients hardly change between H{\sc ii} regions and DIG, we observe no dependence on stellar mass in any of the panels of Fig.~\ref{oh12gradients}.   We use this figure to also show the impact of the adopted metallicity calibration on the measured gradients.  Specifically, for the SF-only gas, we show the comparison between metallicity gradients measured with our fiducial calibration and with the M13 calibration.  The M13 calibration yields substantially shallower metallicity gradients, and occasionally even positive metallicity gradients (in some high-mass galaxies), due to a drop in metallicity in the inner 0.5$R_{e}$.  \citet{Sanchez-Menguiano2017} also reported a drop in metallicity in the inner 0.5Re when using M13, not found when using the DOP16 calibration (see their Appendix C). Our aforementioned results regarding the metallicity gradients of H{\sc ii} regions are consistent with those of \citet{Sanchez-Menguiano2017}, although their sample is three times as large and has in some cases larger spatial coverage.

\begin{figure}
\begin{center}
 \includegraphics[width=70mm]{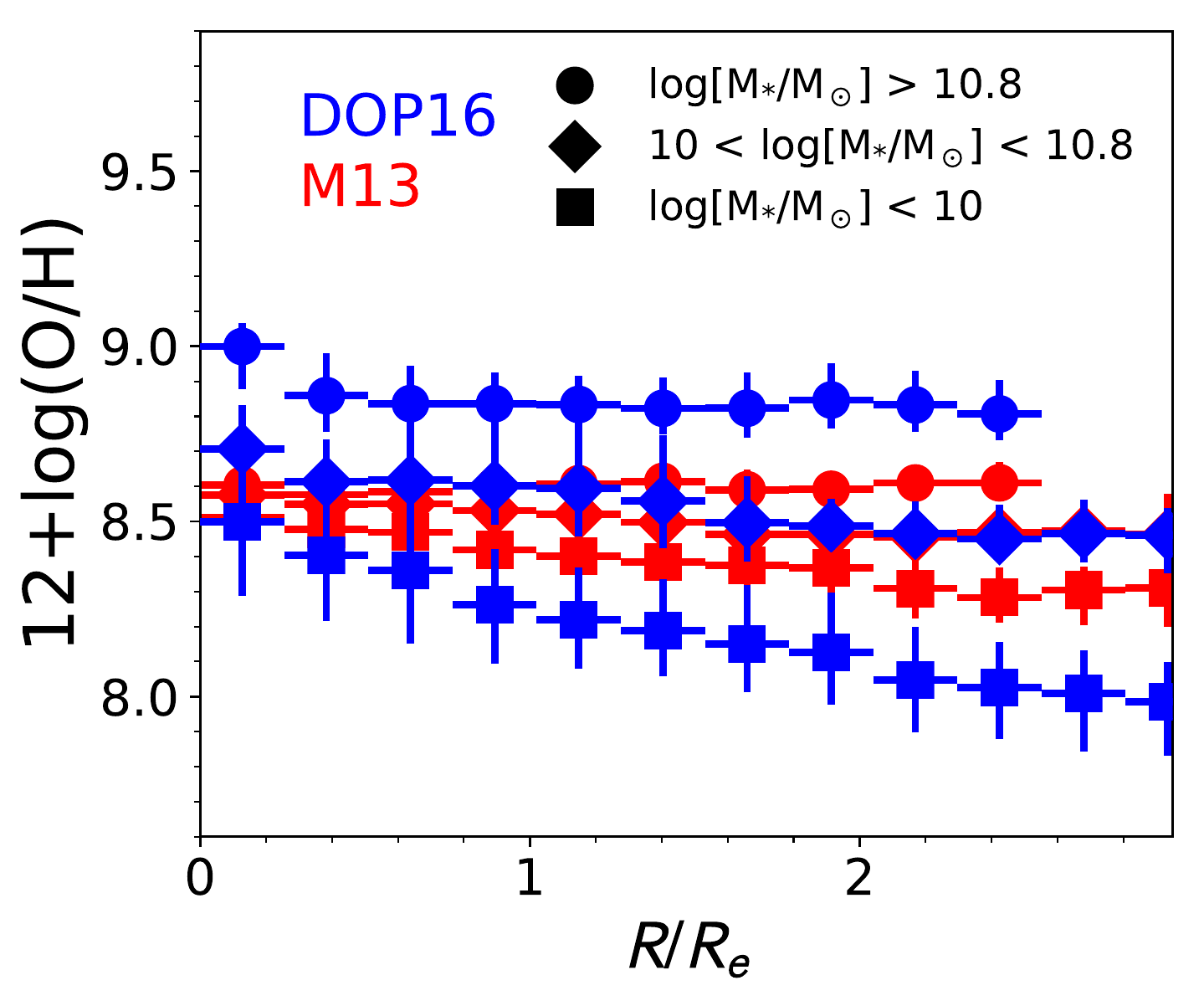}
\caption{Median metallicity radial profiles measured using DOP16  (blue) and O3N2-M13 (red) metallicity calibrators, grouped in three mass bins: squares for low-mass galaxies ($ M_{\star} < 10^{10} M_{\sun}$), diamonds for intermediate-mass galaxies ($10^{10} M_{\sun} < M_{\star} < 10^{10.8} M_{\sun}$) and dots for high-mass galaxies ($M_{\star} >10^{10.8} M_{\sun}$). There are 12-13 galaxies in each mass bin.}
\label{oh12gradients_bymass}
\end{center}
\end{figure}

Fig.~\ref{oh12gradients_bymass} shows the radial metallicity profiles of the SF regions dividing the sample in three mass bins with a roughly similar number of galaxies per bin: low-mass galaxies ($ M_{\star}\,<\,10^{10}\,M_{\sun}$), intermediate-mass galaxies ($10^{10}\,M_{\sun}\,<\,M_{\star}\,<\,10^{10.8}\,M_{\sun}$) and high-mass galaxies ($ M_{\star}\,>10^{10.8}\,M_{\sun}$). Both  DOP16 and M13 calibrations have been used to compute the metallicities.  At fixed galactocentric distance, the median metallicity values increase with increasing total mass (at all radii for both metallicity calibrators).   It is evident, however, that the shape of these profiles depends on the adopted calibration: the metallicity profiles decrease for all mass bins for our fiducial calibration DOP16, whereas for M13 they are much flatter, even rising for the high-mass bin.

Significant information concerning the enrichment histories of galactic disks that is lost when computing the azimuthal averages that produce the metallicity gradients; this produces the scatter in each radial bin that is shown in the previous figures, which is on average $\sim0.2$ dex. This typical scatter that we find across our sample has also been reported for  studies  of individual galaxies based on data of similar resolution to ours, e.g., in HCG~91c by \citet{Vogt2017}. The gradients result from averaging at any given radius substantially different metallicities in H{\sc ii} regions, DIG, spiral arms, bars, non-gravitational (inflow/outflow) components, galactic rings, and other small-scale substructure which stands out in the 2D maps. As the DIG  have lower metallicities than the H{\sc ii} regions, the median metallicities reported in the radial (azimuthally-averaged) gradients are in between the two extremes produced by these two gas components. In general, morphological substructure stands out clearly in the 2D metallicity maps: the nuclear rings in NGC~5806 and NGC~7552 have very different metallicity than their surroundings; the H{\sc ii} regions in the spiral arms in NGC~289 and NGC~4030 have different metallicity compared to the interarm regions. The latter may result from inflow of low-metallicity gas along the arms, either from the diffuse circum-galactic medium or due to the accretion of a low-metallicity satellite; alternatively, we may be seeing an increase in metallicity due to enhanced star-formation in the gas-compression zones generated by the spiral arms. In general however the H{\sc ii} regions have typical metallicities that are higher than those found in the interarm regions. In Appendix~\ref{App:AppendixA}, we describe the results for each individual galaxy, relating the scatter and features in the metallicity maps to the other gas diagnostics presented in this paper. In Appendix~\ref{App:AppendixC} we compare the 2D metallicity distribution with its azimuthally-averaged metallicity gradient.

%\section{The sample-averaged gas properties of SFMS disks on $\sim$100 pc scales}
\subsection{The spatially resolved  Mass-Metallicity relation on $\sim$100 pc scales}
\label{Sect_MZR}
%An important question is whether the ``scatter'' (i.e., azimuthal variations) in gas metallicity that we report above is physically driven (in a quantitative manner) by the local  (e.g., local SFR or $\Sigma_{\star}$) or by the  global (e.g., total SFR or total stellar mass) properties of the galactic disks.

An important question is whether the global relationships reported above arise from local ones. The high-spatial-resolution MAD data enable to explore relationships between local physical properties such as local SFR, local metallicities or $\Sigma_{\star}$.

\begin{figure}
\begin{center}
 \includegraphics[width=90mm]{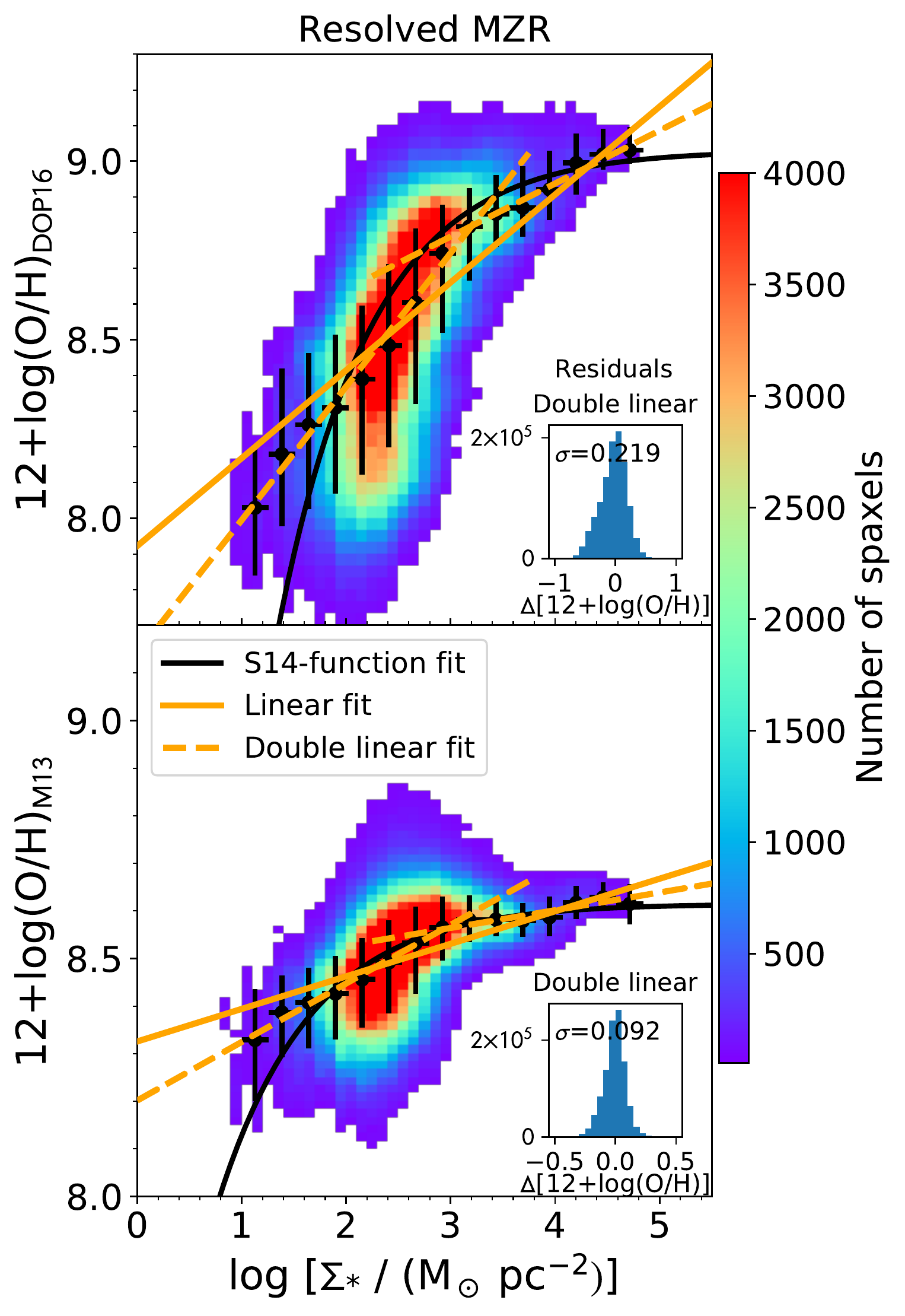}
\caption{Spatially resolved stellar surface mass density-gas metallicity relation (i.e., the RMZR) using DOP16  (top) and M13 (bottom) calibrations, colour-coded by the number of spaxels in each x-y bin. In these plots, there are $\sim1070000$ SF spaxels from the 38 galaxies. The orange solid lines show the linear fit to all the median values of $Z_{\rm gas}$. Two additional fits are performed to the data above and below the apparent threshold in surface mass density at $\sim10^3 M_{\sun}/pc^{2}$ (orange dashed lines). The black solid line is the fit to the S14-function previously used in the literature.}
\label{fits_A}
\end{center}
\end{figure}

In Fig.~\ref{fits_A} we present the distribution of the oxygen abundance as function of $\Sigma_{\star}$ for all the SF spaxels provided by our sample galaxies. The oxygen abundances are derived using DOP16 (top) and M13 (bottom) calibrations. This figure is based on the $\sim$\, 1070000 SF spaxels  out of the $\sim$\, 1330000 spaxels from the  entire  sample of 38 galaxies. The number of spaxels in each x-y bin is represented by a colour scale. These plots show a clear relationship between $\Sigma_{\star}$ and $Z_{\rm gas}$, i.e., a resolved MZR (RMZR) for $\sim$\,100 pc local scales. Specifically, $Z_{\rm gas}$ increases with increasing $\Sigma_{\star}$ up to a stellar surface mass density of $\sim10^3 M_{\sun}/{\rm pc}^{2}$. Interestingly, $Z_{\rm gas}$ continues to increase beyond this threshold value, but more gently than at lower $\Sigma_{\star}$.  Using the M13 calibration instead of our fiducial DOP16 calibration produces a more dramatic flattening of the relation above $\sim10^3 M_{\sun}/{\rm pc}^{2}$. In contrast, with the DOP16  the RMZR keeps increasing up to the highest stellar surface mass densities that we probe with our data. With our smaller spatial scales, we extend the RMZR to  $\Sigma_{\star}$ of $\sim10^5 M_{\sun}/{\rm pc}^{2}$, i.e., one and 1.5 magnitude(s) higher than those reached with MaNGA and CALIFA, respectively.

In order to explain the shape of the RMZR, we search for a mathematical formula which can reproduce both the linear behaviour at lower  $\Sigma_{\star}$ and the flattening (asymptotic behaviour) at higher  $\Sigma_{\star}$. We fit the median metallicity values inside each bin of $\Sigma_{\star}$ using three different approaches: (i) a linear function based on two free variables, i.e. slope and y-intercept, (ii) a three-variable function, $y=a+b(x-c)e^{-(x-c)}$ \citep{Sanchez2014}, S14-function thereafter and (iii)  double linear function for the two regimes below and above the $\sim10^3 M_{\sun}/{\rm pc}^{2}$ threshold which comprises four free variables. The threshold at $\sim10^3 M_{\sun}/{\rm pc}^{2}$ has been calculated as the optimal value that describes the break in the two linear regimes. In Table~\ref{MZR_fits} the coefficients of the fitted functions for both DOP16 and M13 calibrations are summarized. The residuals are computed as $ \Delta {\rm 12+log(O/H)} =y-\overline{y}$ for all y (metallicity) values from all the SF spaxels. We compute the scatter of the RMZR as the standard deviation of the residuals after subtracting the fitted relations to the RMZR. The distribution of the residuals as well as the scatter in this distribution are presented in a subplot inside  Fig.~\ref{fits_A}.
 
In order to assess which function represents the data better, we analyse the residuals, $\chi^2$ and $p$-value of the aforementioned fits. From Table~\ref{MZR_fits} we see that the double linear fit presents less scatter than the others for both metallicity calibrators. Additionally, the $\chi^2$ for the double linear fit is also the lower one for both metallicity calibrators. Even if $\chi^2$ determines the goodness of the fits, including more degrees of freedom (more free variables) to our fitted functions may result in a lower $\chi^2$ parameter. Thus we study the $p$-value to see the significance of our results, finding that the higher $p$-value corresponds to the double linear function for both metallicity calibrators, confirming that this function represents the data distribution better than the other two. In any case, both the double linear fit and the S14 functions can explain the saturation in metallicity at the highest densities. This saturation has been suggested to be consequence of the maximum yield of oxygen in spiral galaxies \citep{Pilyugin2007},  low specific SFR in the inner regions of the galaxies  or efficiency of cooling in high-metallicity and low-temperature H{\sc ii} regions \citep{Rosales-Ortega2012}.

\begin{table}
\caption{Best fits to the RMZR using three different functions: (i) linear fit \textit{y=mx+n}, (ii)  fit using the S14-function $y=a+b(x-c)e^{-(x-{\rm c})}$ and (iii) double linear fit, performing two linear fits before and after $10^3 M_{\sun}/{\rm pc}^{2}$ .  The fits have been done using both DOP16 and M13 metallicity calibrators. The scatter has been measured as the standard deviation of the residuals after subtracting the best fit to the data.}
 \label{MZR_fits}
\centering
\begin{tabular}{|c|c|clc|}
  &  & {\textit{DOP16}}&{\textit{M13}}\\
 \hline
\multicolumn{4}{|c|}{Linear \textit{y=mx+n}} \\
 \hline 
& \textit{m}(dex/log$\Sigma_*$) & 0.25 $\pm$ 0.02 &0.07 $\pm$ 0.01\\ 
 & \textit{n}(dex) &7.92 $\pm$ 0.07 & 8.33 $\pm$ 0.03\\ 
Residuals&  $ \sigma$(dex)  &0.227 & 0.096\\  
$\chi^2$ &    &4.526 & 2.846\\  
$p-$value &    &0.033 & 0.092\\  
 \hline 
\multicolumn{4}{|c|}{S14-function $y=a+b(x-c)e^{-(x-c)}$} \\ 
 \hline 
& \textit{a}(dex) & 9.03 $\pm$ 0.05 & 8.61 $\pm$ 0.05\\ 
& \textit{b}(dex/log$\Sigma_*$) & 0.000 $\pm$ 0.001 & 0.000 $\pm$ 0.001\\ 
& \textit{c}(dex) & 10.00 $\pm$ 16.04 & 10.00 $\pm$ 16.04\\ 
Residuals& $ \sigma$(dex) &0.283 & 0.108\\      
$\chi^2$ &    &3.215 & 1.331\\  
$p-$value &    &0.200 & 0.514\\  
 \hline 
\multicolumn{4}{|c|}{Double Linear \textit{y=mx+n}}\\ 
 \hline 
x< $10^3 M_{\sun}/{\rm pc}^{2}$ & \textit{m}(dex/log$\Sigma_*$) & 0.37 $\pm$ 0.02 & 0.12 $\pm$ 0.01\\ 
& \textit{n}(dex) & 7.62 $\pm$ 0.04 & 8.20 $\pm$ 0.01\\ 
x> $10^3 M_{\sun}/{\rm pc}^{2}$ & \textit{m}(dex/log$\Sigma_*$) & 0.15 $\pm$ 0.01 & 0.04 $\pm$ 0.01\\ 
& \textit{n}(dex) & 8.34 $\pm$ 0.06 & 8.45 $\pm$ 0.04\\ 
Residuals& $ \sigma$(dex) &0.219 & 0.092\\  
$\chi^2$ &    &0.416 & 1.411\\  
$p-$value &    &0.937 & 0.703\\    
 \hline     
\end{tabular}
\end{table}

\begin{figure*}
\begin{center}
 \includegraphics[width=165mm]{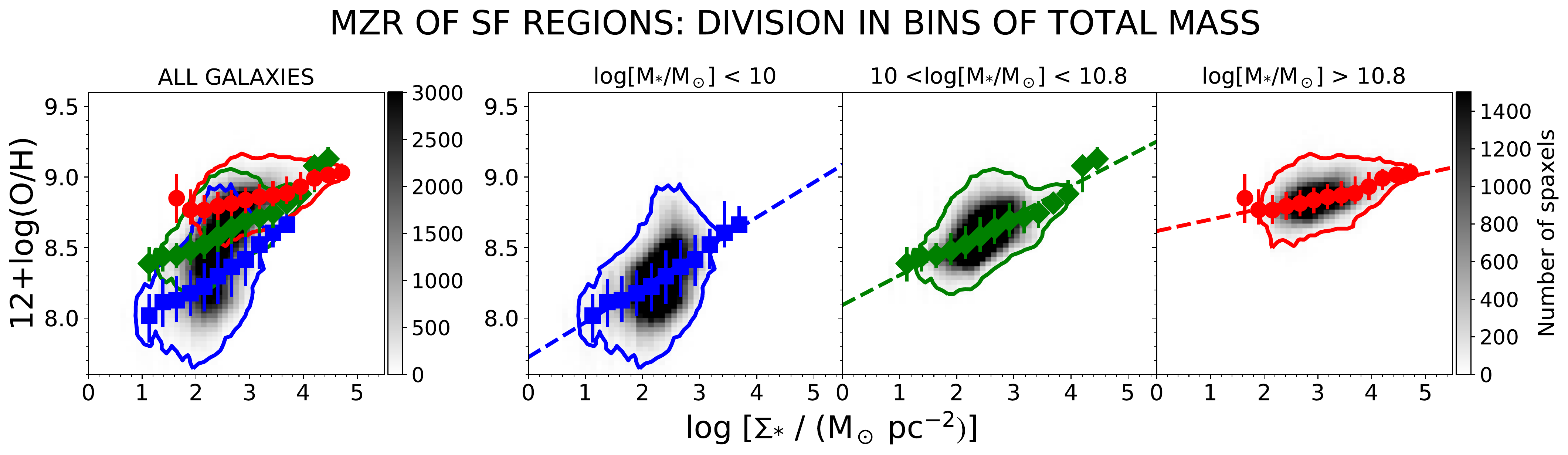}
\caption{RMZR  for all SF spaxels from the whole sample (left), followed by subsamples defined according to the total stellar mass of the galaxies  (low, intermediate and high stellar mass moving from left to right). Each plot is colour-coded in grey scale by the number of spaxels in each x-y bin. The blue, green and red contours for the low-, intermediate- and high-mass galaxies enclose the loci where the data-points lie. The corresponding dashed blue, green and red lines indicate the linear fit to the median values of $Z_{\rm gas}$ for each mass bin. The same contours  are reported in the left-most panel, to ease the comparison.}
\label{A_number_color_noAGN}
\end{center}
\end{figure*}

In order to understand the physical interpretation of the chosen formula, we divide the galaxies of our sample in three mass bins with a roughly similar number of galaxies per bin. The second, third and fourth  panels of Fig.~\ref{A_number_color_noAGN} show the $\Sigma_{\star}$ vs $Z_{\rm gas}$ relation respectively for  the lower (< $10^{10} M_{\sun}$), intermediate ($10^{10} M_{\sun} < M_{\star} < 10^{10.8} M_{\sun}$) and high  (> $10^{10.8} M_{\sun}$) stellar mass bins.  For each mass bin, we use a different colour for the contour of the RMZR relation; the contours enclose the loci where the data-points lie: we use blue for the low-mass galaxies, green for the intermediate-mass and red for the more massive galaxies.  These contours are superimposed in the left-most panel (i.e., the one containing the whole sample of 38 galaxies) using the same colour-coding. Specifically, the slope of the RMZR relation is rather steep in the lowest mass bin and flattens substantially towards  the higher masses (slopes $ \alpha=0.25 \pm 0.01, 0.21 \pm 0.01$ and $0.08 \pm 0.01$ for the low-, intermediate- and high-mass bins, respectively). Contrary to the behaviour of considering all the spaxels from all the galaxies, the median metallicities at each $\Sigma_{\star}$ of the three mass bins are very well fitted by a line, and no threshold in $\Sigma_{\star}$ is found. This points toward a scenario where the maximum oxygen yield that produces the saturation at high densities depends on the total stellar mass, and the saturation observed is the result of combining all the stellar mass bins in the same plot. In other words, the double linear fit is the best formula which fits to the data and shows the saturation at high masses. However, a single linear correlation explains the RMZR at each bin of total stellar mass.

We also note that a wide range of $Z_{\rm gas}$ is measured at a given $\Sigma_{\star}$ from the right plots of Fig.~\ref{A_number_color_noAGN}: at fixed values of surface mass density which are found in galaxies of all masses in our sample,  the range in metallicities measured within the disks slightly increases with the total stellar mass of the galaxies.   For example, at a surface mass density of $10^{2.5} M_{\sun}/{\rm pc}^{2}$, the lowest stellar mass bin has gas metallicities within the range 7.8<$Z_{\rm gas}$<8.9, the intermediate-mass galaxies have a range 8.3<$Z_{\rm gas}$<9.1, and the most massive bin a range of 8.4<$Z_{\rm gas}$<9.3. This does not imply, however, that the metal enrichment has a global origin rather than a local one, as denser (less dense) regions are found in more (less) massive galaxies. The dependence of the RMZR with the total stellar mass will be discussed in Sect.~\ref{sectiondiscussion}.

\begin{figure*}
\begin{center}
 \includegraphics[width=165mm]{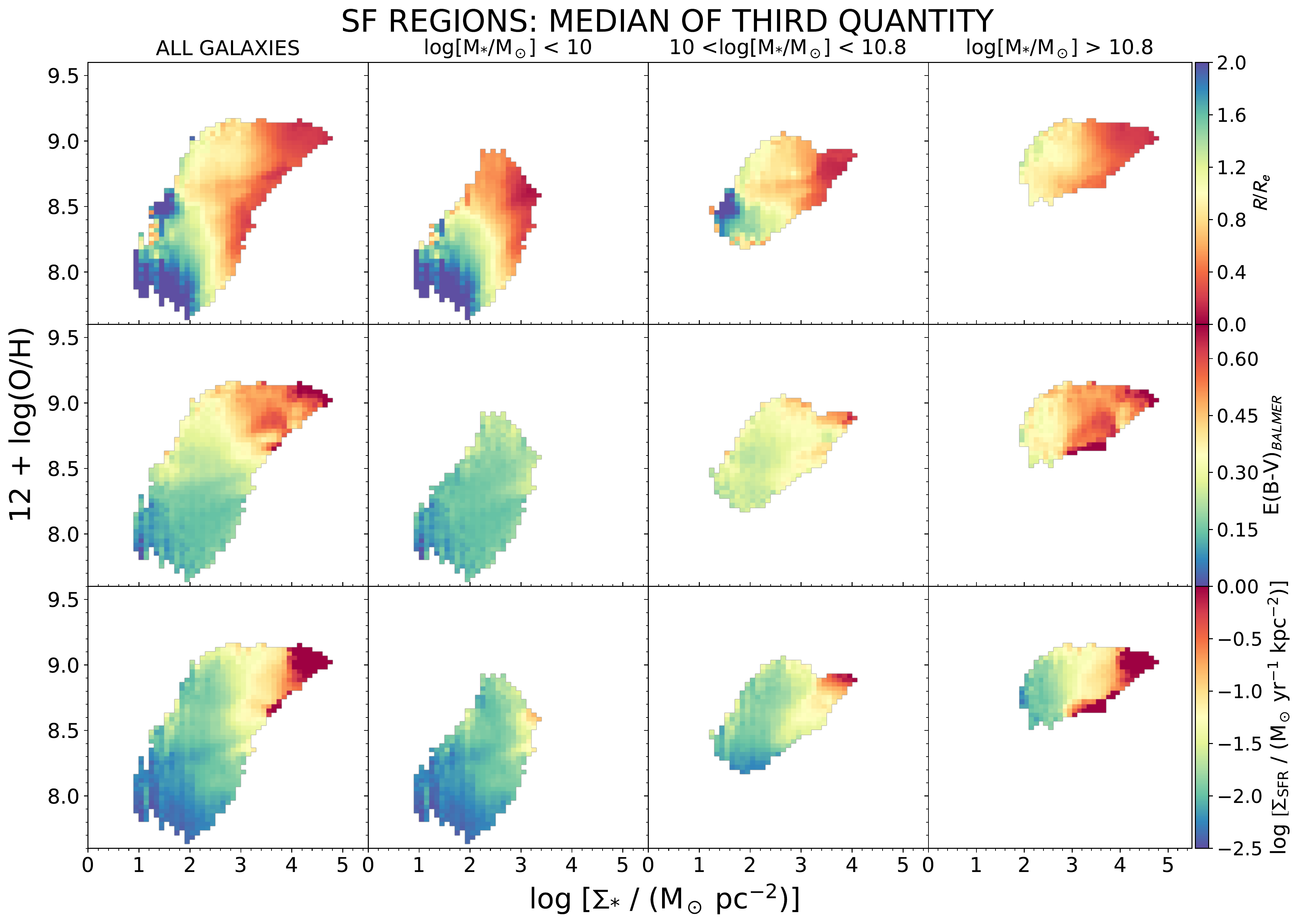}
\caption{RMZR colour-coded by the median value of a third parameter: from top to bottom, galactocentric distance, dust extinction from the Balmer decrement  and SFR surface density. As in Fig.~\ref{A_number_color_noAGN}, the plots show the results for all SF spaxels from the whole sample (left column), followed by the subsamples defined according to the total stellar mass of the galaxies  (low, intermediate and high stellar mass from left to right).}
\label{supergalaxyA}
\end{center}
\end{figure*}

We furthermore explore  how other important parameters depend on $\Sigma_{\star}$ and $Z_{\rm gas}$  by adopting a ``supergalaxy view" in Fig.~\ref{supergalaxyA}.  By ``supergalaxy view" we mean a diagnostic diagram in which, at each (x,y) point, the value of the third variable (colour-coded) represents the median of the distribution of values of that third variable obtained from the spaxels with those specific (x,y) values. We present two complementary figures in  Appendix~\ref{App:AppendixD}, which show respectively the standard deviation $ \sigma $ of the distribution of values for the third (colour-coded) parameter, and the number of galaxies contributing to each x-y bin. These complementary figures are important to establish how meaningful and statistically-representative the  median values are in each (x,y) location of the diagrams.

The four parameters that we consider are, from top to bottom of the figure, galactocentric distance (normalized by the effective radius),  $E(B-V)$ computed from the Balmer decrement measured in our spectra and $\Sigma_{\rm SFR}$. The top row of Fig.~\ref{supergalaxyA} shows how galactocentric distance (normalized by the effective radius) varies with $Z_{\rm gas}$ and $\Sigma_{\star}$. As expected from an inside-out scenario, $\Sigma_{\star}$ of the disks decreases with increasing galactocentric distance (the median position of denser spaxels -high $\Sigma_*$- is at low galactocentric distance, i.e., the central regions). In the three panels showing the different mass bins, denser regions are found in the more massive galaxies and vice versa. The middle row shows the variation of  the dust extinction with $Z_{\rm gas}$ and $\Sigma_{\star}$. \citet{Garn2010} and \citet{Zahid2012} found that globally, dust extinction  increases with SFR, metallicity and stellar mass, but as stellar mass is the dominant factor, the relationships between dust and metallicity or SFR and extinction and SFR are secondary effects \citep{Garn2010}. We find locally that, when all the galaxies are included, $E(B-V)$ increases with increasing $\Sigma_{\star}$ and increasing $Z_{\rm gas}$. However, the different mass bins show that $E(B-V)$ does not strongly correlate with $\Sigma_{\star}$ or $Z_{\rm gas}$ but increases with the total stellar mass at a given $\Sigma_{\star}$.

The last row shows the dependence of  $\Sigma_{\rm SFR}$ with $Z_{\rm gas}$ and $\Sigma_{\star}$. When considering all masses together, we find a  complex dependence of  $\Sigma_{\rm SFR}$ with  $Z_{\rm gas}$ and $\Sigma_{\star}$. As expected from the intrinsic local relation between $\Sigma_{\star}$ and $\Sigma_{\rm SFR}$ (RSFMS, see following section), low SFR densities are found in regions of low $Z_{\rm gas}$ and high SFR densities are found in regions of high metallicities, but a wide range of $\Sigma_{\rm SFR}$ is found at intermediate metallicities and $\Sigma_{\star}$. For intermediate-mass galaxies,  $\Sigma_{\rm SFR}$ increases with  $\Sigma_{\star}$ and $Z_{\rm gas}$. For low and high-mass galaxies; however, there is not a clear correlation between $\Sigma_{\rm SFR}$ and $Z_{\rm gas}$ at fixed $\Sigma_{\star}$. 
It is therefore important to further discuss the secondary dependence of the RMZR with $\Sigma_{\rm SFR}$ (Sect.~\ref{sectiondiscussion}).

\subsection{The  stellar mass density versus SFR  density relation on MAD scales}

\begin{figure}
\begin{center}
 \includegraphics[width=90mm]{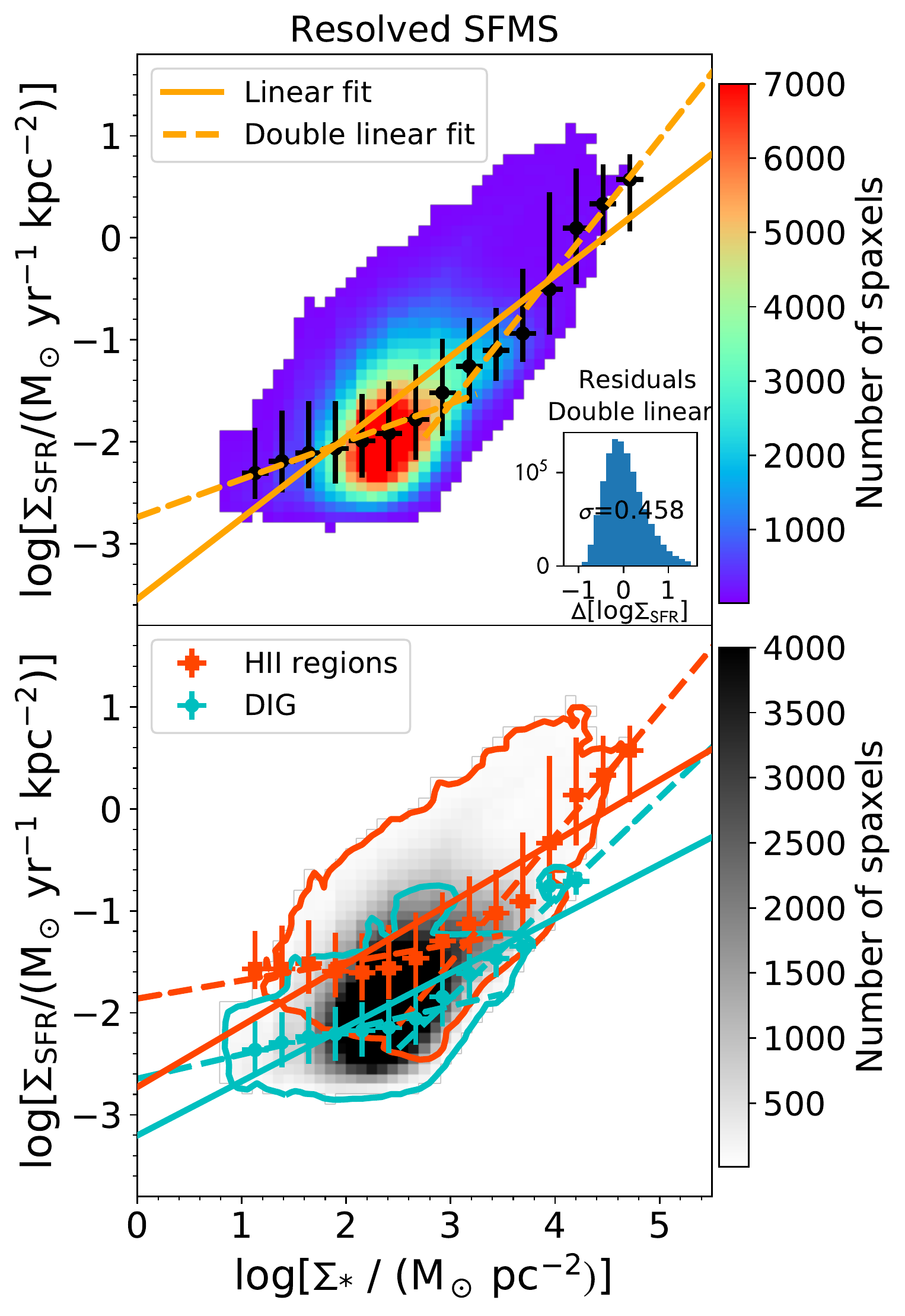}
\caption{\textit{Top:} Spatially resolved stellar mass surface density-star formation rate density relation (SFMS) at $\sim$100 pc scales for all the SF spaxels from all the galaxies in the sample, colour coded by the number of spaxels in each x-y bin. The orange solid line shows the linear fit to all the median values of $\Sigma_{\rm SFR}$. Two additional fits are performed to the data above and below the apparent threshold in surface mass density at $\sim10^3 M_{\sun}/pc^{2}$, plotted as orange dashed lines. \textit{Bottom:} Resolved SFMS, with overplotted contours from the H{\sc ii} regions (red) and DIG (cyan). Again, a single and double linear fits have been done to the median values of $\Sigma_{\rm SFR}$ in bins of  $\Sigma_{\star}$ for both H{\sc ii} regions (red squares) and DIG (cyan circles).}
\label{B_number_comp}
\end{center}
\end{figure}

As discussed in the introduction, the almost linear correlation  between  total $M_{\star}$  and total  SFR, i.e., the so-called SFMS, is well known at low (e.g \citealt{Brinchmann2004}; \citealt{Renzini2015}) and high redshifts (e.g., \citealt{Noeske2007}; \citealt{Daddi2007};  \citealt{Wuyts2013}; \citealt{Tacchella2015}). Here we address the question of whether this relation holds on the $\sim$100 pc  scales probed by MAD, when looking at the individual regions inside disk galaxies, i.e. a resolved SFMS. We first focus on the relationship for all the SF regions in the top panel of Fig.~\ref{B_number_comp}, where the colour-coding shows the number of spaxels in each x-y bin. We find that $\Sigma_{\rm SFR}$ is positively correlated with $\Sigma_{\star}$ when considering all galaxies together, although showing some scatter. This correlation spans five orders of magnitude in $\Sigma_{\star}$ and four in $\Sigma_{\rm SFR}$. Just as we observed in the RMZR, the local $\Sigma_{\rm SFR}$-$\Sigma_{\star}$ relation is characterized by a break around $10^{3} M_{\odot}/{\rm pc}^2$ above and below which the relation is steeper and flatter. As in the previous subsection, we use different functions to fit the median values of $\Sigma_{\rm SFR}$ in bins of  $\Sigma_{\star}$: (i) single linear fit and (ii) two linear fits before and after the threshold at $\Sigma_{\star}$\,=10$^{3}\, M_{\sun}\, {\rm pc^{-2}}$. Table~\ref{SFMS_fits} collects the slopes, intercepts and scatter for each function. A double linear fit reduces the scatter again, although not significantly with respect to the single linear one. The $ \chi^2 $ is lower and the $p$-value is higher for the double-linear fit, which shows that it is a better fit than the single-linear. We note a sharp cut at lower $\Sigma_{\rm SFR}$, which may cause the relation to flatten at lower values of $\Sigma_{\star}$. Taking into account a possible incompleteness of our sample at lower values of $\Sigma_{\rm SFR}$, the single line fit represents better the contours of the distribution.

\citet{Cano-Diaz2016} found a RSFMS using SF regions of galaxies of all morphological types from the CALIFA sample (0.5-1.5 kpc spatial resolution). They found a slope of $\gamma=d$~log($\Sigma_{\star})/d$~log($\Sigma_{{\rm H} \alpha}))=0.72 \pm 0.04$ and zeropoint of -7.63 (in units of log($M_{\sun}yr^{-1}kpc^{-2}$)]. Our results at higher spatial resolution agree, finding a slope of $\gamma=0.714 \pm 0.064$ and a zeropoint of -7.564 in their units. Using MaNGA ($\sim$1 kpc spatial resolution) data for star-forming regions in star-forming galaxies, \citet{Hsieh2017} found a slope of $\gamma=d$~log($\Sigma_{\star})/d$~log($\Sigma_{{\rm H} \alpha}))=0.715 \pm 0.001$ and zeropoint of 33.204 erg\,s$^{-1}$\,kpc$^{-2}$.  In our data at higher spatial resolution we find a slope of $\gamma=0.714 \pm 0.064$ and a zeropoint of 33.704 erg\,s$^{-1}$\,kpc$^{-2}$. The aforementioned results indicate that the relationship between $\Sigma_{\star}$ and $\Sigma_{\rm SFR}$ remains unchanged  for star-forming regions over several order of magnitudes of spatial scales, from kpc scales down to the  $\sim$100 pc scales probed with MUSE in the MAD galaxies.

The bottom panel of Fig.~\ref{B_number_comp} shows the RSFMS distinguishing the H{\sc ii} regions (red contours) and DIG (cyan contours)  over the SFMS for all the SF spaxels, now grey colour-coded. We perform the same fits to the distinct components (H{\sc ii} regions and DIG), finding similar slopes for both single and double linear fits. The median $\Sigma_{\rm SFR}$ values for the H{\sc ii} regions have higher $\Sigma_{\rm SFR}$ on average (i.e., higher $y$-intercept on the fits). In other words, we find a ``retired'' RSFMS for the DIG component. As our study focuses on the regions ionized by hot/young stars, this result helps to understand the origin of the DIG at sub-kpc scales and agrees with \cite{Weilbacher2017}, where they found that the leaking UV photons from H{\sc ii} regions can explain the ionization of the DIG. Our study cannot rule out the possibility that the hot old stars can ionize part of the DIG because we do not study the LINER-like ionized regions. In this regard, \citet{Hsieh2017} found a ``retired" RSFMS for the LINER spaxels in their SF galaxies.

\begin{table}
\caption{Best fits to the resolved SFMS using two different functions: (i) linear fit using the equation \textit{y=mx+n}, (ii) double linear fits before and after $10^3 M_{\sun}/{\rm pc}^{2}$. The scatter has been measured as the standard deviation of the residuals after subtracting the best fit to the data.}
 \label{SFMS_fits}
\centering
\begin{tabular}{|c|c|c|}
\multicolumn{3}{|c|}{Linear \textit{y=mx+n}} \\
 \hline 
& \textit{m}(dex/log$\Sigma_*$) & 0.79 $\pm$ 0.07 \\ 
 & \textit{n}(dex/log$\Sigma_{\rm SFR}$) &-3.54 $\pm$ 0.20 \\ 
Residuals&  $ \sigma$(dex/log$\Sigma_{\rm SFR}$)  &0.472 \\  
$\chi^2$ &    &6.424 \\  
$p-$value &    &0.011 \\  
 \hline 
\multicolumn{3}{|c|}{Double Linear \textit{y=mx+n}}\\ 
 \hline 
x< $10^3 M_{\sun}/{\rm pc}^{2}$ & \textit{m}(dex/log$\Sigma_*$) & 0.37 $\pm$ 0.04\\ 
& \textit{n}(dex/log$\Sigma_{\rm SFR}$) & -2.74 $\pm$ 0.08 \\ 
x> $10^3 M_{\sun}/{\rm pc}^{2}$ & \textit{m}(dex/log$\Sigma_*$) & 1.30 $\pm$ 0.09 \\ 
& \textit{n}(dex/log$\Sigma_{\rm SFR}$) & -5.52 $\pm$ 0.35 \\ 
Residuals& $ \sigma$(dex/log$\Sigma_{\rm SFR}$) &0.458 \\  
$\chi^2$ &    &0.604 \\  
$p-$value &    &0.895 \\  
 \hline 
\end{tabular}
\end{table}

\begin{figure*}
\begin{center}
 \includegraphics[width=165mm]{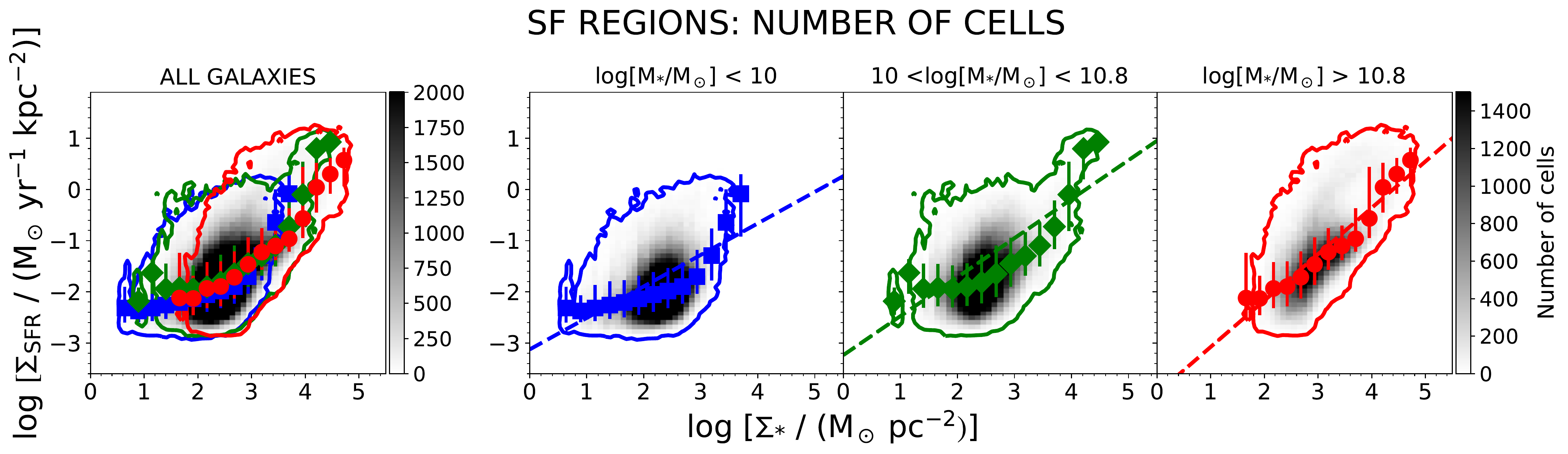}
\caption{Spatially resolved stellar mass surface density-star formation rate density relation (RSFMS) at $\sim$100 scales for all the SF spaxels from all the galaxies (left), followed by the separation of the sample in three mass bins as in Fig.~\ref{supergalaxyA}.}
\label{B_number_color_noAGN}
\end{center}
\end{figure*}

When dividing the sample into stellar mass bins (Fig.~\ref{B_number_color_noAGN}), the $\Sigma_{\rm SFR}$-$\Sigma_{\star}$ relations of the respective bins nearly coincide. The break in the $\Sigma_{\rm SFR}$-$\Sigma_{\star}$ relation moves between  $10^{3}$ and $10^{4} M_{\odot}/{\rm pc}^2$ with increasing total stellar mass . As a result, the slopes of the relation do increase with total mass ($\gamma=0.612 \pm 0.108$ for the low-mass galaxies, $0.762  \pm 0.118$ for galaxies with intermediate stellar masses and $0.906 \pm 0.068$ for the high-mass galaxies), but this variation is small compared to the 3 orders of magnitude variation in $\Sigma_{\rm SFR}$ over the range of $\Sigma_{\star}$. We note however that the scatter in $\Sigma_{\rm SFR}$ at a constant  $\Sigma_{\star}$ is quite large; this scatter does not  change substantially across the range of total  stellar mass that we sample (with the  exception of the very centres of intermediate and high-mass galaxies, most likely due to the superposition of bulge and disk data). Sect.~\ref{sectiondiscussion} will further study whether the RSFMS has a local or global origin.

\begin{figure*}
\begin{center}
 \includegraphics[width=165mm]{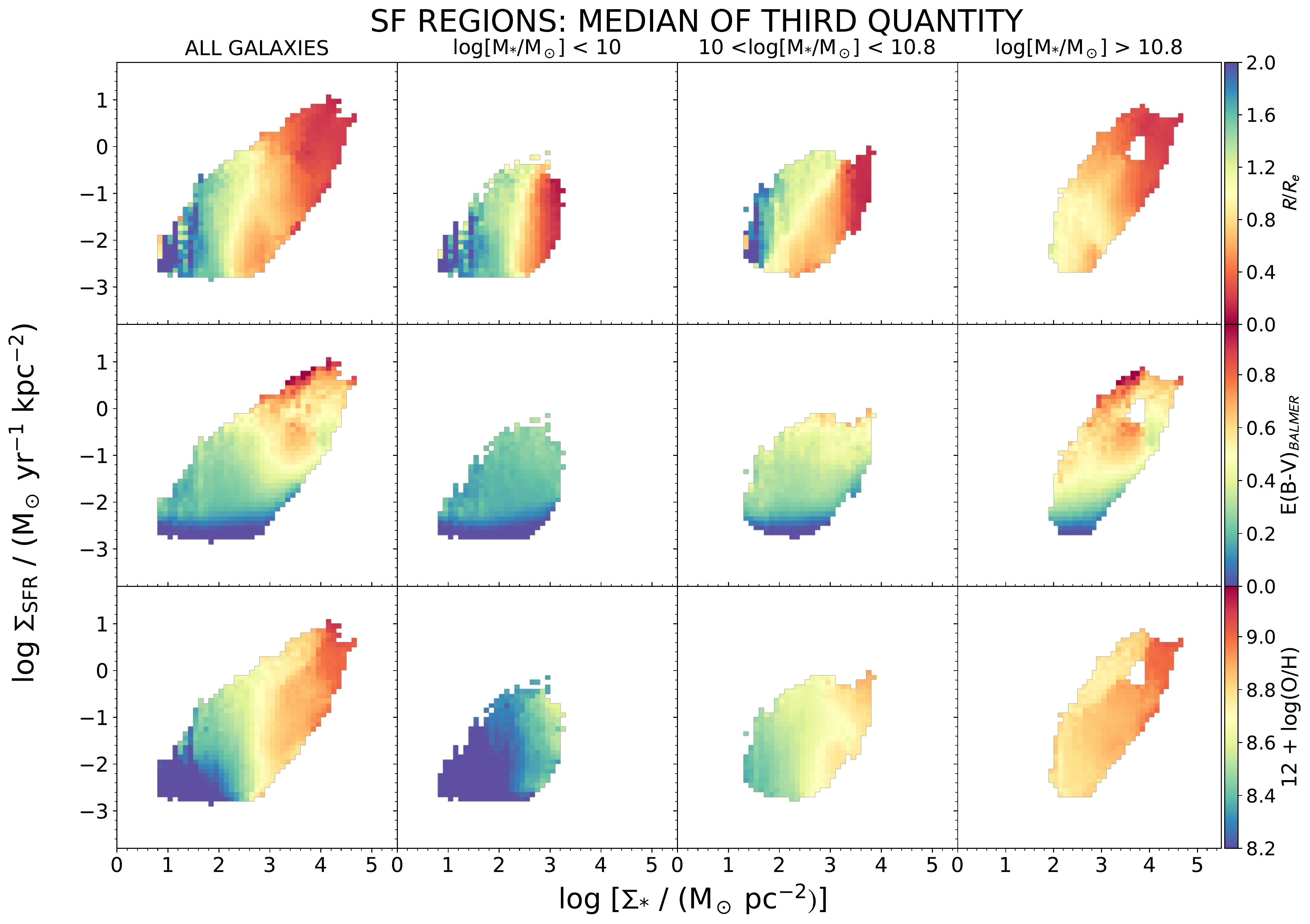}
\caption{Supergalaxy view of the spatially resolved SFMS for all the SF spaxels from all the galaxies (left), followed by the separation of the sample in three mass bins as in Fig.~\ref{supergalaxyA}. Each row shows different colour-coding according to the median value of a third quantity: galactocentric distance (top row), dust extinction from the Balmer decrement (middle row) and $Z_{\rm gas}$ (bottom row).}
\label{supergalaxyB}
\end{center}
\end{figure*}

We investigate the $\Sigma_{\rm SFR}$-$\Sigma_{\star}$ relation further by mapping the dependence of third parameters on this relation.  These are the same as those analysed in the previous section and are represented by the colour coding in Fig.~\ref{supergalaxyB}. Galactocentric distance does not vary significantly with $\Sigma_{\rm SFR}$ but increases with decreasing $\Sigma_{\star}$ as expected from an inside-out scenario.  This is true in all stellar mass bins, indicating that $\Sigma_{\star}$ is directly correlated with the galactocentric distance regardless of the value of total stellar mass.  Dust extinction increases with increasing $\Sigma_{\rm SFR}$ but does not depend on $\Sigma_*$.  This indicates that the apparent correlation between dust extinction and $\Sigma_{\star}$  observed in the RMZR (middle row of Fig.~\ref{supergalaxyA}) is a consequence of the correlation between $\Sigma_{\star}$ and $\Sigma_{\rm SFR}$.  Thus dust extinction follows SFR and does not seem to depend on either $\Sigma_{\star}$ or metallicity. Finally, the last row shows the variation of gas metallicity with $\Sigma_{\star}$  and  $\Sigma_{\rm SFR}$: the only correlations seem to be between $\Sigma_{\star}$ and $Z_{\rm gas}$ and between $\Sigma_{\star}$ and $\Sigma_{\rm SFR}$, but not between $\Sigma_{\rm SFR}$ and $Z_{\rm gas}$. At fixed $\Sigma_{\star}$, the metallicity does not vary with $\Sigma_{\rm SFR}$. These results agree with an scenario in which the dust is formed with newly born stars, irrespectively of the amount (mass) of stars or the local metal enrichment.

%%%%%%%%%%%%%%%%%%%%%%DISCUSSION%%%%%%%%%%%%%%%%%%%%%%%%%%%%%%%%%%

\section{Discussion}
\label{sectiondiscussion}
\subsection{Resolved relations: local or global origin?}

First we explore the possible dependence of the local gas metallicity with the global stellar mass. As found in Sect.~\ref{sectionmetalgradients}, the inner-disk gas metallicities that we measure at a given (normalized) galactocentric distance increase with the total stellar mass of the galaxies. Fig.~\ref{A_number_color_noAGN} showed that the local enrichment varies with $\Sigma_{\star}$ and the total stellar mass of the galaxies. It is therefore important to understand whether this local enrichment varies due to $\Sigma_{\star}$, total stellar mass or both.

\begin{figure*}
\begin{center}
 \includegraphics[width=165mm]{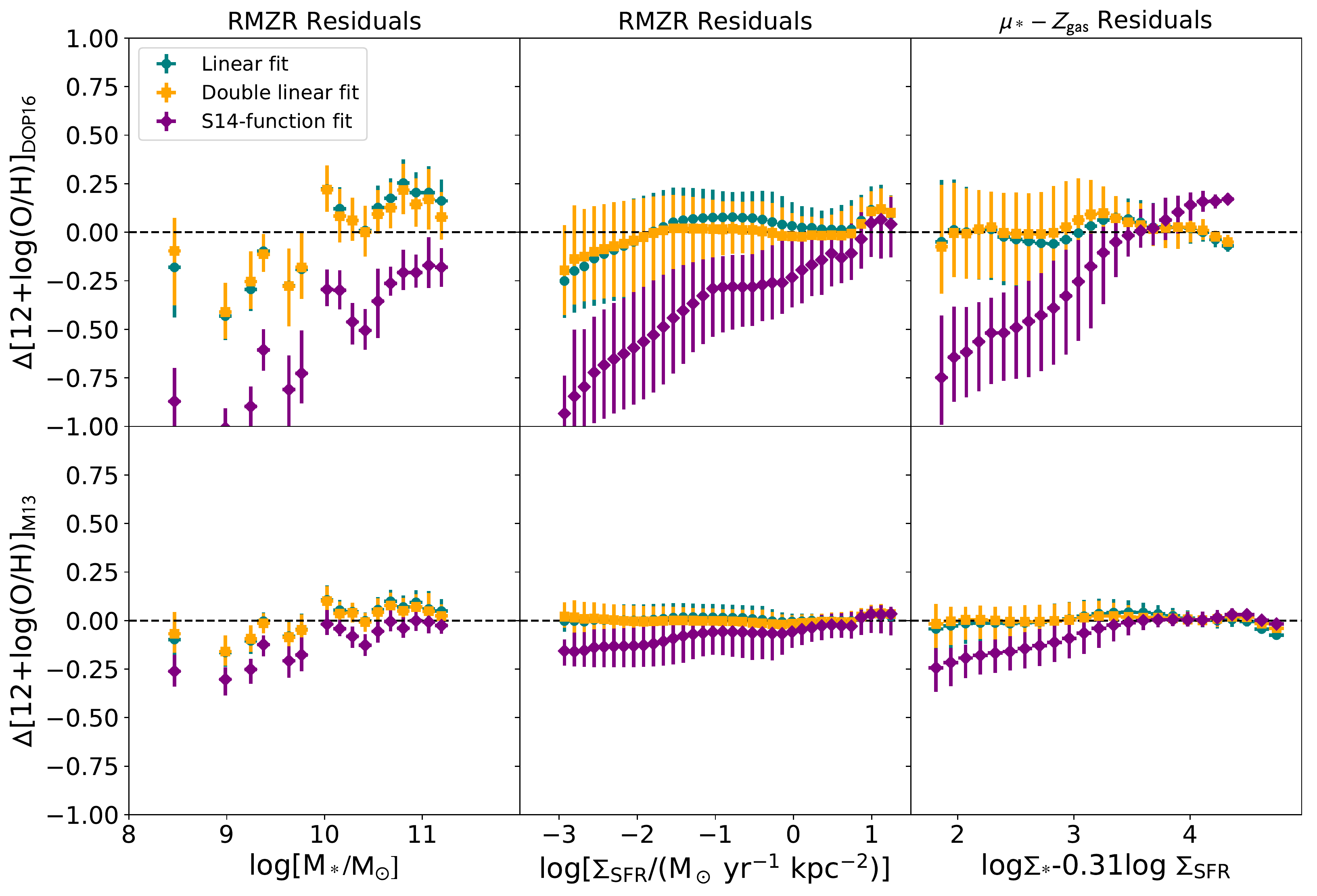}
\caption{Analysis of the residuals $ \Delta {\rm 12+log(O/H)} $ for the DOP16 (top) and M13 (bottom) calibrations, as a function of the total stellar mass (left), log$\Sigma_{\rm SFR}$ (middle) and log$\Sigma_{\star}$ - 0.31 log $\Sigma_{\rm SFR}$. The median  and 1$ \sigma $ deviation (errorbar) from all the SF spaxels in each x-bin have been presented as blue dots, orange squares and purple diamonds for the linear fit, double-linear fit and S14-function fit respectively.}
\label{A_number_residuals}
\end{center}
\end{figure*}

To do so, we study the behaviour of the residuals from the RMZR as a function of the total mass of the galaxy. The median residuals in bins of total stellar mass are computed and presented in the left panel of Fig.~\ref{A_number_residuals} for the three performed fits: single linear, double linear and S14-function. From this figure we see that (i) M13 calibration shows lower residuals due to the lower range in metallicities of the calibrator, (ii) the S14-function fit delivers larger residuals than the single linear and double linear fits for both metallicity calibrators, specially for the low mass galaxies and (iii) the residuals are within the metallicity calibration uncertainties except for the low mass galaxies. The residuals using the linear and double linear fits do not depend on the total stellar mass. However, the bad fits  on the lower $ \Sigma_{\star}$ data using the S-14 function deliver high residuals for the low mass galaxies. We thus conclude that the RMZR does not depend on the total stellar mass, as the residuals when adopting a linear/double linear function do not depend on the total stellar mass. The chemical enrichment occurs at local scales (<100 pc), also observed at larger spatial resolution (e.g., at 1-2 kpc, \citealt{Barrera-Ballesteros2016}).

\begin{figure}
\begin{center}
 \includegraphics[width=80mm]{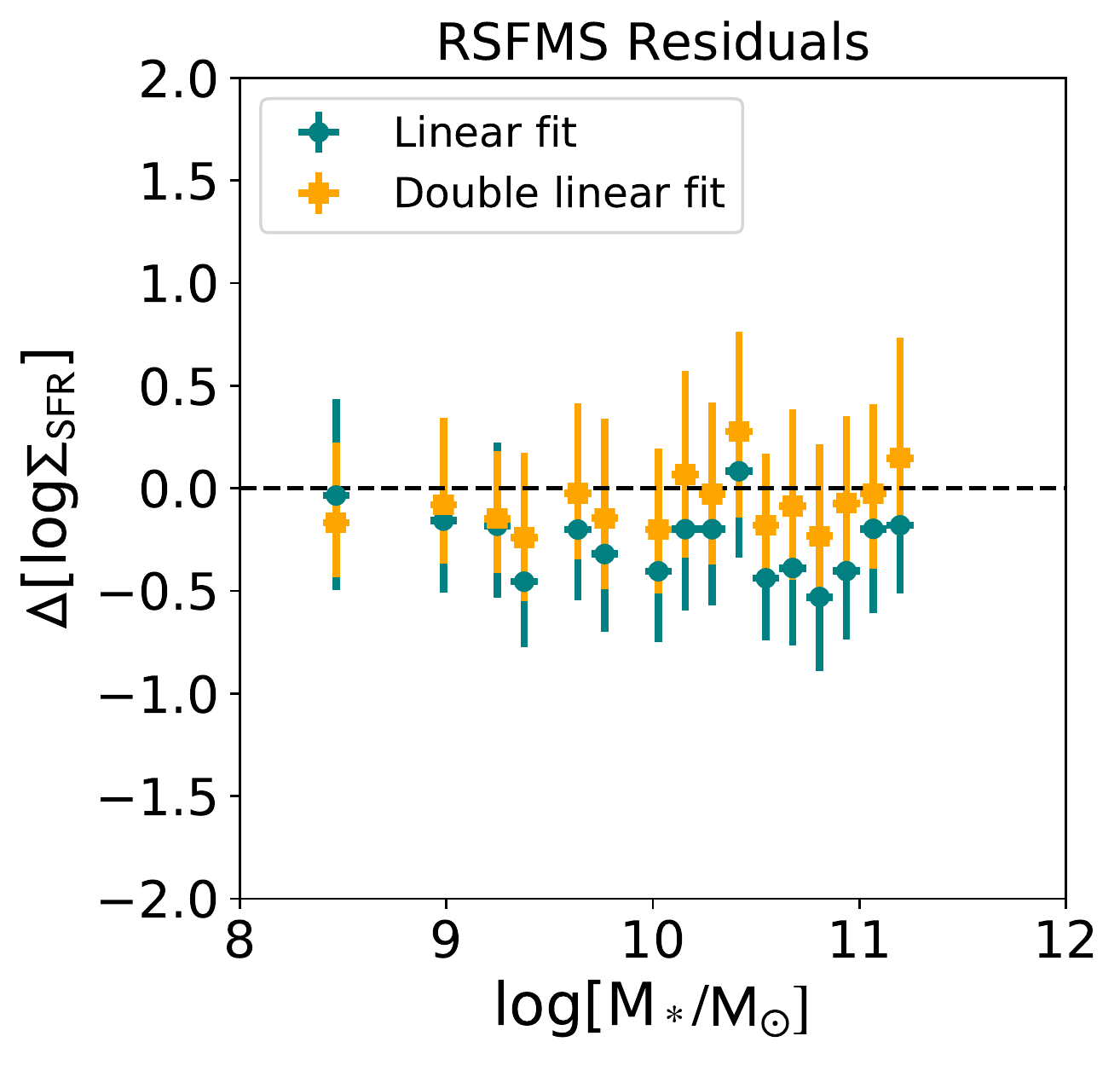}
\caption{Analysis of the residuals $ \Delta{\rm log}\Sigma_{\rm SFR} $ as a function of the total stellar mass. The median  and 1$ \sigma $ deviation from all the SF spaxels in each x-bin have been presented as blue dots and orange squares for the linear and double-linear fits respectively.}
\label{B_number_residuals}
\end{center}
\end{figure}

Similarly, it is important to determine whether the RSFMS has a local or global origin. Fig.~\ref{B_number_color_noAGN} already showed that the shape of the RSFMS barely changed with increasing total stellar mass. Analogously to the study of the residuals of the RMZR, we now analyze the residuals from the RSFMS as a function of the total stellar mass. Fig.~\ref{B_number_residuals} confirms that the resolved SFMS does not depend on the total stellar mass, as the residuals of this relation do not depend on the total stellar mass.

\subsection{Secondary dependence of the RMZR with the SFR}

In global terms, the total SFR has been suggested to be a second parameter on the global MZR (e.g., \citealt{Ellison2008}; \citealt{Mannucci2010}; \citealt{Lara-Lopez2010}). Neither MaNGA \citep{Barrera-Ballesteros2017}, CALIFA \citep{Sanchez2017} nor SAMI \citet{Sanchez2018} have found the corresponding second relation relation in their IFU data. This discrepancy between global and resolved studies is an important matter of study. Equally important is to find the physical scenario that can describe our collection of results.

First, the bottom left panel of Fig.~\ref{supergalaxyA} showed a complex relationship between the RMZR and $\Sigma_{\rm SFR}$: $\Sigma_{\rm SFR}$ increases with both increasing $Z_{\rm gas}$ and $\Sigma_{\star}$. Second, the RSFMS (Fig.~\ref{B_number_comp}) revealed increasing $\Sigma_{\rm SFR}$ with $\Sigma_{\star}$. It is therefore unclear whether the relationship seen in Fig.~\ref{supergalaxyA} between the RMZR and $\Sigma_{\rm SFR}$ is due to the RSFMS or due to the SFR as a secondary parameter in the RMZR.

As in the previous subsection, the analysis of the residuals clarifies the existence of this possible dependence. The second column of Fig.~\ref{A_number_residuals} shows the behaviour of the residuals of the RMZR as a function of $\Sigma_{\rm SFR}$. We see that the residuals do not correlate with $\Sigma_{\rm SFR}$. We also define $ \mu _*$=log$\Sigma_{\star}$-$\alpha$log $\Sigma_{\rm SFR}$  \citep{Mannucci2010} as a possible quantity to minimize the scatter in the RMZR relation. Therefore we explore the $ \mu_{\star} $-$Z_{\rm gas}$ relation, fitting it using three different functions (single linear, double linear and S14-function), and study the residuals  (third column of Fig.~\ref{A_number_residuals}). The minimum scatter in the residuals was found for $ \alpha =0.31$ and are $ \sigma_{\rm DOP16} =$0.237, 0.234 and 0.286 dex and  $ \sigma_{\rm M13} =$0.094, 0.092 and 0.108 dex for the single linear, double linear and S14-function fits respectively. These residuals are larger than those of the original RMZR, confirming that there is no secondary dependence of the RMZR with the  $\Sigma_{\rm SFR}$. Additionally, the unclear relationship between RMZR and $\Sigma_{\rm SFR}$ may well be due to a re-scaling of the stellar-mass axis (as suggested in \citealt{Barrera-Ballesteros2017}).

We thus conclude that there are local (<100 pc) $\Sigma_{\star}$-$Z_{\rm gas}$ and  $\Sigma_{\star}$-$\Sigma_{\rm SFR}$ relations for SF regions in galaxies which lie in the SFMS. This result, where less-dense regions are less chemically evolved than the denser regions (which tend to reside in the centres of galaxies), is consistent with an inside-out growth (i.e., the central regions form earlier and are denser than the outer regions). The global MZR thus emerges from the local one, and the local enrichment does not depend on the total stellar mass. In other words, the chemical enrichment is dominated by local processes (\citealt{Rosales-Ortega2012}; \citealt{Sanchez2014}; \citealt{Barrera-Ballesteros2016,Barrera-Ballesteros2017}). The local SFR seems to depend only on the local potential, being directly correlated with the local $\Sigma_{\star}$ on top of which the local star formation is taking place, but this relation is independent of the total stellar mass. Dust attenuation appears to follow SFR alone.  It is independent of local metallicity, $\Sigma_{\star}$  and total stellar mass.

%%%%%%%%%%%%%%%%%%%%%%CONCLUSIONS%%%%%%%%%%%%%%%%%%%%%%%%%%%%%%%%%%

\section{Conclusions}

We have presented the first results for the 38 star-forming disk galaxies along the $z=0$  SFMS that have been observed so far in the MAD survey. The novelty of MAD is the ability to probe the emission line fluxes on spatial scales of $\sim$100 pc. After correcting the emission line fluxes from dust extinction using the local Balmer decrements measured in the MUSE spectra, we have computed the relevant emission line ratios required to measure accurate  SFR, BPT diagrams, diffuse ionized gas components and gas metallicities. To parametrize the local BPT properties within the disks we  defined a continuous variable $\eta$ that unambiguously identifies the location of the spectrum on the BPT diagram, and thus  the source of the gas ionization. Our main conclusions are as follows:

\begin{enumerate}
\item There is a correlation between the mean gas metallicity inside 0.5~$R_e$ and the total stellar mass (mass-metallicity relation). This correlation  holds for all star forming components, i.e., when H{\sc ii} regions and DIG are considered separately. 
\item Adopting as our fiducial gas metallicity calibrator the theoretically-based parameter introduced by DOP16, we find a universal trend for negative gas metallicity gradients in the inner regions of the local disks which are sampled in our MAD data. This result is consistent   with the cosmologically-motivated inside-out scenario for the growth of galactic disks, although it is not a sufficient condition to prove such a scenario. Radially-declining metallicities, i.e., negative metallicity gradients, are found also when separately considering   H{\sc ii} regions and the DIG. Outside 0.5 $R_e$, the metallicity decreases with radius more rapidly for low mass galaxies than for high-mass galaxies. 
\item The 2D gas metallicity distributions reveal a large amount of structure and scatter at any galactocentric radius, which is lost when deriving the  azimuthal averages to produce the gradients. Generally, the H{\sc ii} regions are on average $\sim0.2$ dex more metal enriched than the DIG at any radius. 
\item We find that the RMZR holds at the $\sim$100 pc scales sampled by MAD for about four orders of magnitude in $\Sigma_{\star}$ and over a broad range of metallicities, 7.8<$ Z_{gas} $<9.3. The relation  flattens above a threshold of   $\Sigma_{\star}\gtrsim\,10^3\,M_{\sun}{\rm pc}^{-2}$, although $Z_{\rm gas}$ continues increasing for increasing $\Sigma_{\star}$ with a shallower slope. We find that the global MZR arise from the local MZR at these scales (RMZR), as the residuals of the RMZR do not depend on the total stellar mass. We find that there is no secondary dependence of the RMZR with the $\Sigma_{\rm SFR}$, as introducing this as a secondary parameter does not reduce the scatter of the RMZR.
\item We find a correlation between $\Sigma_{\star}$ and $\Sigma_{\rm SFR}$ on the explored $\sim$100 pc scales, spanning five orders of magnitude in $\Sigma_{\star}$ and four in $\Sigma_{\rm SFR}$. This relation holds for both H{\sc ii} regions and DIG, being the $\Sigma_{\rm SFR}$ in the H{\sc ii} regions $\sim$1 dex higher than that of the DIG (we find a ``retired" RSFMS for the DIG). The slope of this relation depends only weakly with total stellar mass. The residuals of the resolved SFMS do not correlate with the total stellar mass.
\item Dust extinction seems to follow $\Sigma_{\rm SFR}$ only.  It is independent of $\Sigma_{\star}$ and metallicity. 
\end{enumerate}
\label{section7}

%%%%%%%%%%%%%%%%%%%%%%%ACKNOWLEDGMENTS%%%%%%%%%%%%%%%%%%%%%%%%%%%%%%%%%

\section*{Acknowledgements}

We thank Amanda Bluck for her useful comments during the preparation of this manuscript. We thank the referee for their comments, which improved the quality of the paper. This work is based on observations taken at ESO/VLT in Paranal and we would like to thank the ESO staff for their assistance and support during the MUSE GTO campaigns. We acknowledge the usage of the HyperLeda database (http://leda.univ-lyon1.fr). This research has made use of the NASA/IPAC Extragalactic Database (NED) which is operated by JPL, Caltech, under contract with NASA. SEF and CMC acknowledge support from the Swiss National Science Foundation.  AMI acknowledges support from the Spanish MINECO through project AYA2015-68217-P. JS acknowledges VICI grant 639.043.409 from the Netherlands Organization for Scientific Research (NWO). V.P.D. is supported by STFC Consolidated grant ST/M000877/1 and acknowledges the personal support of George Lake, and of the Pauli Center for Theoretical Studies, which is supported by the Swiss National Science Foundation (SNF), the University of Z\"urich, and ETH Z\"urich during a sabbatical visit in 2017.

%%%%%%%%%%%%%%%%%%%%%%%%%%%%%%%%%%%%%%%%%%%%%%%%%%



\bibliographystyle{mnras}
\bibliography{references} % if your bibtex file is called example.bib

%\begin{thebibliography}{99}
%\end{thebibliography}

\section*{SUPPORTING INFORMATION}

Additional Supporting Information may be found in the online version of this article:

\textbf{Appendix A}. Notes on individual galaxies

\textbf{Appendix B}. Determination of the BPT-parameter $ \eta $

\textbf{Appendix C}. Beyond the radial gradients: 2D metallicity distributions

\textbf{Appendix D}. Complementary figures to the resolved Mass-Metallicity relation

\textbf{Appendix E}. Complementary figures to the resolved Mass-SFR relation

%%%%%%%%%%%%%%%%%%%%%%%%%%

%\bsp
\clearpage

\appendix

\section{Notes on individual galaxies}
\label{App:AppendixA}

In one page per galaxy, the maps of the emission line fluxes and gas diagnostics analysed in this paper are presented, together with notes on each individual galaxy. 

The top eight maps show the dust-corrected emission line fluxes from the strong lines, in units of $\times 10^{-20}$ erg s$^{-1}$ cm$^{-2}$. From left to right and top to bottom, the presented emission lines are: H$ \beta $, [O{\sc iii}]$ \lambda $4959, [O{\sc iii}]$ \lambda $5007, [N{\sc ii}]$ \lambda $6548, H$\alpha$, [N{\sc ii}]$ \lambda $6583, [S{\sc ii}]$ \lambda $6717 and [S{\sc ii}]$ \lambda $6731.

The bottom eight maps show all the diagnostics studied in this paper: a) Stellar mass surface density map. b) Colour-composite RGB image created using narrowband images centred on H$ \alpha $ (from red to blue according to the velocity at each spaxel) and [O{\sc iii}] (green). c) Extinction map computed via the Balmer decrement using the H$ \alpha $/H$ \beta $ ratio from our spectra. d) SFR  surface density map of the SF regions derived from the dust-corrected H$ \alpha $ flux. e) N{\sc ii}-BPT diagram. Each point corresponds to one spaxel, colour coded by the BPT-parameter $\eta$. f) 2-D BPT map of the galaxy, colour-coded as in e. g) Gas metallicity map  of the SF regions derived using the DOP16 calibration. h) In greyscale, the dust-corrected SFR  surface density. In red, the position of the diffuse ionized gas (DIG).

In all the maps, the foreground stars and the spaxels that do not have a S/N of 3 in all the studied lines have been masked. For NGC3783 and NGC4593, the central parts where the broad lines due to the AGN emission resulted in a failure of the fitting code have been also masked. For each galaxy, all the maps have the same orientation (North is up and East is left). The physical scale (1 kpc) is presented in the top left panel.

\clearpage

%\subsection{NGC~4030}

\begin{figure*}
\begin{center}
 \includegraphics[width=165mm]{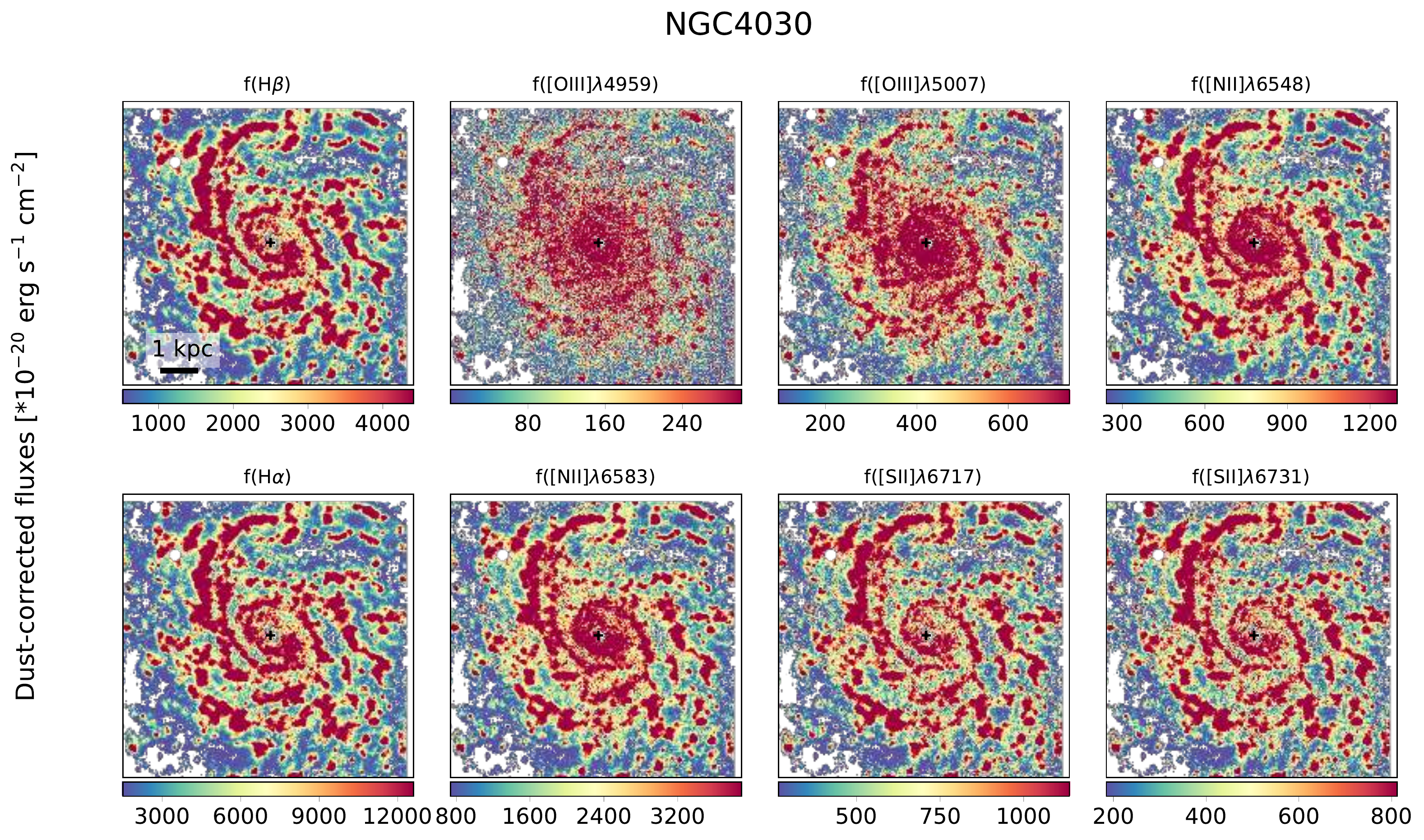}
% \caption{As Fig.~\ref{allfluxes} but for NGC~4030.}
% \label{ngc4030fluxes}
% \end{center}
% \end{figure*}

% \begin{figure*}
% \begin{center}
 \includegraphics[width=165mm]{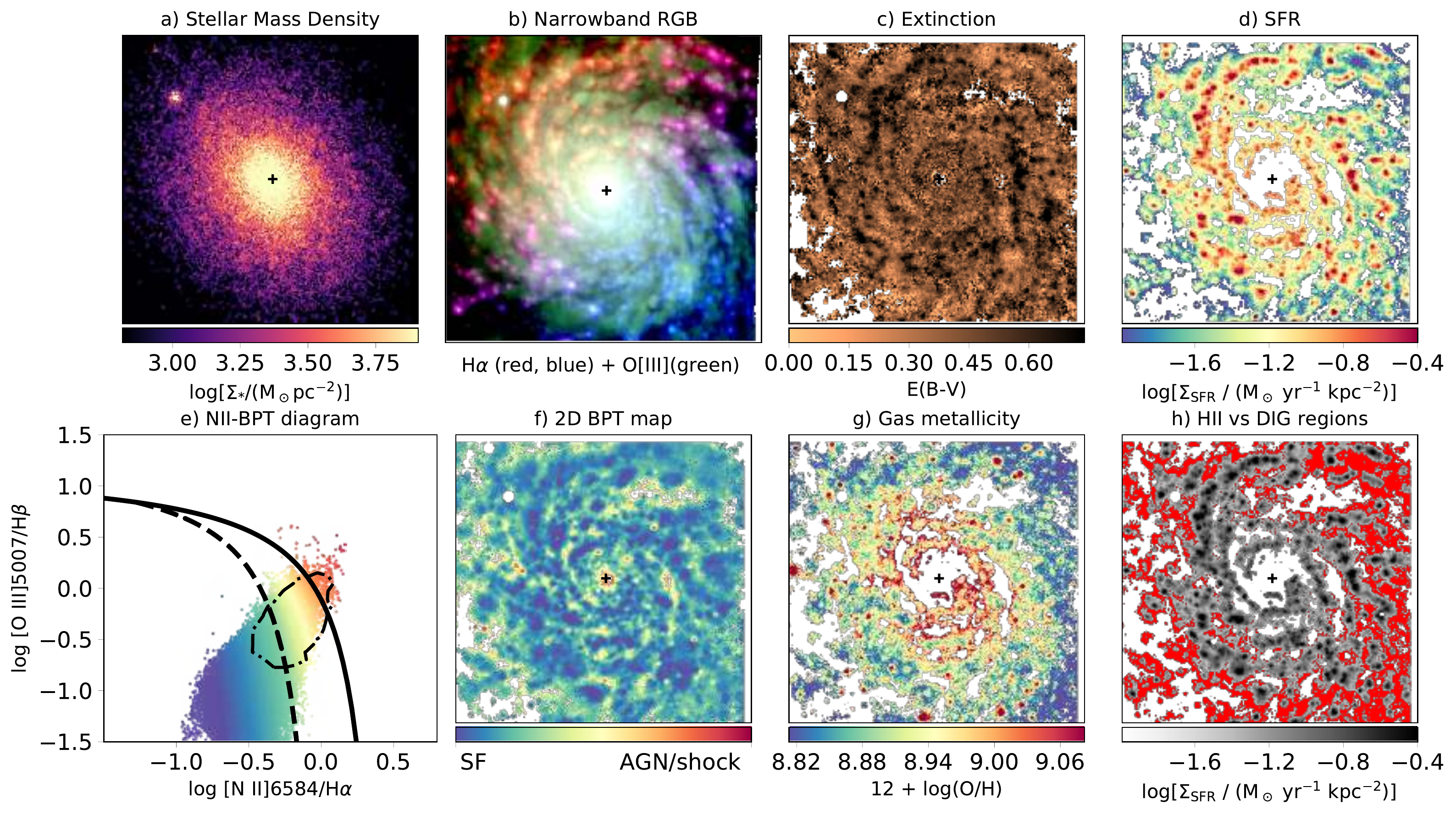}
\caption{NGC~4030 is a massive, multi-armed, unbarred spiral with a small bulge. The spiral arms are composed of multiple H{\sc ii} regions. The interarm region is well identified in both the BPT and DIG maps. There are no asymmetries in the ionized emission of the galaxy, and only narrow lines have been observed (no signs of outflows are found). As in NGC~3521, the metallicity gradient for the H{\sc ii} regions is flatter than that for the DIG gas, although both are negative with radius. The H{\sc ii} regions in the centre and those in the outer spiral deviate from the linear fit by $\sim$0.25 dex, whereas the DIG in the interarm regions are $\sim$ 0.15 dex lower than the average.}
\label{ngc4030plots}
\end{center}
\end{figure*}

\clearpage
%\subsection{NGC~3521}   

\begin{figure*}
%\begin{center}
\includegraphics[width=165mm]{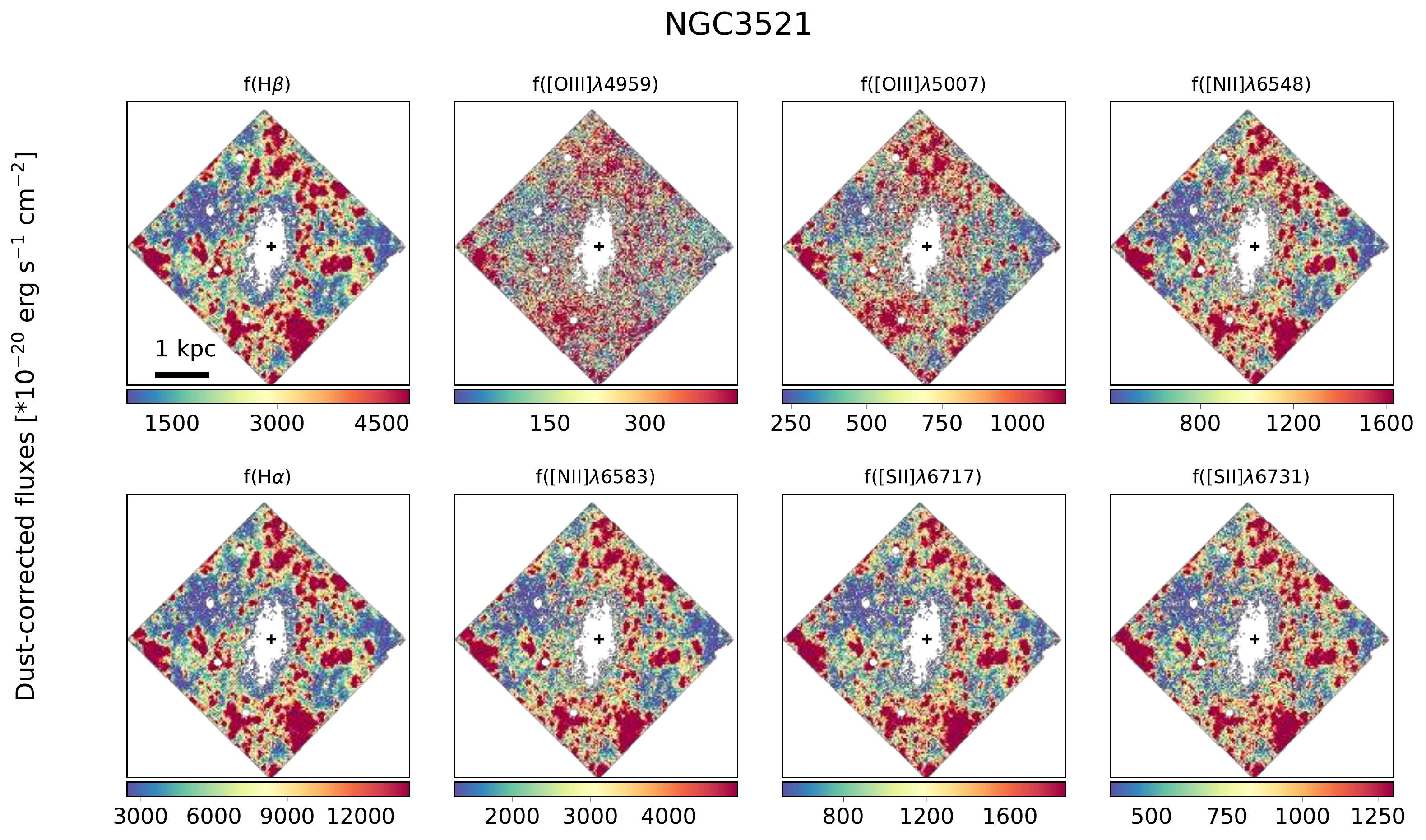}
% \caption{As Fig.~\ref{allfluxes} but for NGC~4030.}
% \label{ngc3521fluxes}
% \end{center}
% \end{figure*}

% \begin{figure*}
% \begin{center}
 \includegraphics[width=165mm]{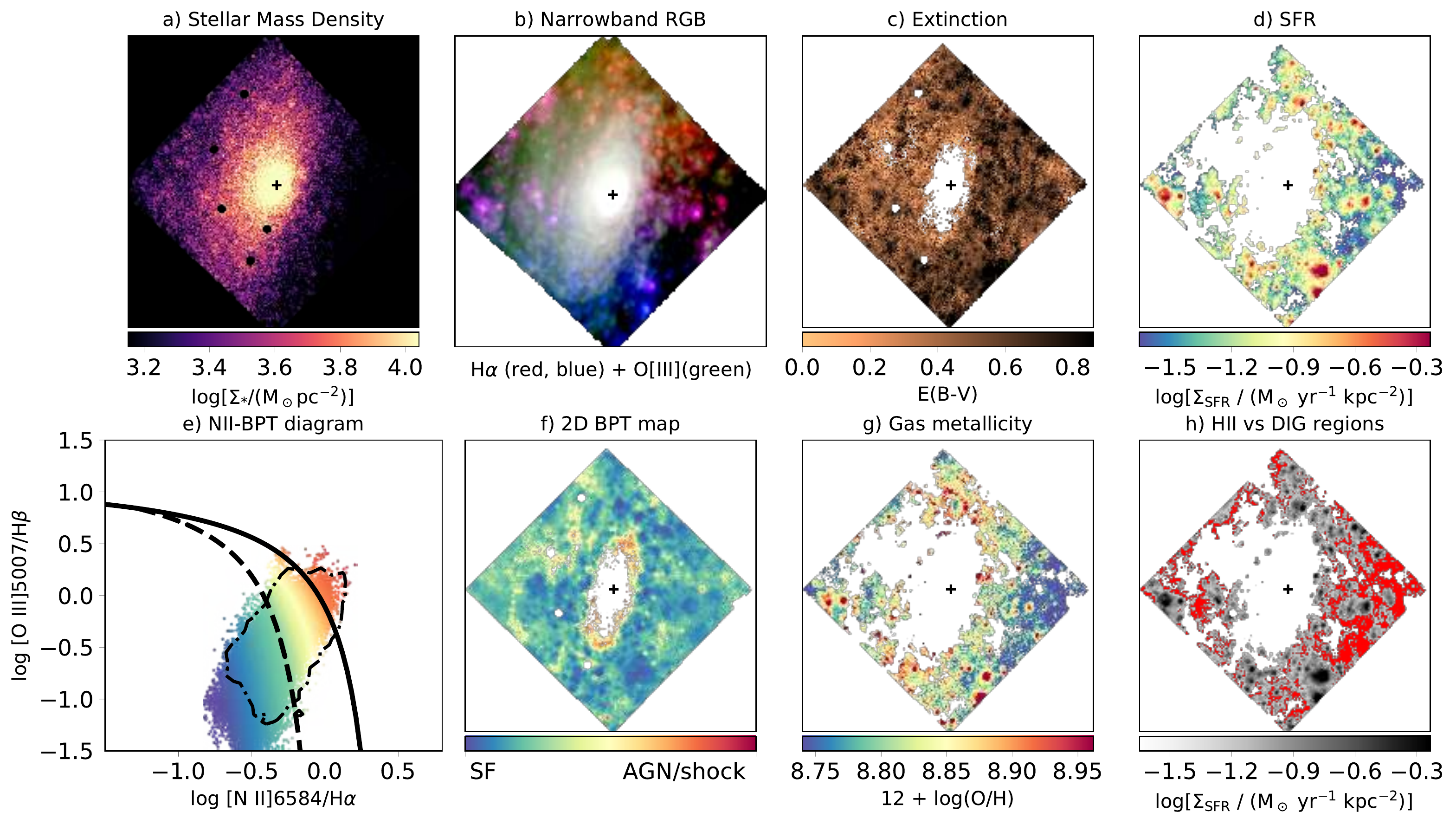}
\caption{This galaxy is one of the most massive ones in our sample ($M_{\star}$=10$^{11.2} M_{\sun}$). It shows a very bright and dense bulge, devoid of star formation. There are, however, many H{\sc ii} regions surrounding the central part. There are two distinct components in the SFR map: one quenched central bulge part and a disk that is actively forming stars. We see that the emission of this galaxy is asymmetric, showing a denser, metal enriched eastern area with lower extinction and lower SFR values than the western part. These asymmetries cannot be explained by interactions, as NGC~3521 is isolated \citep{Karachentseva1973}. The central part and central H{\sc ii} regions have higher metallicity than the outer regions, deviating from the linear fit about $\sim$0.2 dex. The metallicity in the DIG gas in the western part deviate -0.15 dex from the linear fit. The gradient for the H{\sc ii} regions is mostly flat, whereas that for the DIG gas is steeper.
}
\label{ngc3521plots}
%\end{center}
\end{figure*}

\clearpage
%\subsection{NGC~3256}   

\begin{figure*}
\begin{center}
 \includegraphics[width=165mm]{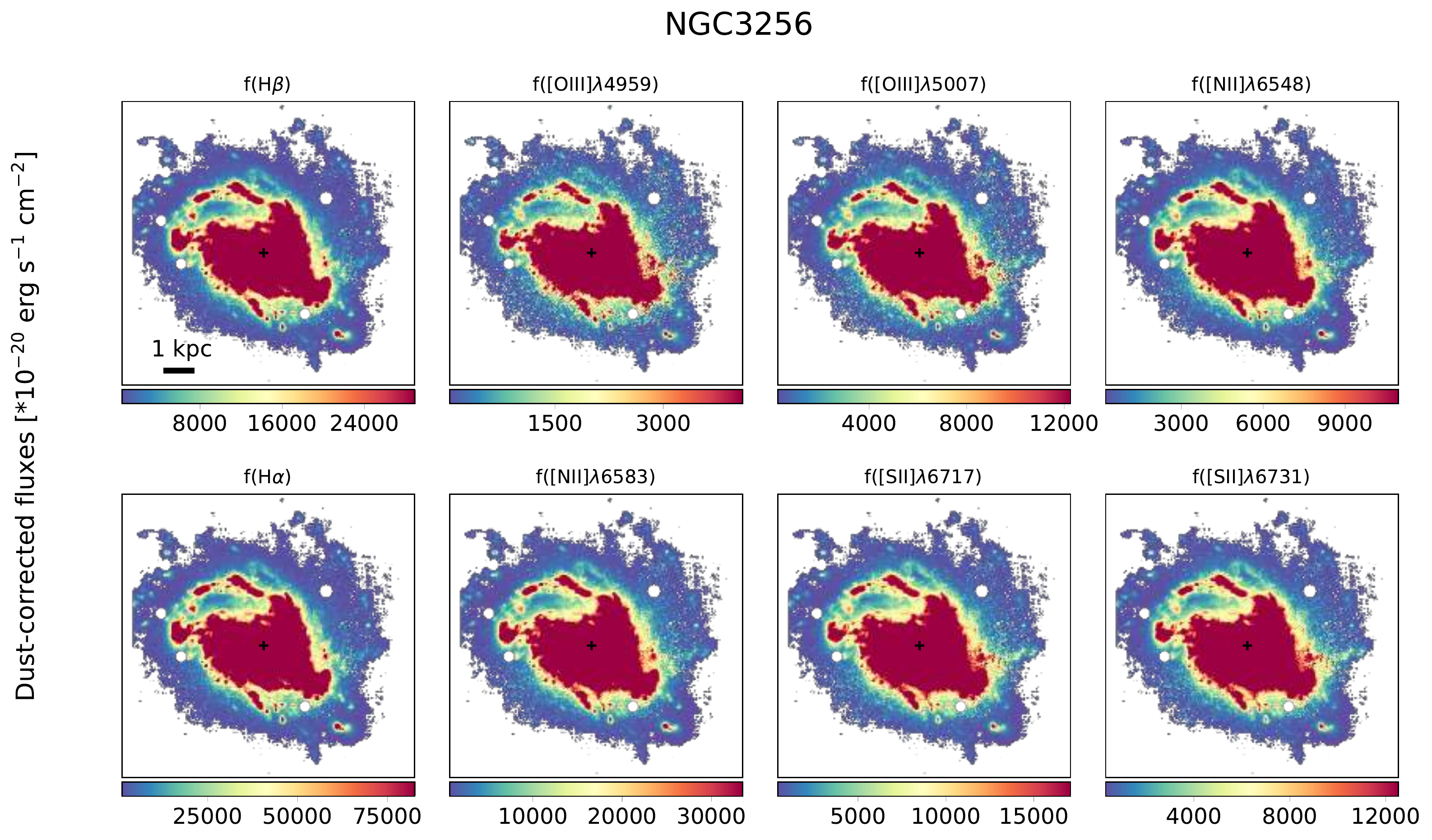}
% \caption{As Fig.~\ref{allfluxes} but for NGC~4030.}
% \label{ngc3256fluxes}
% \end{center}
% \end{figure*}

% \begin{figure*}
% \begin{center}
 \includegraphics[width=165mm]{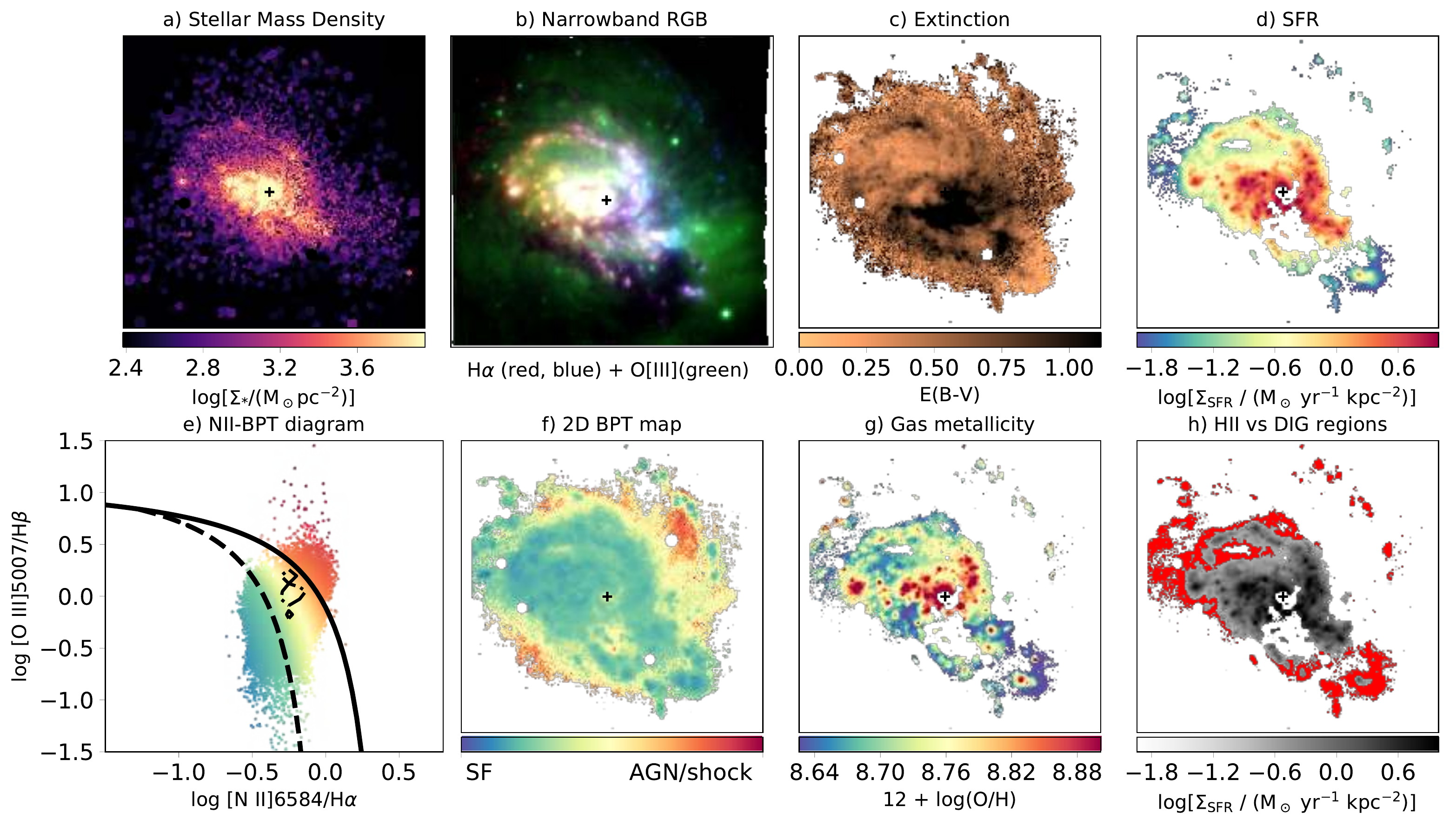}
\caption{NGC~3256 is a merger relic, with two clear components in extinction and gas metallicity maps. Two bright point sources can also be identified in the continuum (probably two distinct nuclei). The extinction map traces a substantial dust content in the southern part of the galaxy. The ionization in the outer parts are closer to AGN/shock than to the chaotic-central with star-forming ionization. Compared to the UV maps from GALEX, where there is not that much star formation, the SFR map from H$\alpha$ indicates very high values of SF, indicating that the SF is very recent (in the last tens of Myr). We see wings and broad lines in the diffuse ionized gas located in the northern part of the galaxy. This means that the gas velocity dispersions are very high, probably because of the high turbulence. There are probably double components in the lines, as they appear asymmetric in many of the central regions as well, indicating the possible presence of outflows. The metallicity gradient is flat due to the average of both the high and low metallicities from the two merging components. It is interesting to see in the metallicity plots of H{\sc ii} vs DIG (Fig.~\ref{oh12scatter_DIG}) that the metallicity gradient is negative for H{\sc ii} regions and positive for the DIG. The highest deviations from the average gradient are found in the H{\sc ii} regions in the centre ($\sim$0.2 dex) and outer DIG in the SW ($\sim$-0.2 dex).
}
\label{ngc3256plots}
\end{center}
\end{figure*}

\clearpage
%\subsection{NGC~4603}   

\begin{figure*}
\begin{center}
 \includegraphics[width=165mm]{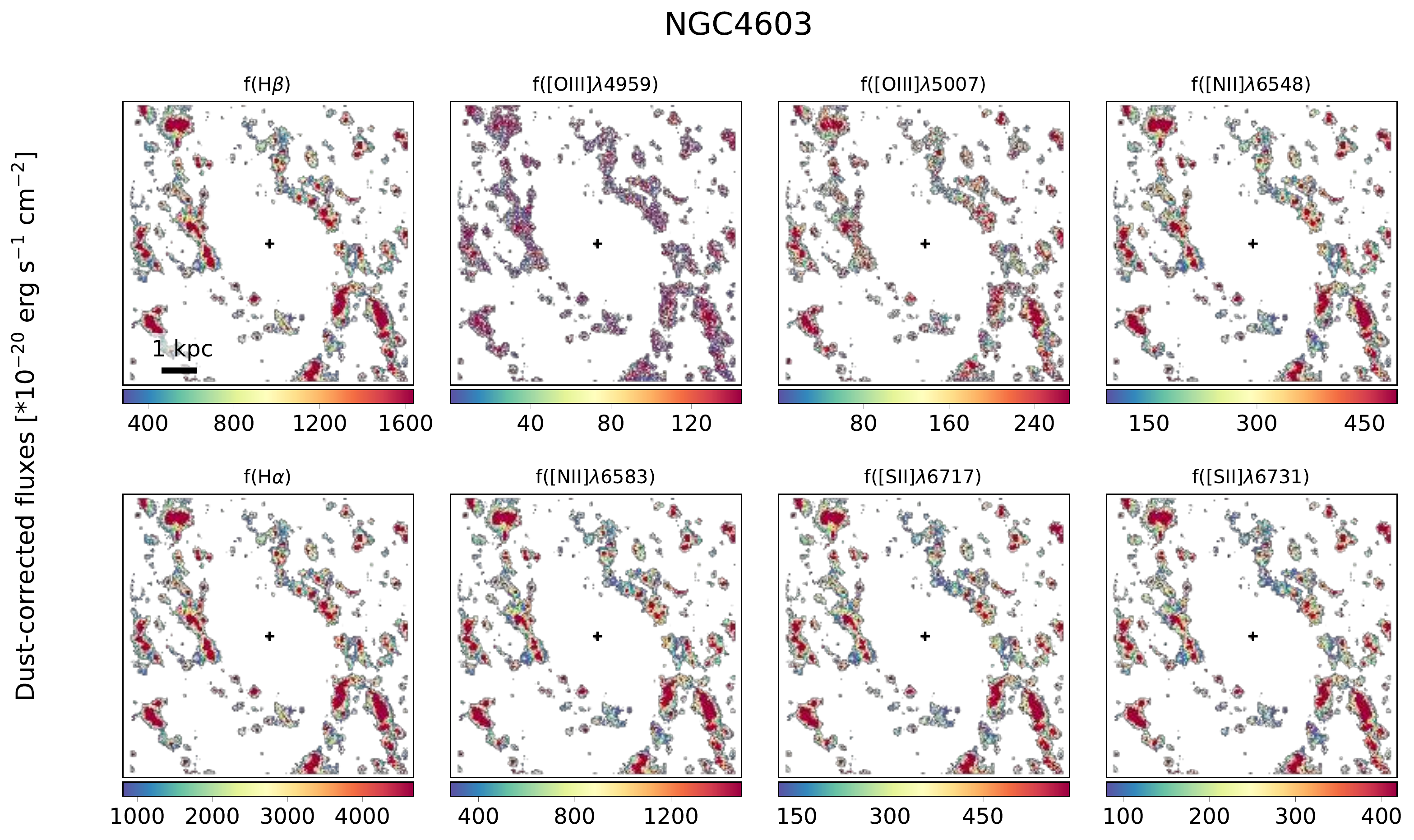}
% \caption{As Fig.~\ref{allfluxes} but for NGC~4603.}
% \label{ngc4603fluxes}
% \end{center}
% \end{figure}

% \begin{figure}
% \begin{center}
 \includegraphics[width=165mm]{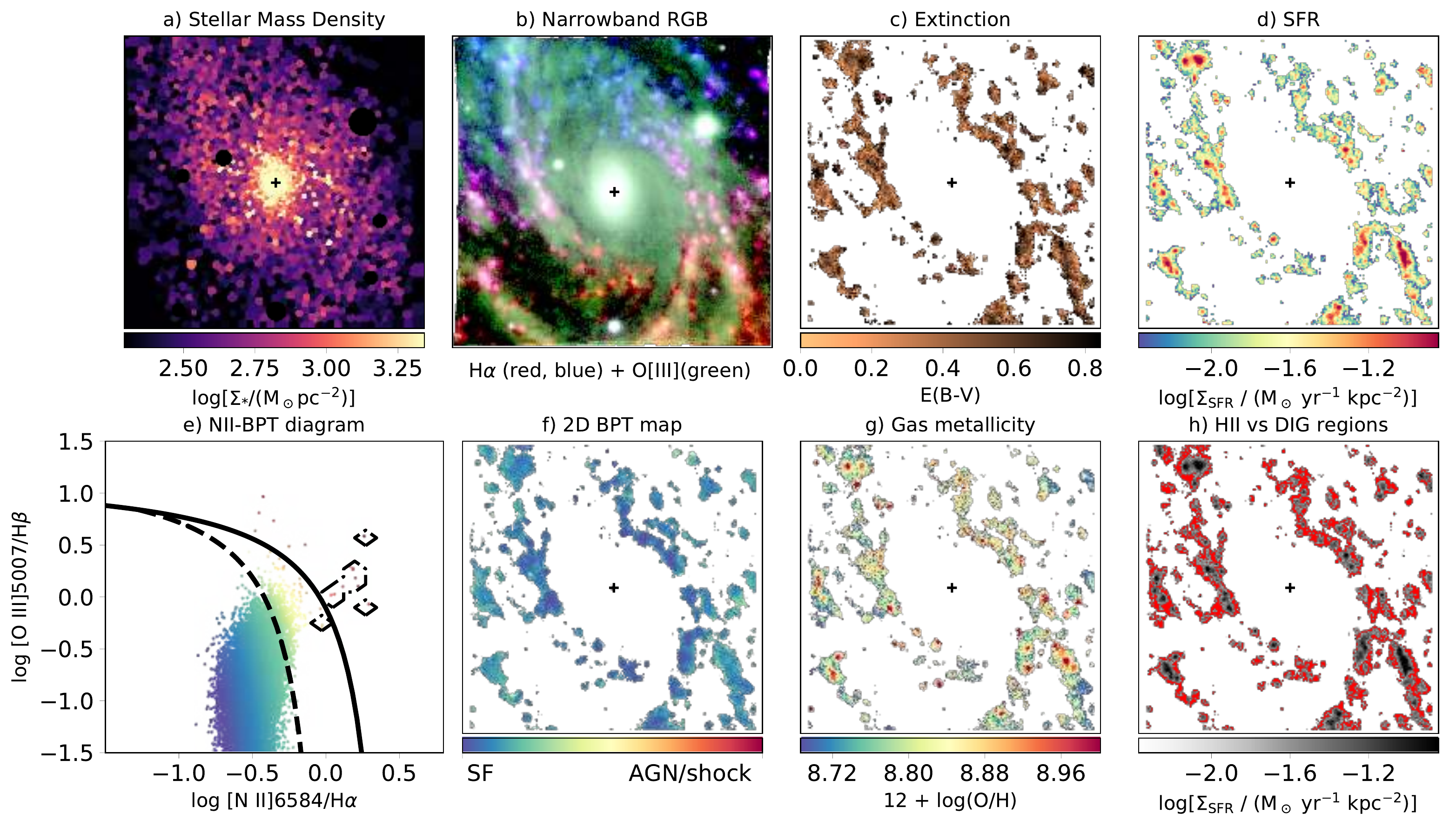}
\caption{This galaxy shows very low levels of star formation in its centre, without an apparent bar. This central part has AGN/shocked ionization, and presents a bulge-like stellar component without an ionized gas counterpart. Two arms can be identified as traced by the H{\sc ii} regions. The gas metallicity decreases throughout the inner disk, with very high values that stand out in the centre (deviating $\sim$0.2 dex from the linear fit), and also larger deviations for the H{\sc ii} regions in the outer parts.
}
\label{ngc4603plots}
\end{center}
\end{figure*}

\clearpage
%\subsection{NGC~3393}   
\begin{figure*}
\begin{center}
 \includegraphics[width=165mm]{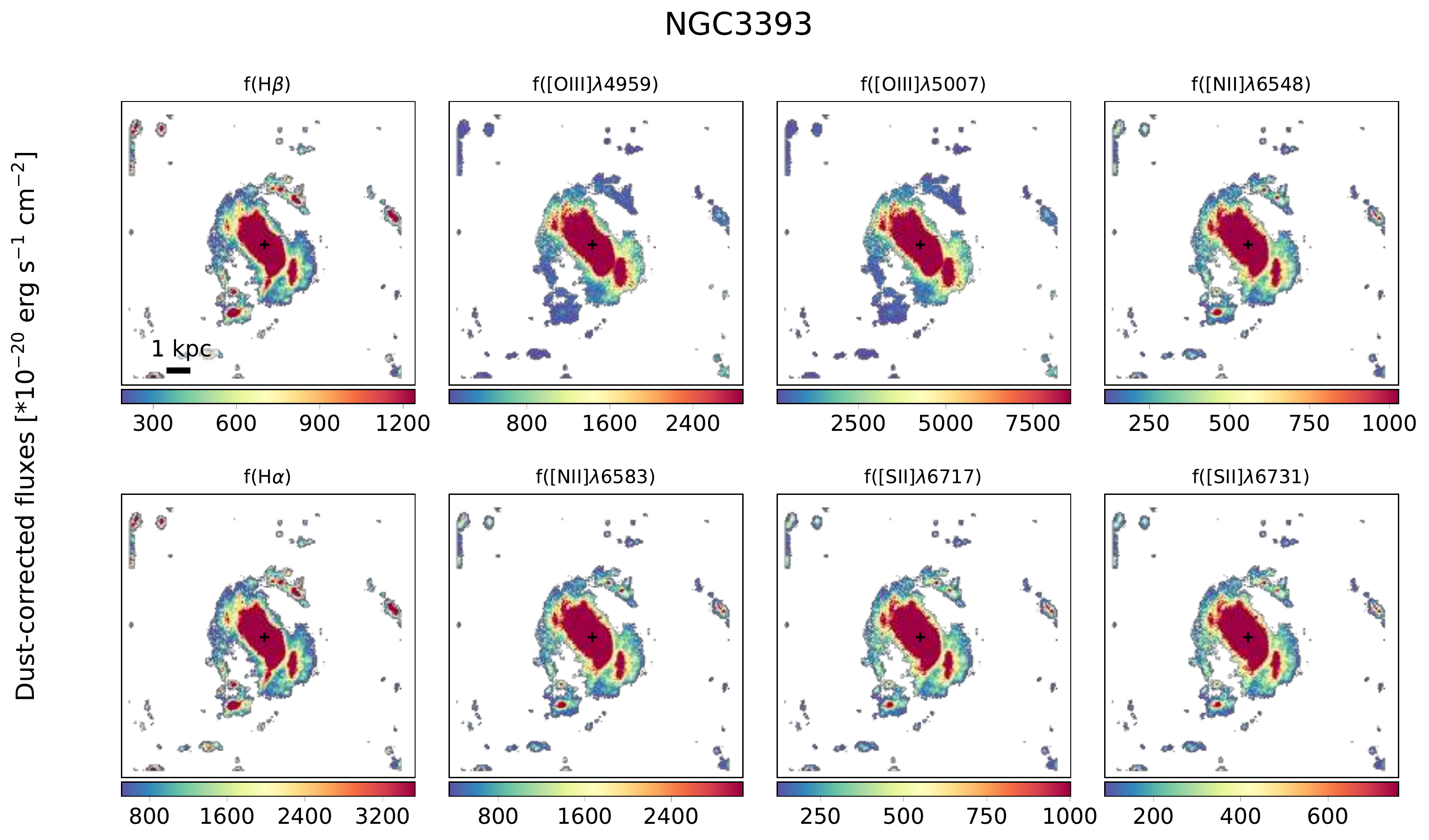}
% \caption{As Fig.~\ref{allfluxes} but for NGC~4030.}
% \label{ngc3393fluxes}
% \end{center}
% \end{figure*}

% \begin{figure*}
% \begin{center}
 \includegraphics[width=165mm]{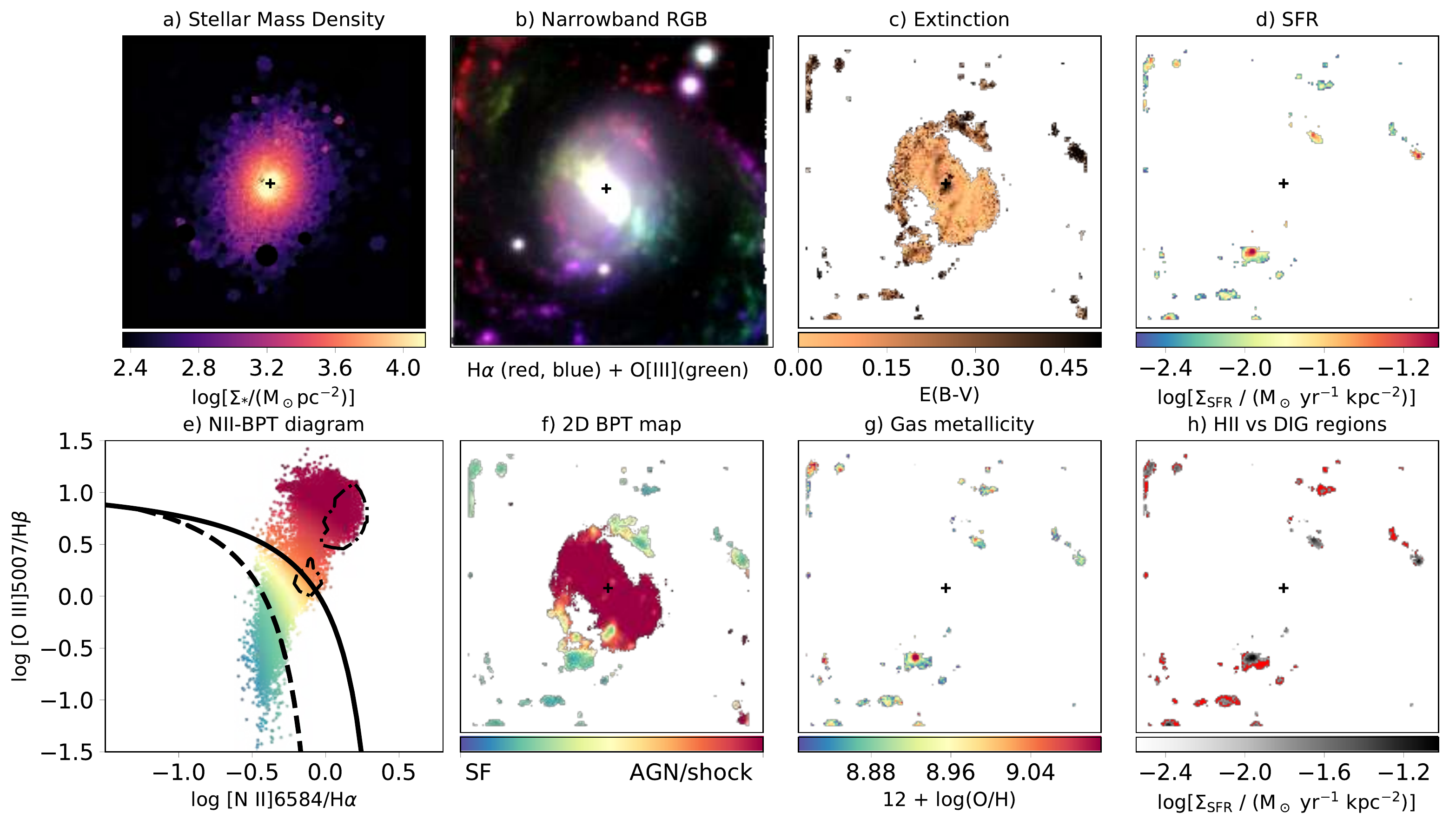}
\caption{NGC~3393 is a Sy2 galaxy and most of the ionization source is AGN/shock/post AGB stars from the BPT. Even with bright nuclear emission, it is possible to distinguish an inner spiral emission in all the lines inside the ionization cone (also seen in \citealt{Pogge1997} and \citealt{Cooke2000}), and translated into an inner spiral in the SFR map. There are not so many H{\sc ii} regions, being present only around two faint spiral arms near the edges of the CCD. 
}
\label{ngc3393plots}
\end{center}
\end{figure*}

\clearpage

%\subsection{NGC~1097}   

\begin{figure*}
\begin{center}
 \includegraphics[width=165mm]{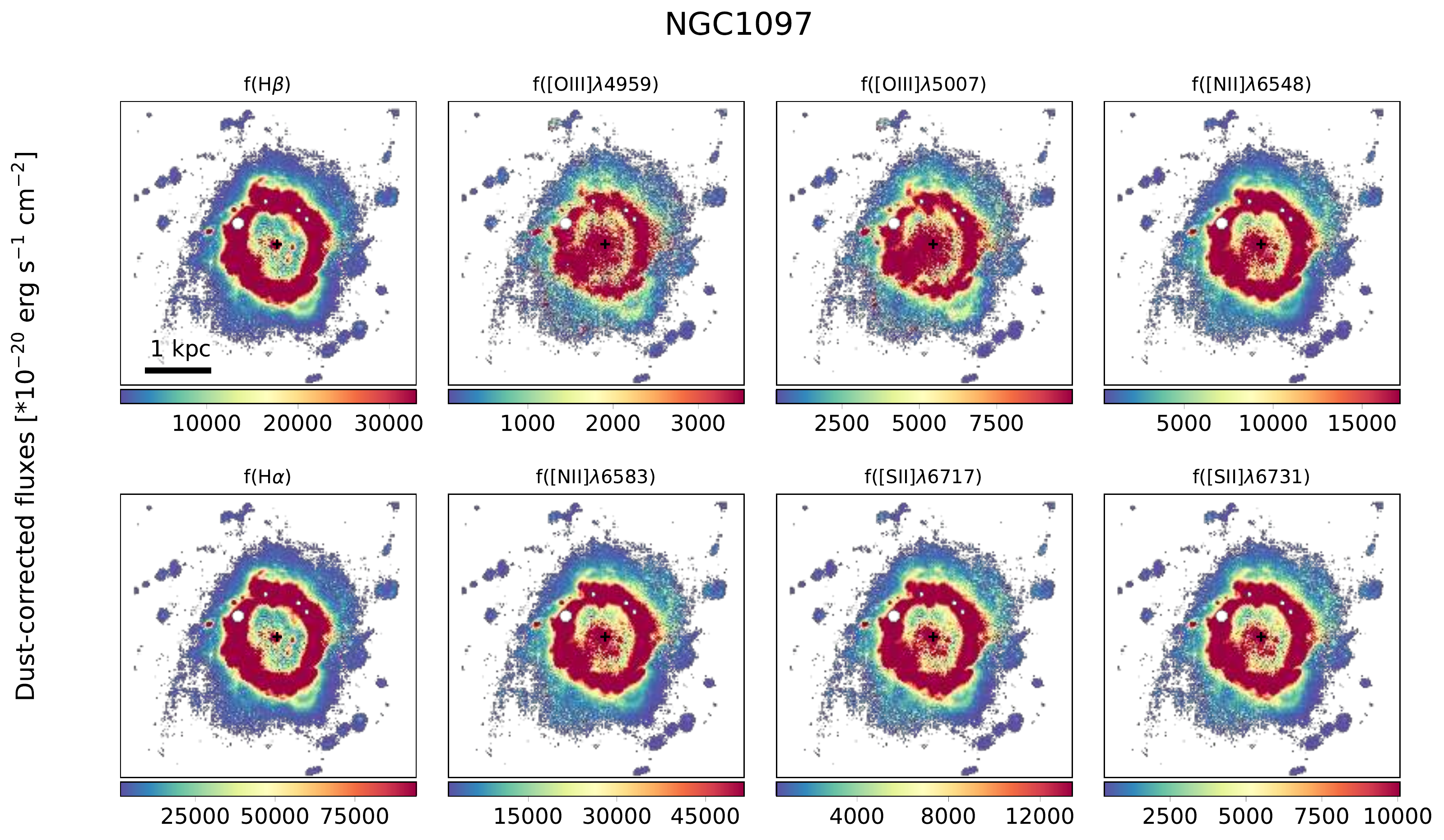}
% \caption{As Fig.~\ref{allfluxes} but for NGC~4030.}
% \label{ngc1097fluxes}
% \end{center}
% \end{figure*}

% \begin{figure*}
% \begin{center}
 \includegraphics[width=165mm]{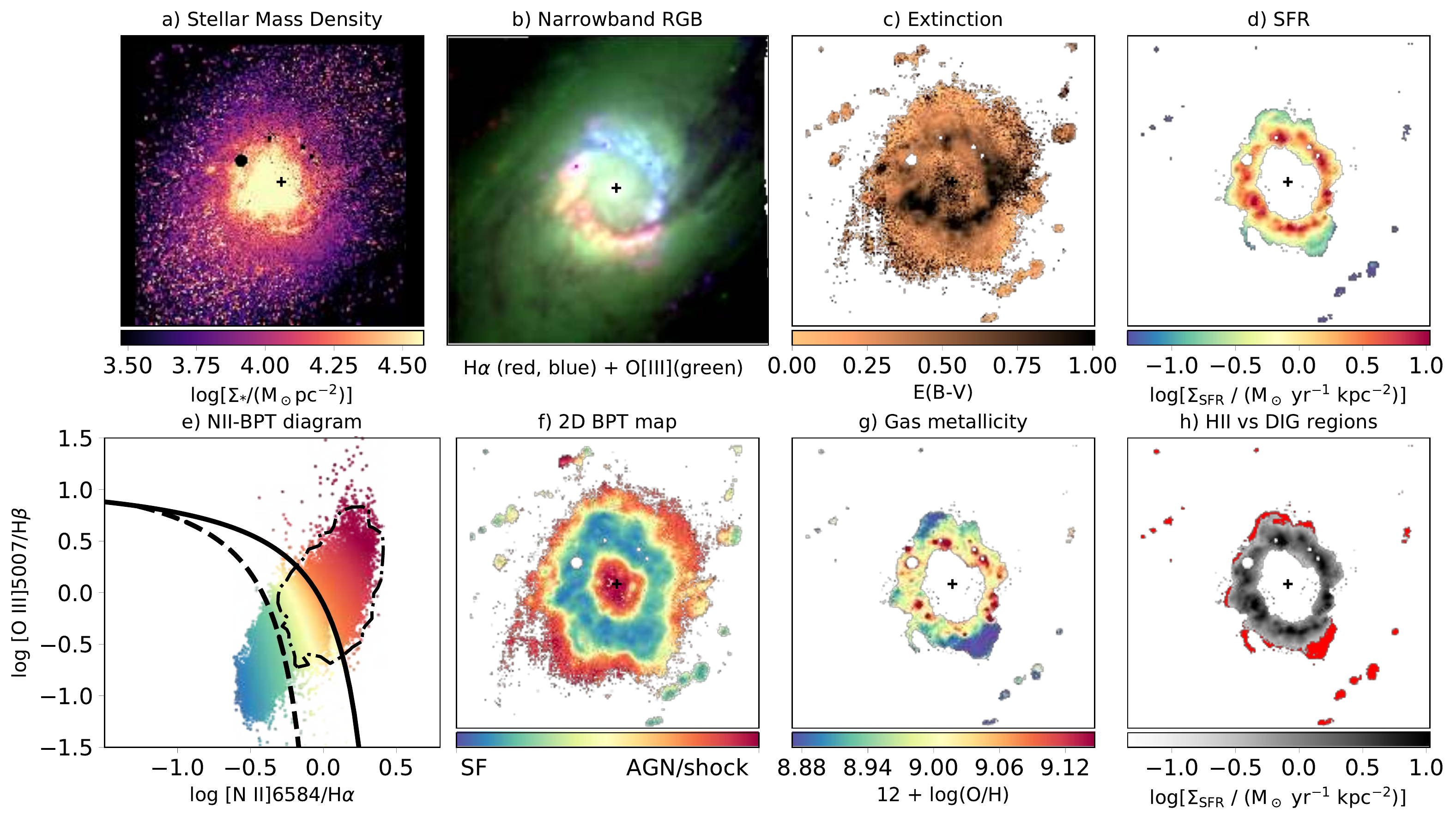}
\caption{MAD observations of NGC~1097 focus on the central part, where a nuclear star-forming ring is found. There is intense line emission in the H{\sc ii} regions of this ring, and small gas clouds outside. However, most of the star formation is located in the ring. There is a strong dust lane following the bar, starting at the end of the bulge, clearly seen in the extinction map.
}
\label{ngc1097plots}
\end{center}
\end{figure*}
\clearpage
%\subsection{NGC~289 }   
\begin{figure*}
\begin{center}
 \includegraphics[width=165mm]{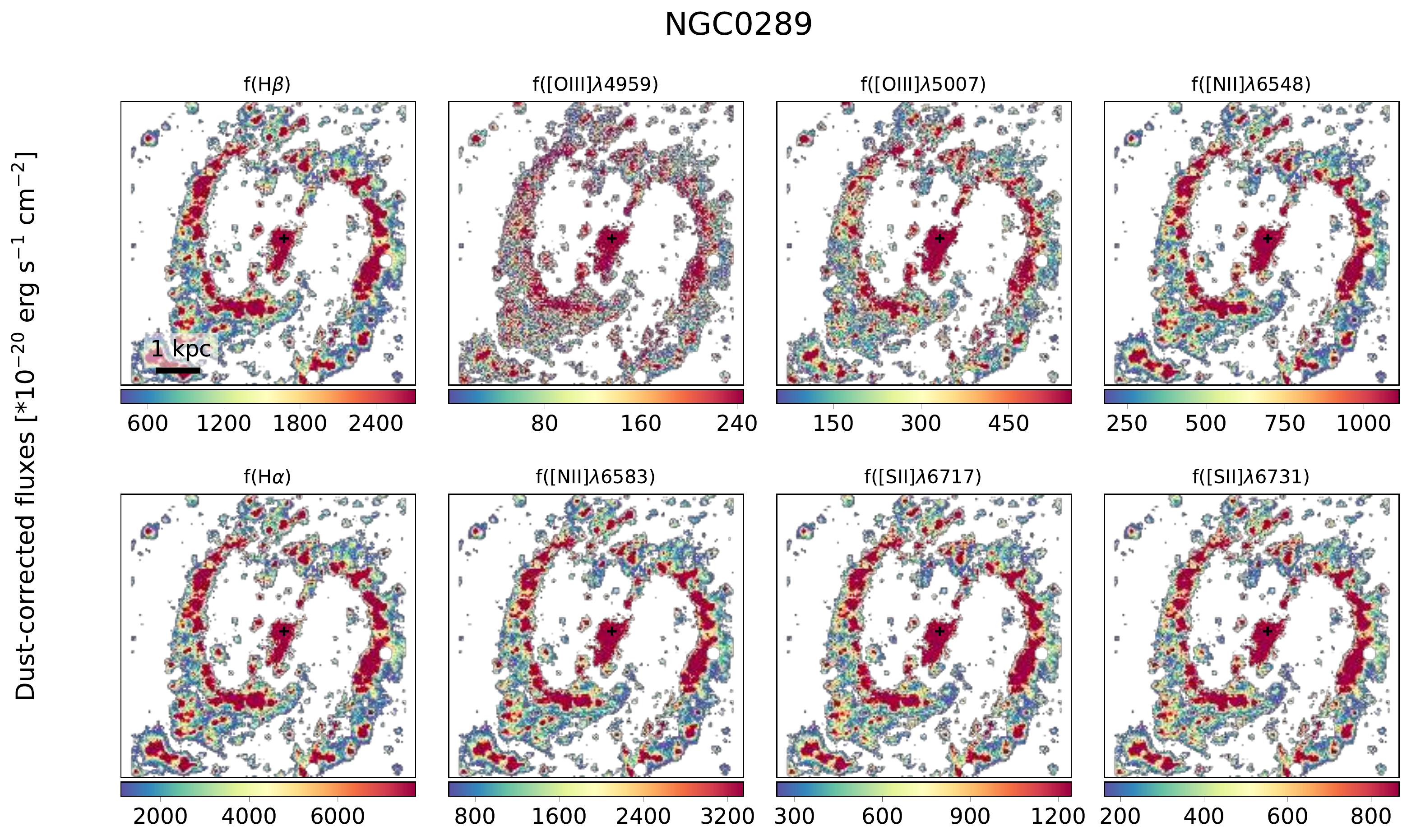}
% \caption{As Fig.~\ref{allfluxes} but for NGC~4030.}
% \label{ngc0289fluxes}
% \end{center}
% \end{figure*}

% \begin{figure*}
% \begin{center}
 \includegraphics[width=165mm]{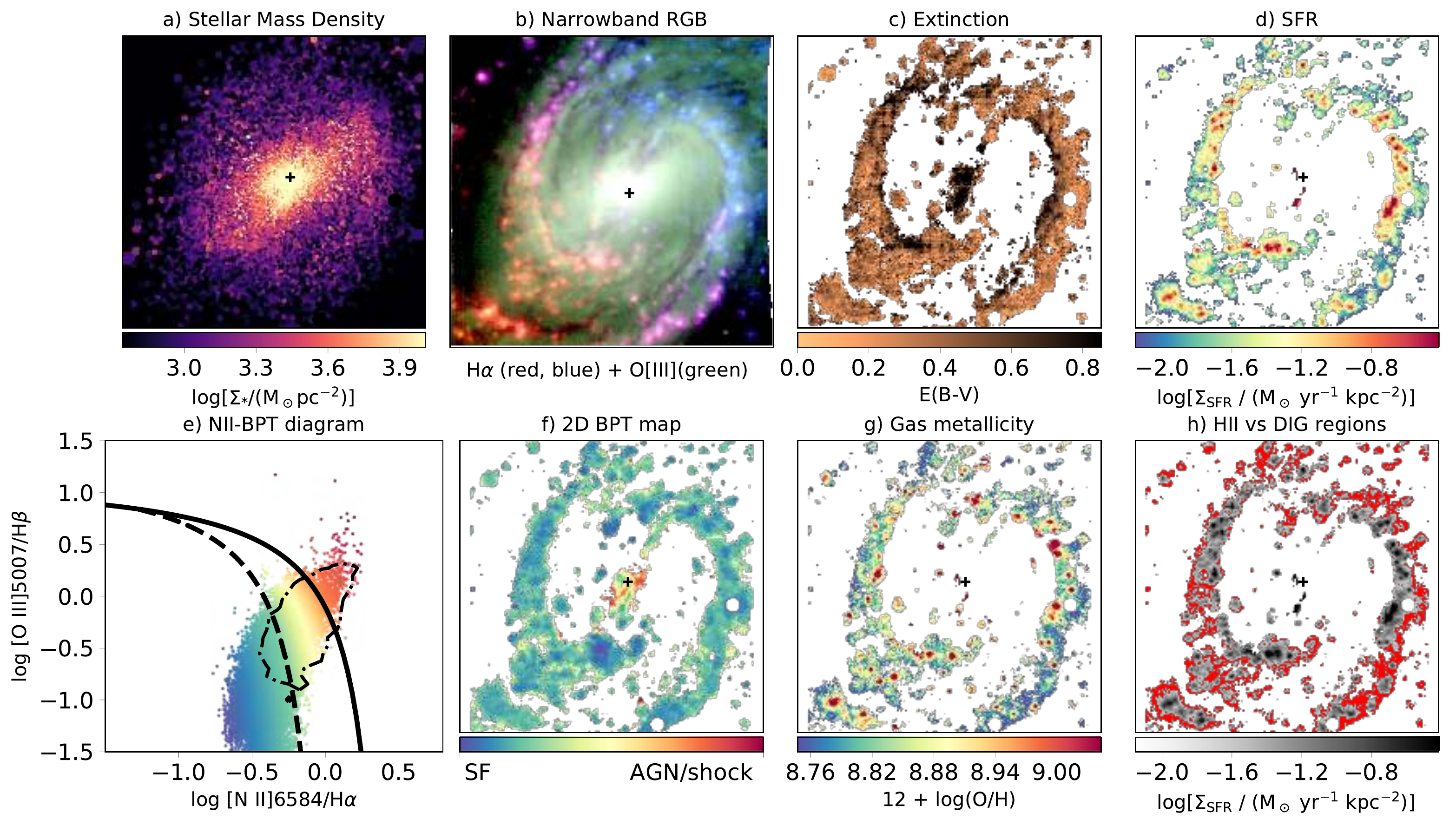}
\caption{The stellar mass density plot shows the bar region. The inner structure of this galaxy is characterized by the two well-defined spiral arms that finish at the centre of the galaxy (in contrast with other barred galaxies such as NGC~864, where the spiral arms start at the ends of the bar, see \citealt{Erroz-Ferrer2012}). The inner spiral structure is traced by H{\sc ii} regions and dusty regions. The bar shows very low levels of star formation in the form of diffuse ionized gas, with ionization in the AGN/shocked area of the BPT (probably caused by the shocks due to the potential of the bar). This barred region presents higher metallicities than the two spiral arms, with values deviating from the linear fit $\sim$0.3 dex. The H{\sc ii} regions in the central parts of the inner spiral do not, however, deviate from the linear fit.
}
\label{ngc0289plots}
\end{center}
\end{figure*}

\clearpage
%\subsection{NGC~4593}   
\begin{figure*}
\begin{center}
 \includegraphics[width=165mm]{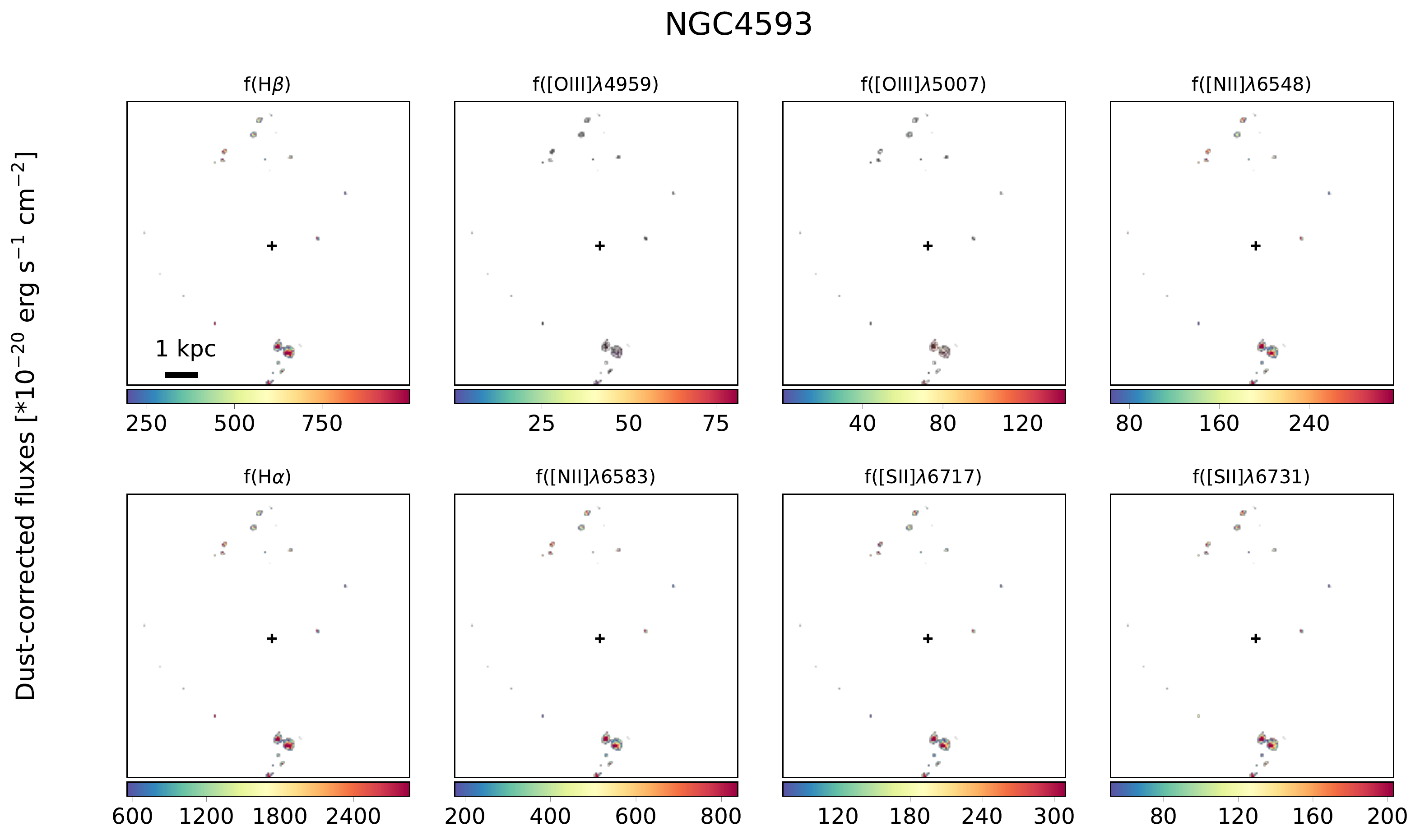}
% \caption{As Fig.~\ref{allfluxes} but for NGC~4030.}
% \label{ngc4593fluxes}
% \end{center}
% \end{figure*}

% \begin{figure*}
% \begin{center}
 \includegraphics[width=165mm]{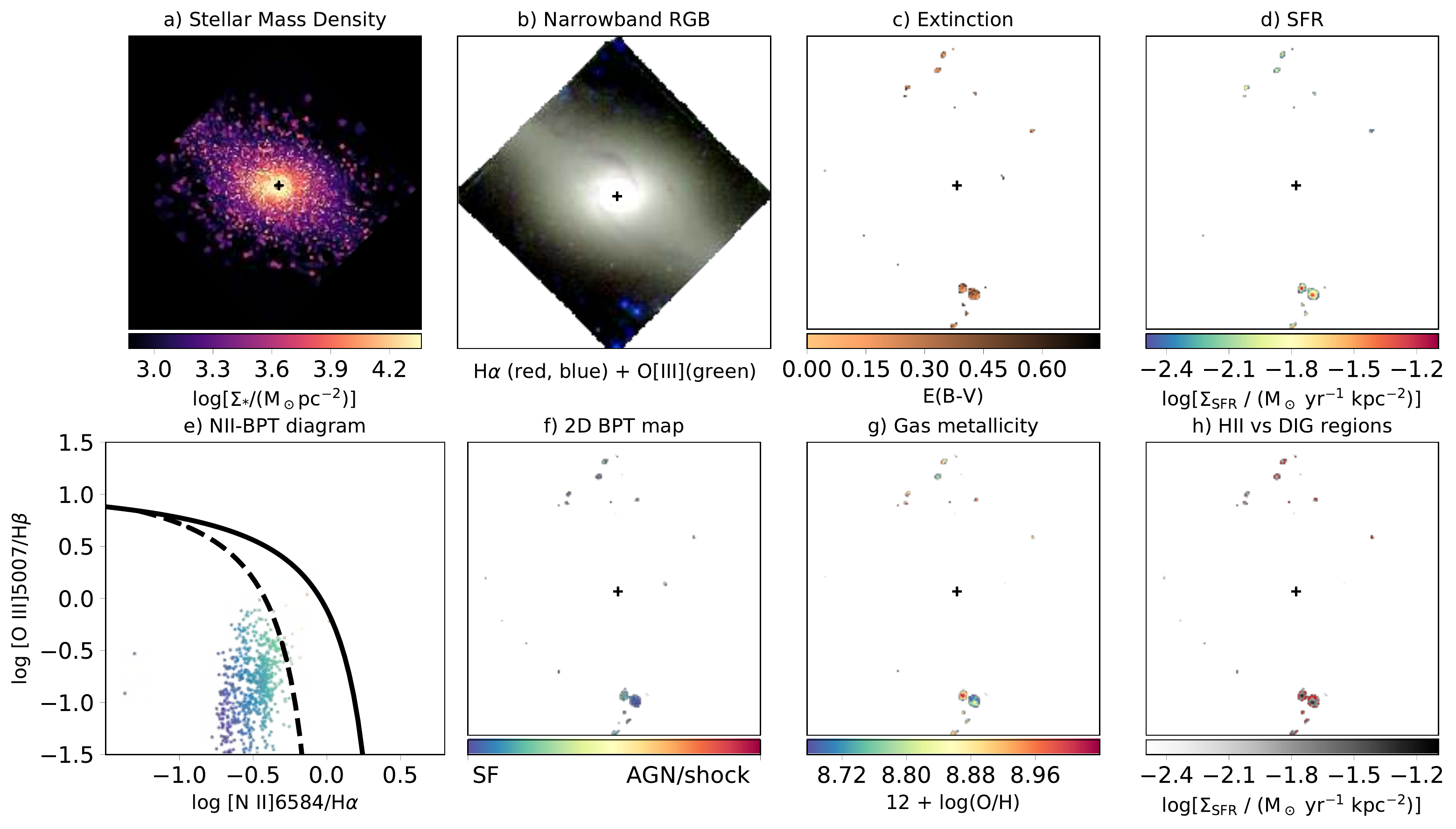}
\caption{NGC~4593 is a Sy1 galaxy, showing very broad Balmer lines in the central part (masked due to the failure in the fitting analysis). NGC~4593 has a large scale bar which does not show SF. The diagnostic maps for this galaxy are mostly empty due to the lack of ionized gas in the bar region. There are a few H{\sc ii} regions inside the MUSE FoV.}
\label{ngc4593plots}
\end{center}
\end{figure*}

\clearpage
%\subsection{IC~2560 }   
\begin{figure*}
\begin{center}
 \includegraphics[width=165mm]{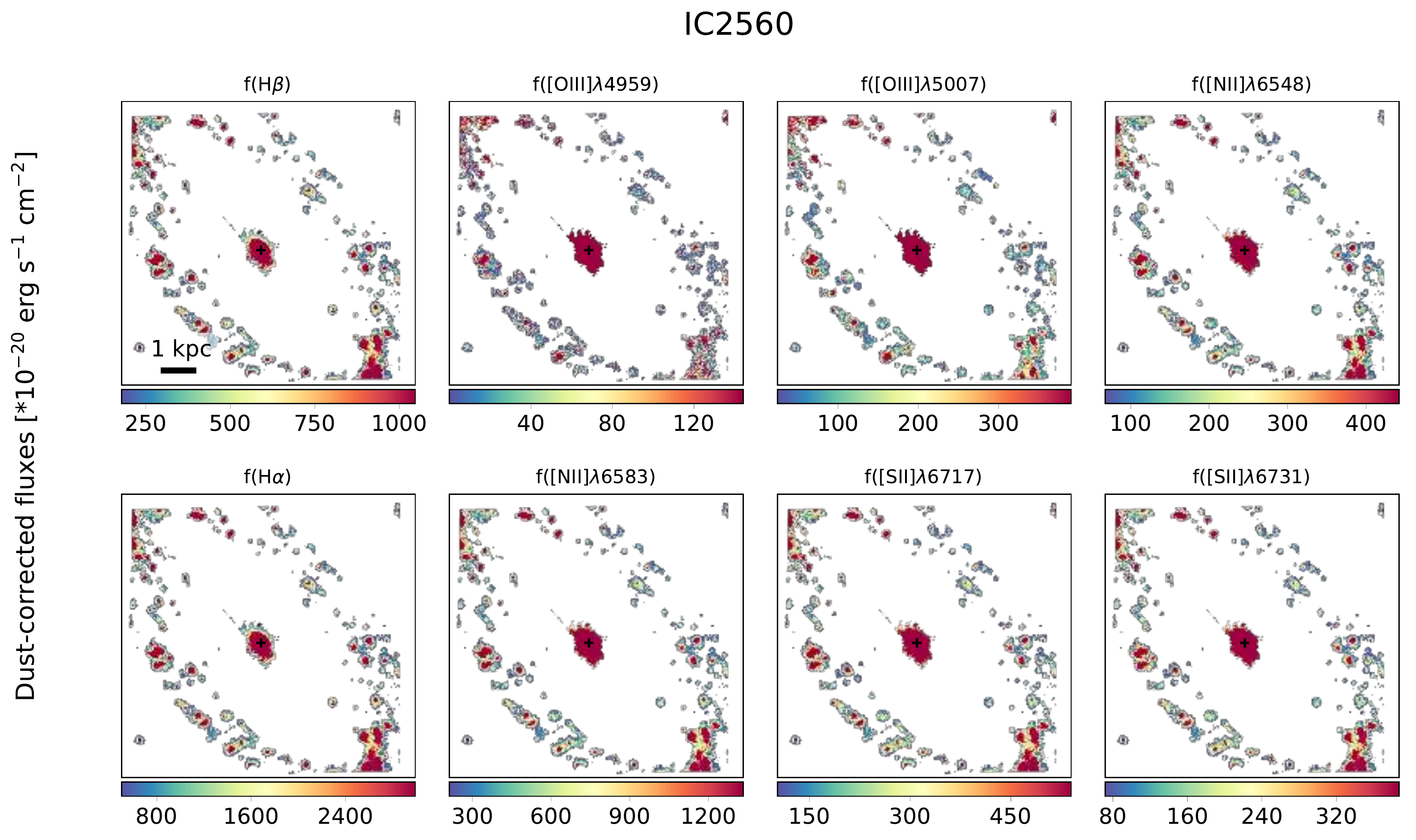}
% \caption{As Fig.~\ref{allfluxes} but for NGC~4030.}
% \label{ic2560fluxes}
% \end{center}
% \end{figure*}

% \begin{figure*}
% \begin{center}
 \includegraphics[width=165mm]{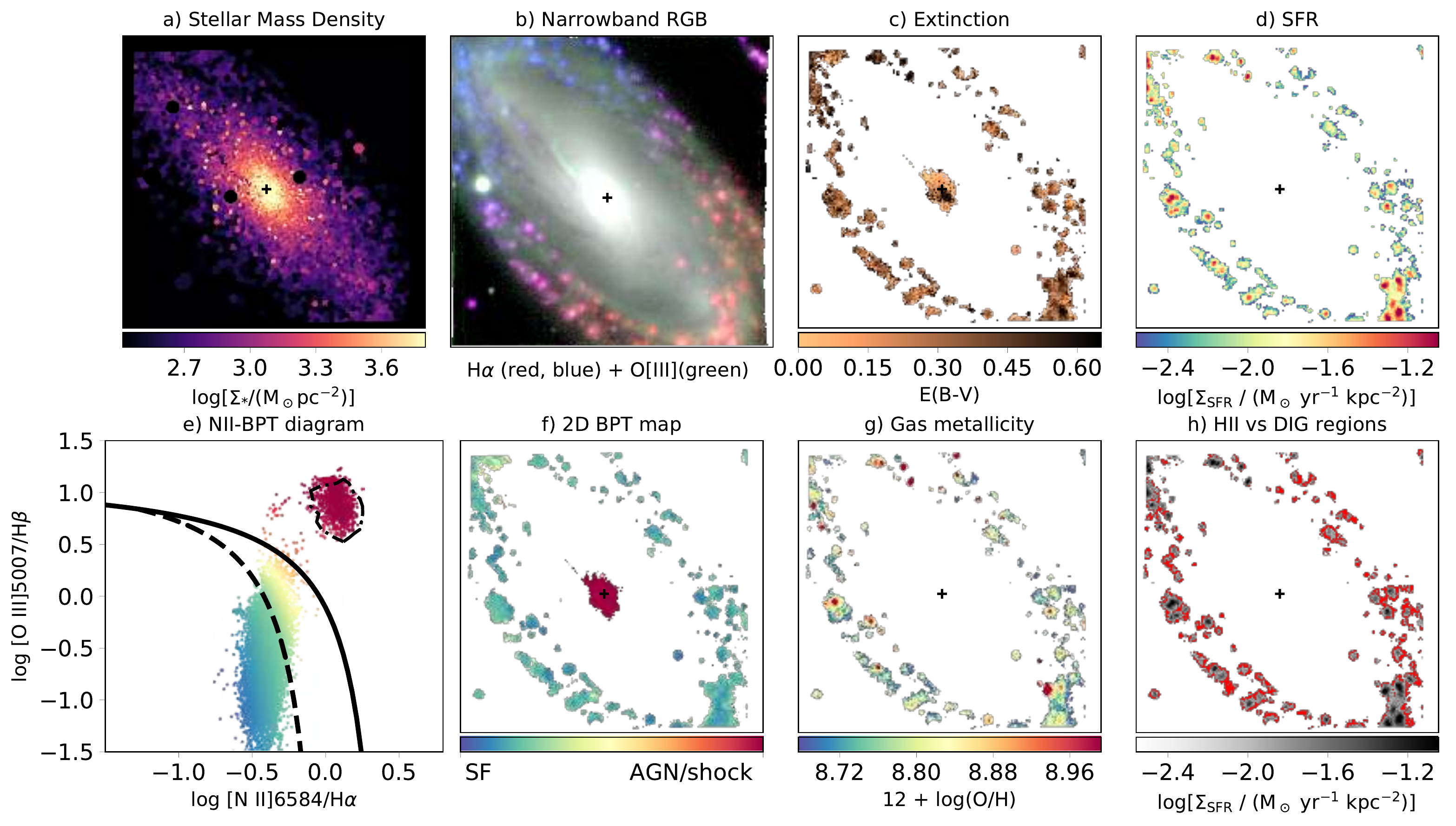}
\caption{This barred galaxy is classified as Sy2 \citep{VeronCetty2006}, with prominent [O{\sc iii}] emission in the central parts. The bar region is mostly devoid of star formation, except in the dust lanes that form a spiral to the centre. The bar region shows AGN/shock ionization, as expected. The two spiral arms close into a ring structure, where most of the star formation is localized. The largest deviations in gas metallicity from the linear fit are found on the major axis of the bar (with values $\sim$0.3 dex deviating from the linear fit) and the centres of the H{\sc ii} regions inside the ring (with values $\sim$0.3 dex deviating from the linear fit).
}

\label{ic2560plots}
\end{center}
\end{figure*}

\clearpage
%\subsection{NGC~5643}   
\begin{figure*}
\begin{center}
 \includegraphics[width=165mm]{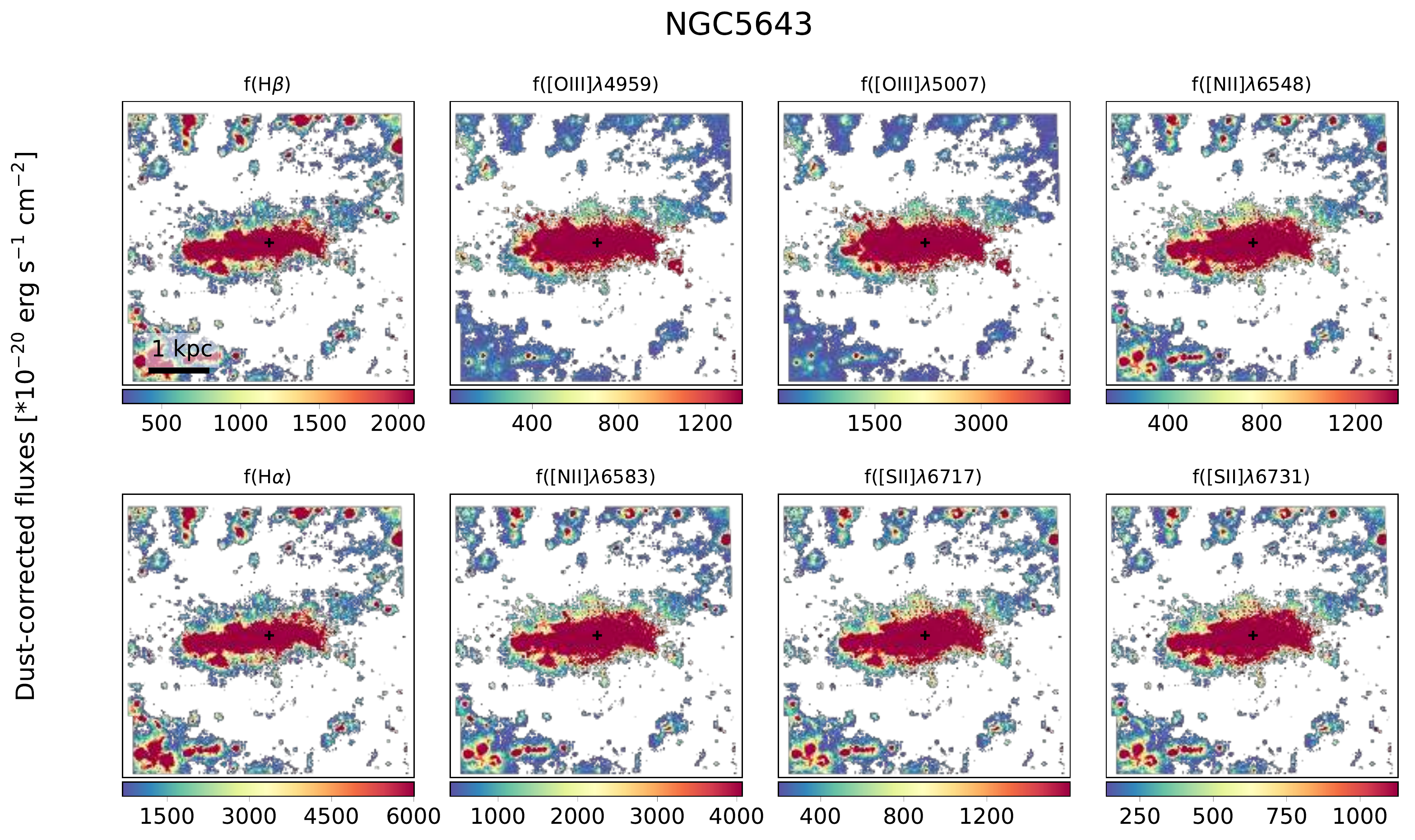}
% \caption{As Fig.~\ref{allfluxes} but for NGC~4030.}
% \label{ngc5643fluxes}
% \end{center}
% \end{figure*}

% \begin{figure*}
% \begin{center}
 \includegraphics[width=165mm]{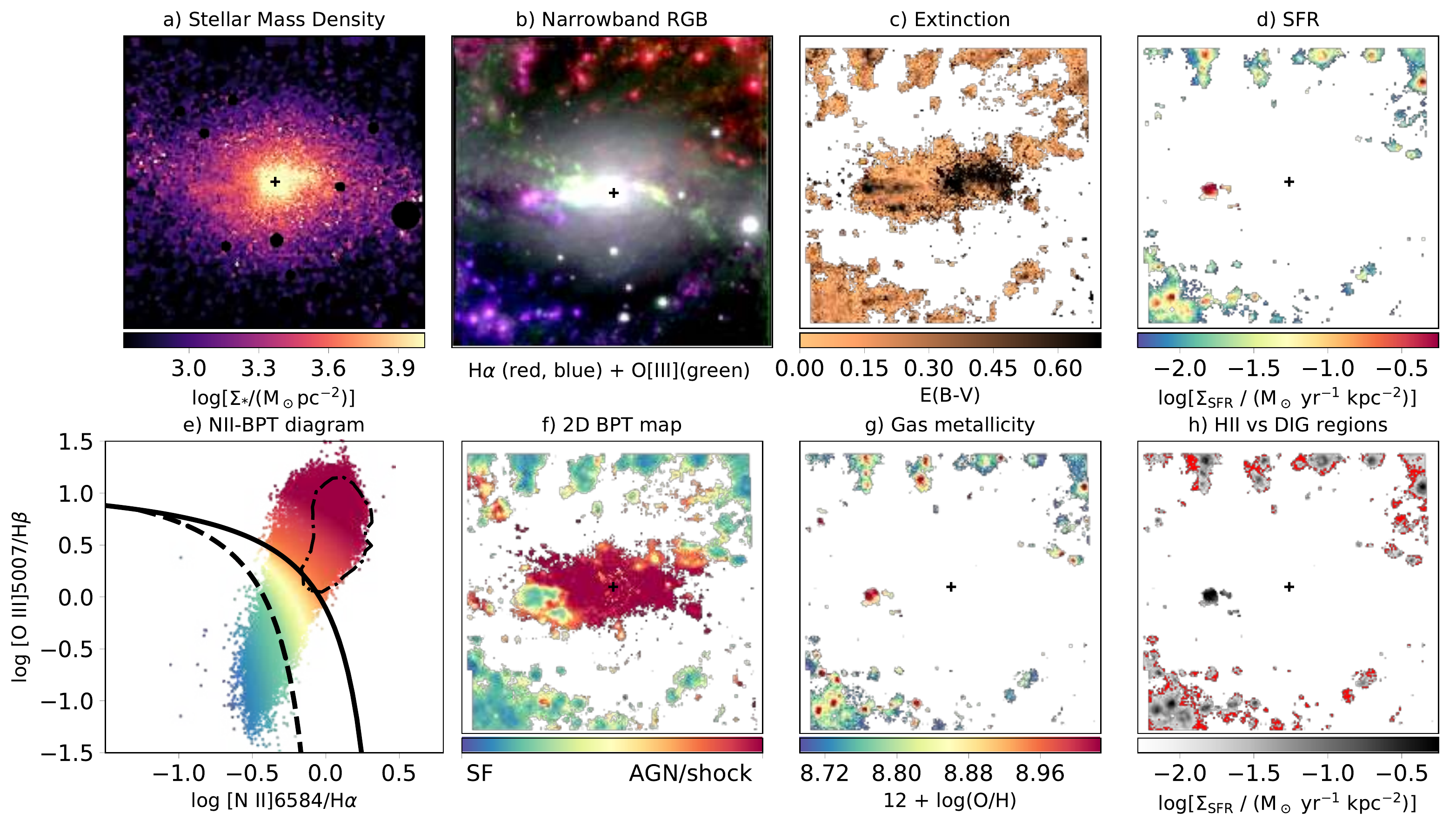}
\caption{The first MUSE observations of the galaxy were obtained in science verification mode, and the first results were presented in \citet{Cresci2015}. This galaxy was re-observed by our team with longer exposure times and therefore deeper observations. This barred galaxy shows a well-defined [O{\sc iii}] structure outflowing from the central part, where a Sy2 is located. The bar is prominent in H$\alpha$ emission, although this emission may not be coming from H{\sc ii} regions. There are, however, well identified H{\sc ii} regions in the bar that are in the ionization cone of the AGN, and claimed to be due to positive feedback in  \citet{Cresci2015}. Apart from this, the ionization of most of the central part is due to AGN/shocks. The bar region shows some H{\sc ii} regions, and the multi-spiral pattern starts at the edges of the MUSE FoV. The metallicity is higher in the H{\sc ii} regions in the central part of the bar and where the outflow is found. Then, the metallicity decreases with radius for both H{\sc ii} and DIG.
}
\label{ngc5643plots}
\end{center}
\end{figure*}

\clearpage
%\subsection{NGC~3081}   
\begin{figure*}
\begin{center}
 \includegraphics[width=165mm]{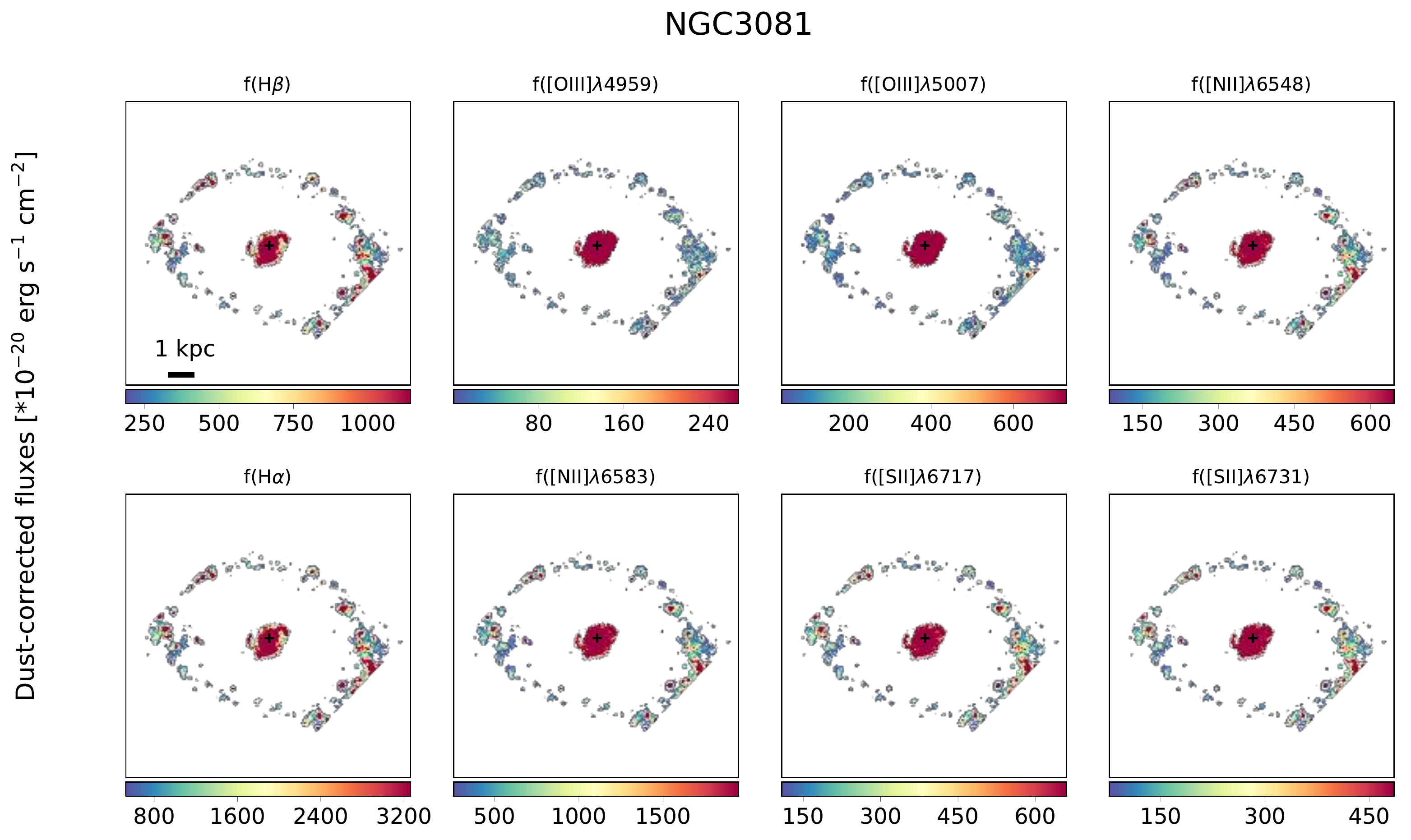}
% \caption{As Fig.~\ref{allfluxes} but for NGC~4030.}
% \label{ngc3081fluxes}
% \end{center}
% \end{figure*}

% \begin{figure*}
% \begin{center}
 \includegraphics[width=165mm]{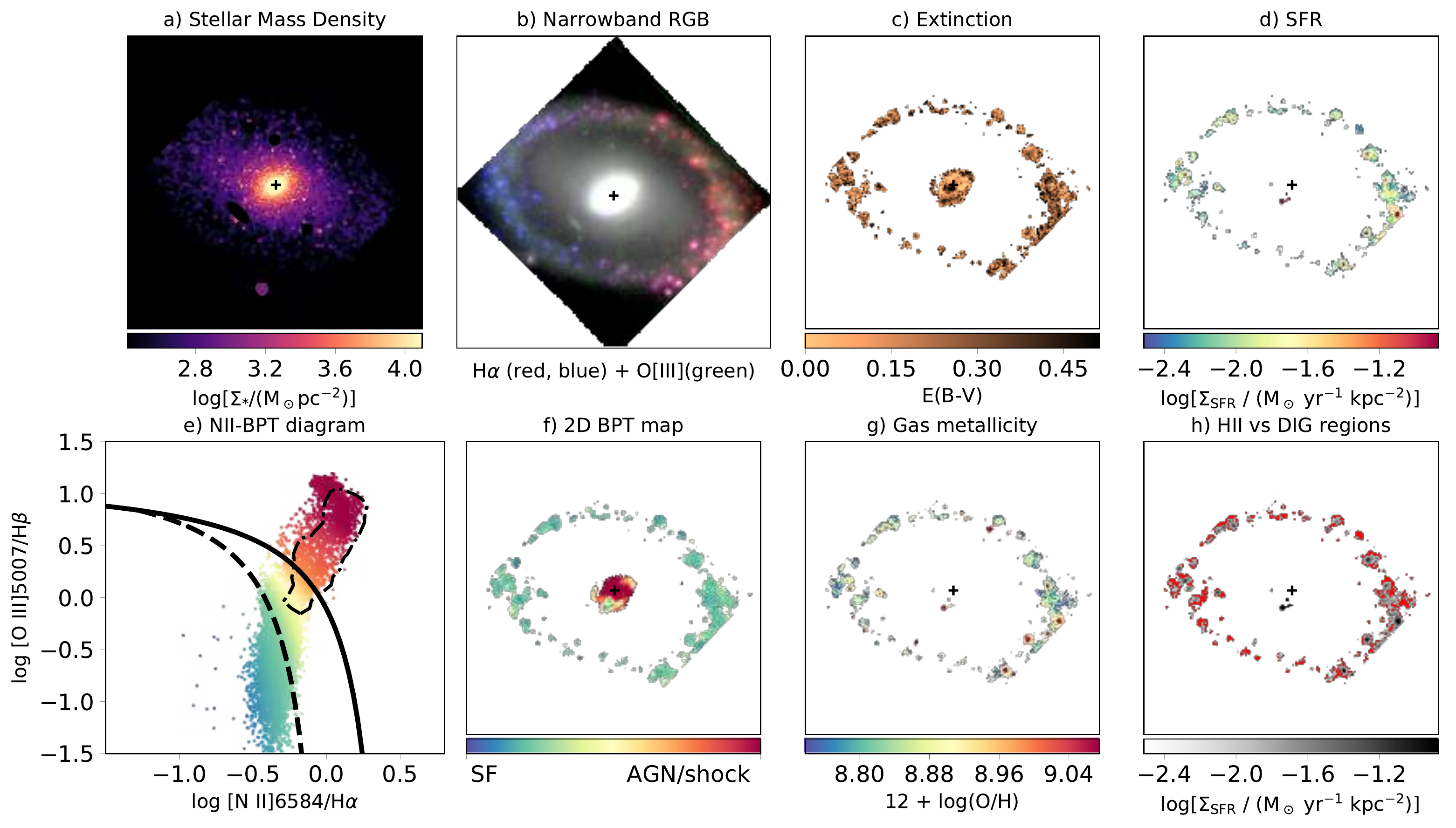}
\caption{This Sy2 galaxy presents a very bright nuclear part with many emission lines throughout the spectra, but neither the Balmer lines nor the He lines are very broad. The central emission is very interesting, as the emission lines form a inner spiral structure, whereas the extinction map shows a dust ring that is not aligned with the outer ring. This galaxy has two bars (e.g., \citealt{Buta1998}), and the star formation is very low in the large-scale one. The emission line ratios are mostly those for the intermediate and AGN/shock regions of the BPT diagram, except for the H{\sc ii} regions in the nuclear and outer rings. The metallicity decreases with radius, where most of the deviation from the linear fit are found in the centres of the H{\sc ii} regions of the outer ring ($\sim$0.3 dex) as the metallicity of the H{\sc ii} regions is similar at all radii.
}

\label{ngc3081plots}
\end{center}
\end{figure*}

\clearpage
%\subsection{NGC~4941}   
\begin{figure*}
\begin{center}
 \includegraphics[width=165mm]{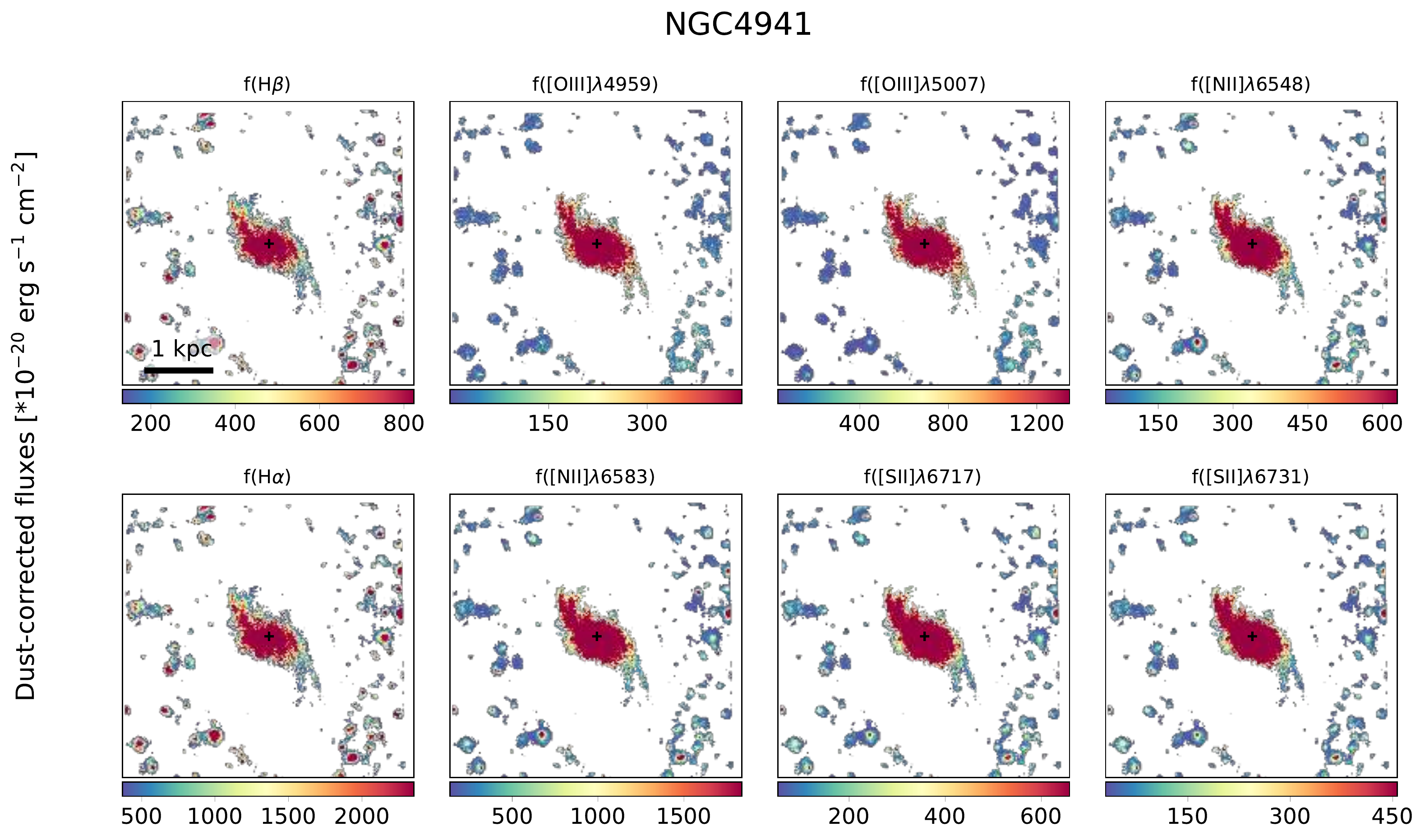}
% \caption{As Fig.~\ref{allfluxes} but for NGC~4030.}
% \label{ngc4941fluxes}
% \end{center}
% \end{figure*}

% \begin{figure*}
% \begin{center}
 \includegraphics[width=165mm]{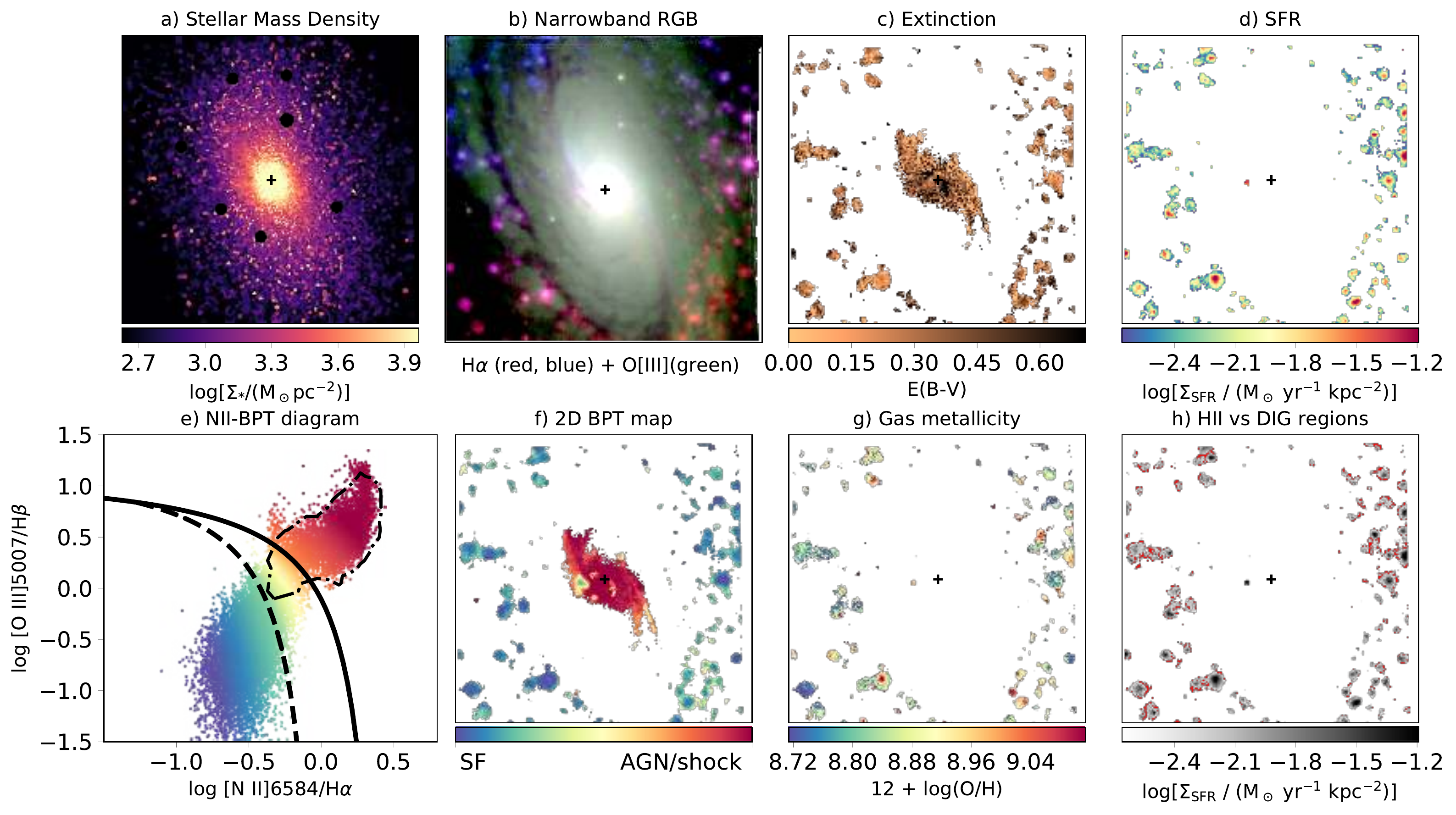}
\caption{This unbarred galaxy presents an inner ring without a clear outer spiral structure. The central two kpc are dominated by AGN/shocked emission, and exhibit an inner spiral structure. The stellar mass content of the galaxy is misaligned with the gaseous properties: the bulk of the stellar mass does not show a bar-like structure, whereas the BPT diagram shows a boxy structure at a position angle (PA) $\sim45\degr$ (from North to East), ending in two spirals that follow the dust lanes. There are also higher values of SF with higher values of the metallicity in the H{\sc ii} regions in the centre and following the dust lanes. The H{\sc ii} regions are mostly located in the ring, and between this ring and the central emission the levels of SF are very low. The largest deviations in metallicity are found in the central high metallicities ($\sim$0.2 dex), not in the H{\sc ii} regions in the outer parts.
}

\label{ngc4941plots}
\end{center}
\end{figure*}

\clearpage
%\subsection{NGC~5806}   

\begin{figure*}
\begin{center}
 \includegraphics[width=165mm]{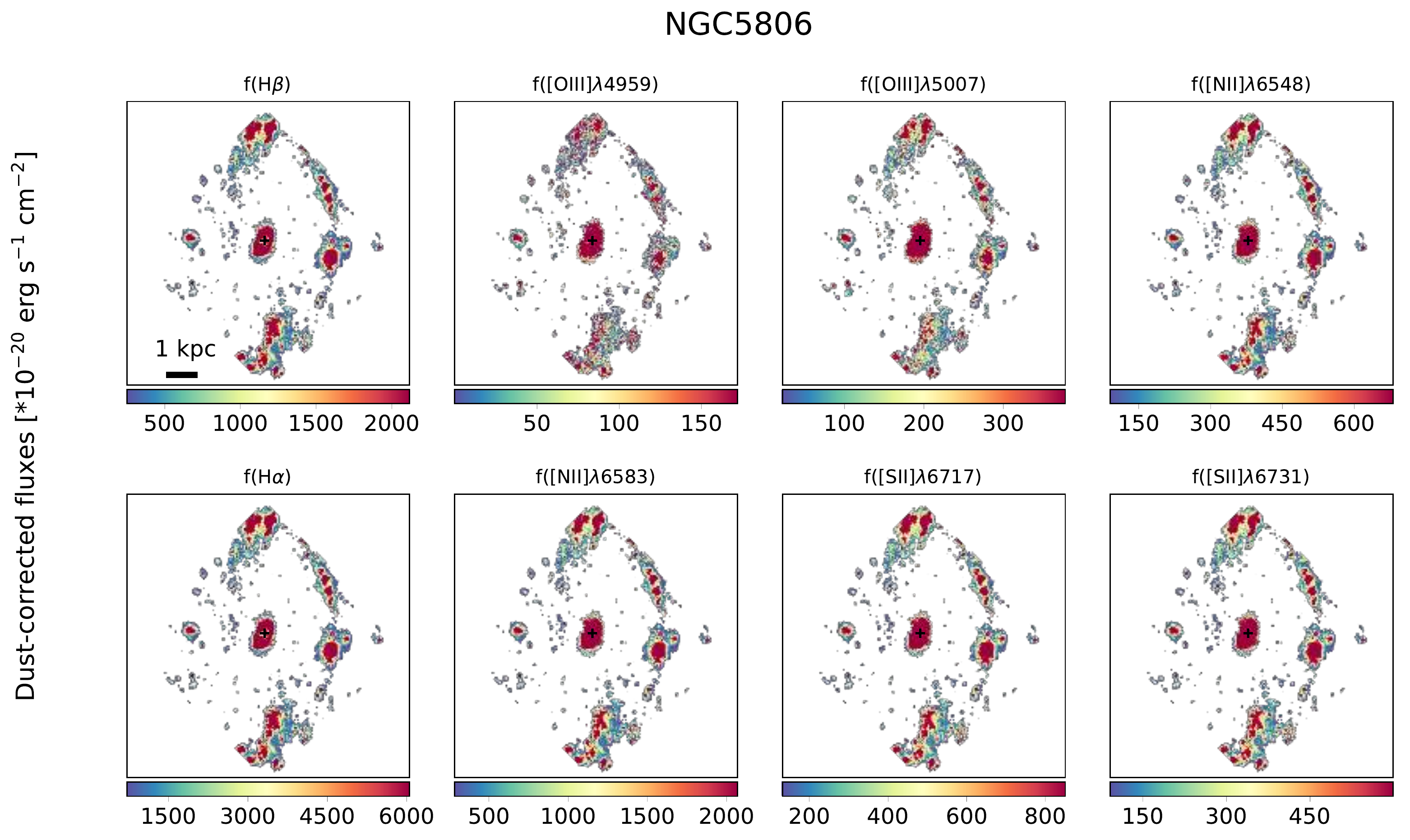}
% \caption{As Fig.~\ref{allfluxes} but for NGC~4030.}
% \label{ngc5806fluxes}
% \end{center}
% \end{figure*}

% \begin{figure*}
% \begin{center}
 \includegraphics[width=165mm]{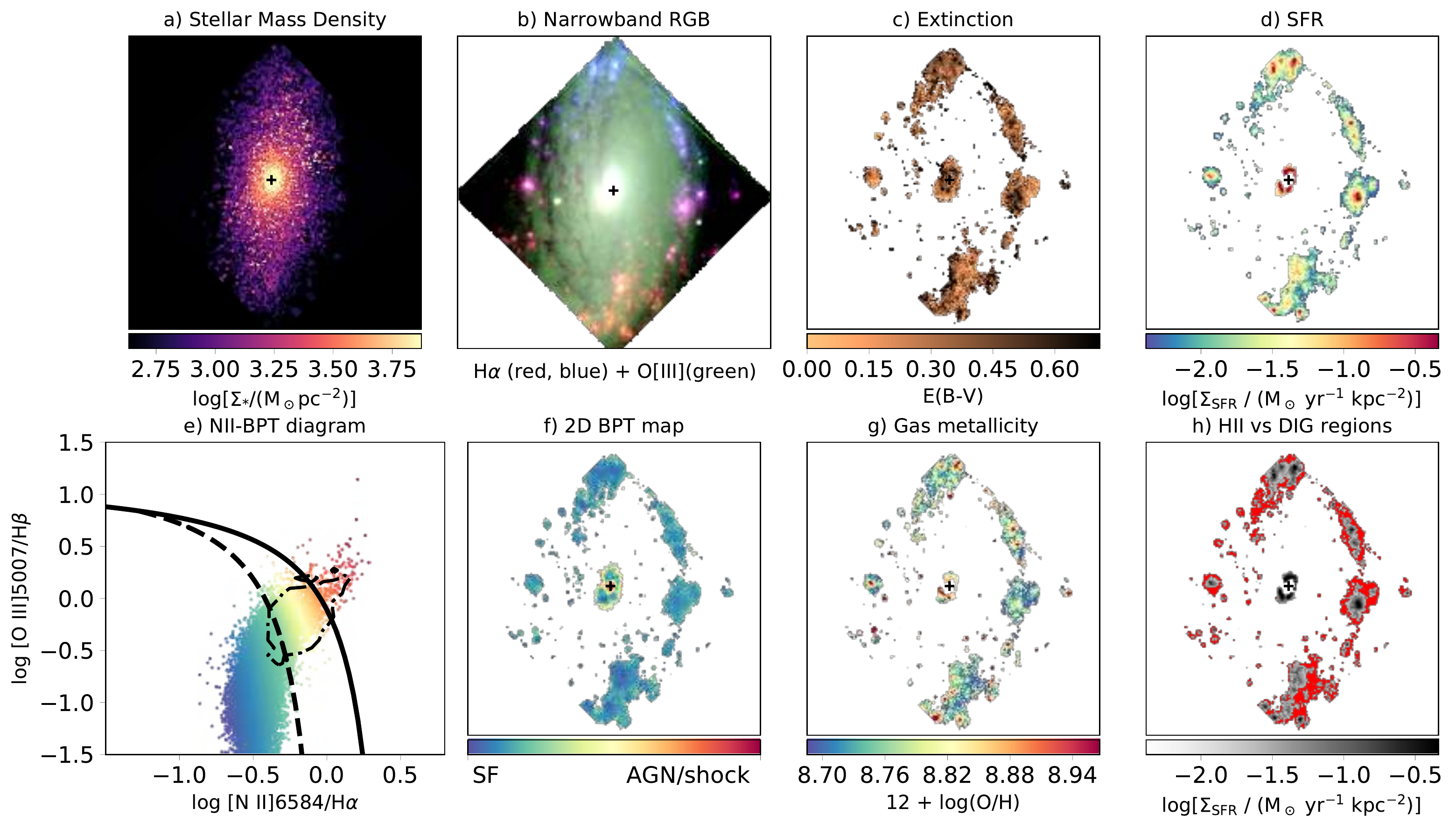}
\caption{This barred galaxy shows low levels of star formation inside the bar, where AGN/shock ionization is present. There is, however, a very prominent star formation ring, as seen in the SFR map. This star forming ring has lower metallicities and lower stellar velocity dispersion (MAD3) than its surroundings. The spiral arms form an outer ring with higher levels of star formation. Although it is classified as Sy2, we do not find broad lines in the centre. However, the emission line ratios do classify the central spaxels in the AGN/shock region of the BPT. Although the metallicity decreases with radius, there a obvious deviations from the linear fit in the nuclear ring ($\sim$-0.10 dex) and central DIG ($\sim$0.2 dex).
}

\label{ngc5806plots}
\end{center}
\end{figure*}

\clearpage

%\subsection{NGC~3783}   

\begin{figure*}
\begin{center}
 \includegraphics[width=165mm]{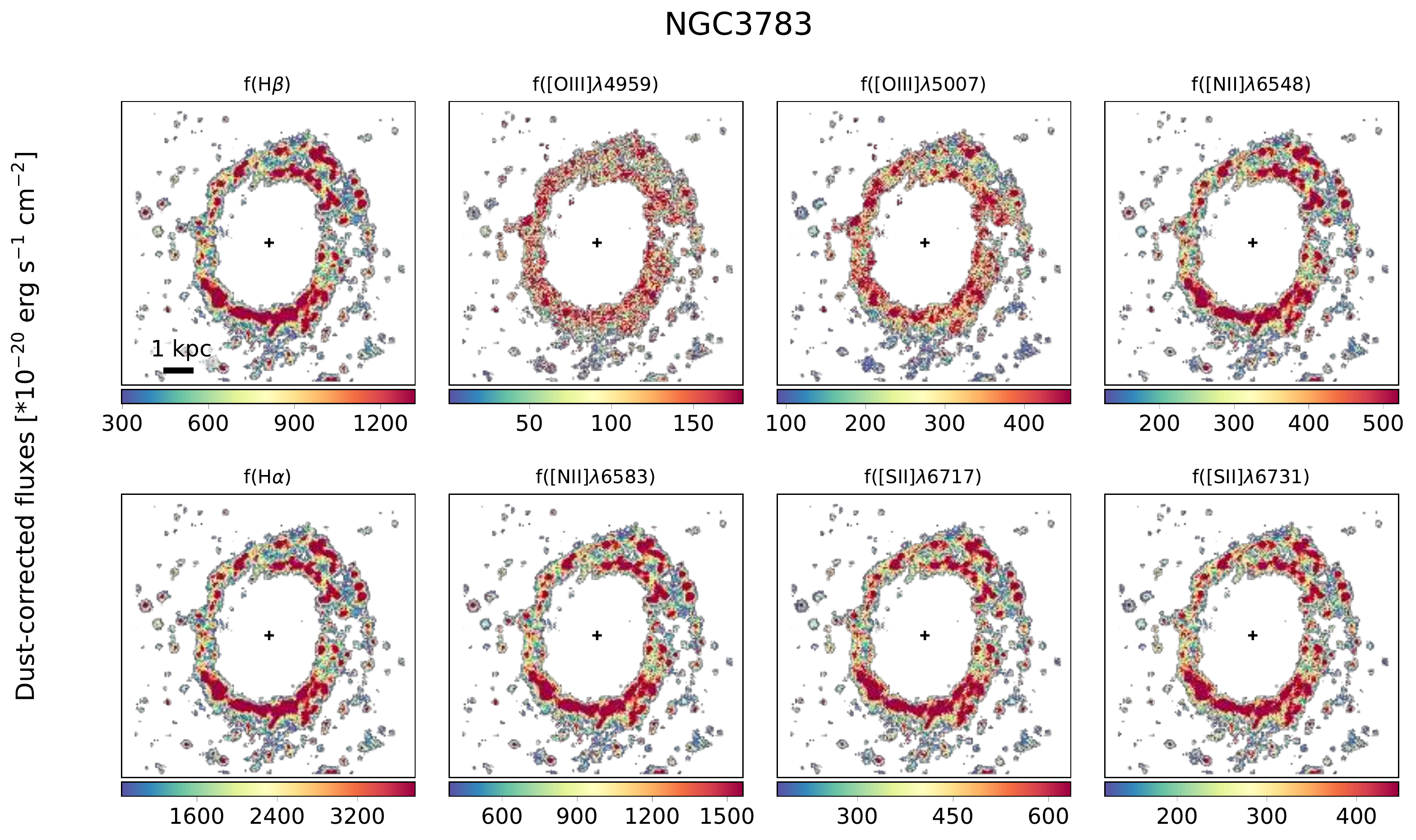}
% \caption{As Fig.~\ref{allfluxes} but for NGC~4030.}
% \label{ngc3783fluxes}
% \end{center}
% \end{figure*}

% \begin{figure*}
% \begin{center}
 \includegraphics[width=165mm]{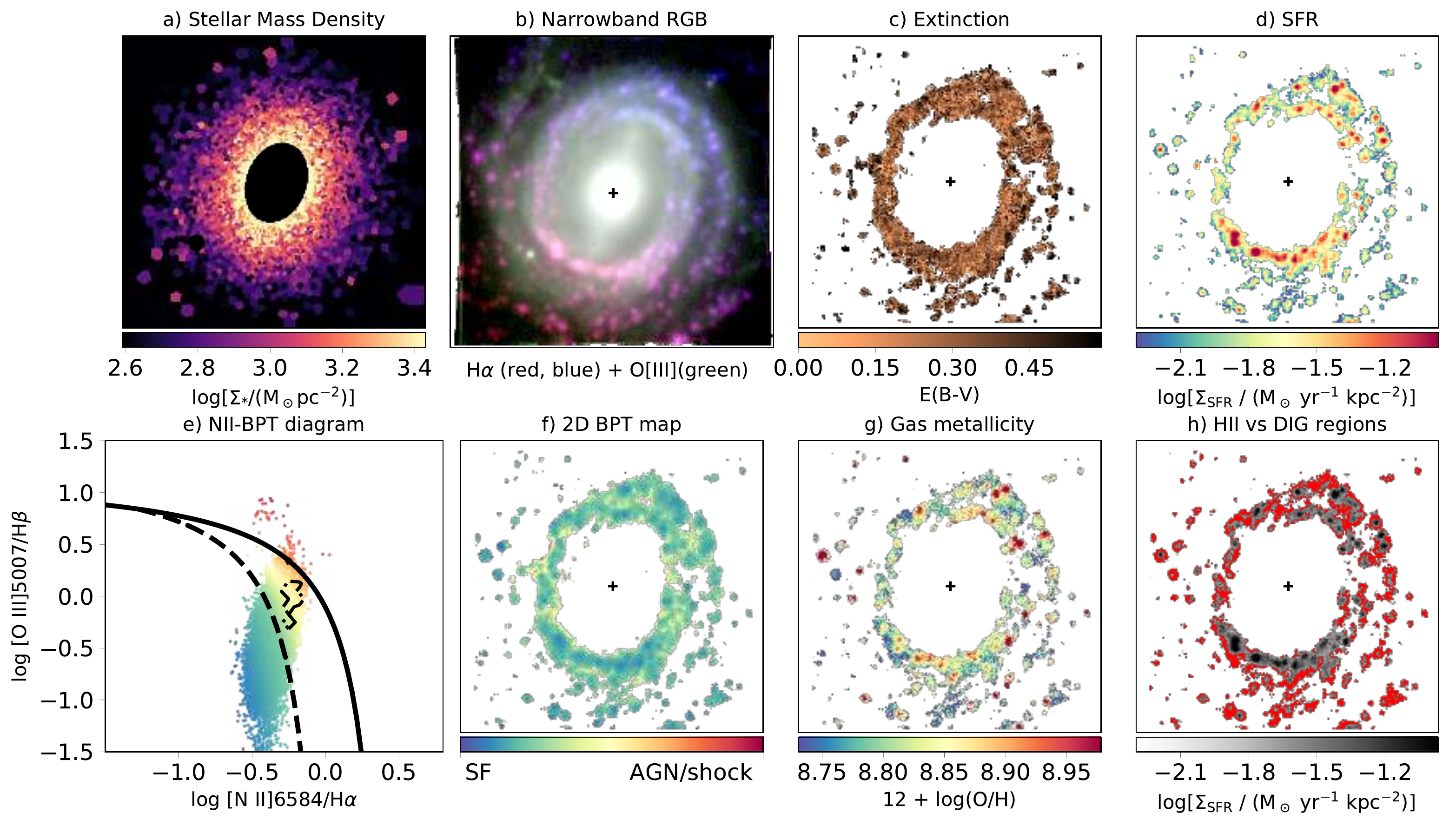}
\caption{This barred galaxy presents a Sy1 nucleus \citep{Raban2008} with very broad emission line features. The central part of the galaxy has been masked as the lines there are very broad due to the AGN, and our emission line fitting is not reliable there. There is an inner ring with most of the SFR of the galaxy. There is emission from a fainter and outer arm that continues the ring and disappears in the NE part of the image. The metallicity decreases with galactocentric distance. The largest deviation in metallicity from the  linear fit is found in the H{\sc ii} regions of the H{\sc ii} ring, with values $\sim$ 0.2 dex higher than the linear fit.
}

\label{ngc3783plots}
\end{center}
\end{figure*}

\clearpage
%\subsection{NGC~5334}   

\begin{figure*}
\begin{center}
 \includegraphics[width=165mm]{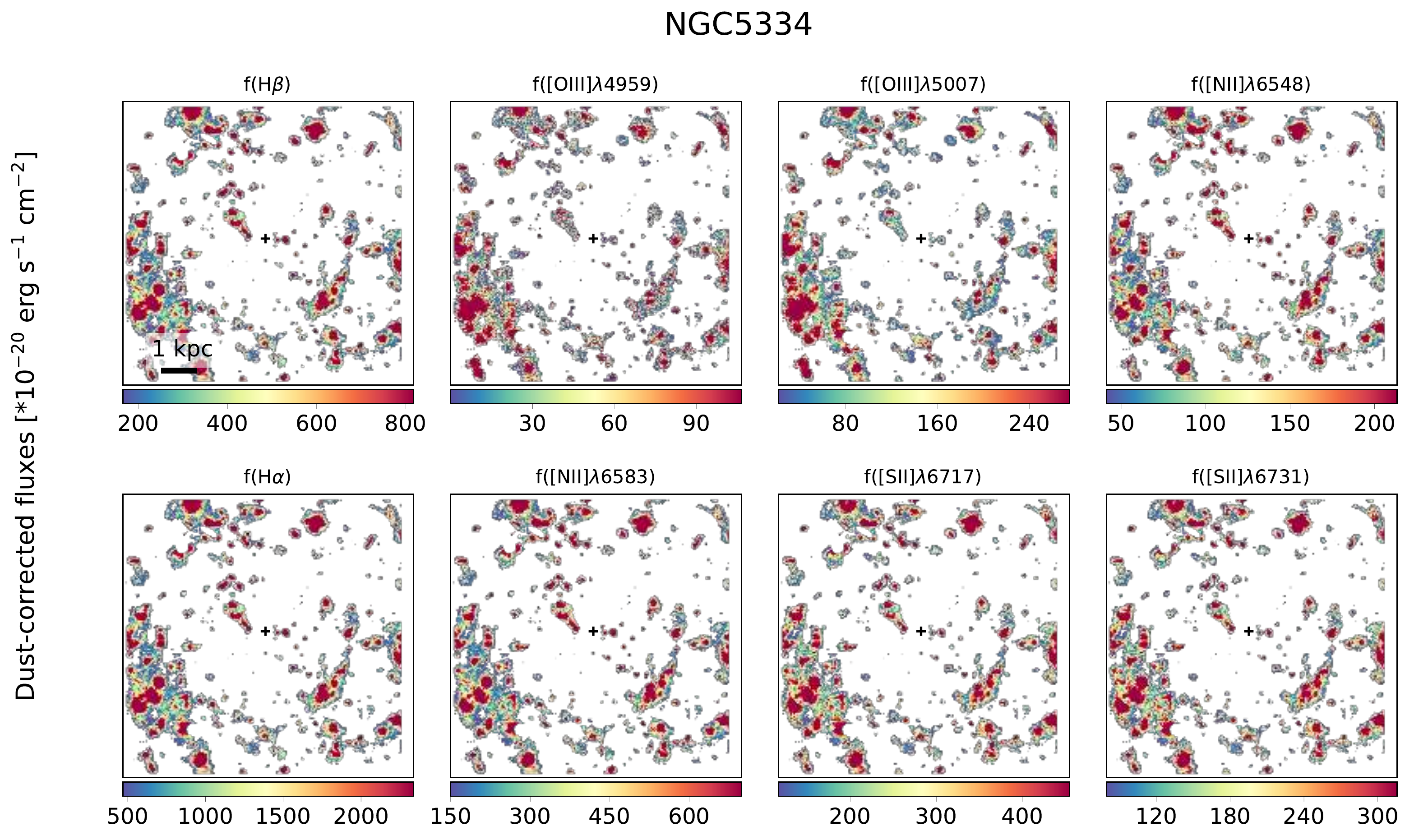}
% \caption{As Fig.~\ref{allfluxes} but for NGC~4030.}
% \label{ngc5334fluxes}
% \end{center}
% \end{figure*}

% \begin{figure*}
% \begin{center}
 \includegraphics[width=165mm]{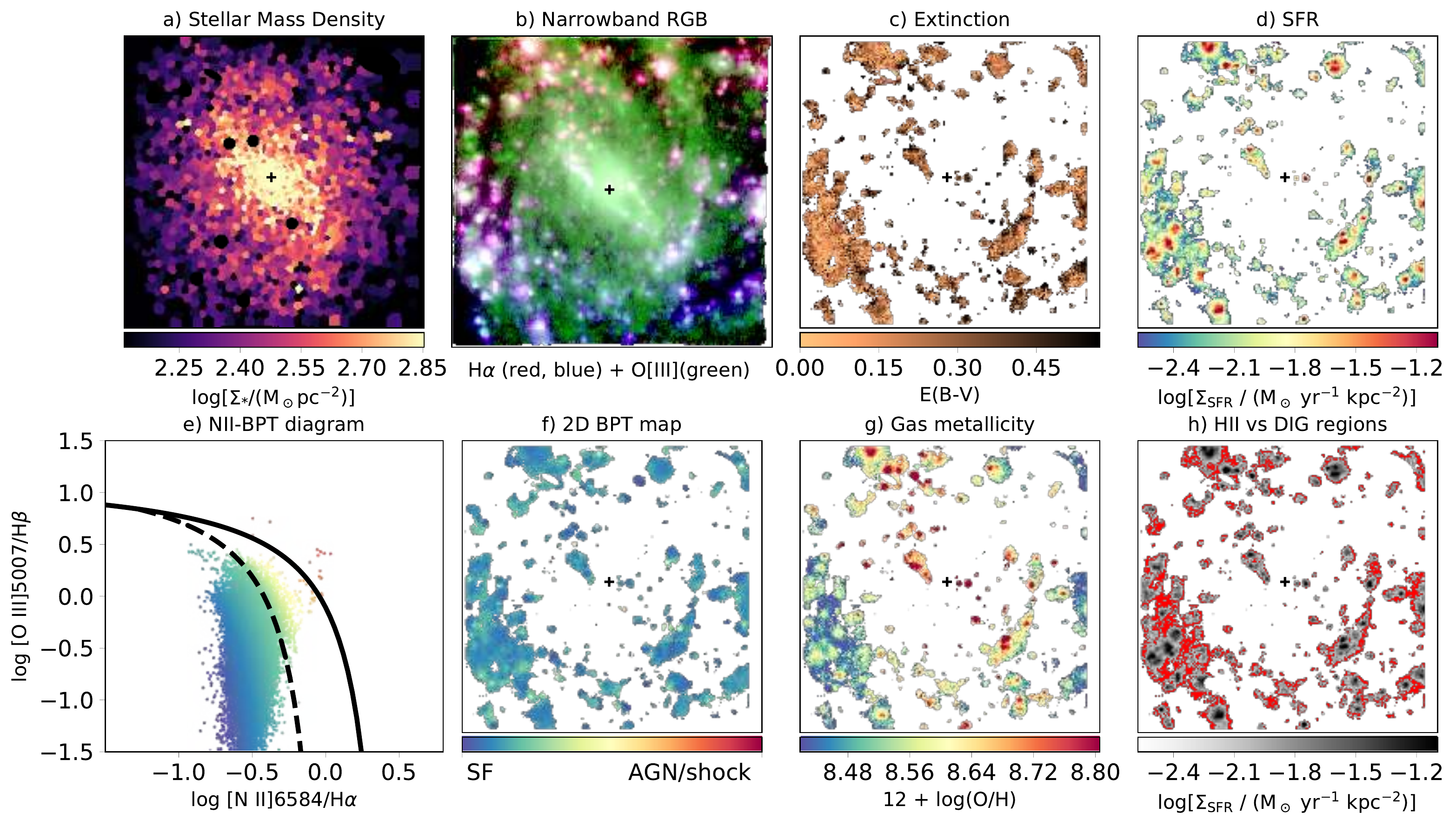}
\caption{NGC~5334 presents a prominent bar in the stellar surface density map, which does not produce much emission in H$ \alpha $ (as seen from the SFR map). The metallicity map shows that the central H{\sc ii} regions are more metal enriched than those in the spiral arms and than the DIG ($\sim$0.2 dex higher). There are many H{\sc ii} regions in the spiral arms, although our FoV does not allow us to see the extent and type of these arms.
}

\label{ngc5334plots}
\end{center}
\end{figure*}

\clearpage
%\subsection{NGC~7162}   

\begin{figure*}
\begin{center}
 \includegraphics[width=165mm]{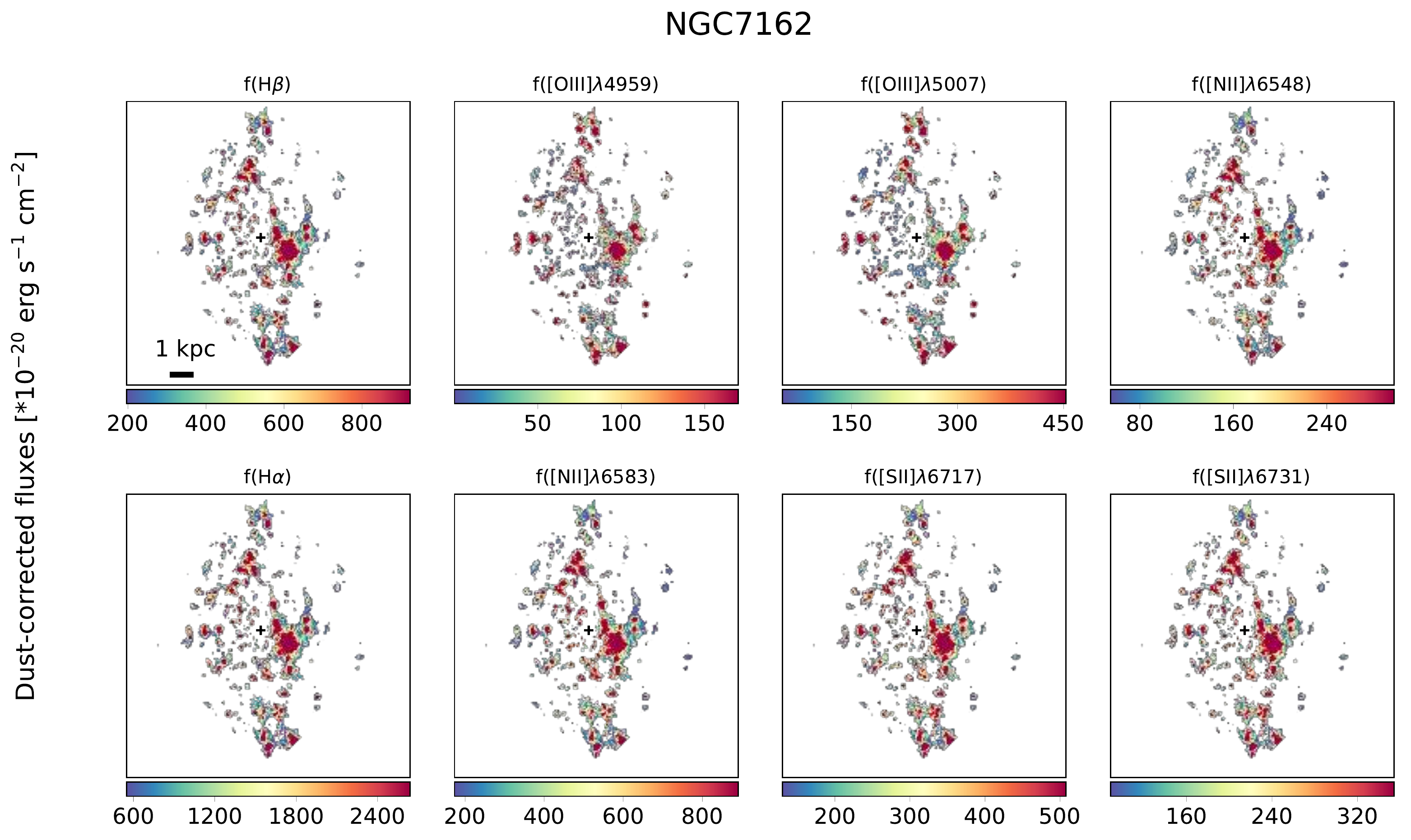}
% \caption{As Fig.~\ref{allfluxes} but for NGC~4030.}
% \label{ngc7162fluxes}
% \end{center}
% \end{figure*}

% \begin{figure*}
% \begin{center}
 \includegraphics[width=165mm]{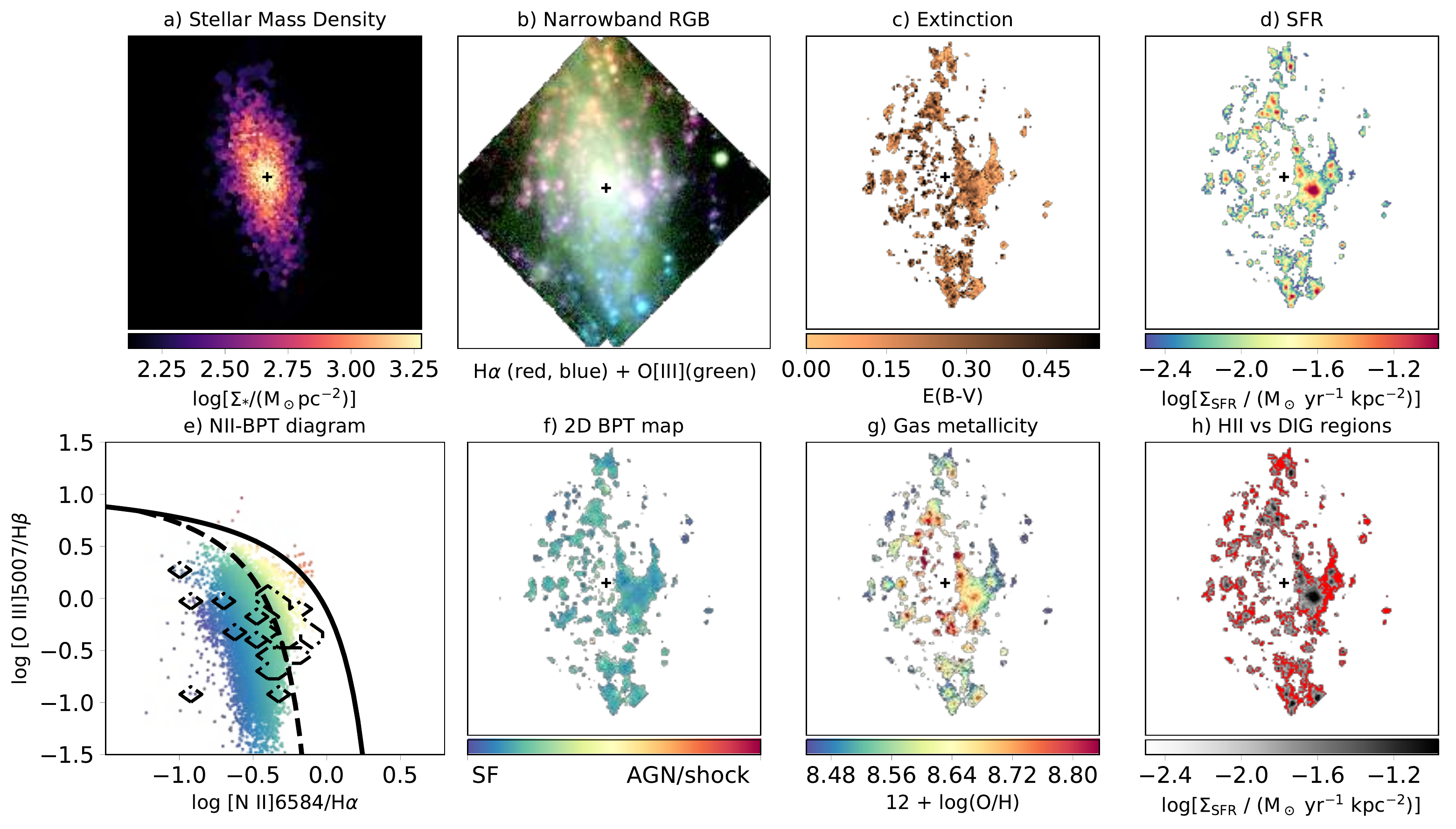}
\caption{NGC~7162 is an unbarred spiral with moderate inclination and clear asymmetric structure, probably due to the interaction with NGC~7166 (at an angular separation of 11 arcmin). As in NGC~3521, the most luminous part in the stellar structure coincides with the side with higher metallicities and lower SFRs. Indeed, the most luminous H{\sc ii} region is located $\sim$1 kpc SW of the centre, but that region is not the most metal enriched. The metallicity is clearly decreasing with radius, with some scatter ($\sim$0.2 dex) in the centres of the H{\sc ii} regions compared to the linear fit.
}

\label{ngc7162plots}
\end{center}
\end{figure*}

\clearpage

%\subsection{NGC~1084}   

\begin{figure*}
\begin{center}
 \includegraphics[width=165mm]{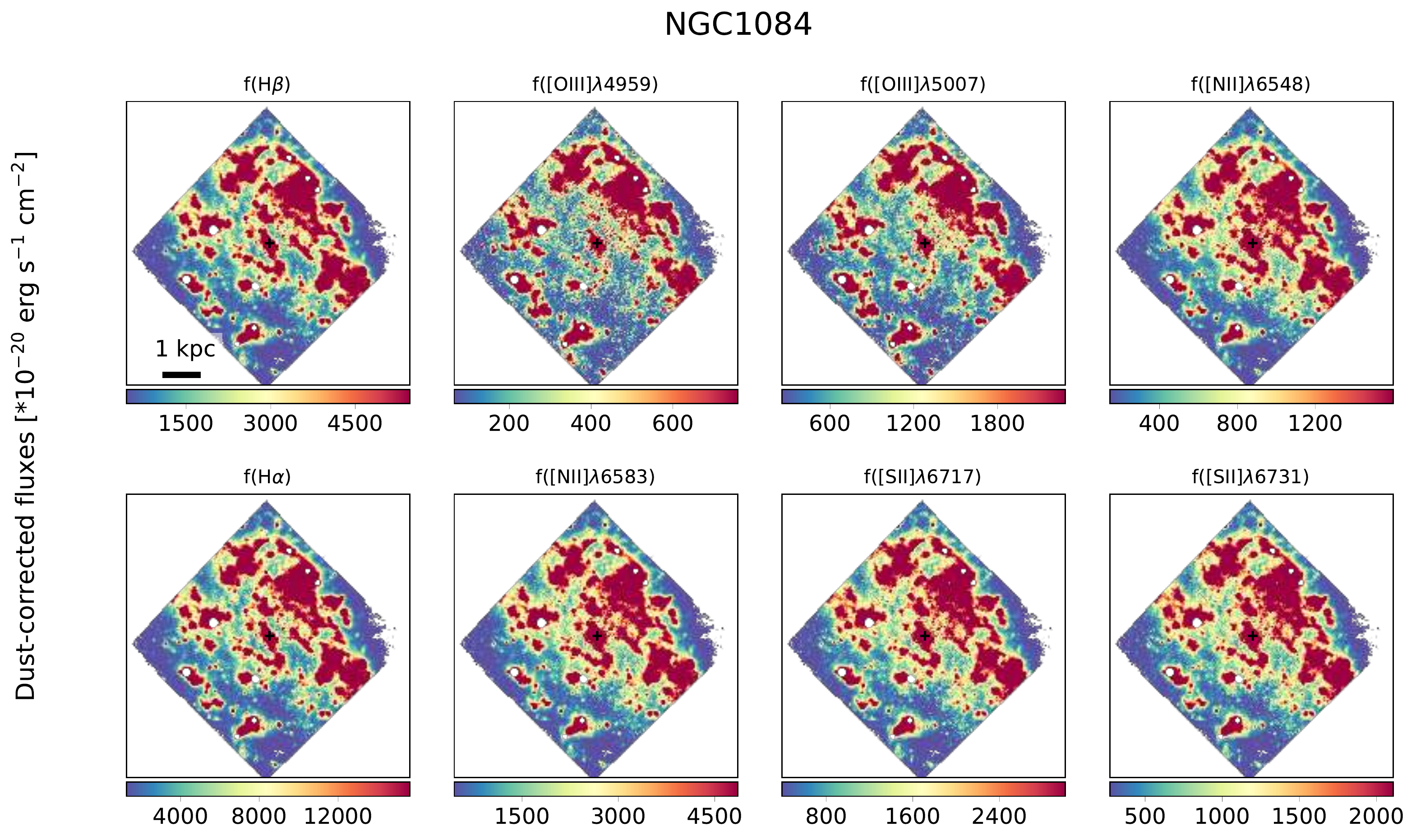}
% \caption{As Fig.~\ref{allfluxes} but for NGC~4030.}
% \label{ngc1084fluxes}
% \end{center}
% \end{figure*}

% \begin{figure*}
% \begin{center}
 \includegraphics[width=165mm]{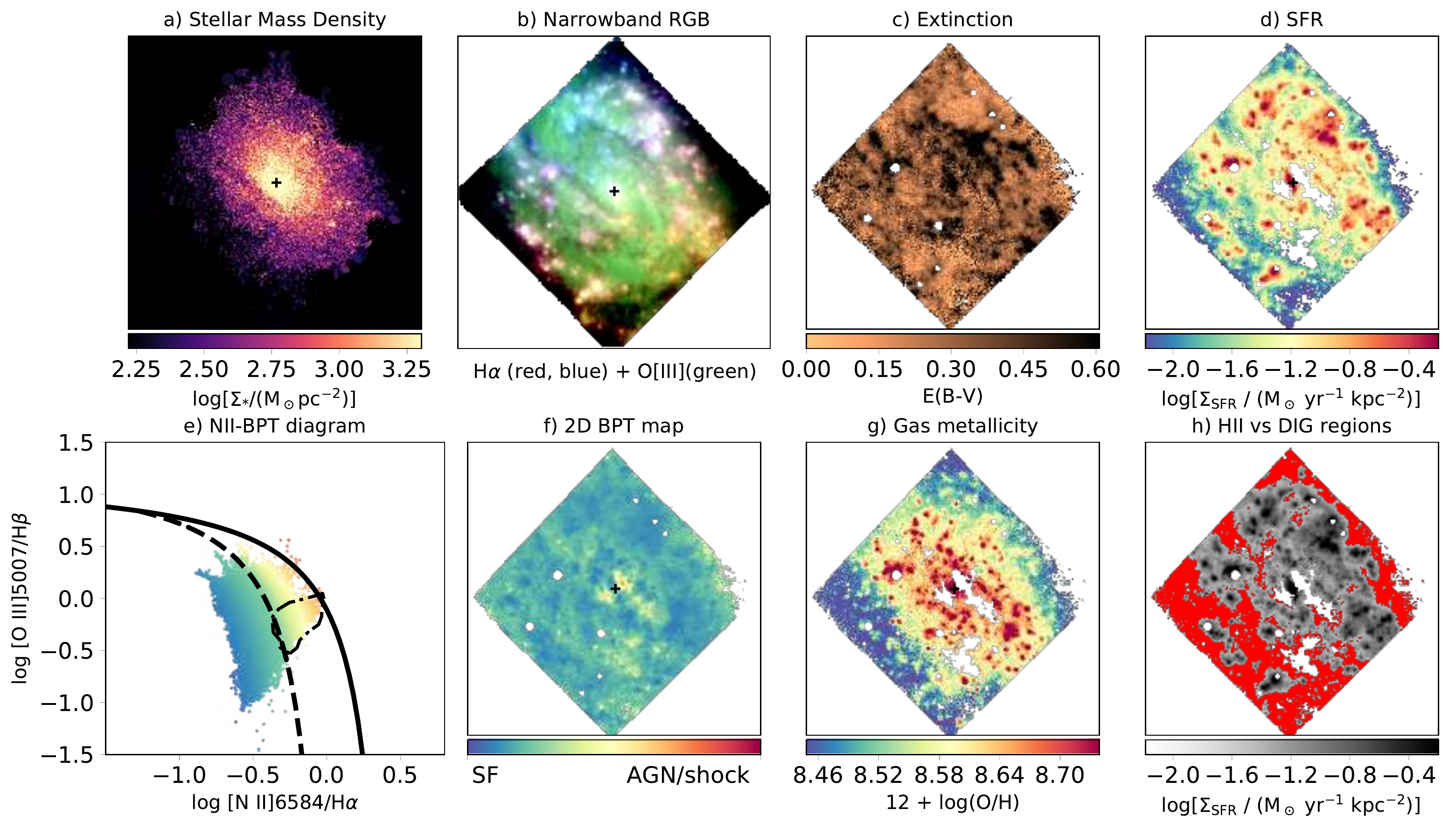}
\caption{Both the emission line maps and the narrowband image of NGC~1084 appear asymmetric, with more emission in the northern part. The metallicity, however, seems to be equally distributed with respect to the central part. Most of the regions show SF ionization source.
}
\label{ngc1084plots}
\end{center}
\end{figure*}
\clearpage
%\subsection{NGC~1309}   

\begin{figure*}
\begin{center}
 \includegraphics[width=165mm]{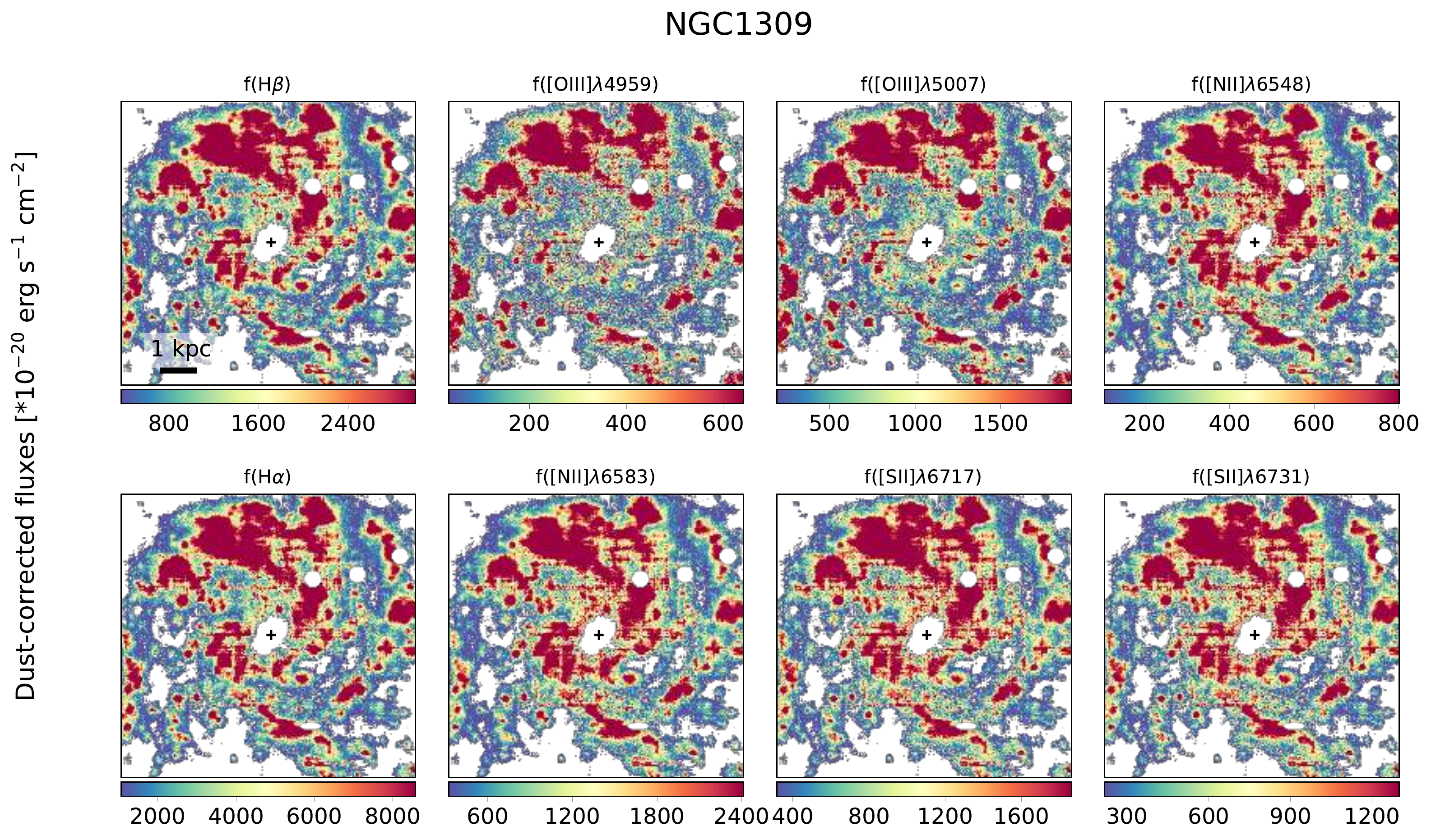}
% \caption{As Fig.~\ref{allfluxes} but for NGC~4030.}
% \label{ngc1309fluxes}
% \end{center}
% \end{figure*}

% \begin{figure*}
% \begin{center}
 \includegraphics[width=165mm]{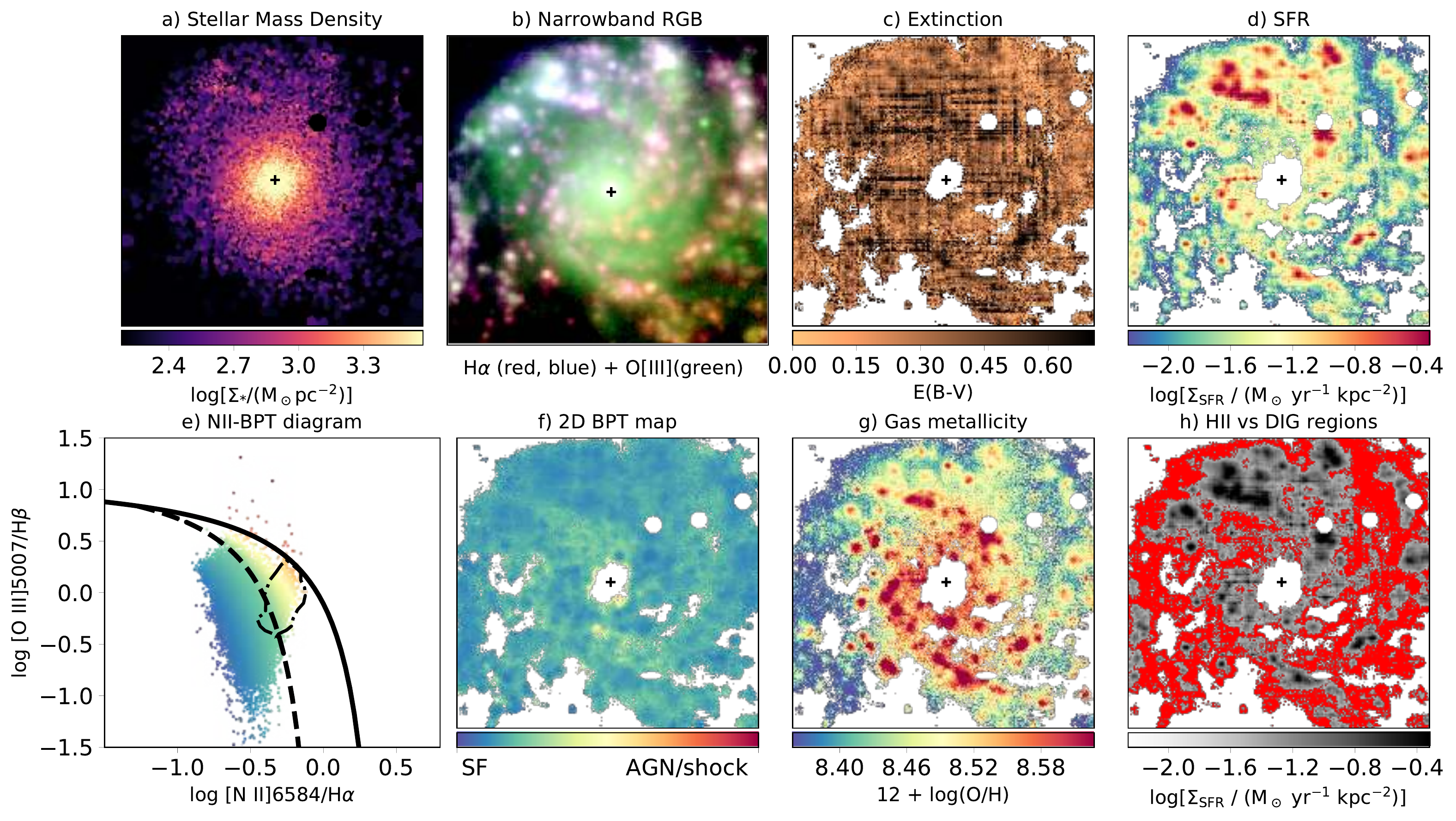}
\caption{This galaxy shows multiple spiral arms. These arms do not reach the very central part, where a bulge component without SF lies. On the one hand, the galaxy appears asymmetric in the SFR map, as most of the H$\alpha$ emission is located in the northern part of the spiral disk. But on the other hand, the gas metallicity map looks symmetric. It shows a clear gradient, with decreasing values towards the outer parts. Most of the deviations from the linear fit are found in the H{\sc ii} regions in the spiral arms.
}
\label{ngc1309plots}
\end{center}
\end{figure*}

\clearpage
%\subsection{NGC~5584}   

\begin{figure*}
\begin{center}
 \includegraphics[width=165mm]{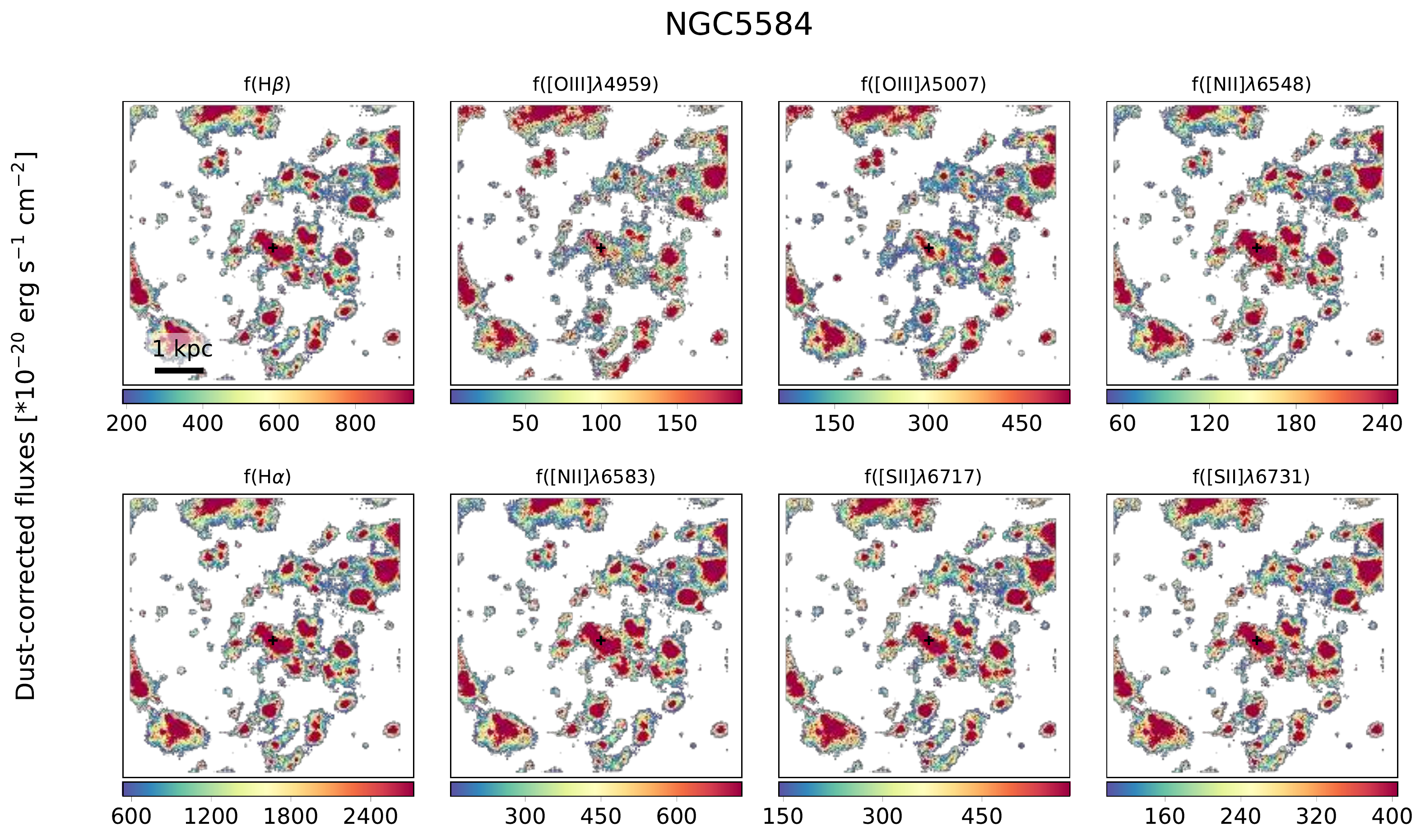}
% \caption{As Fig.~\ref{allfluxes} but for NGC~4030.}
% \label{ngc5584fluxes}
% \end{center}
% \end{figure*}

% \begin{figure*}
% \begin{center}
 \includegraphics[width=165mm]{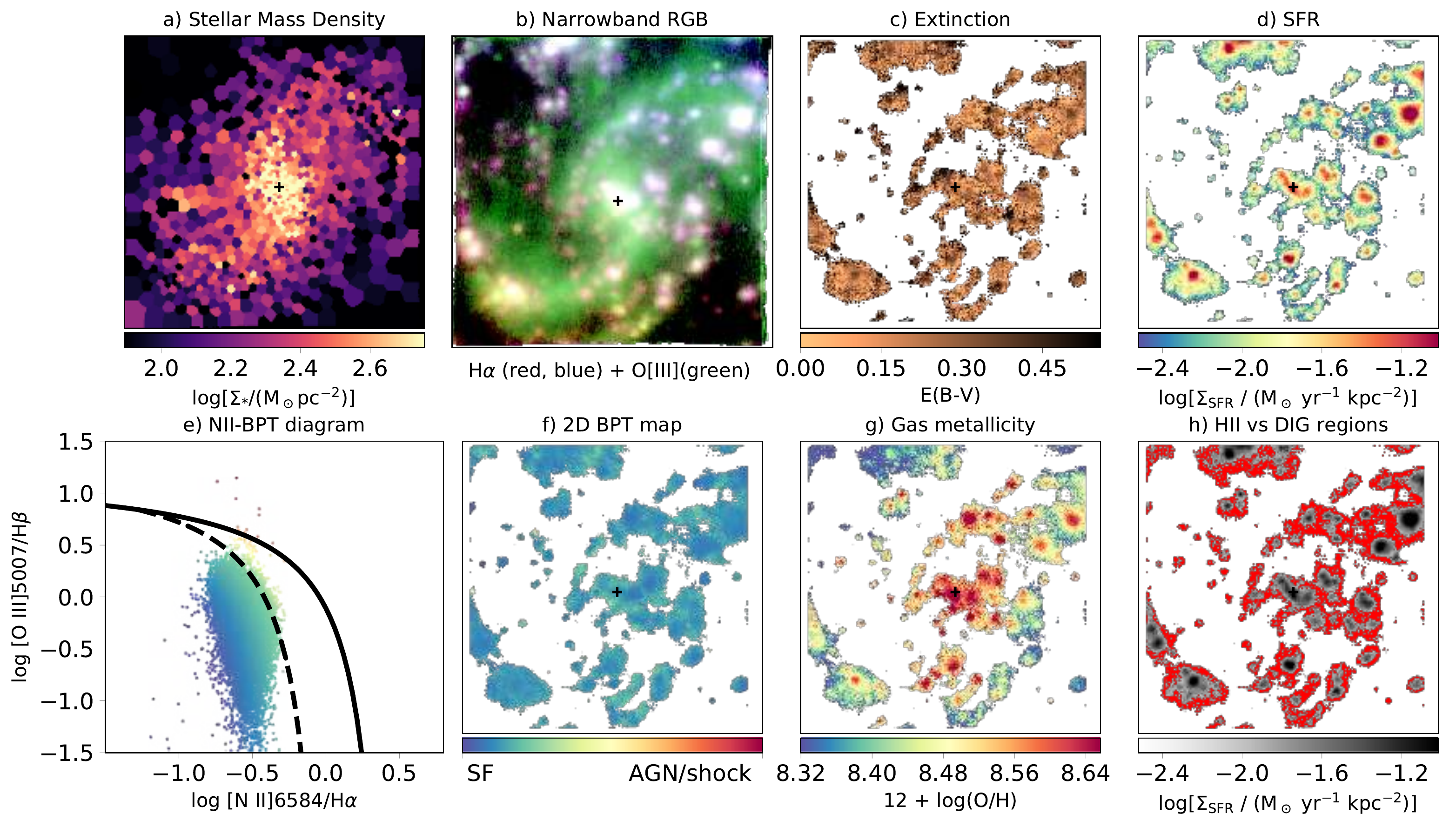}
\caption{NGC~5584 has a prominent bar with two spiral arms arising at its ends. There are, however, many H{\sc ii} regions not located in these two main arms. There is a clear metallicity gradient, with metal enriched gas in the central part, declining as we move outwards. The gas metallicity map also shows the structure of the galaxy, with the H{\sc ii} regions in the bar and inner spiral arms being more metal enriched than the DIG and regions in the outer interarms. The largest deviations in metallicity from the linear fit are found in the outer parts, where the H{\sc ii} regions are $\sim$0.2 dex higher than the linear fit and the DIG has metallicities $\sim$0.2 dex lower than the linear fit. This galaxy does not seem to present any AGN/shocked ionization.
}
\label{ngc5584plots}
\end{center}
\end{figure*}

\clearpage
%\subsection{NGC~4900}   

\begin{figure*}
\begin{center}
 \includegraphics[width=165mm]{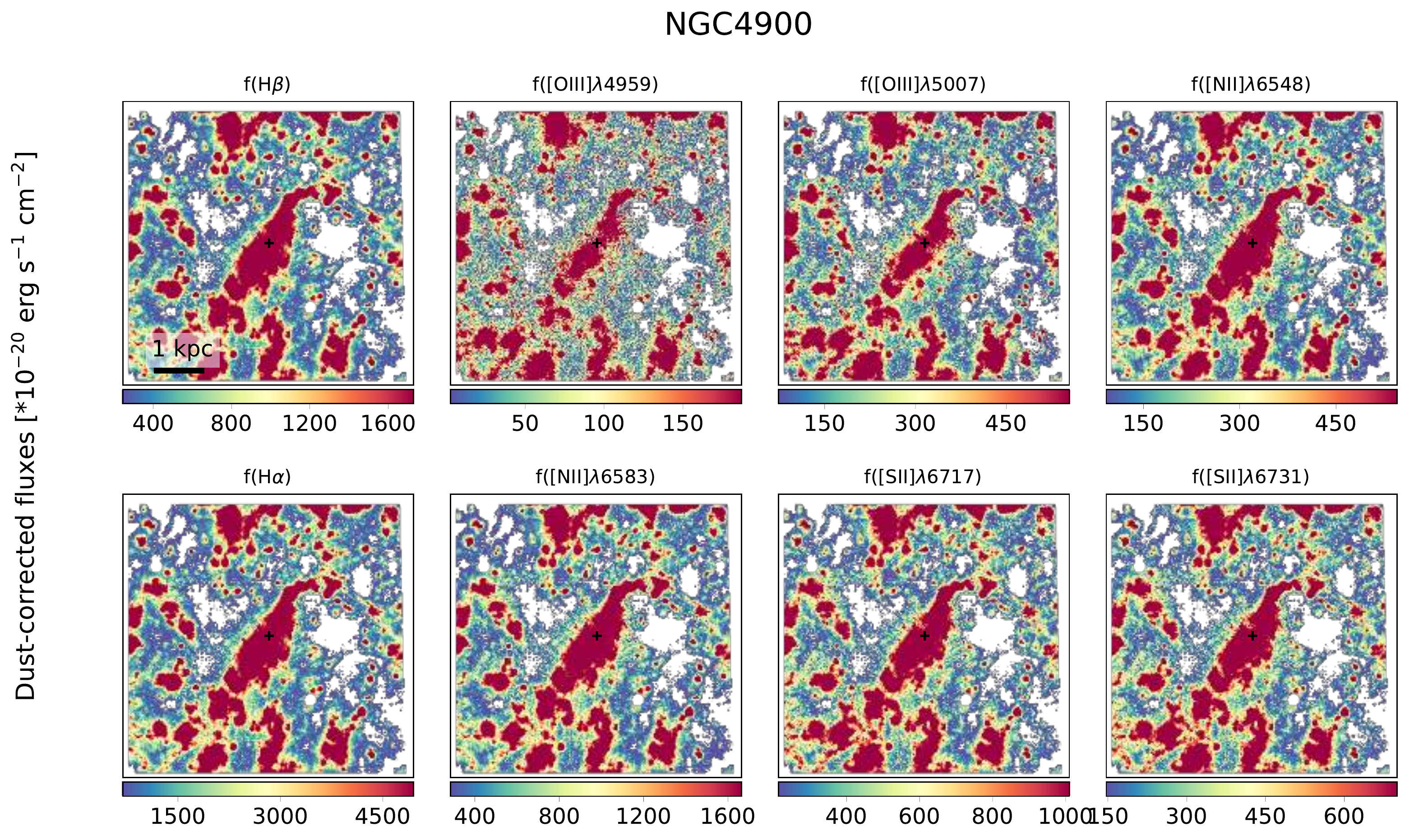}
% \caption{As Fig.~\ref{allfluxes} but for NGC~4030.}
% \label{ngc4900fluxes}
% \end{center}
% \end{figure*}

% \begin{figure*}
% \begin{center}
 \includegraphics[width=165mm]{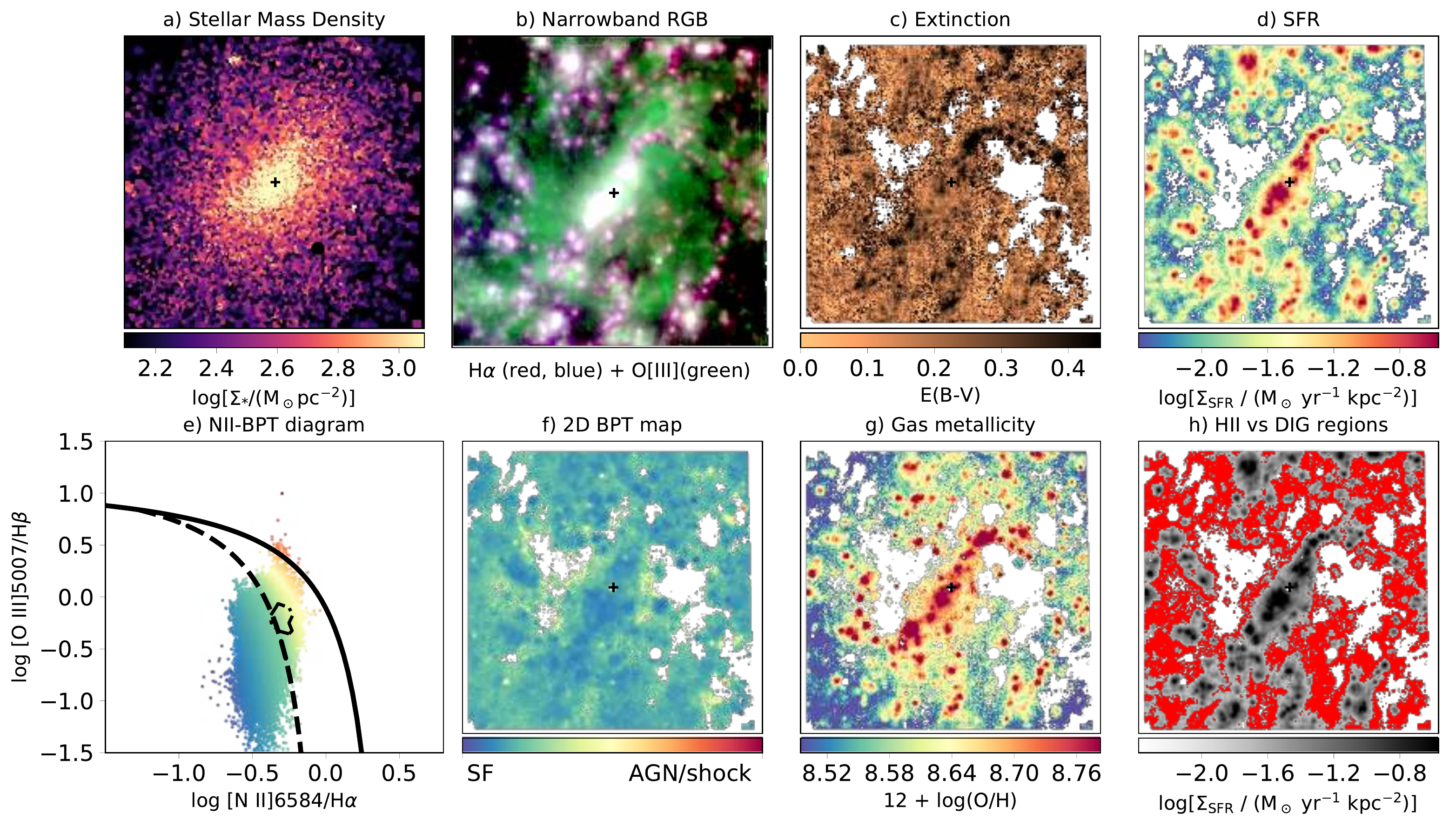}
\caption{Most of the star formation in the H{\sc ii} regions of NGC~4900 is localized in the spiral arms, although not forming filamentary structures, but rather patchy emission. The bar presents high values of star formation as well, with metal enriched gas. The centre itself presents a lower level of star formation than the surroundings. The H{\sc ii} regions in the spiral arms present lower metallicity than those in the bar and its vicinity. The peak of the star formation and metallicity is located in a blob SE from the centre. This interesting region next to the nucleus shows many weak emission lines (He{\sc i}$ \lambda $6678, He{\sc i}$ \lambda $7065, [Ar{\sc iii}]$ \lambda $7136, [O{\sc ii}]$ \lambda $7300 doublets, [C{\sc iii}]$ \lambda $8579, [C{\sc iii}]$ \lambda $8500, Pa12 and Pa10). These lines appear very narrow and without an outflow component, so we do not consider this region to be outflowing/inflowing from/towards the galactic centre. The largest metallicity deviations from the linear fit are found for the H{\sc ii} regions, with a scatter of $\sim$0.25 dex at all radii.}
\label{ngc4900plots}
\end{center}
\end{figure*}
\clearpage

%\subsection{NGC~7496}   

\begin{figure*}
\begin{center}
 \includegraphics[width=165mm]{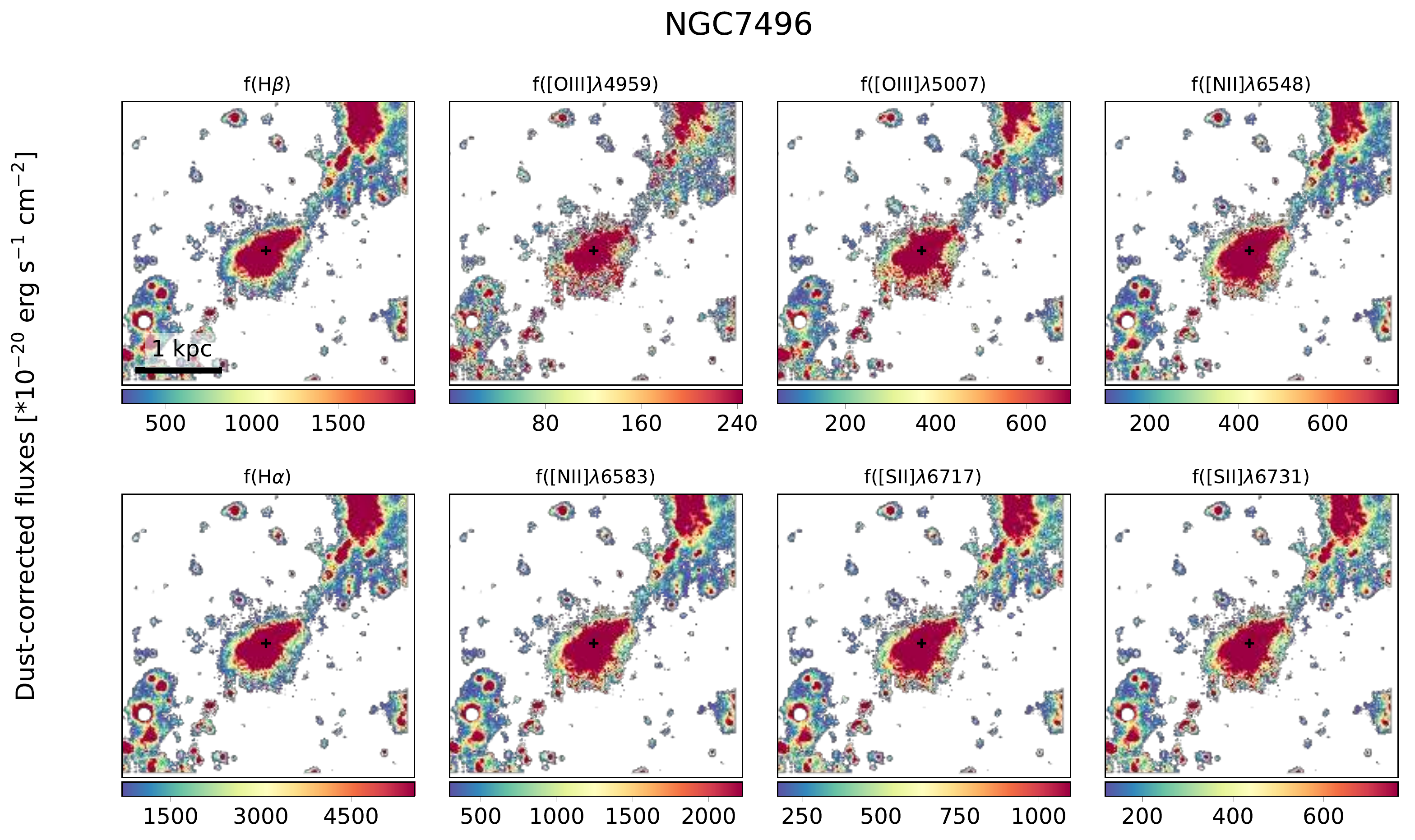}
% \caption{As Fig.~\ref{allfluxes} but for NGC~4030.}
% \label{ngc7496fluxes}
% \end{center}
% \end{figure*}

% \begin{figure*}
% \begin{center}
 \includegraphics[width=165mm]{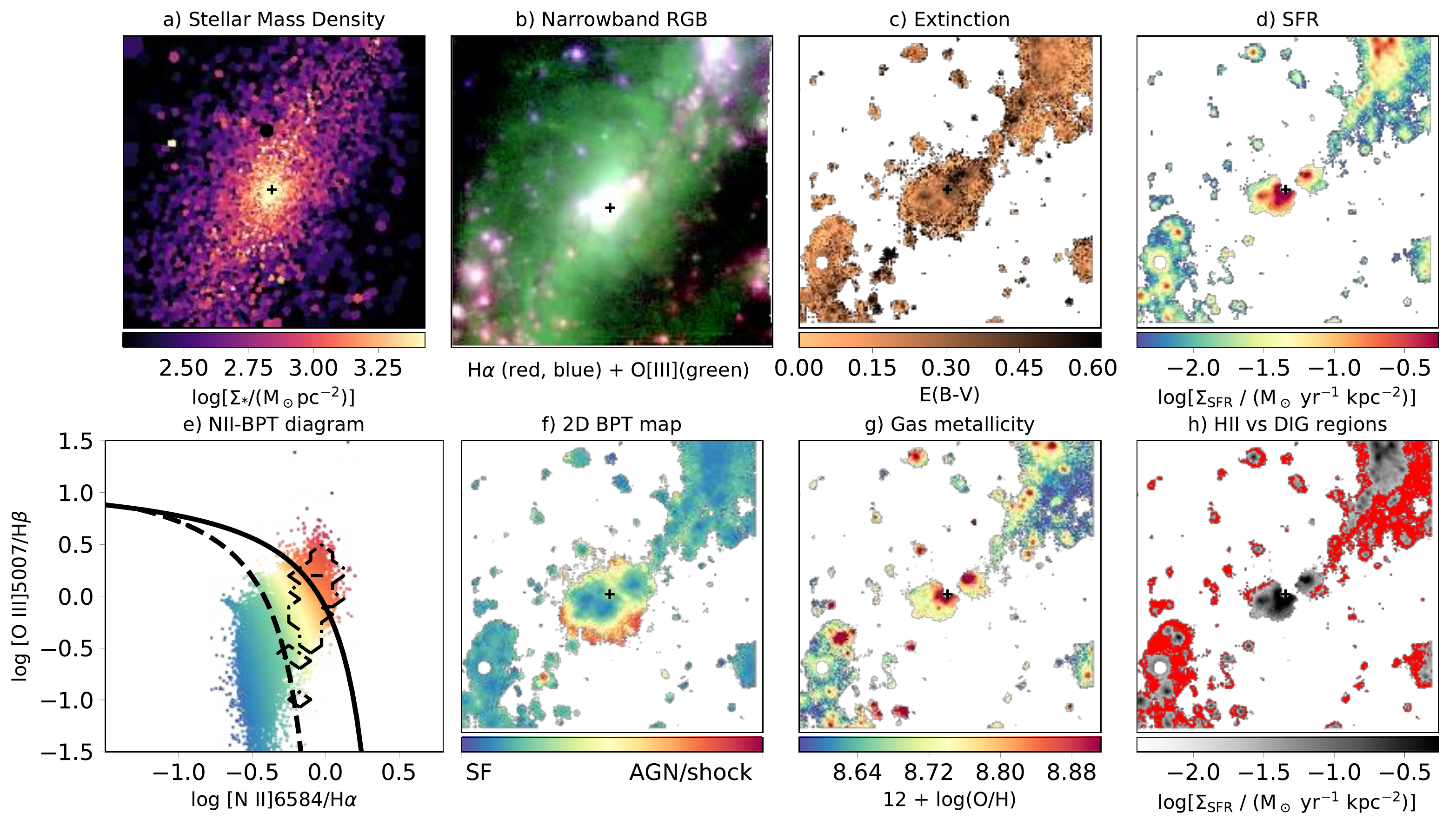}
\caption{This barred galaxy shows a dense, bulge-like structure. There are H{\sc ii} regions aligned in two well defined spiral arms that end in the central part. The emission lines are symmetric and narrow inside the H{\sc ii} regions but broad in the centre, compatible with the Sy2 classification by \citet{VeronCetty2006}. The barred region shows AGN/shock ionization except in the centre, where the ionization is dominated by SF. The very centre is metal enriched, more than its surroundings. Again, the DIG presents, on average, lower levels of metallicity than the H{\sc ii} regions.
}
\label{ngc7496plots}
\end{center}
\end{figure*}
\clearpage
%\subsection{NGC~7552}   

\begin{figure*}
\begin{center}
 \includegraphics[width=165mm]{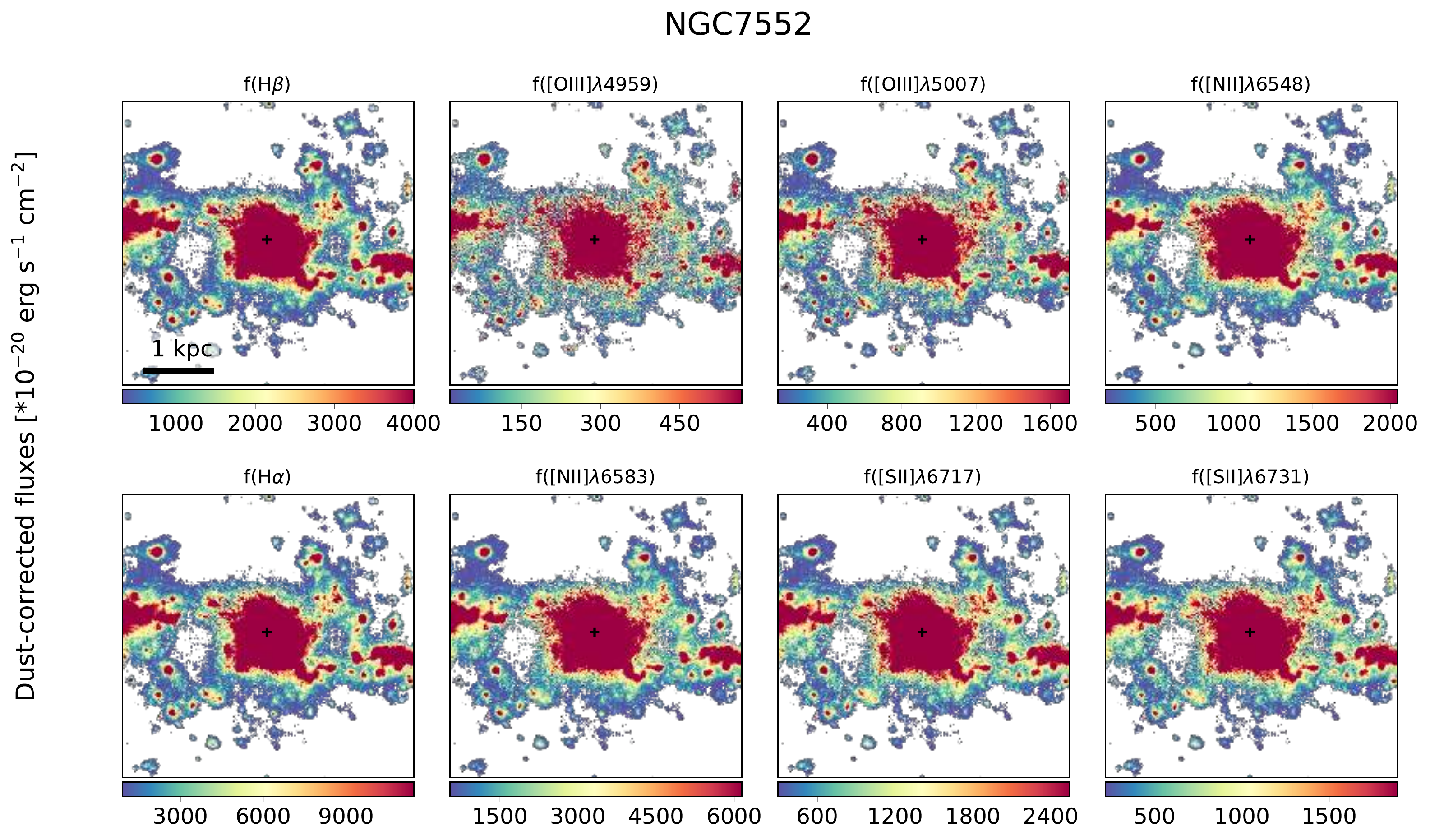}
% \caption{As Fig.~\ref{allfluxes} but for NGC~4030.}
% \label{ngc7552fluxes}
% \end{center}
% \end{figure*}

% \begin{figure*}
% \begin{center}
 \includegraphics[width=165mm]{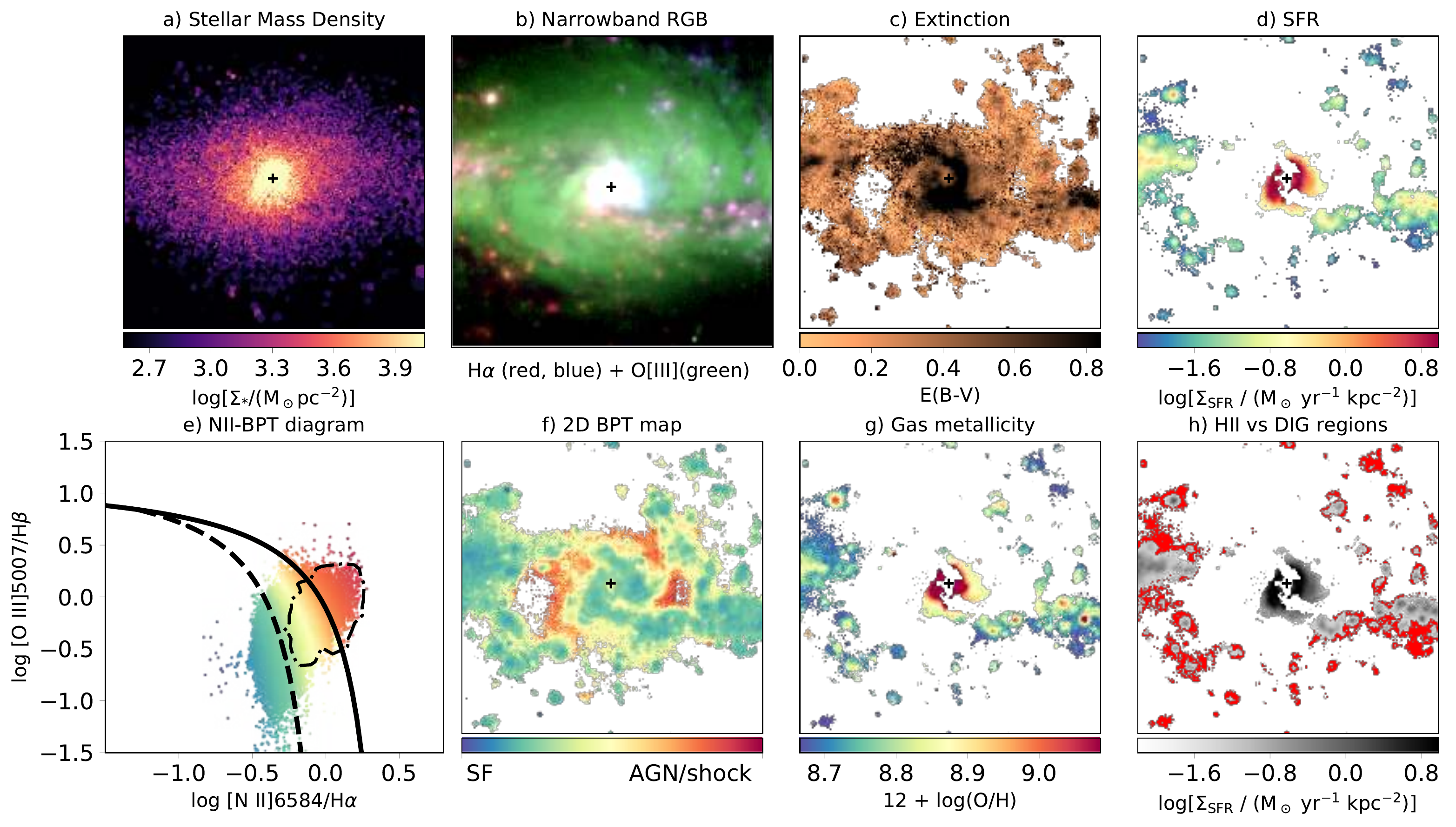}
\caption{NGC~7552 is a barred galaxy with two well defined dust lanes in the extinction map and narrowband image. Most of the H{\sc ii} regions are located along these two dust lanes. The two dust lanes terminate at a nuclear star forming ring, highly enriched in metallicity. The emission lines show a double component (blue-shifted outflow) in the eastern side of the centre, that is enhanced in metallicity and shows the largest deviation from the linear fit ($\sim$0.3 dex).
}
\label{ngc7552plots}
\end{center}
\end{figure*}
\clearpage

%\subsection{NGC~1512}   

\begin{figure*}
\begin{center}
 \includegraphics[width=165mm]{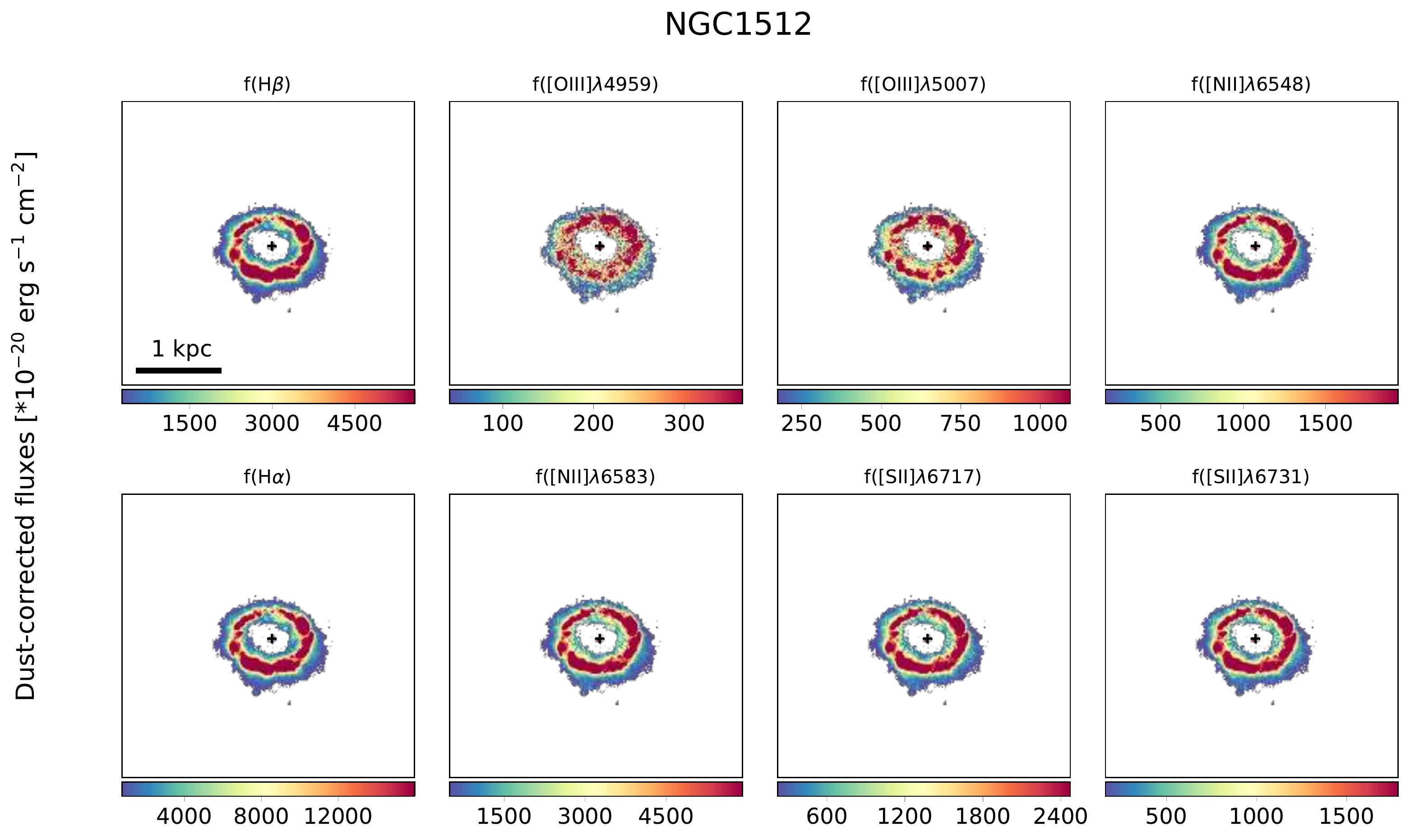}
% \caption{As Fig.~\ref{allfluxes} but for NGC~4030.}
% \label{ngc1512fluxes}
% \end{center}
% \end{figure*}

% \begin{figure*}
% \begin{center}
 \includegraphics[width=165mm]{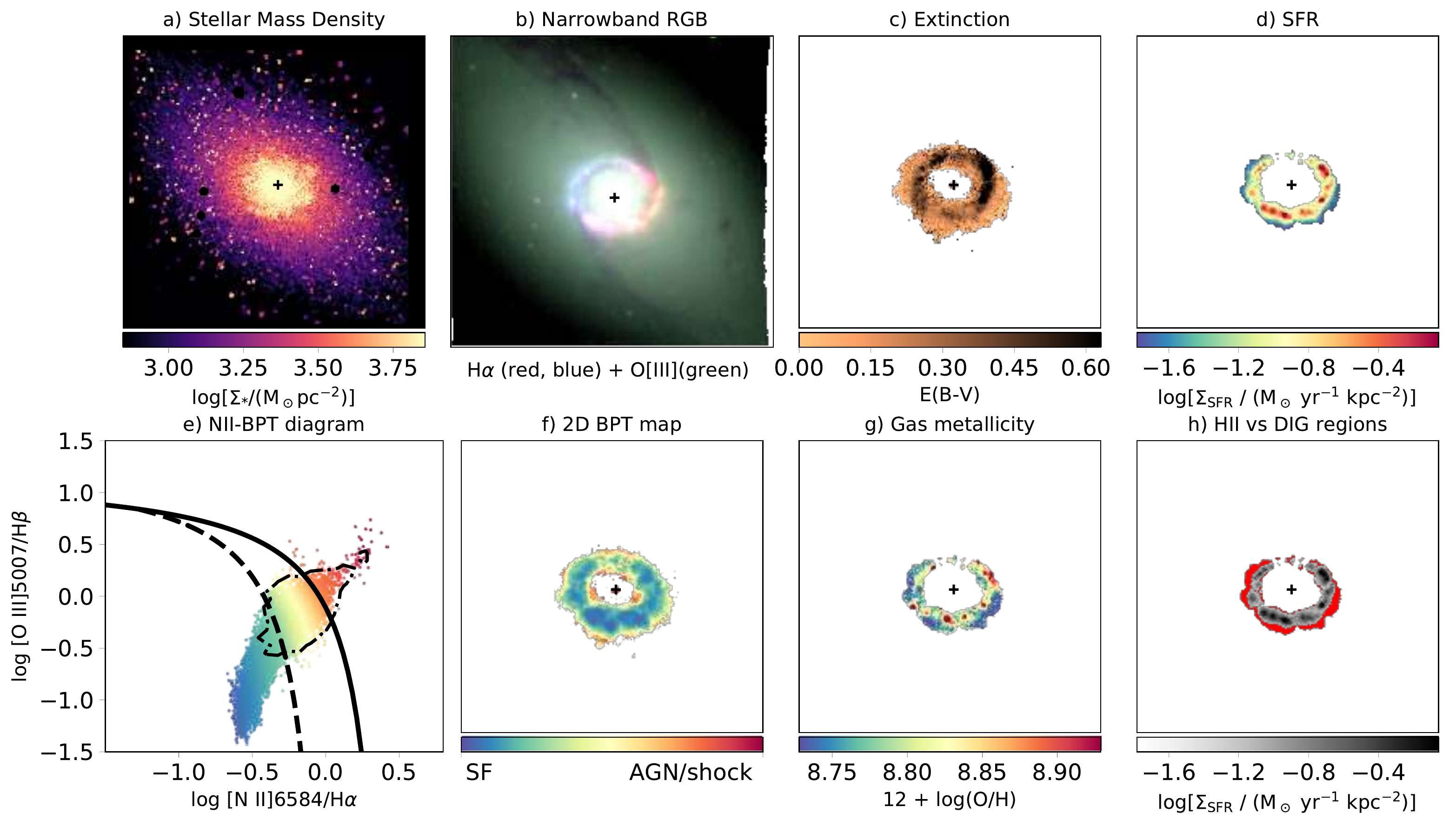}
\caption{As in NGC~1097, MAD observations of NGC~1512 focuses on the central part of this barred galaxy, where a nuclear star-forming ring has been found. The narrowband image shows two clear dust lanes arising from the central bulge. The bulge and bar are devoid of star formation.
}
\label{ngc1512plots}
\end{center}
\end{figure*}
\clearpage

%\subsection{NGC~7421}   

\begin{figure*}
\begin{center}
 \includegraphics[width=165mm]{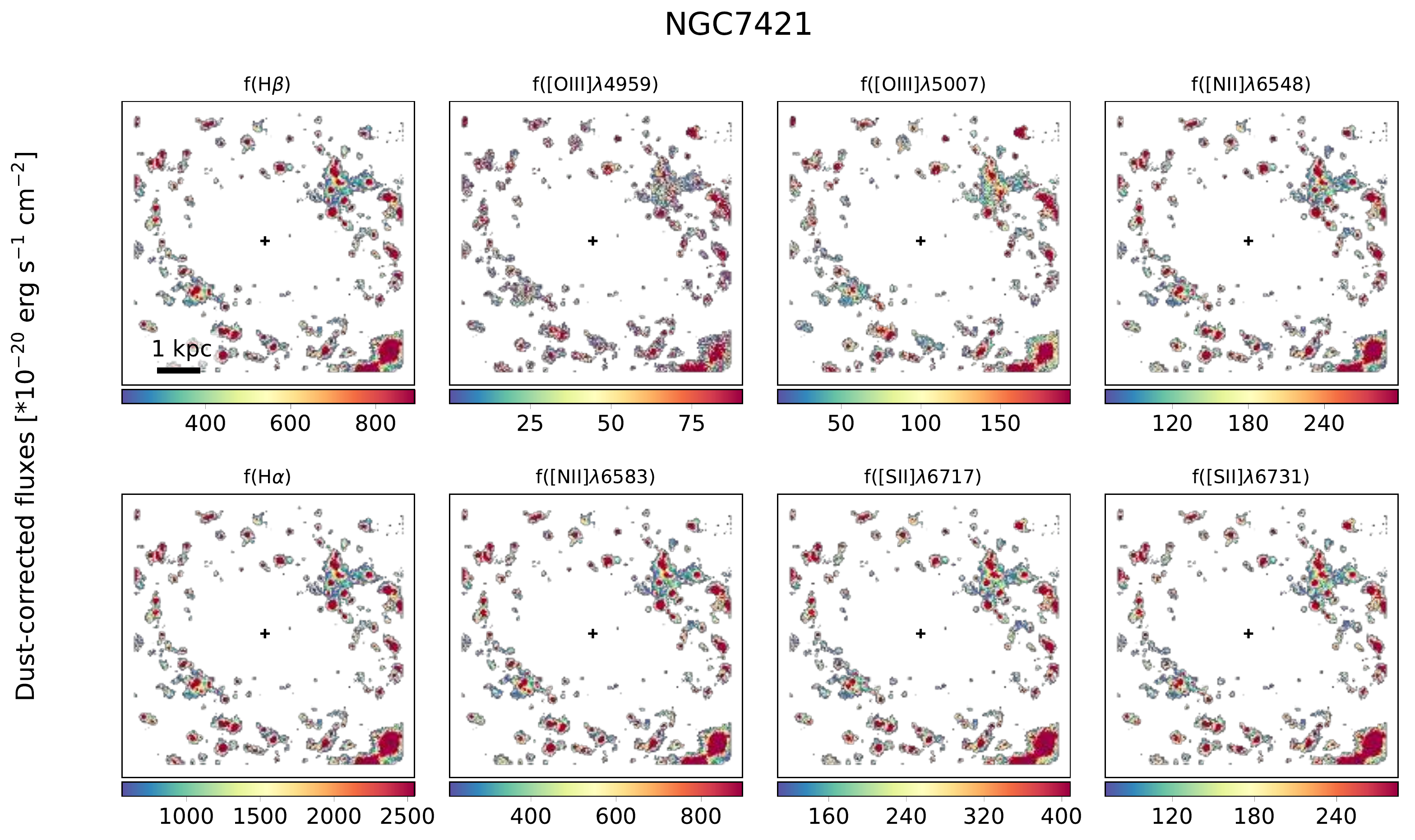}
% \caption{As Fig.~\ref{allfluxes} but for NGC~4030.}
% \label{ngc7421fluxes}
% \end{center}
% \end{figure*}

% \begin{figure*}
% \begin{center}
 \includegraphics[width=165mm]{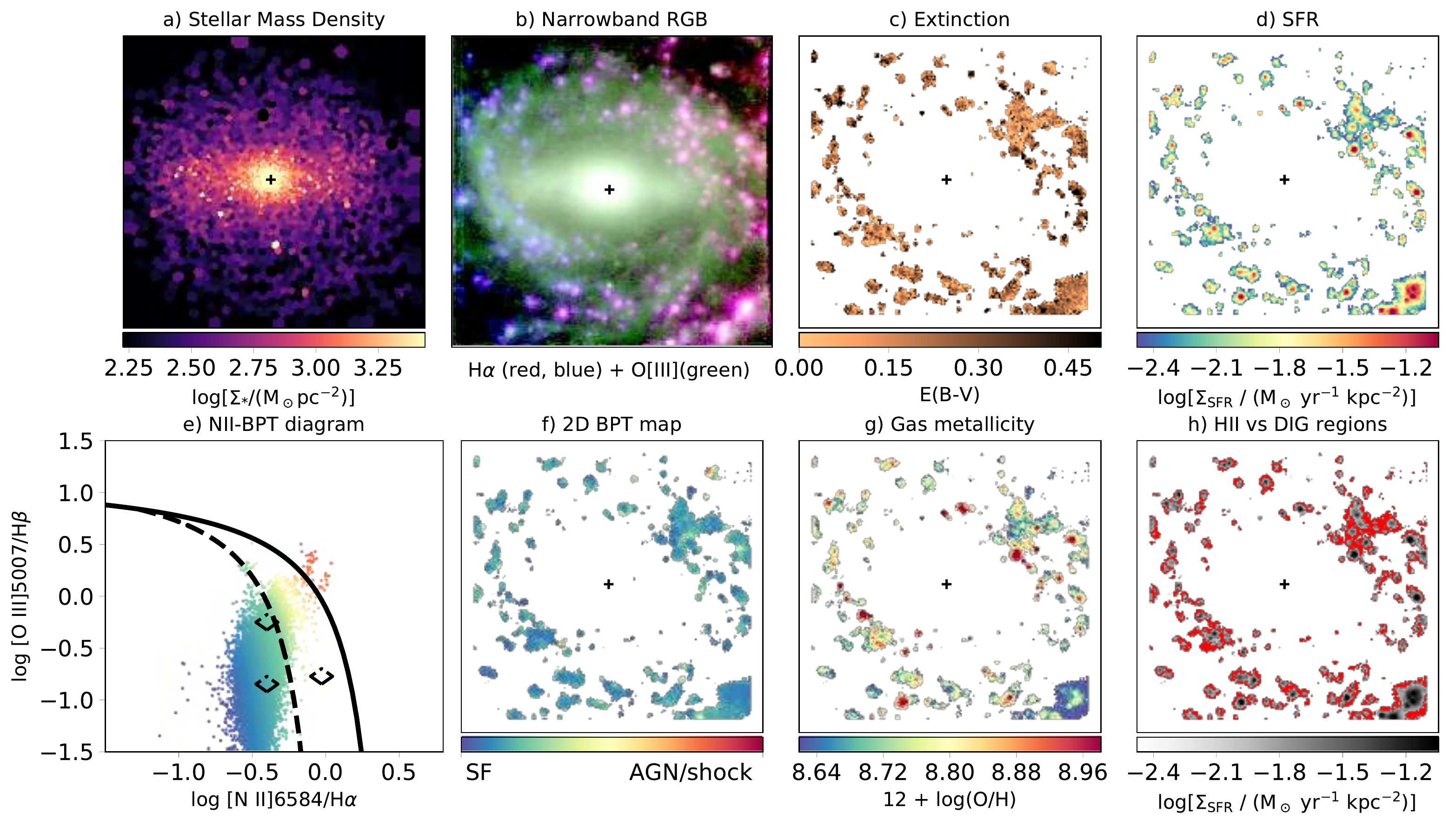}
\caption{Although this barred galaxy forms a pair with NGC~7418 at 19.5 arcmin, the inner properties do not seem to be asymmetric or lopsided. When looking at data with higher spatial coverage (e.g., S$ ^{4} $G data), we see that the bar is offset from the outer disk morphology. The spiral arms close towards a ring in the centre, leaving a region with low levels of SF. 
}
\label{ngc7421plots}
\end{center}
\end{figure*}
\clearpage
%\subsection{ESO~498-G5} 

\begin{figure*}
\begin{center}
 \includegraphics[width=165mm]{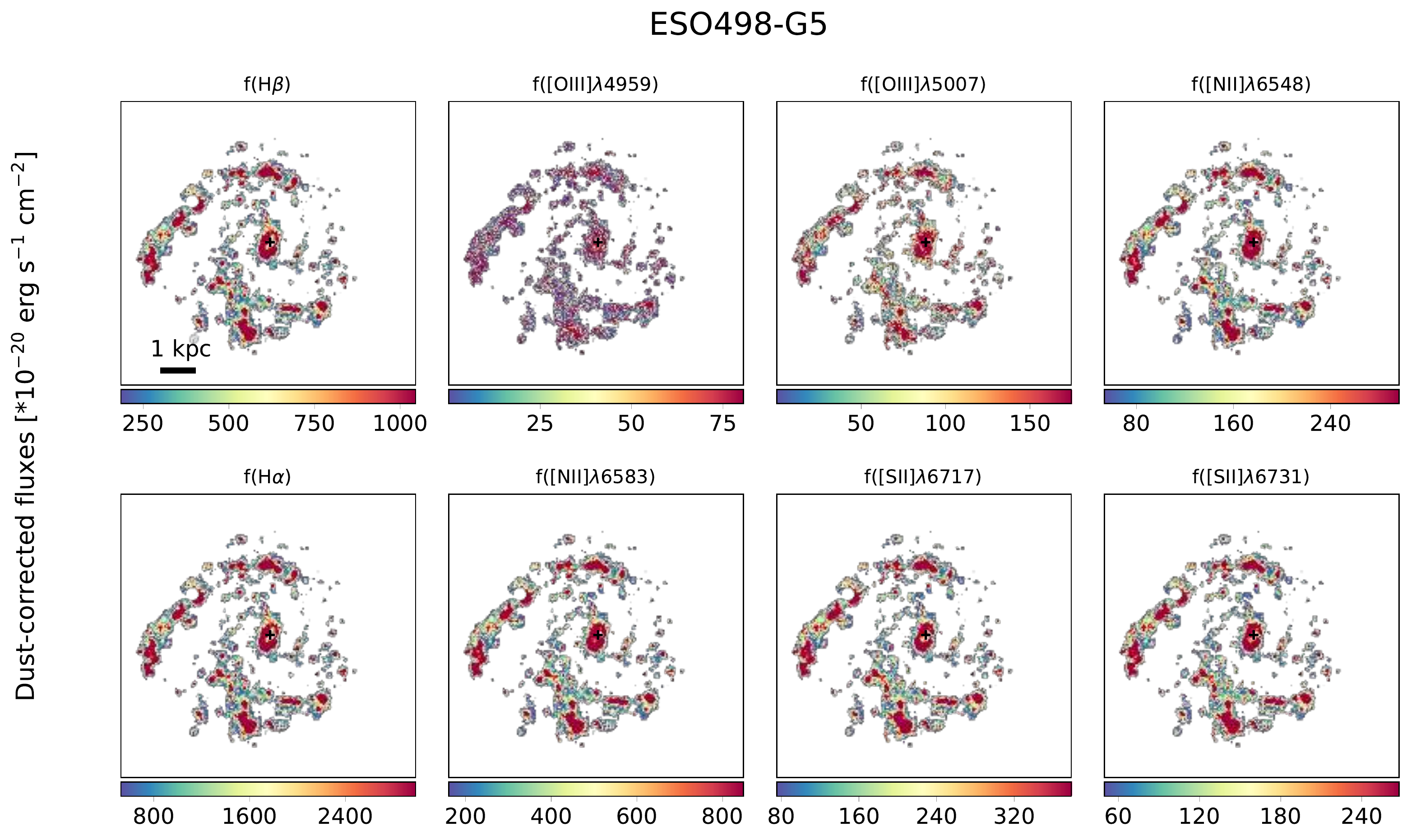}
% \caption{As Fig.~\ref{allfluxes} but for NGC~4030.}
% \label{eso498-g5fluxes}
% \end{center}
% \end{figure*}

% \begin{figure*}
% \begin{center}
 \includegraphics[width=165mm]{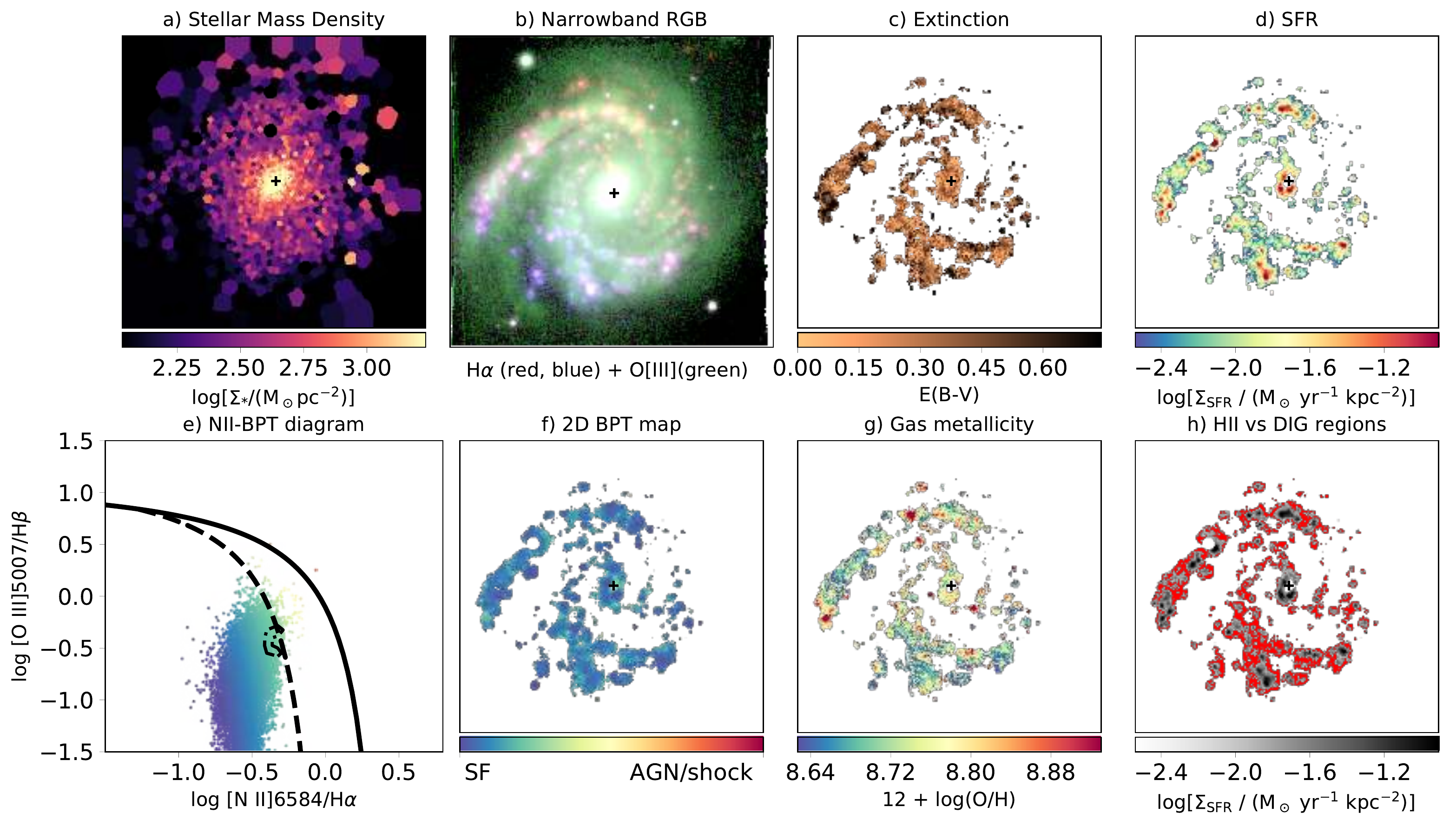}
\caption{ESO~498-G5 presents flocculent spiral arms structure, with no AGN/shock ionization sources. The extinction is rather low and mainly present in the spiral arms. The central part shows higher levels of star formation, probably indicating the formation of a small bulge with young stars. The H{\sc ii} regions in the arms show star formation with lower gas metallicity values than the interarm regions. The metallicities in the interarm regions deviate up to $\sim$0.2 dex from the linear fit.
}
\label{eso498-g5plots}
\end{center}
\end{figure*}
\clearpage
%\subsection{NGC~1042}   

\begin{figure*}
\begin{center}
 \includegraphics[width=165mm]{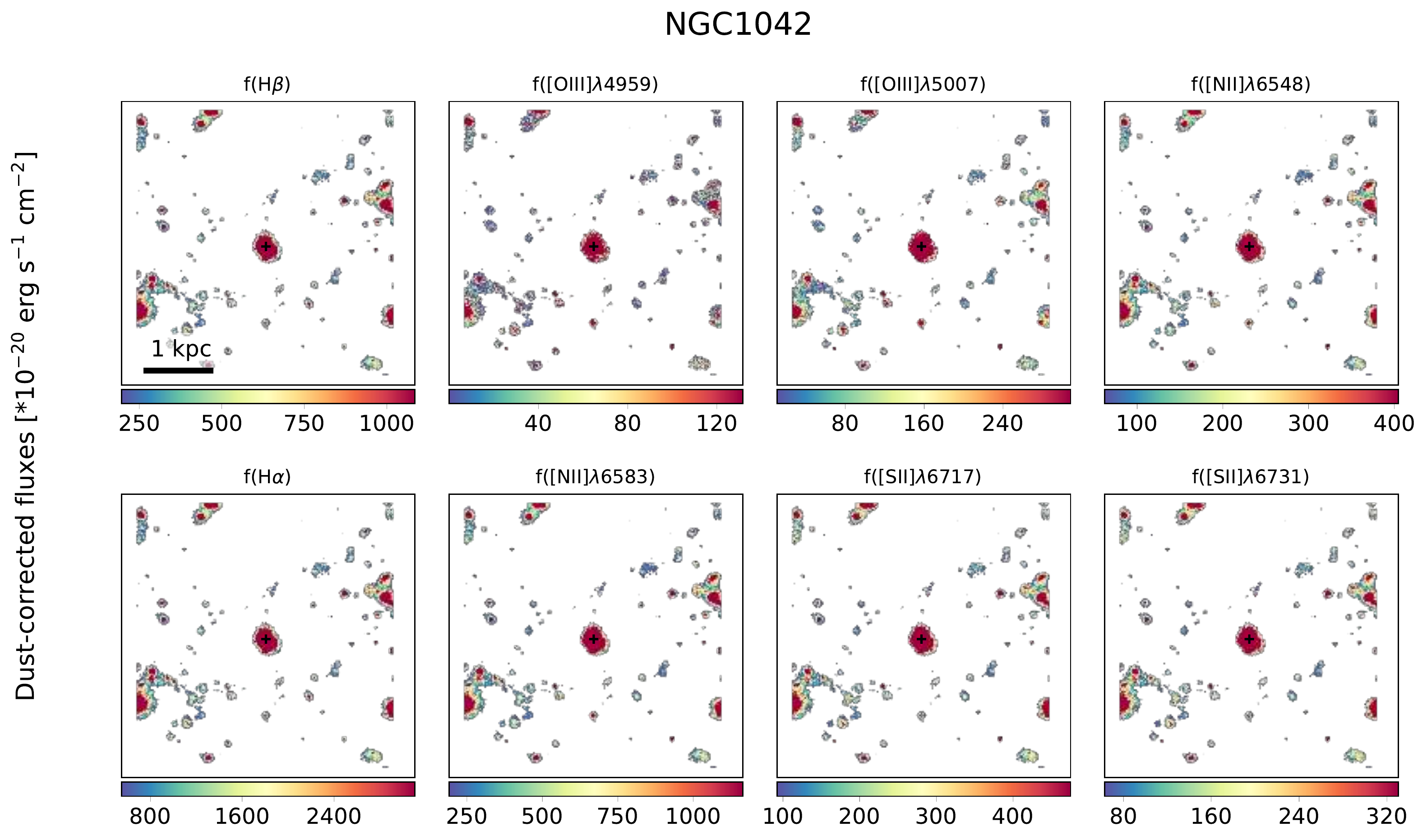}
% \caption{As Fig.~\ref{allfluxes} but for NGC~4030.}
% \label{ngc1042fluxes}
% \end{center}
% \end{figure*}

% \begin{figure*}
% \begin{center}
 \includegraphics[width=165mm]{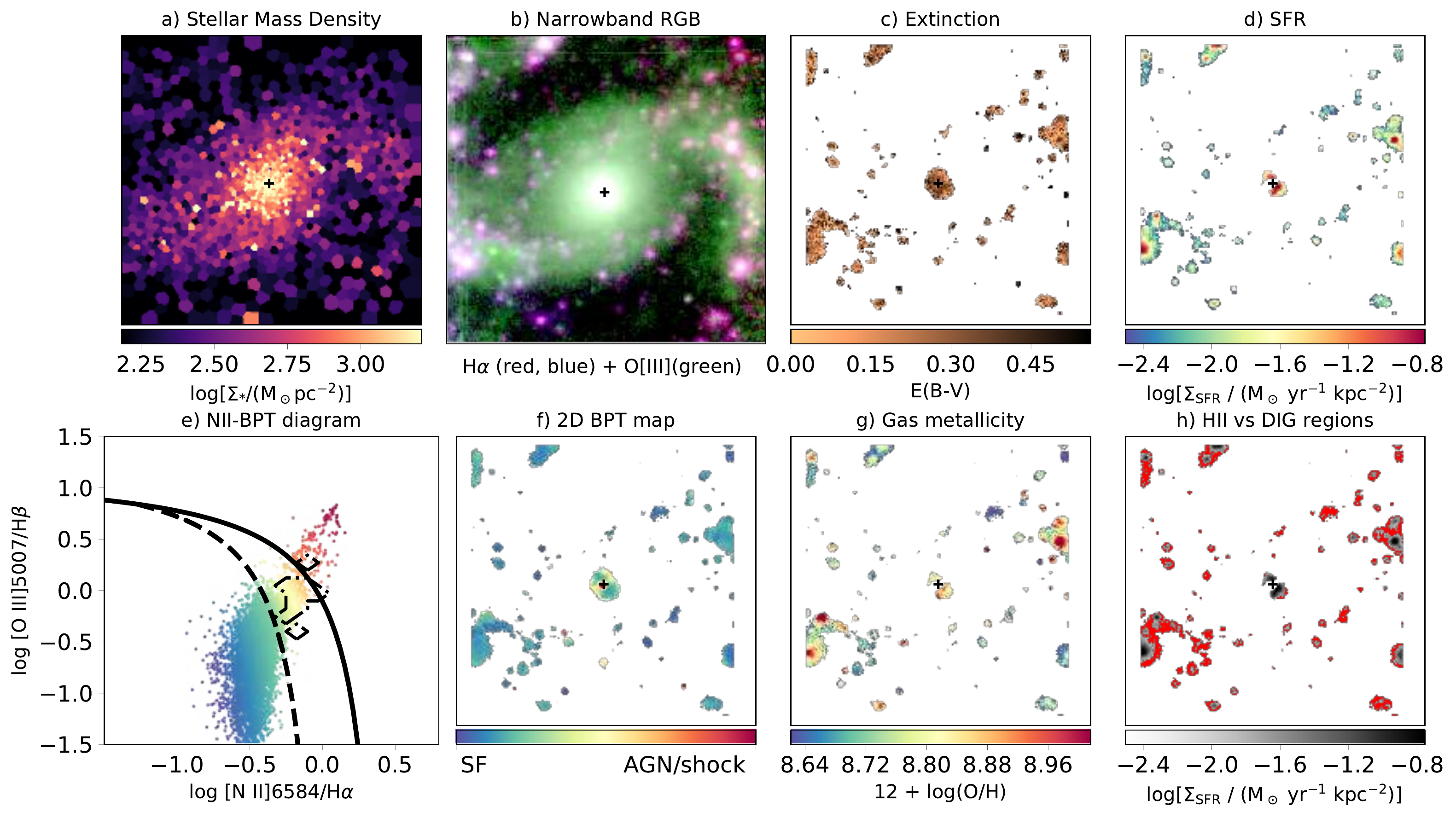}
\caption{The SF in NGC~1042 is mainly localized in the H{\sc ii} regions of the two, symmetric spiral arms which start at the ends of the bar. This bar is devoid of star formation. The central part shows some levels of star formation (traced by the SFR map and the BPT diagram), but with a bar-like structure which goes from NW to SE and is $\sim$200 pc long seen in the BPT diagram. This inner structure has high metallicity values and high SFR. The highest deviations from the linear fit are found in the H{\sc ii} regions at the beginning of the spiral arms ($\sim0.3$dex).
}
\label{ngc1042plots}
\end{center}
\end{figure*}
\clearpage

%\subsection{IC~5273}   

\begin{figure*}
\begin{center}
 \includegraphics[width=165mm]{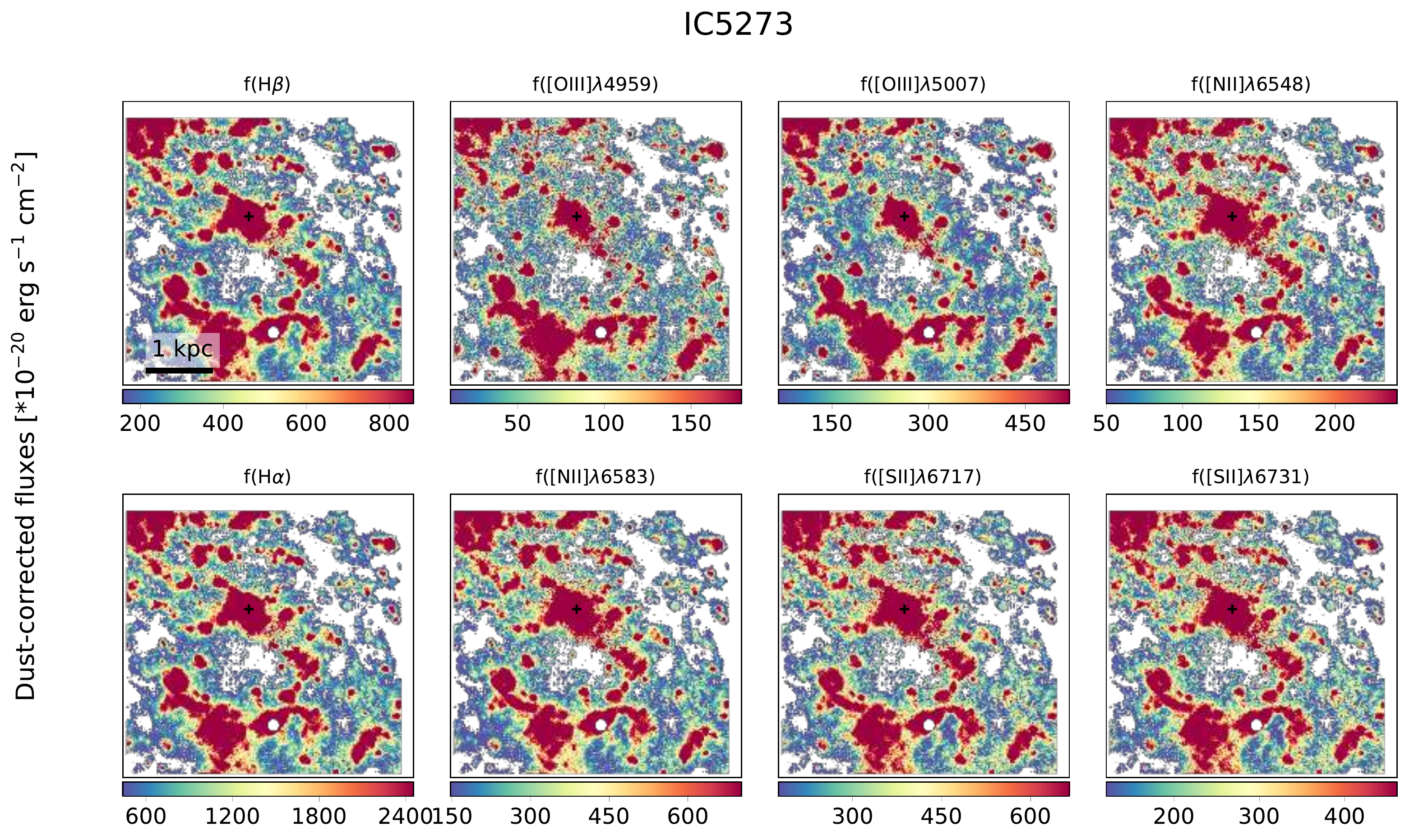}
% \caption{As Fig.~\ref{allfluxes} but for IC~4030.}
% \label{ic5273fluxes}
% \end{center}
% \end{figure*}

% \begin{figure*}
% \begin{center}
 \includegraphics[width=165mm]{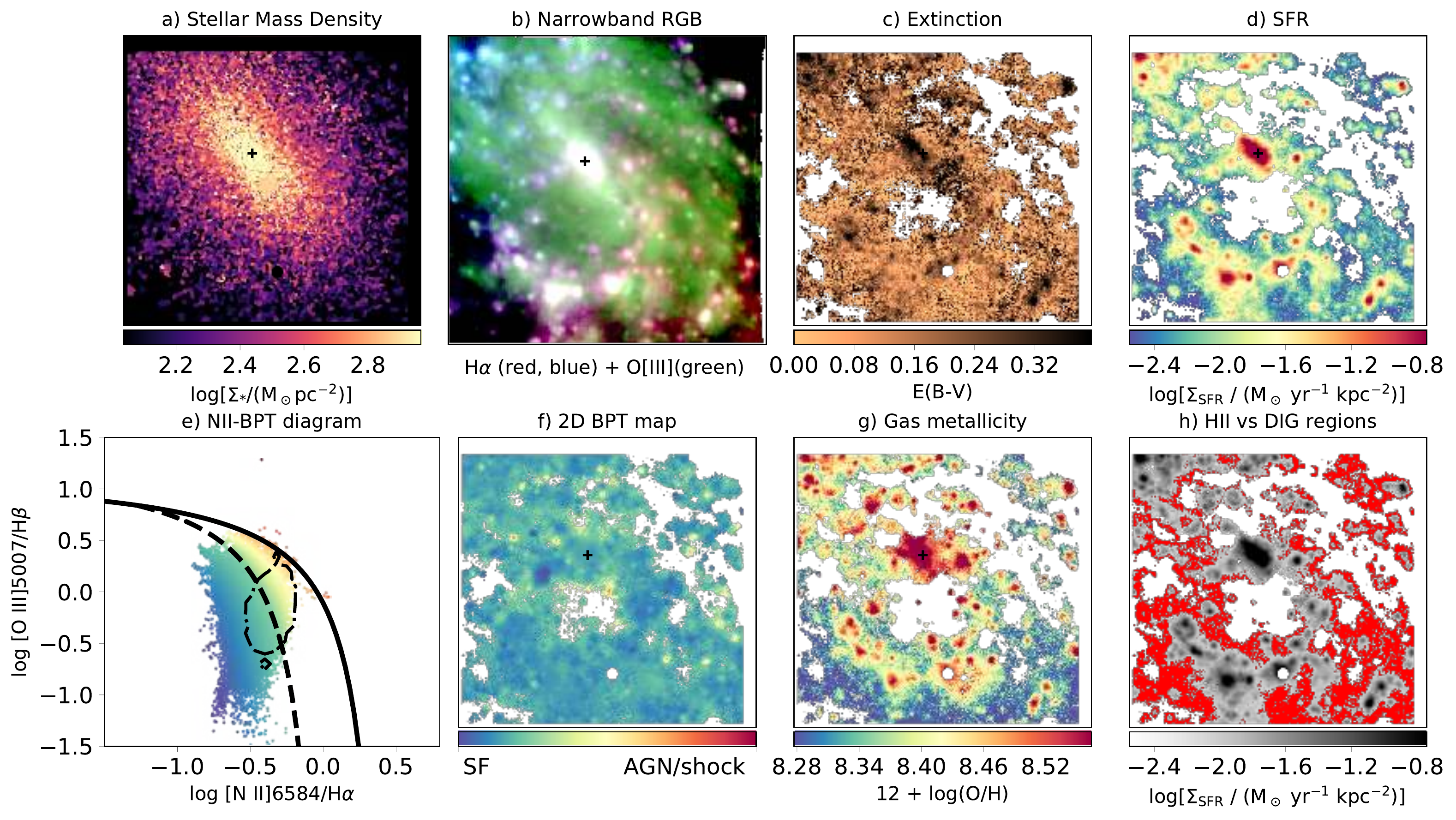}
\caption{IC~5273 is a barred galaxy that shows strong line emission. The main source of ionization is star formation, and the H{\sc ii} regions are located in the two spiral arms that start at the ends of the bar.
}
\label{ic5273plots}
\end{center}
\end{figure*}
\clearpage
%\subsection{NGC~1483}   

\begin{figure*}
\begin{center}
 \includegraphics[width=165mm]{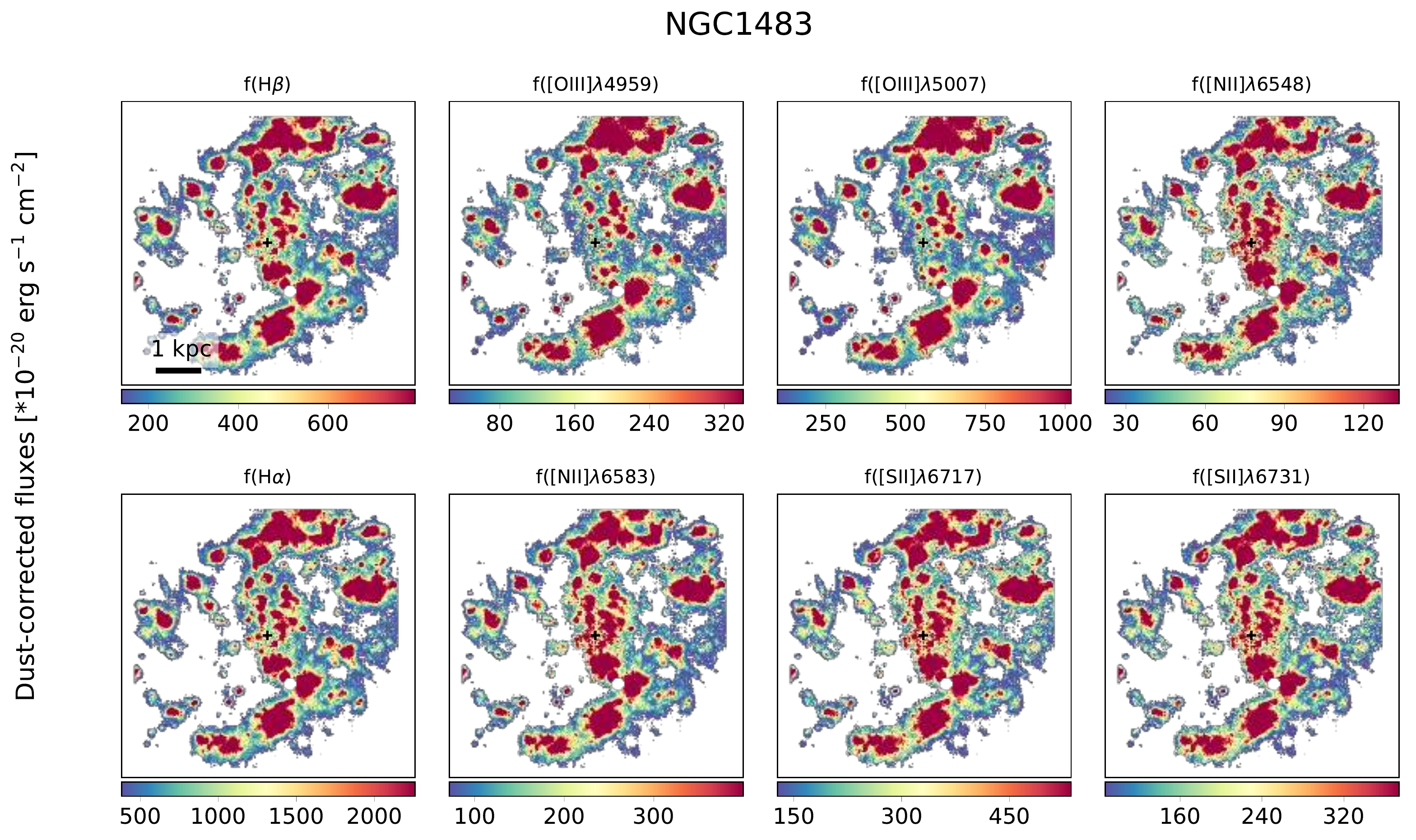}
% \caption{As Fig.~\ref{allfluxes} but for NGC~4030.}
% \label{ngc1483fluxes}
% \end{center}
% \end{figure*}

% \begin{figure*}
% \begin{center}
 \includegraphics[width=165mm]{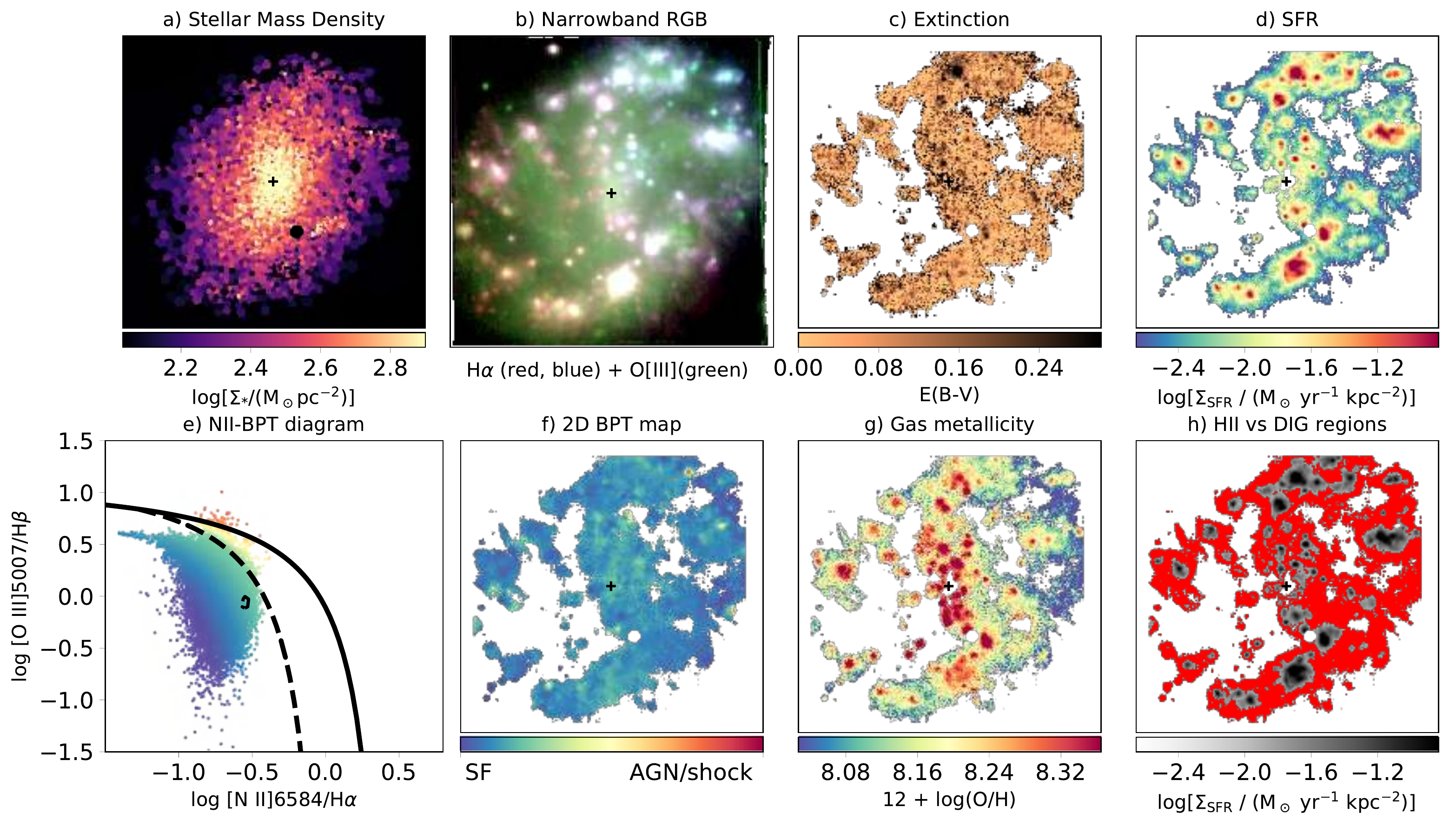}
\caption{This barred galaxy presents SF not only in the spiral arms, but also in H{\sc ii} regions located in the bar and bright H{\sc ii} regions outside the arms. Even with a strong bar, there are no distinct regions of shock-like ionization. The metallicity decreases with radius, being systematically higher for H{\sc ii} regions ($\sim$0.2 dex higher than the linear fit) than for the DIG ($\sim$0.2 dex lower than the linear fit) at all radii.
}
\label{ngc1483plots}
\end{center}
\end{figure*}
\clearpage
%\subsection{NGC~2835}   

\begin{figure*}
\begin{center}
 \includegraphics[width=165mm]{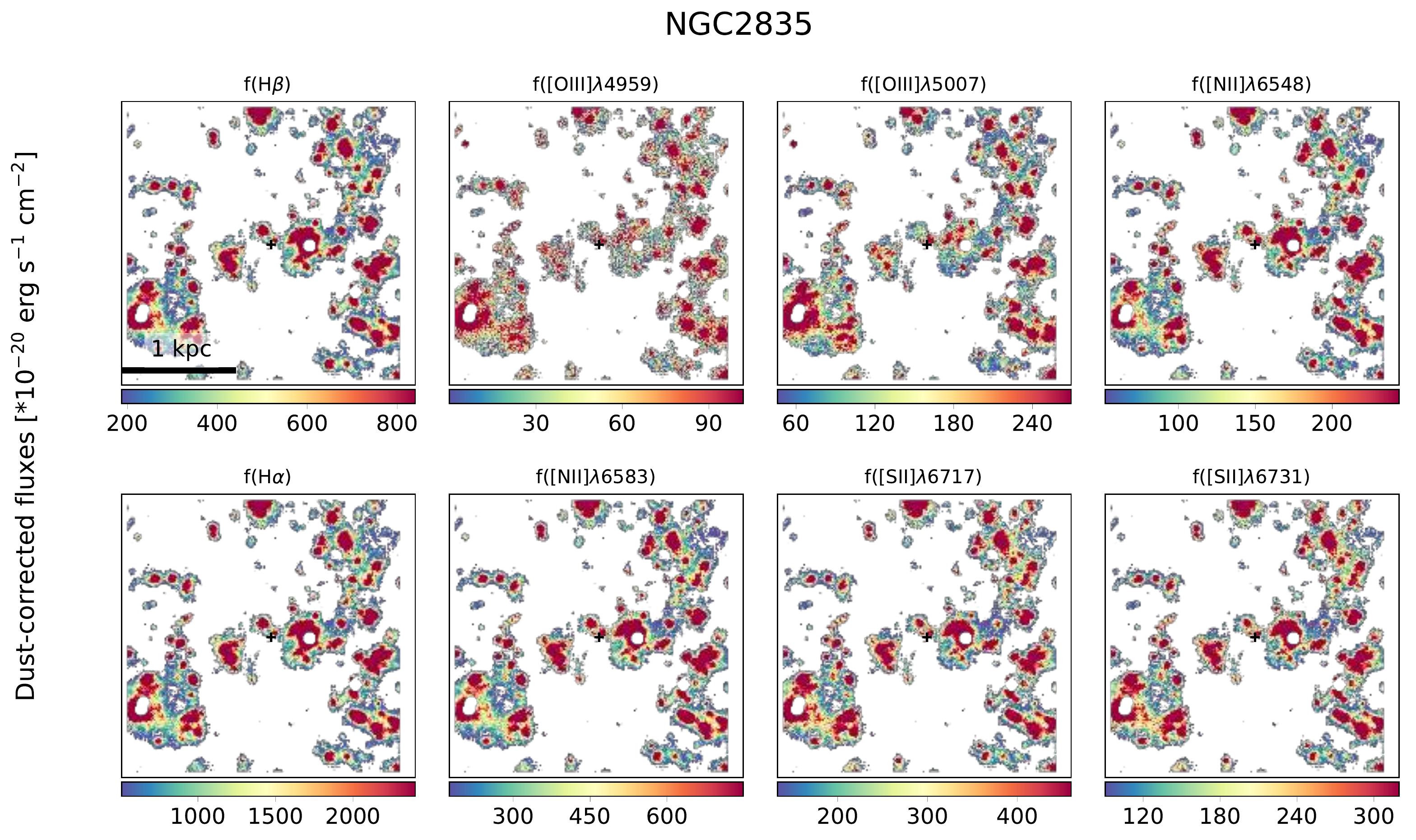}
% \caption{As Fig.~\ref{allfluxes} but for NGC~4030.}
% \label{ngc2835fluxes}
% \end{center}
% \end{figure*}

% \begin{figure*}
% \begin{center}
 \includegraphics[width=165mm]{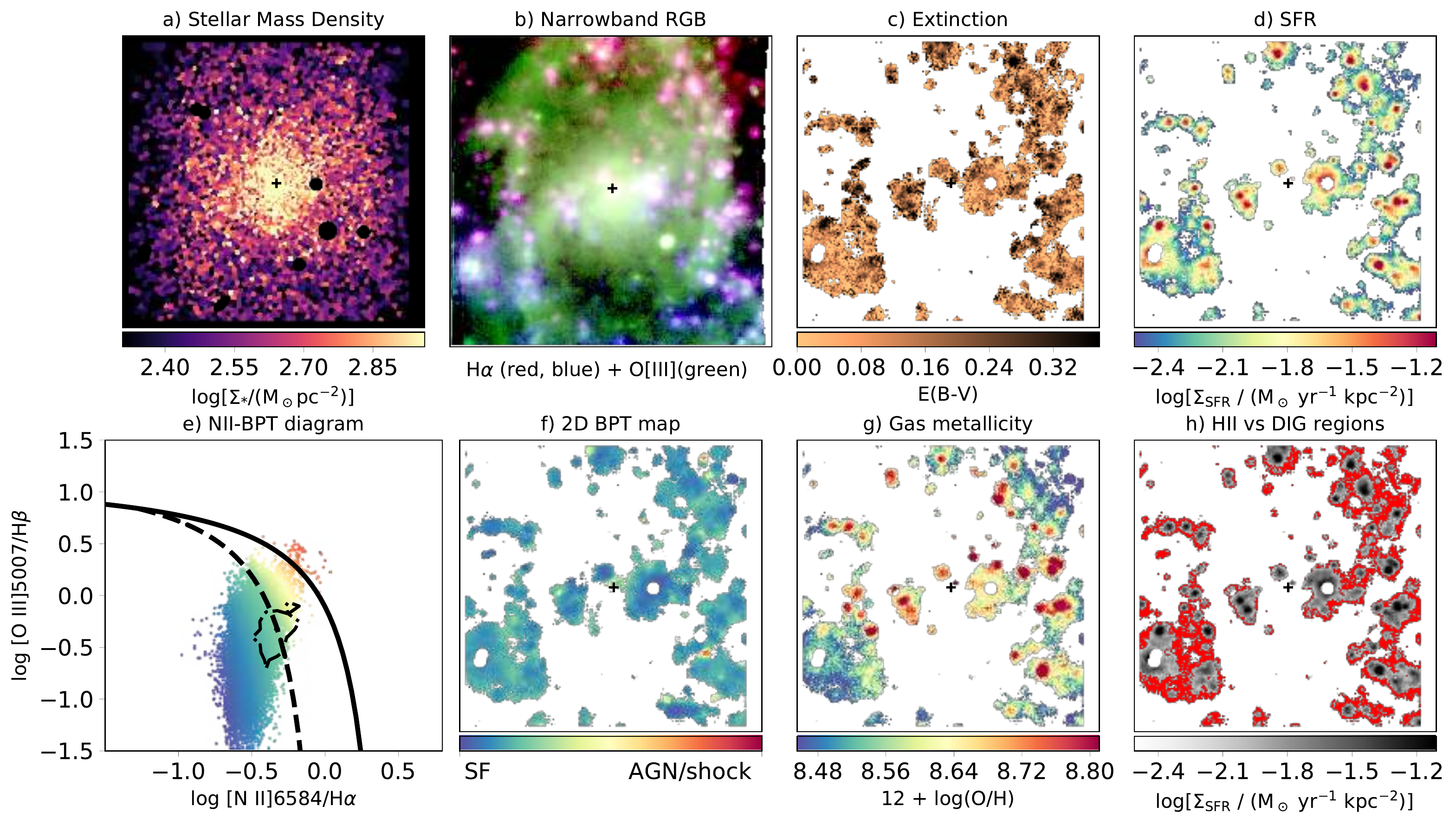}
\caption{Although it is not clear  from the stellar mass density map whether NGC~2835 has a bar, the two main spiral arms start at the ends of a bar-like structure formed by H{\sc ii} regions and higher extinction. However, this bar-like structure has a star-forming ionization. The central part does not show strong emission, meaning that there is little ionized gas in those regions. The metallicity is generally declining towards the outer parts. However, the metallicity in the outer H{\sc ii} regions is $\sim$0.3 dex higher than the linear fit.
}
\label{ngc2835plots}
\end{center}
\end{figure*}
\clearpage
%\subsection{PGC~3853}   

\begin{figure*}
\begin{center}
 \includegraphics[width=165mm]{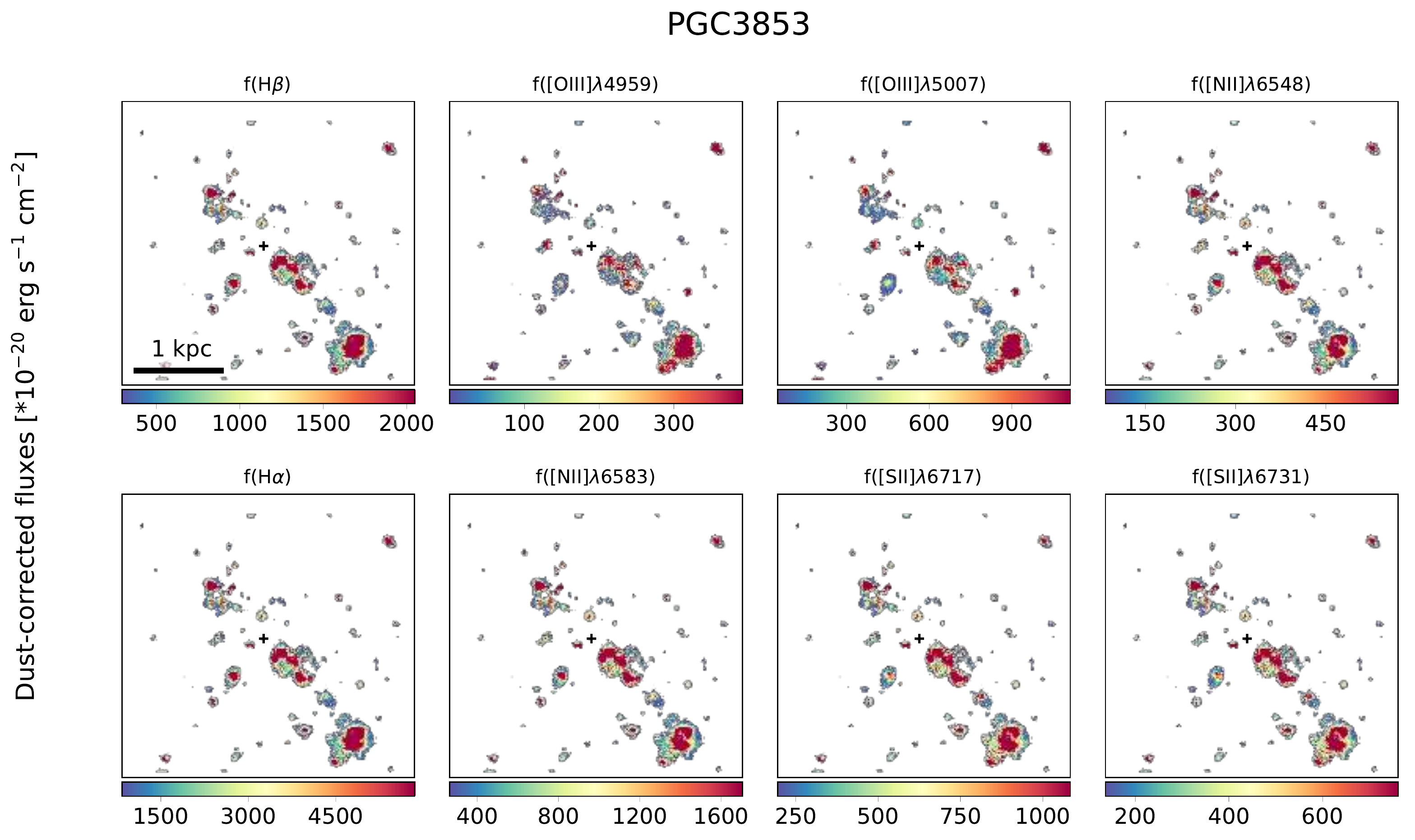}
% \caption{As Fig.~\ref{allfluxes} but for NGC~4030.}
% \label{pgc3853fluxes}
% \end{center}
% \end{figure*}

% \begin{figure*}
% \begin{center}
 \includegraphics[width=165mm]{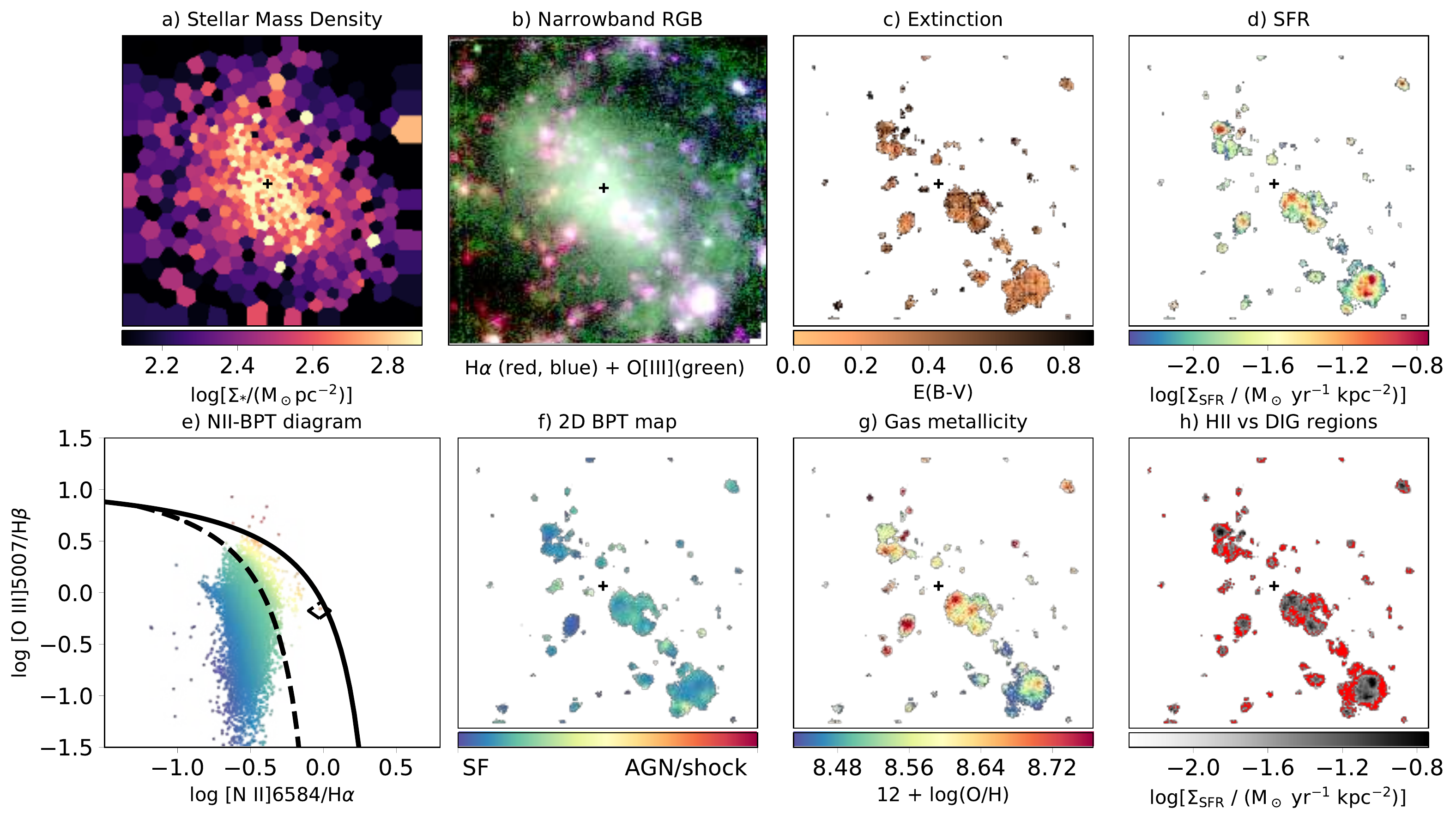}
\caption{Most of the SF in this barred galaxy is found in a structure formed by H{\sc ii} regions which is misaligned by 10 degrees from the position angle of the bar. The gas metallicity declines with radius, and shows higher values in the H{\sc ii} regions than the DIG. The largest deviations from the linear fit are found in the H{\sc ii} regions ($\sim$0.2 dex).
}
\label{pgc3853plots}
\end{center}
\end{figure*}
\clearpage
%\subsection{NGC~337 }   

\begin{figure*}
\begin{center}
 \includegraphics[width=165mm]{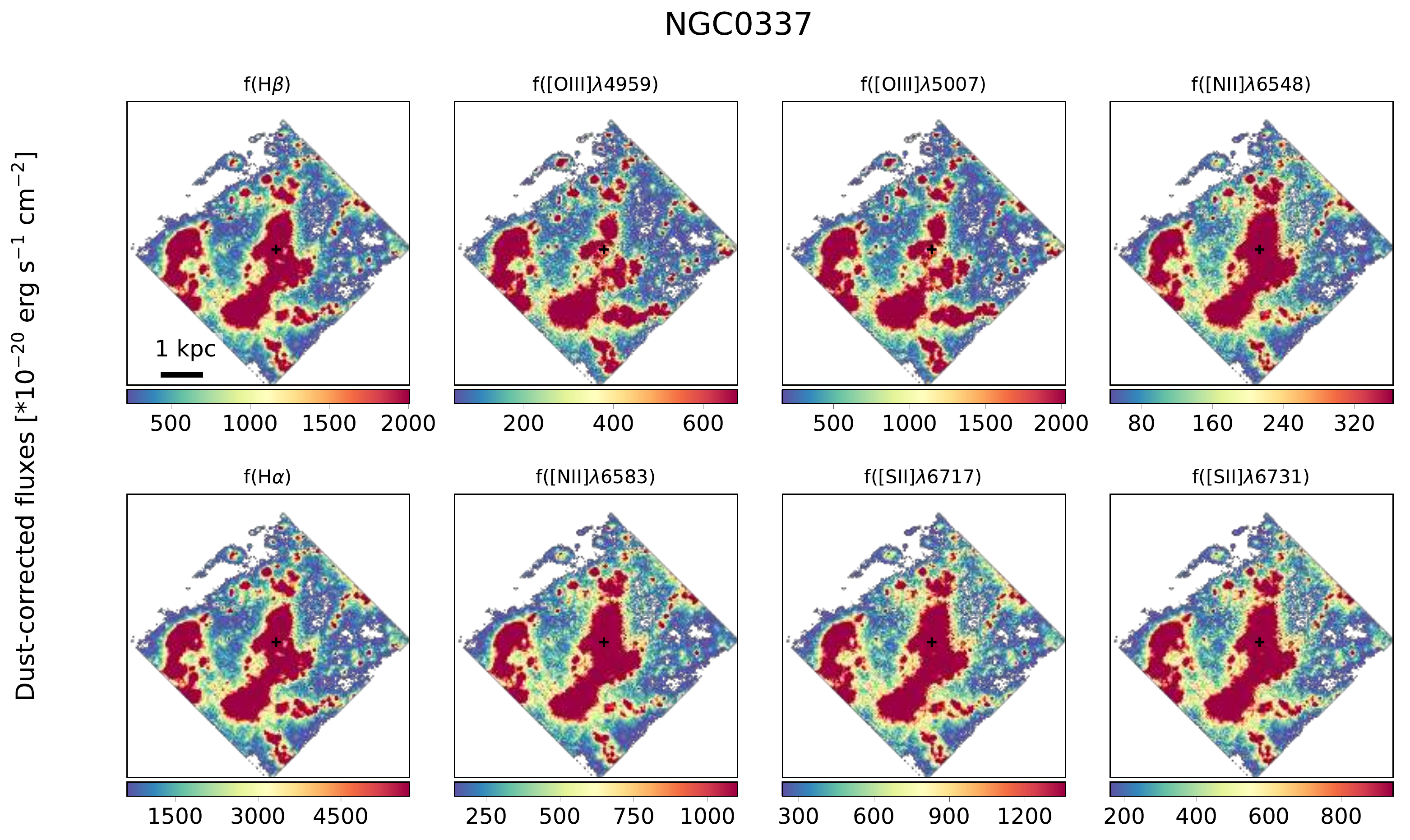}
% \caption{As Fig.~\ref{allfluxes} but for NGC~4030.}
% \label{ngc0337fluxes}
% \end{center}
% \end{figure*}

% \begin{figure*}
% \begin{center}
 \includegraphics[width=165mm]{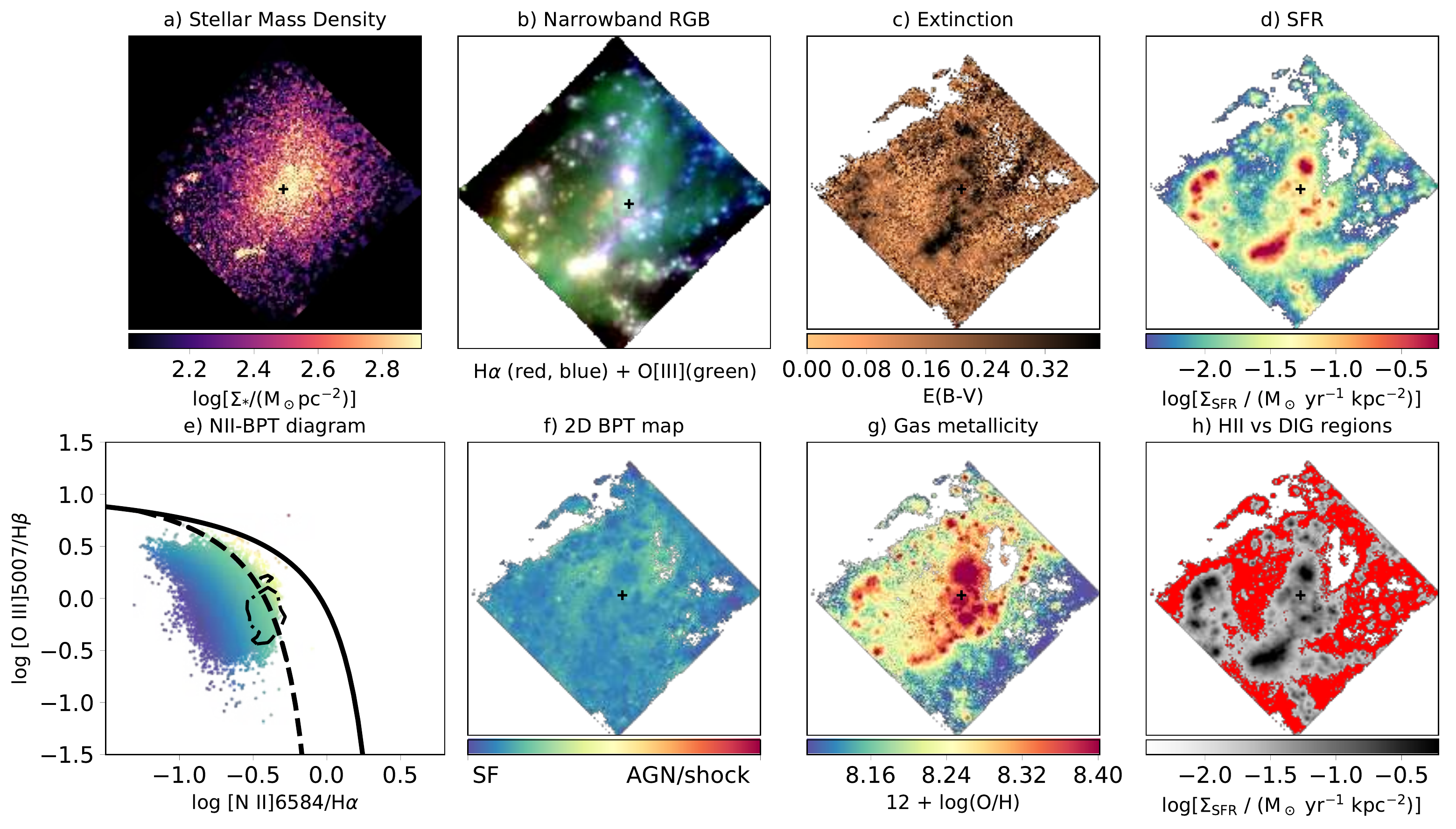}
\caption{The SFR is higher than the average for its total mass, and the spaxels lie in the SF area of the BPT diagram. The SF is localized in large H{\sc ii} regions situated in the spiral arms and also in the stellar bar. The spiral structure is fragmentary and flocculent. The gas metallicity map shows that the H{\sc ii} regions in the bar present higher metallicity than the H{\sc ii} regions outside the bar and the DIG. The largest deviation ($\sim$0.3 dex) from the linear fit are found in the bright H{\sc ii} region $\sim$0.6 kpc NW from the centre. Apart from that, the centres of the H{\sc ii} regions outside the bar deviate around $\sim$0.2 dex from the linear fit.
}
\label{ngc0337plots}
\end{center}
\end{figure*}
\clearpage
%\subsection{NGC~4592}   

\begin{figure*}
\begin{center}
 \includegraphics[width=165mm]{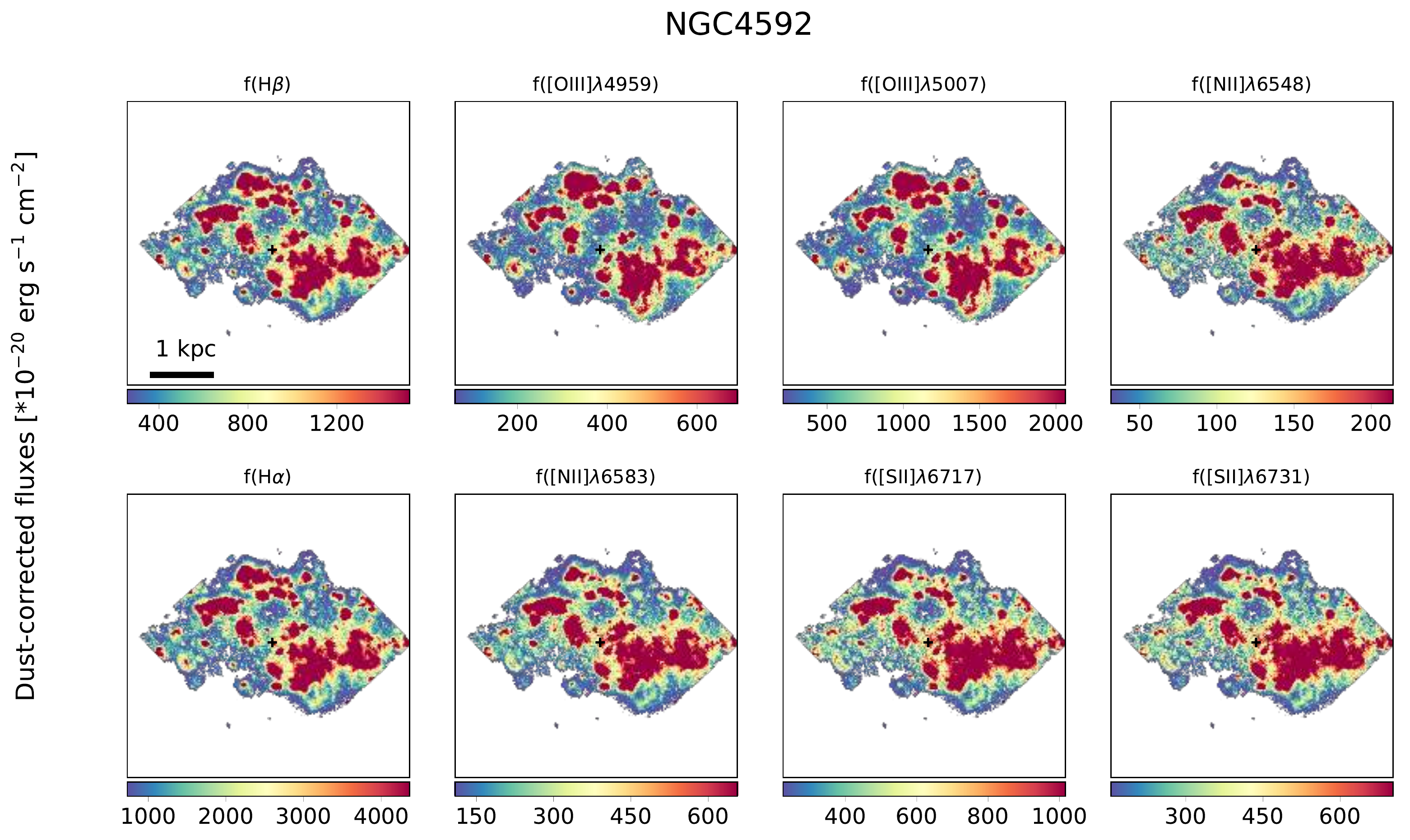}
% \caption{As Fig.~\ref{allfluxes} but for NGC~4030.}
% \label{ngc4592fluxes}
% \end{center}
% \end{figure*}

% \begin{figure*}
% \begin{center}
 \includegraphics[width=165mm]{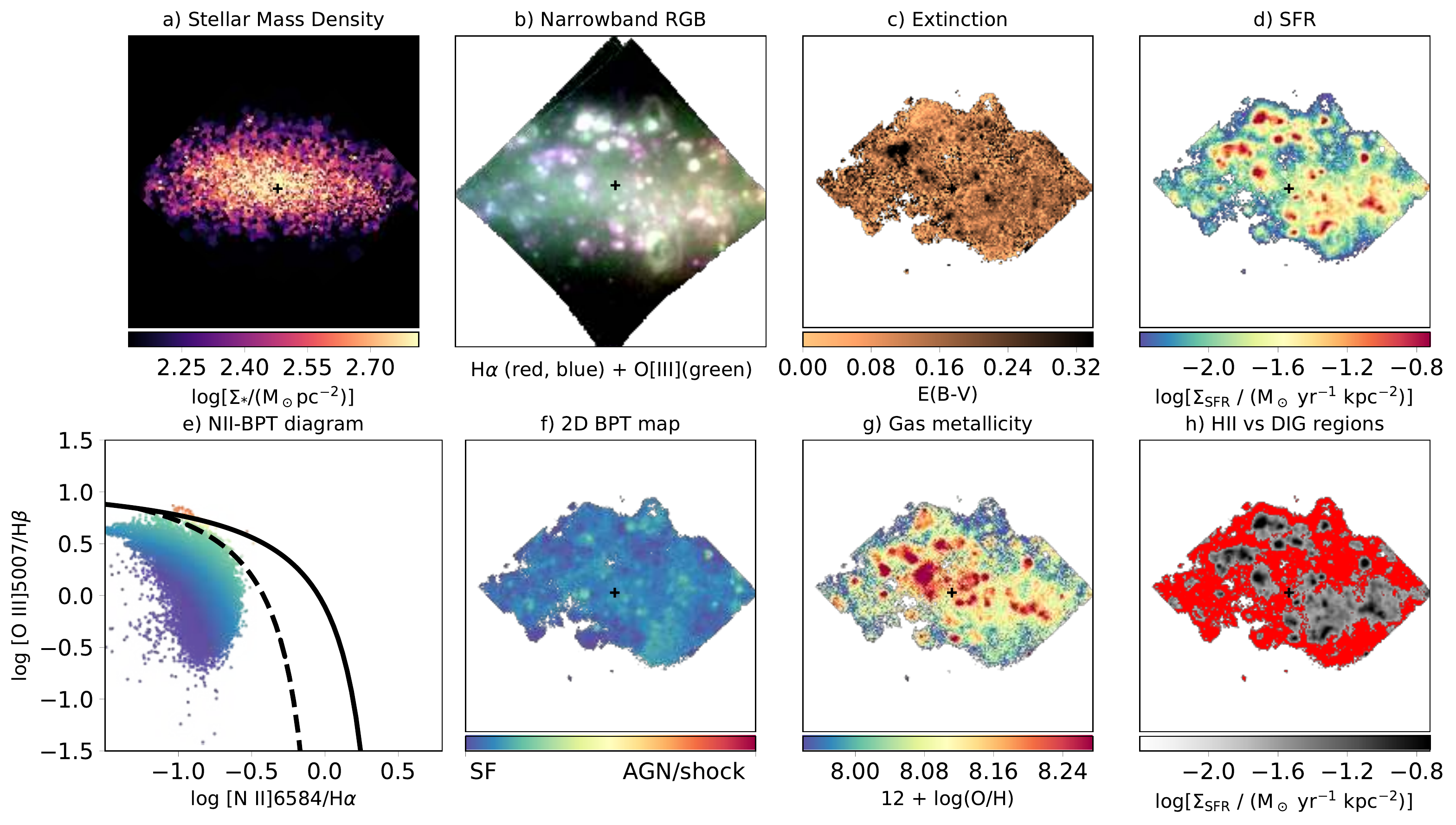}
\caption{This unbarred low mass galaxy is quite inclined, presenting many H{\sc ii} regions which do not form any spiral structure. Although the BPT diagram shows almost only SF, the gas metallicity map shows different substructure, with some H{\sc ii} regions which are metal-enriched in the central parts compared with those in the outer parts. The metallicity of both the H{\sc ii} regions and the DIG decrease with the same gradient. However, the absolute value is $\sim0.1$ dex higher for the H{\sc ii} region than for the DIG.
}
\label{ngc4592plots}
\end{center}
\end{figure*}
\clearpage
%\subsection{NGC~4790}   

\begin{figure*}
\begin{center}
 \includegraphics[width=165mm]{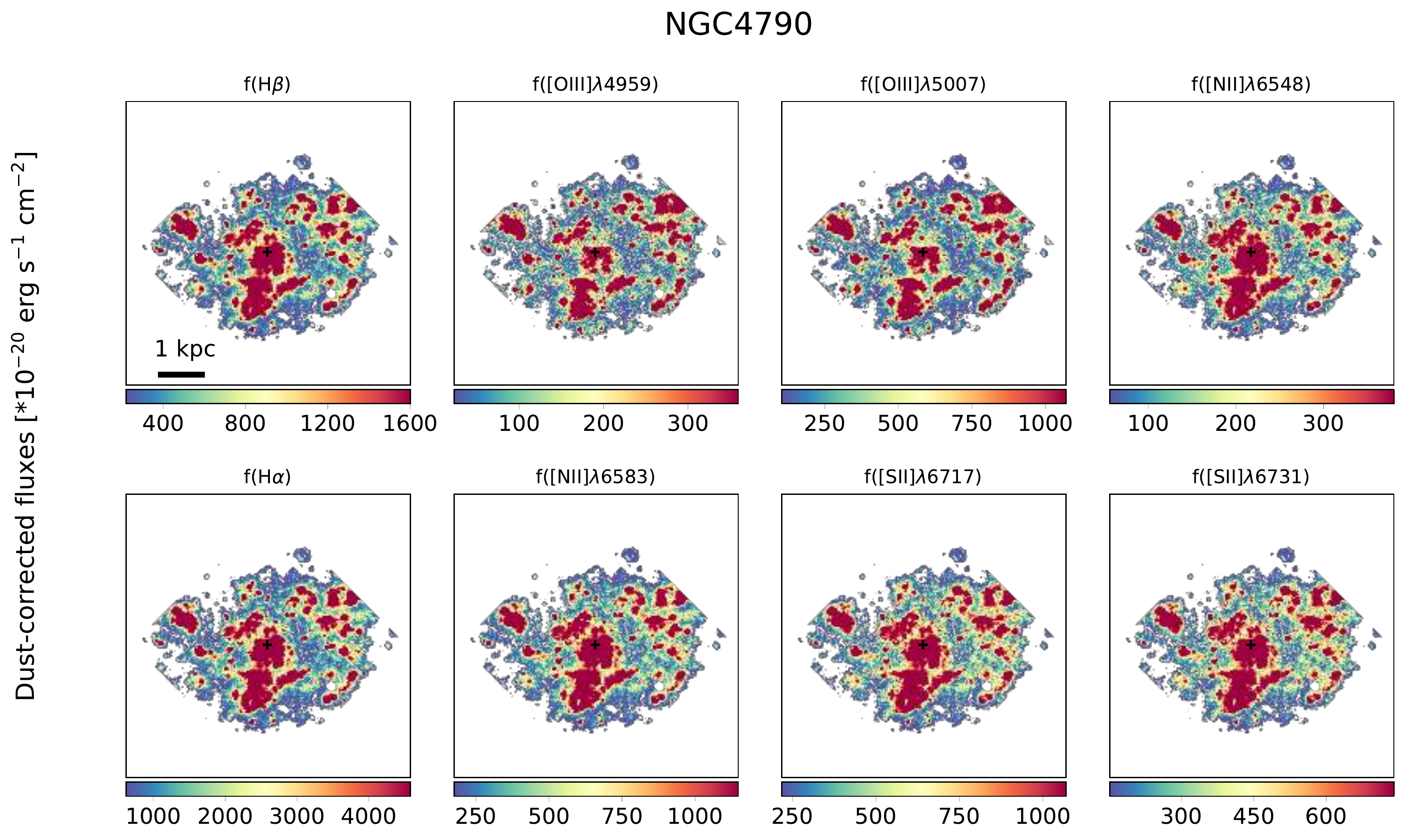}
% \caption{As Fig.~\ref{allfluxes} but for NGC~4030.}
% \label{ngc4790fluxes}
% \end{center}
% \end{figure*}

% \begin{figure*}
% \begin{center}
 \includegraphics[width=165mm]{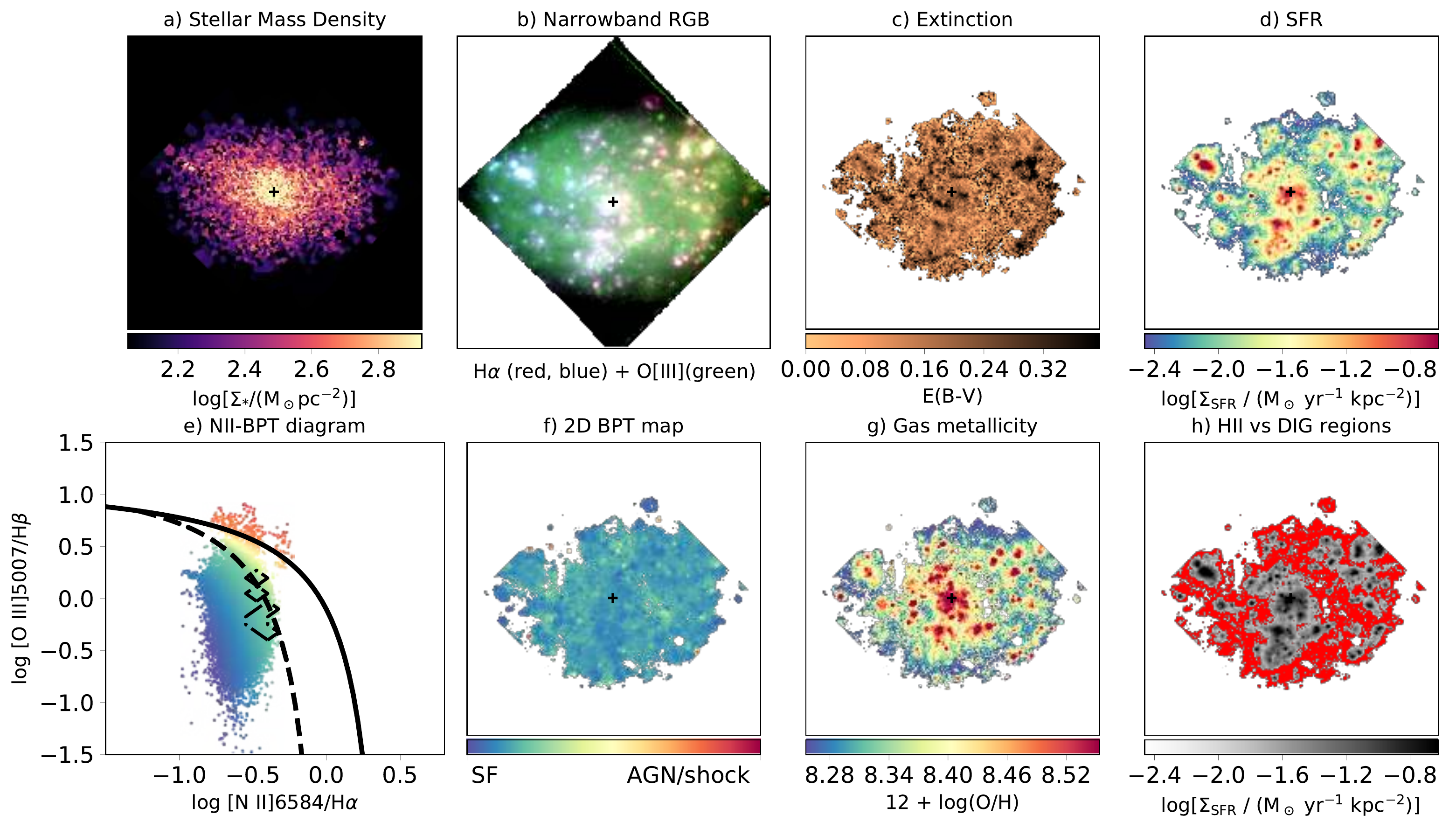}
\caption{NGC~4790 is a galaxy with moderate inclination which is dominated by star formation ionization (BPT diagram). There are spiral arms or fragments, also traced by the dust lanes. The H$ \alpha $ emission is rather asymmetric, although the gas metallicity is symmetric. The metallicities decrease with galactocentric distance, both for H{\sc ii} and for the DIG, with the same gradient. However, the metallicities are $\sim$0.1 dex higher for the H{\sc ii} regions than for the DIG.
}
\label{ngc4790plots}
\end{center}
\end{figure*}
\clearpage
%\subsection{NGC~3513}   

\begin{figure*}
\begin{center}
 \includegraphics[width=165mm]{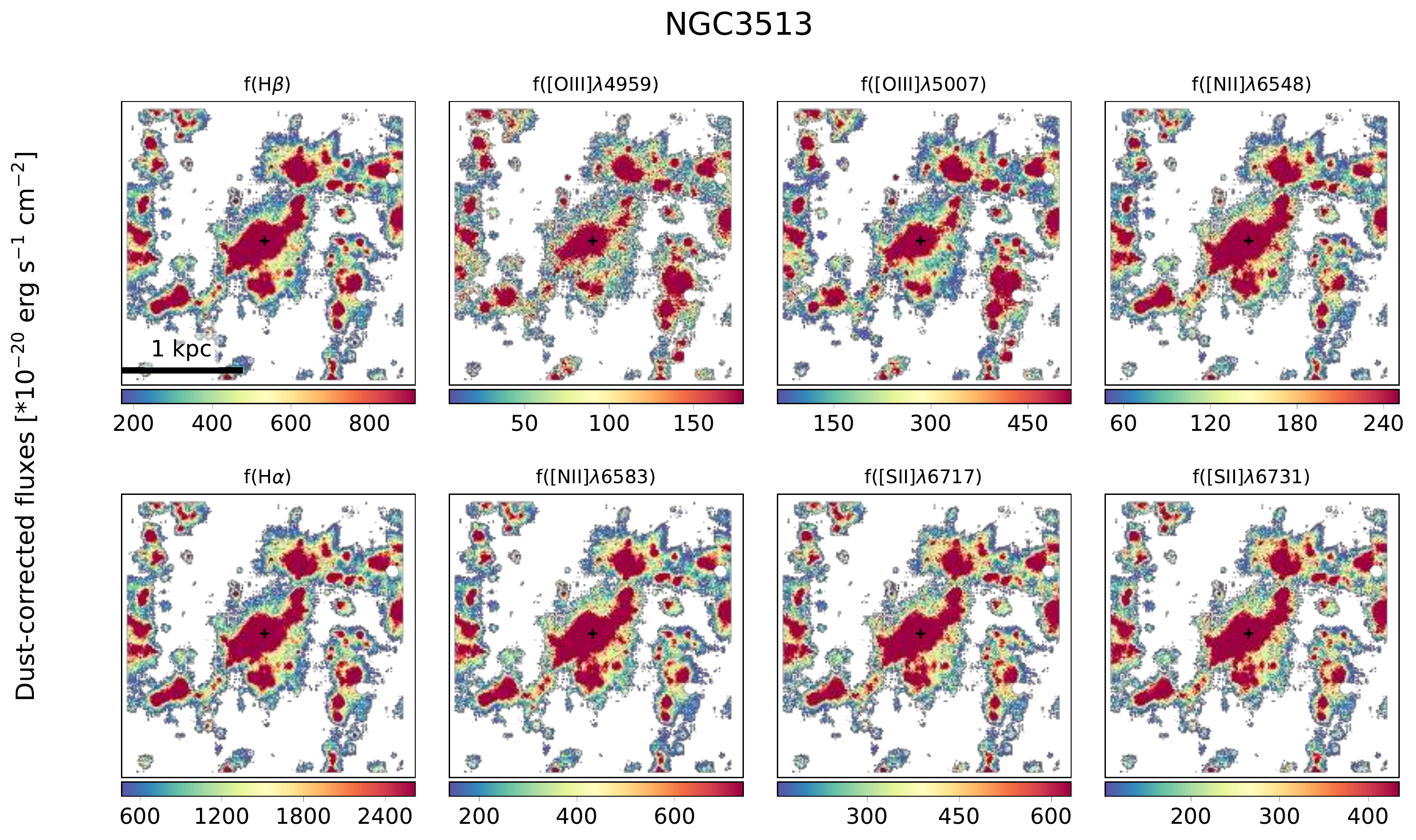}
% \caption{As Fig.~\ref{allfluxes} but for NGC~4030.}
% \label{ngc3513fluxes}
% \end{center}
% \end{figure*}

% \begin{figure*}
% \begin{center}
 \includegraphics[width=165mm]{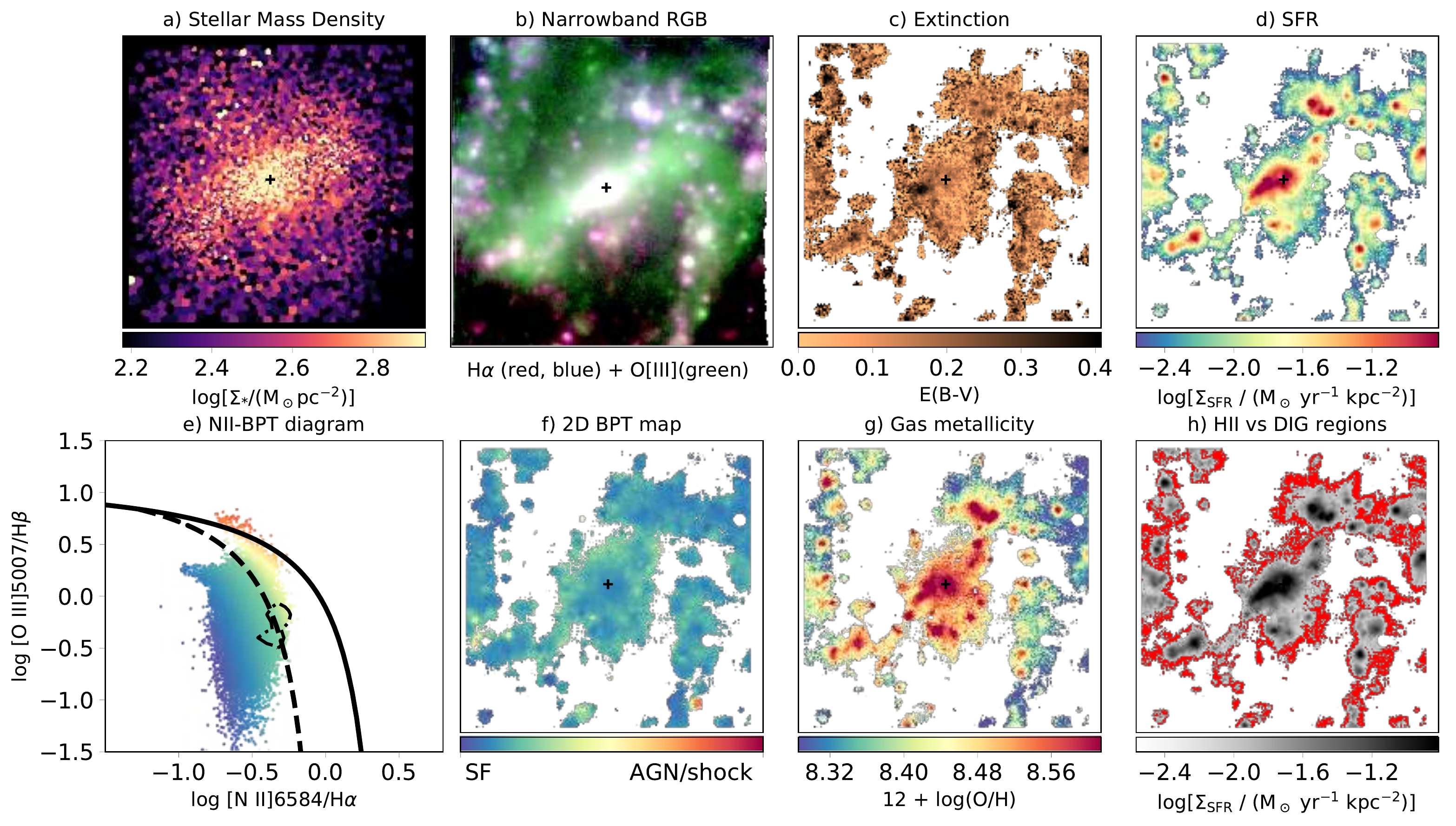}
\caption{NGC~3513 shows a bright bar that ends in two well defined spiral arms. The bar shows levels of star formation in some H{\sc ii} regions located at both parts of the bar. As for NGC~2104, there is a region in the southern part that has AGN/shock ionization ratios, and is due to a very high level of [O{\sc iii}] emission without Hydrogen counterparts. The metallicity is higher in the H{\sc ii} regions in the centre, decreasing with radius although the values in the outer H{\sc ii} regions deviate from the linear fit by $\sim$0.2 dex.
}
\label{ngc3513plots}
\end{center}
\end{figure*}
\clearpage
%\subsection{NGC~2104}   

\begin{figure*}
\begin{center}
 \includegraphics[width=165mm]{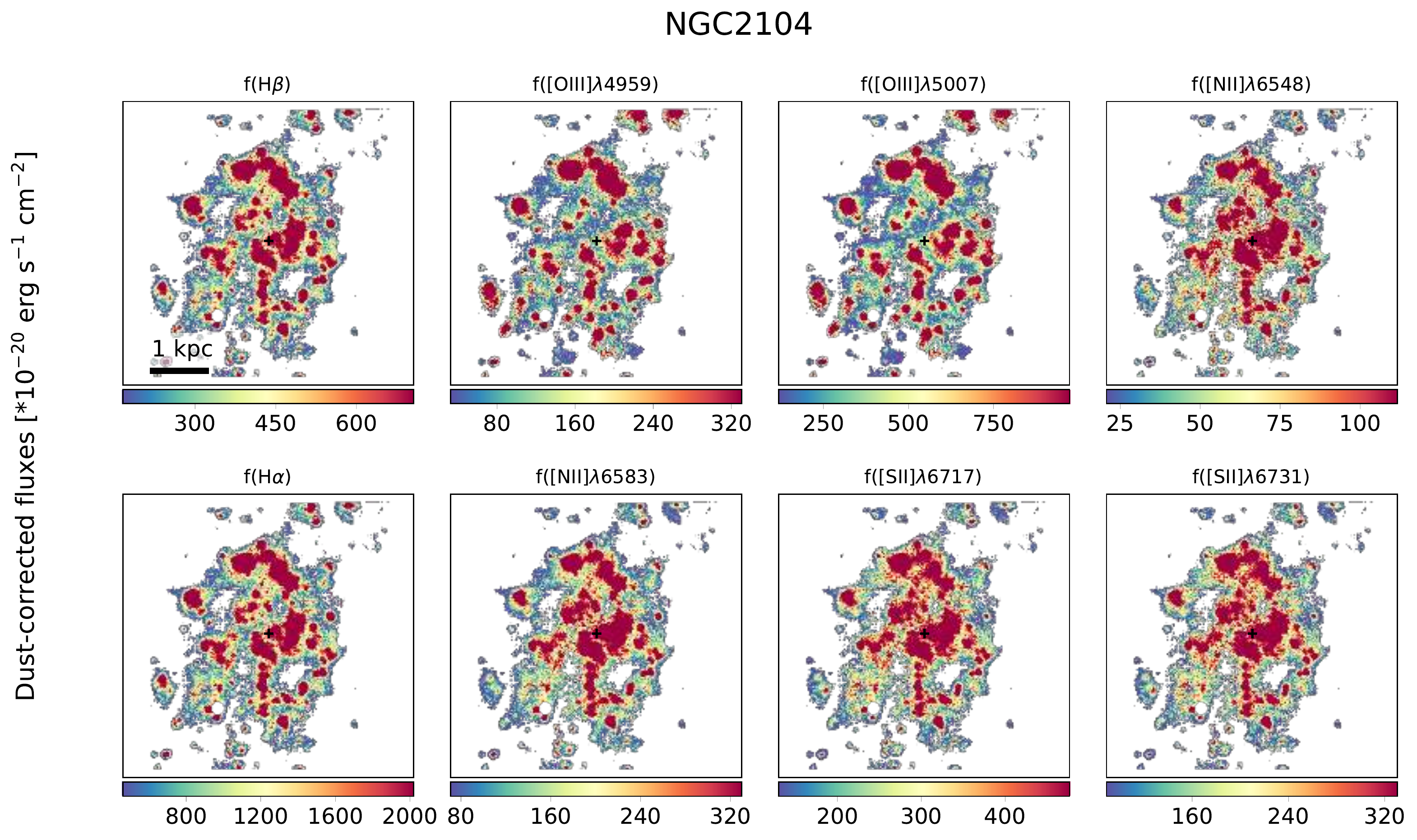}
% \caption{As Fig.~\ref{allfluxes} but for NGC~4030.}
% \label{ngc2104fluxes}
% \end{center}
% \end{figure*}

% \begin{figure*}
% \begin{center}
 \includegraphics[width=165mm]{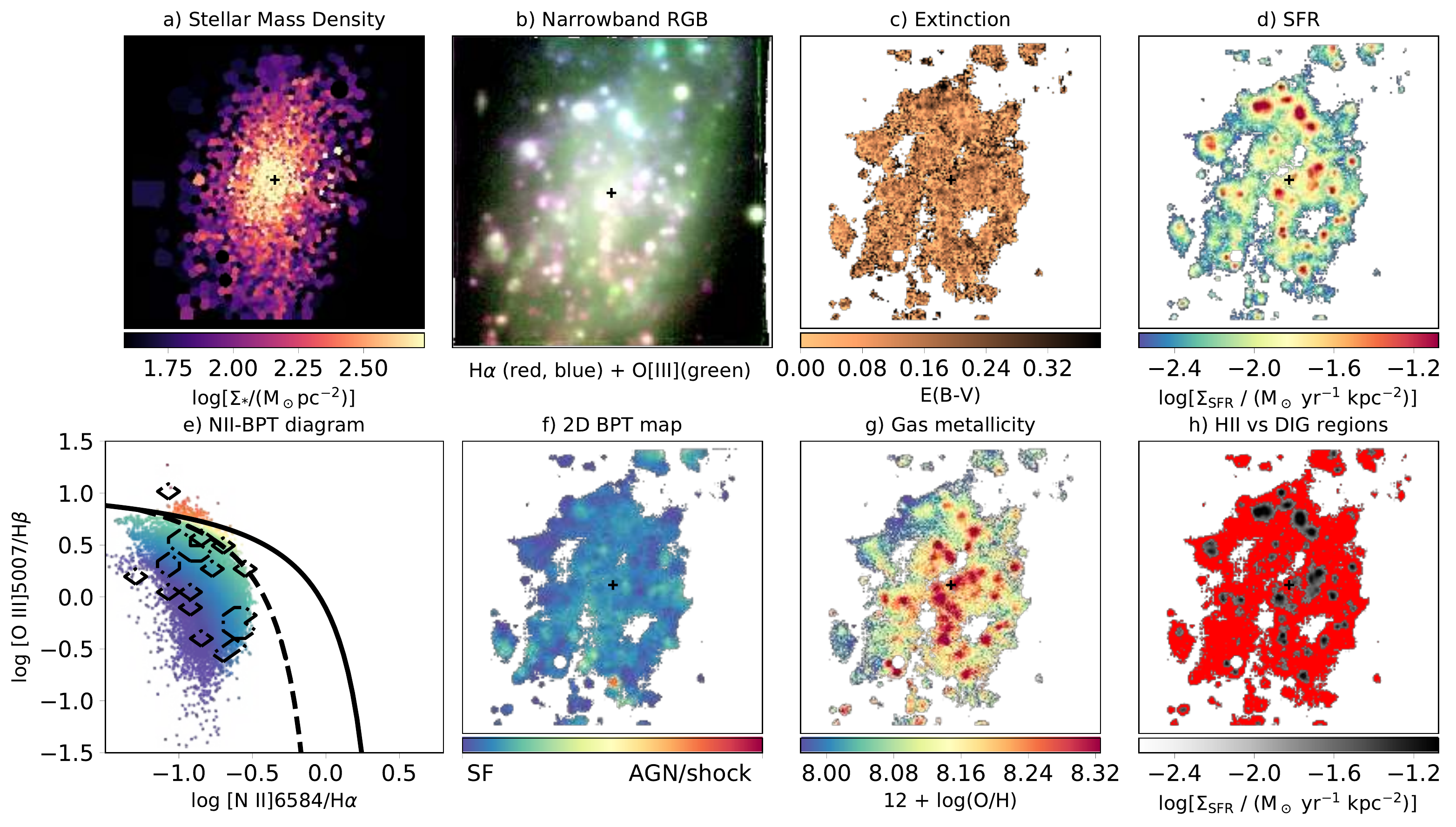}
\caption{This unbarred galaxy does not show a well-defined spiral structure. The H{\sc ii} regions are present throughout the disk, without any well-defined structure. The disk is dominated by star-forming ionization, except for a region in the southern part, with enhanced [O{\sc iii}]$ \lambda $ emission without Balmer lines emission counterparts. The metallicity of the H{\sc ii} regions decreases with radius, having higher values for the H{\sc ii} regions than the diffuse gas.
}
\label{ngc2104plots}
\end{center}
\end{figure*}
\clearpage
%\subsection{NGC~4980}   

\begin{figure*}
\begin{center}
 \includegraphics[width=165mm]{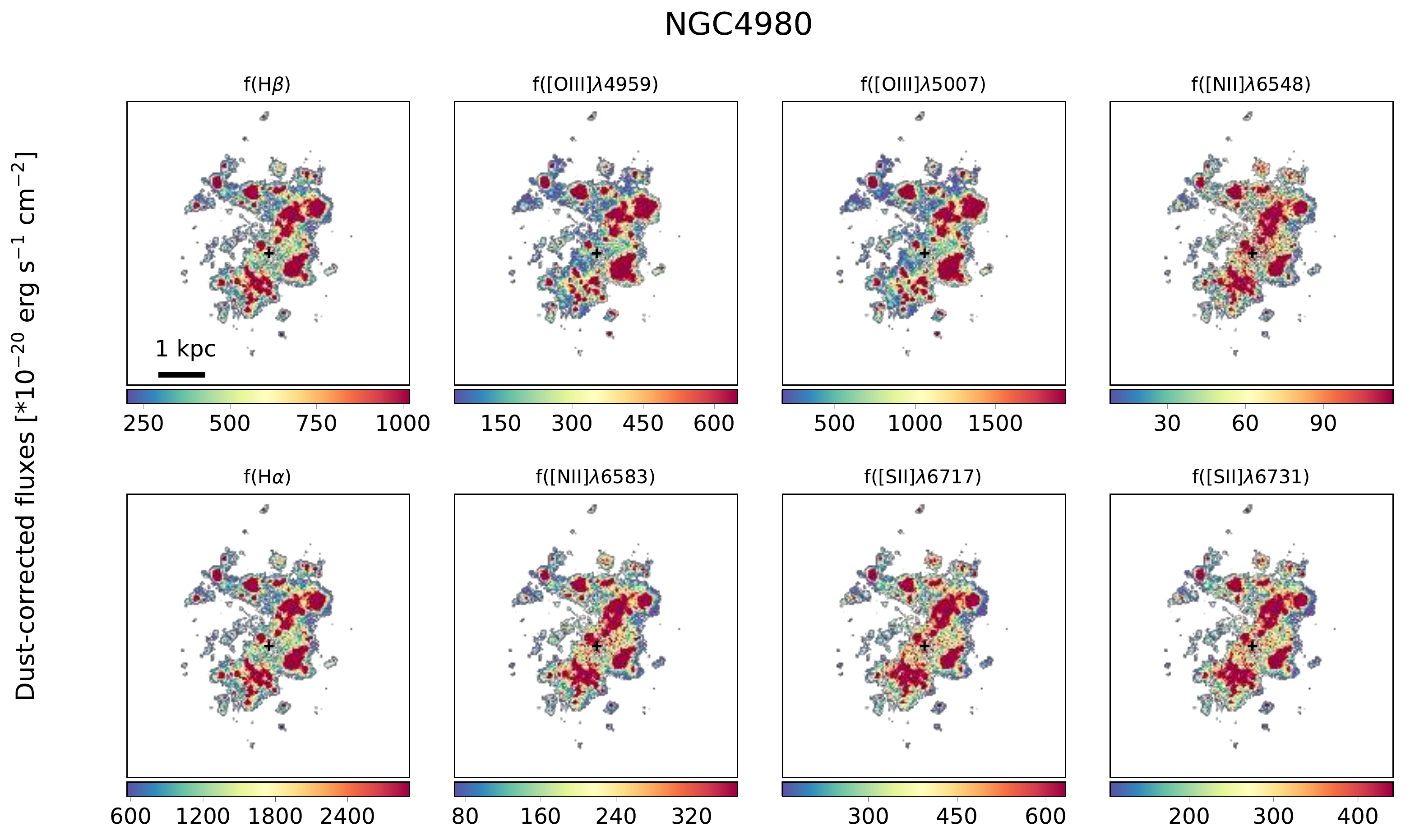}
% \caption{As Fig.~\ref{allfluxes} but for NGC~4030.}
% \label{ngc4980fluxes}
% \end{center}
% \end{figure*}

% \begin{figure*}
% \begin{center}
 \includegraphics[width=165mm]{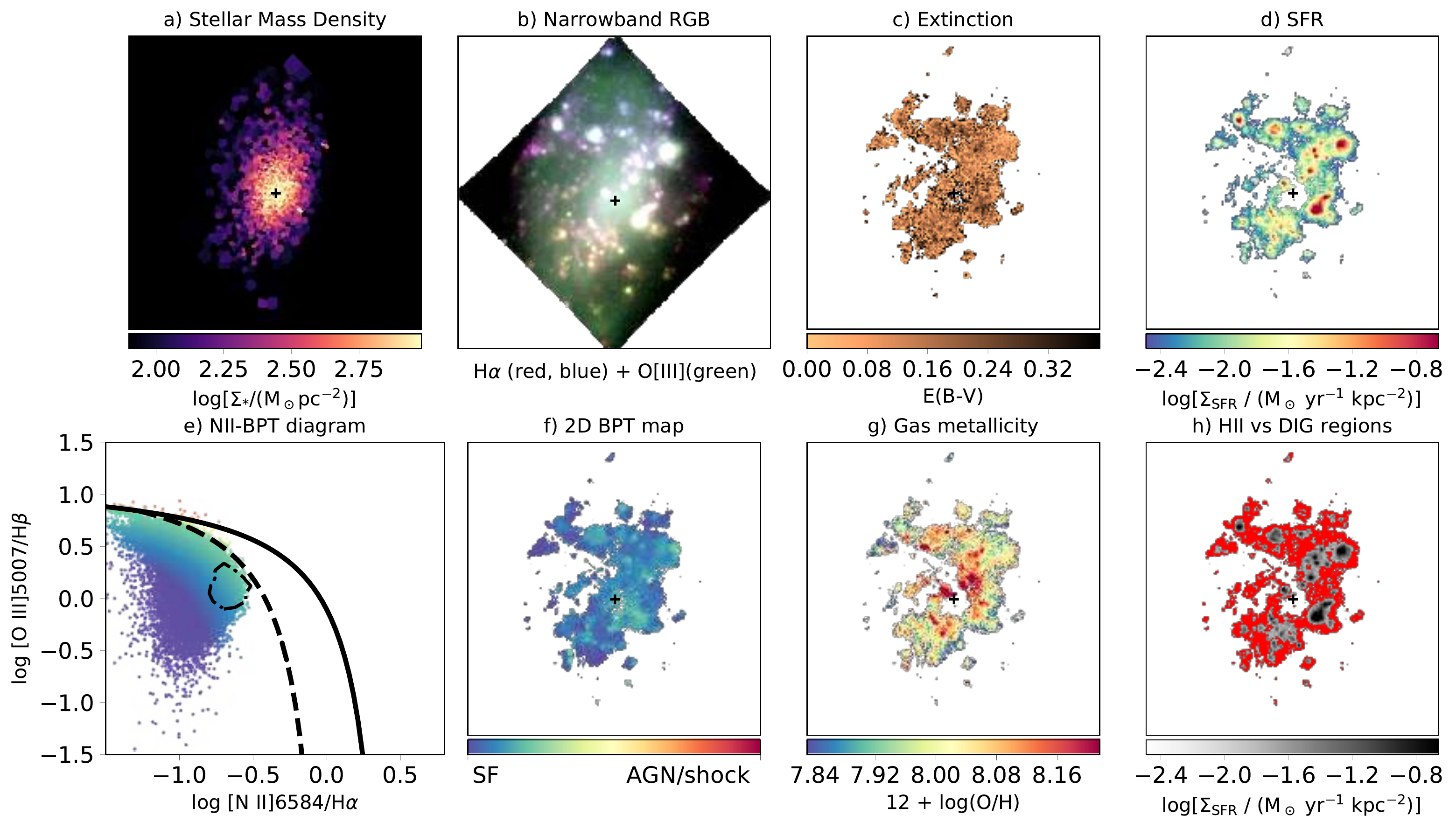}
\caption{The emission of NGC~4980 is asymmetric, in the 3.6 $ \mu $m image from \textit{Spitzer} (from the S$ ^{4} $G survey, \citealt{Sheth2010}), the nuclear and the bar emission is not located in the centre of the galaxy. In other words, there is an offset bar. There is an enhancement in the emission of the eastern part as seen in our images, followed by there is a huge drop in stellar mass and also in the emission of the other lines in the western part. The gas diagnostics also reflect this difference. The western side of the galaxy is metal enriched compared to the eastern part.  The asymmetries in the stellar and gas components may be signs of interaction with other galaxies, although the closest identified galaxy is ESO~444-G2, with a difference in recessional velocity of $ \Delta v\approx210 $km~s$ ^{-1} $ and $ \Delta m_{B}\approx1.43 $mag, which would be in the limit for considering an interacting companion according to \citet{Knapen2014}. The ionization source of this galaxy is star formation. The metallicity map does not clearly correlate with the SFR map: the central regions have higher metallicities, whereas the H$\alpha$ intensity is not decreasing with radius. The largest deviations from the linear fit are in the H{\sc ii} regions in the outer part ($\sim$0.25 dex).
}
\label{ngc4980plots}
\end{center}
\end{figure*}
\clearpage
%\subsection{NGC~4517A}  

\begin{figure*}
\begin{center}
 \includegraphics[width=165mm]{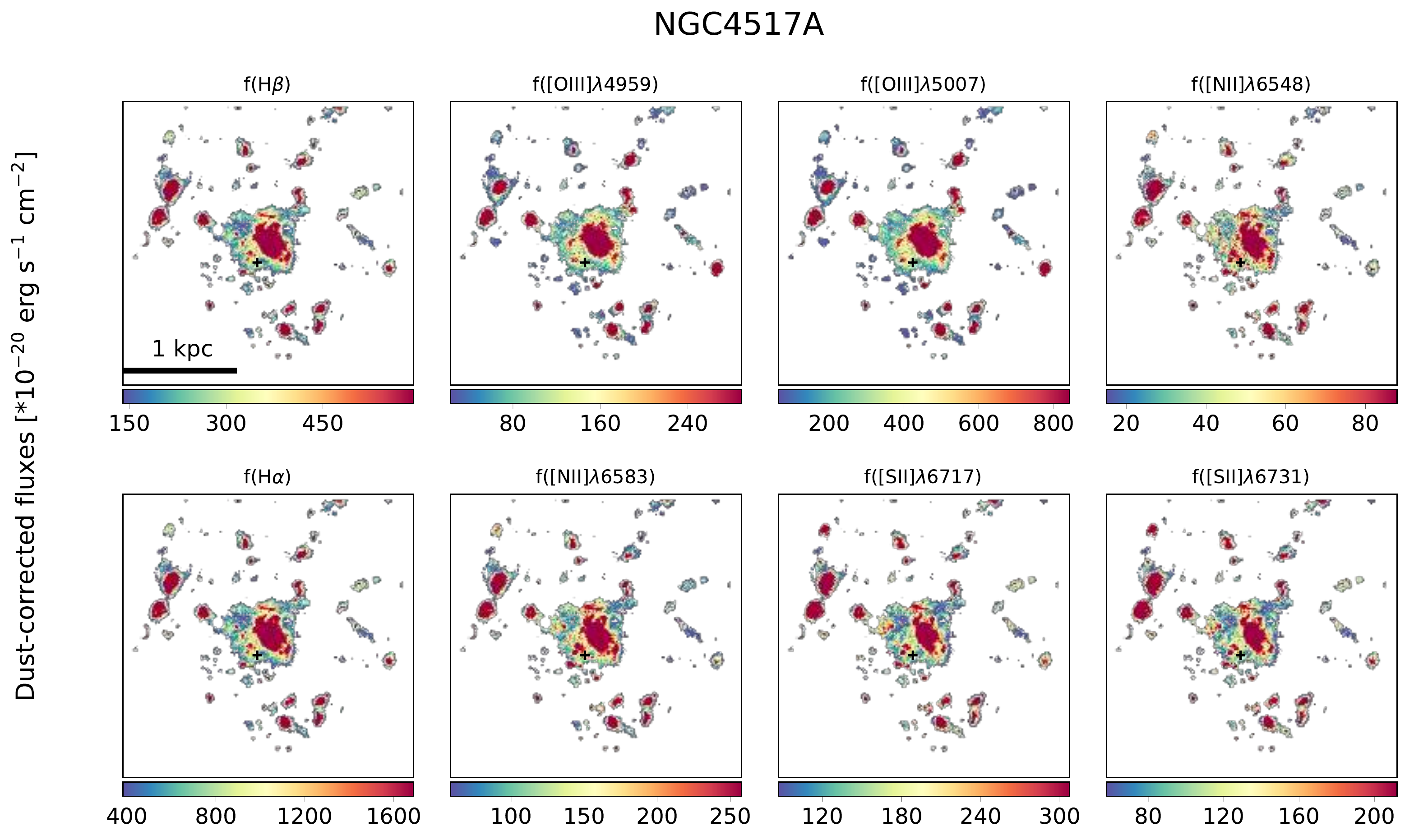}
% \caption{As Fig.~\ref{allfluxes} but for NGC~4030.}
% \label{ngc4517afluxes}
% \end{center}
% \end{figure*}

% \begin{figure*}
% \begin{center}
 \includegraphics[width=165mm]{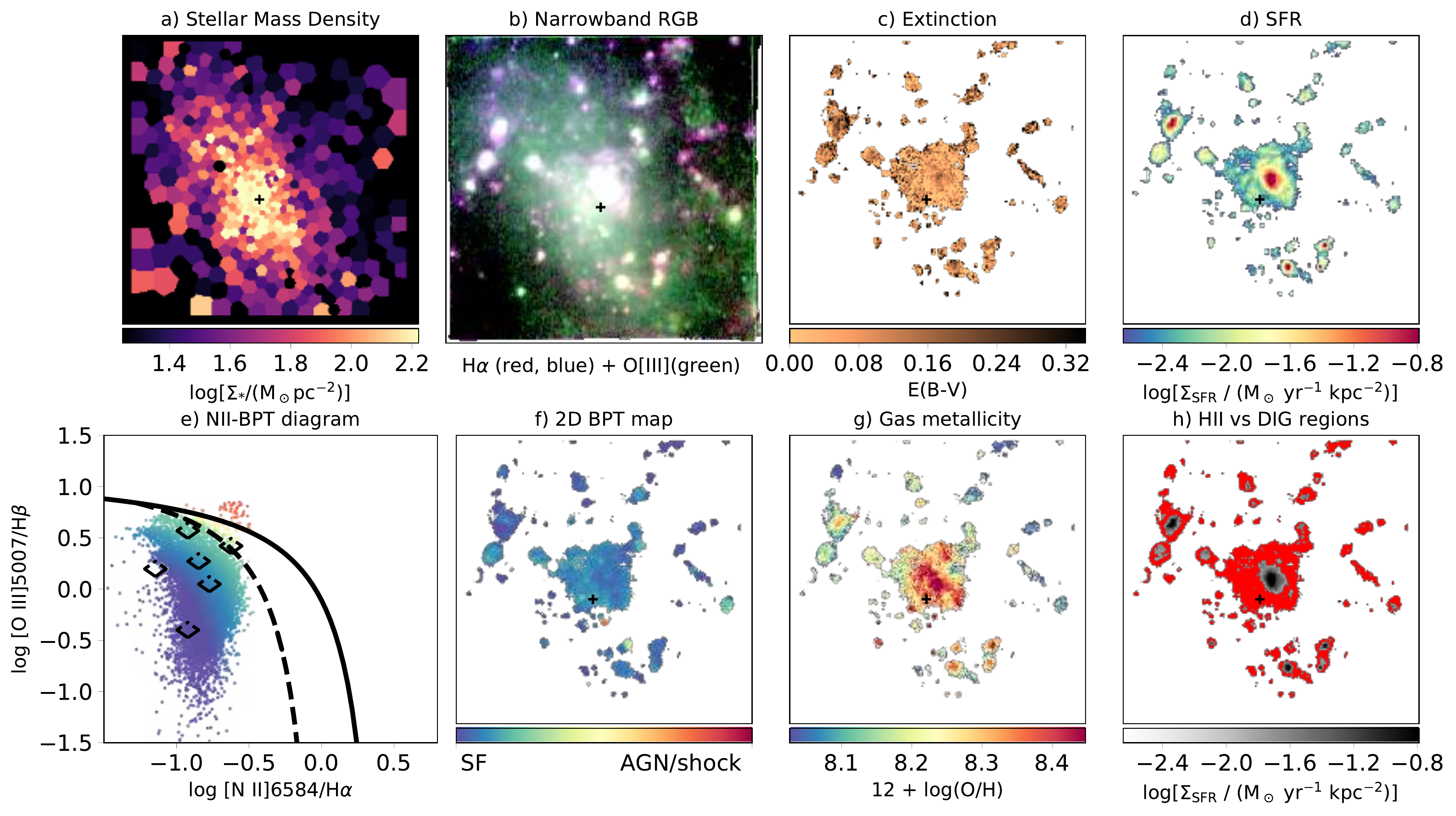}
\caption{This low-mass galaxy has low levels of star formation and very low extinction (as ESO~499-G37). Similarly, there is no spiral structure and the few H{\sc ii} regions are dispersed in the disk. There is also a small bulge-like region in the centre. There is a central star forming region that is offset $\sim$5 arcseconds from the centre of the galaxy. The metallicity decreases with radius, and the largest deviations from the linear fit are in the H{\sc ii} regions in the centre ($\sim$ 0.2 dex) and the DIG ($\sim$- 0.2 dex).
}
\label{ngc4517aplots}
\end{center}
\end{figure*}
\clearpage
%\subsection{ESO~499-G37}

\begin{figure*}
\begin{center}
 \includegraphics[width=165mm]{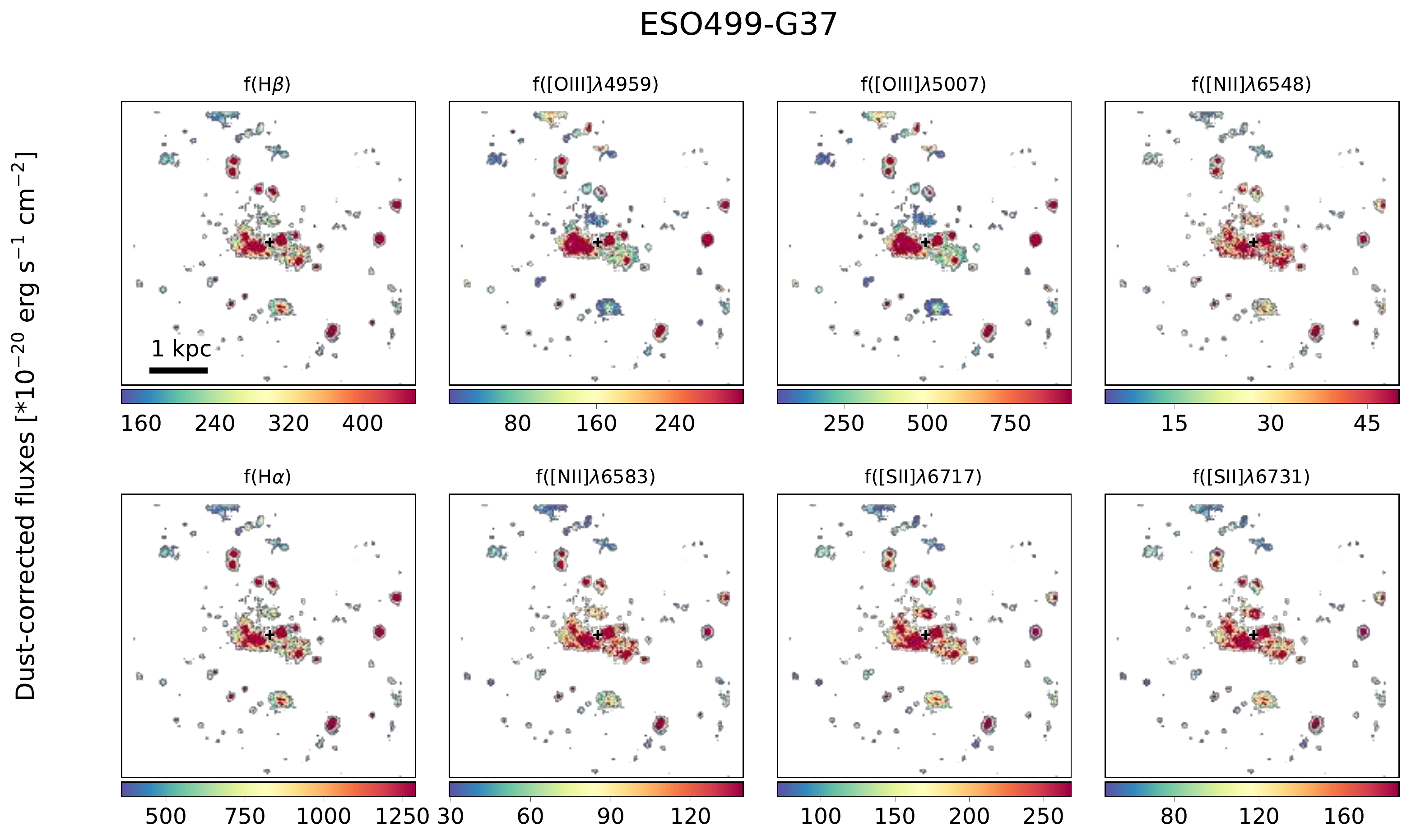}
% \caption{As Fig.~\ref{allfluxes} but for NGC~4030.}
% \label{eso499-g37fluxes}
% \end{center}
% \end{figure*}

% \begin{figure*}
% \begin{center}
 \includegraphics[width=165mm]{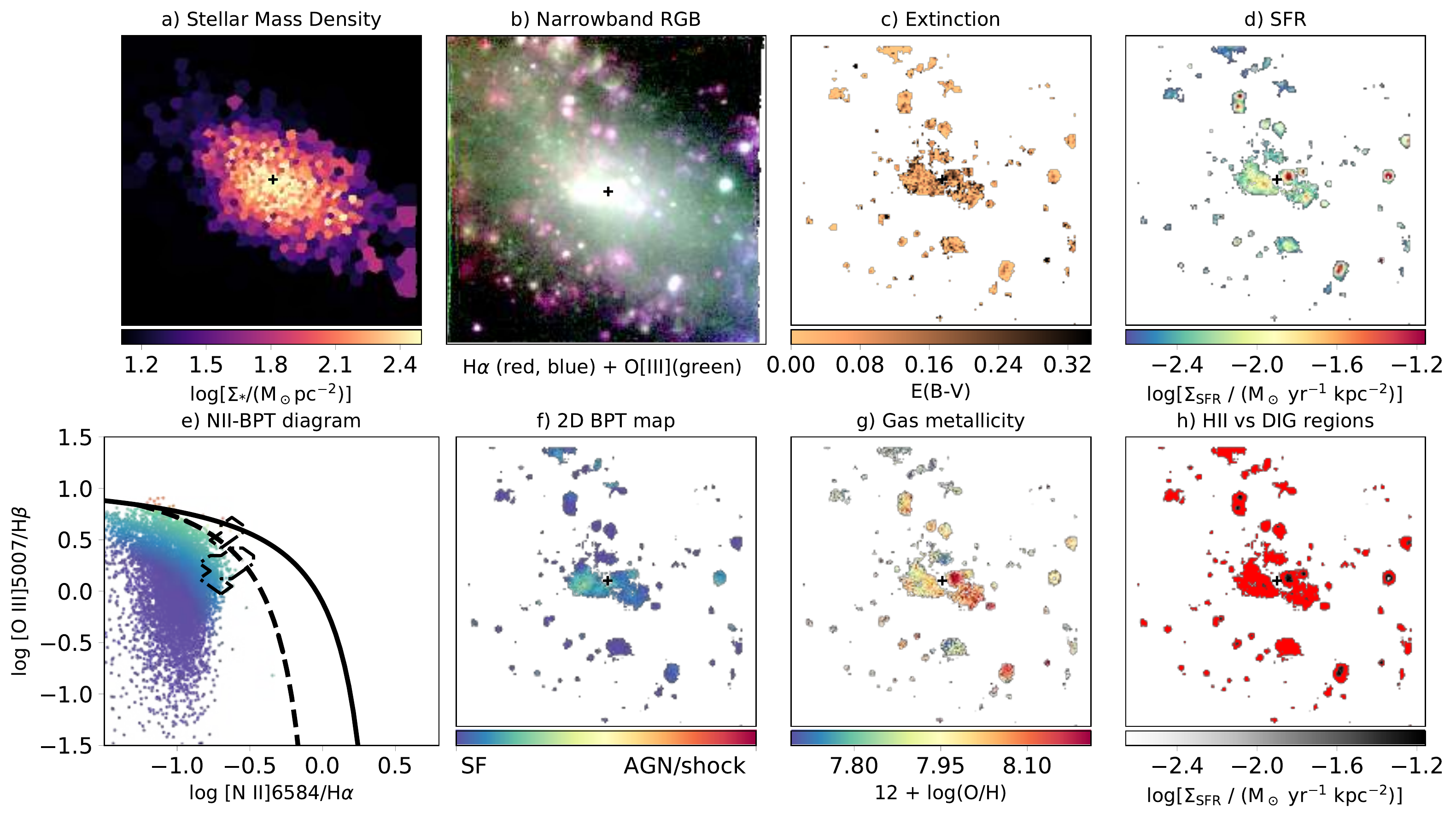}
\caption{ESO~499-G37 is one of the galaxies with lower stellar mass (M$_{\star}$<10$^{9}$M$_{\sun}$). The H$\alpha$ and H$\beta$ emission is very localized in H{\sc ii} regions, and mostly absent in any other place. This ionized gas emission is not filamentary, but patchy in the form of isolated H{\sc ii} regions that do not form a well-defined structure. The average gas metallicity of this galaxy is lower than for higher-mass galaxies and is correlated with the levels of star formation: the large amount of diffuse gas in this galaxy shows lower levels of star formation than the few H{\sc ii} regions, where the metallicity deviates by $\sim$0.4 dex from the  trend.
}
\label{eso499-g37plots}
\end{center}
\end{figure*}
\clearpage

%NGC4030 (SA(b)bc) at z=0.004887 presents a centrally condensed nucleus embedded in a large elliptical bulge. It has a flocculent spiral pattern, with at least three arms which are patchy and wispy, showing signs of dust lanes in the H band (Eskridge et al. 2002). NGC4030 shows a very regular kinematics for both the stellar and gaseous components (Ganda et al. 2006). The major axis rotational velocity reaches $\sim 100 km/s$. The velocity dispersion stays moderately flat within the bulge region and drops off outside (Fabricius et al. 2012). The $H_{\beta}$ map seems to be asymmetric with respect to the centre, since the western side displays higher values than the eastern one. The stellar populations are young over the whole field and the metallicity decreases moving outwards (Ganda et al. 2007).

\section{Determination of the BPT-parameter $\eta$}
\label{App:AppendixB}

For a point P$_1=(x_1, y_1$) in the SF region of the BPT (see Fig.~\ref{eta}), we compute the distance $d_1$ between the point P$_1$ and the dashed curve described by equation $f(x)=0.61/(x-0.05)+1.3$) as:

\begin{align}\label{equation5}
d_1=|\vec{d}_1|=d(\mathrm{P}_{1},f(x))=d((x_1,y_1),f(x))=~~~~~~~~~~\nonumber \\
=\sqrt{(x_1-x)^2+\{y_1-[0.61/(x-0.05)+1.3]\}^2}.
 \end{align} 
The minimum distance would correspond to the solution $D(d_1)/dx=0$, or equivalently, the solution to:
\begin{align}
40000x^4-40000x_1x^3-6000x^3+6000x^2x_1+300x^2\nonumber \\
-300xx_1-31725x+5x_1+24400y_1x\nonumber \\
-1220y_1-13298=0.\label{equation6}
\end{align}
Out of the four roots of eq.~\ref{equation6}, we discard the imaginary solutions and keep the solution that leads to the minimum, positive $d_1$. This solution $(x_0)$ would correspond to the $x$-value of the point $(x_0,y_0)$ which is on the curve described by equation $f(x)$ and is closest to P$_1$, so finally
\begin{align}
d_1=d(((x_1,y_1),(x_0,y_0))=~~~~~~~~~~\nonumber \\
=\sqrt{(x_1-x_0)^2+\{y_1-[0.61/(x_0-0.05)+1.3]\}^2}~~~~~~~~~~\nonumber \\
\end{align}
and according to our definition of $\eta$:
\begin{equation}
\eta_{\mathrm{SF}}=-0.5-d_1.
\end{equation}

Similarly for the AGN/shock region of the BPT diagram, the distance $d_2$ between the point P$_2=(x_2,y_2)$ and the solid curve described by equation $g(x)=0.61/(x-0.47)+1.19$), is:

\begin{align}\label{equation9}
d_2=|\vec{d}_2|=d(\mathrm{P}_{2},g(x))=d((x_2,y_2),g(x))=~~~~~~~~~~\nonumber \\
=\sqrt{(x_2-x)^2+\{y_2-[0.61/(x-0.47)+1.19]\}^2}
 \end{align} 
with the minimum when
\begin{align}
D(d_2)/dx=0 \nonumber \\
1000000x^4-1000000x_2x^3-1410000x^3+1410000x^2x_2\nonumber\\
+662700x^2-662700xx_2 - 829723x+103823x_2\nonumber\\
+610000y_2x-286700y_2-30927=0.
\end{align}
and according to our definition of $\eta$:
\begin{equation}
\eta_{\mathrm{AGN}}=0.5+d_2.
\end{equation}

As for the intermediate region, we can compute the distance from the point $\mathrm{P}_3=(x_3,y_3)$ to both the SF and AGN curves as $d_1=d(\mathrm{P}_{3},f(x))$ and $d_2=d(\mathrm{P}_{3},g(x))$. Following our definition of the continuous variable $\eta_{\rm INTER}\in(-0.5,0.5)$, we compute $|\vec{d}_3|$ as the minimum of $d_1$ and $d_2$, normalized by the sum $d_1 + d_2$. This yields: 
\begin{equation}
  \eta_{\mathrm{INTER}}=-0.5+\dfrac{d_1}{d_1+d_2} = 0.5-\dfrac{d_2}{d_1+d_2}
\end{equation}

\section{Beyond the radial gradients: 2D metallicity distributions}
\label{App:AppendixC}

\begin{figure*}
\begin{center}
 \includegraphics[width=165mm]{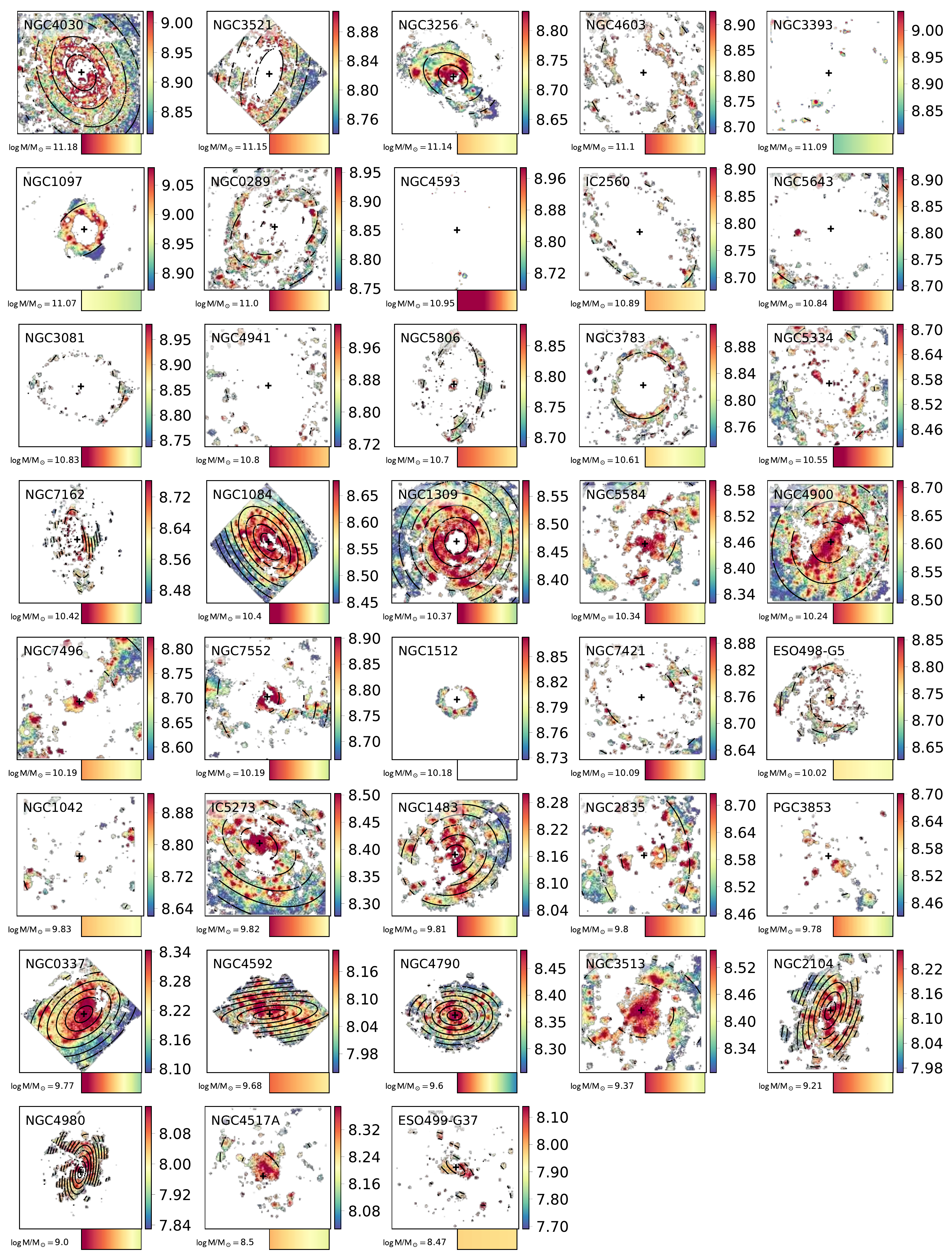}
\caption{2-D gas metallicity maps of the SF regions for the 38 galaxies in this paper, ordered by decreasing stellar mass. In each panel, the azimuthally-averaged metallicity gradient is shown as a colourbar below the map, which clearly makes the point that the metallicity gradients heavily smooth at all radii over very different metallicities, often  associated with clearly identifiable structural components of the disks. The 2D maps and the gradients are based on the identical colour bar (shown on the right). Black contours denote the radial bins in steps of $0.3R/R_e$.}
\label{oh12maps}
\end{center}
\end{figure*}

The 2D metallicity maps shown in Fig.~\ref{oh12maps} explain the origin of the  scatter in metallicity for the 38 galaxies presented in this paper. For an easy comparison, this figure also shows, for each galaxy,  its azimuthally-averaged metallicity gradient a colour-coded strip plotted at the bottom of the 2D maps (using the same colour scheme as the 2-D map).  

\begin{figure*}
\begin{center}
 \includegraphics[width=165mm]{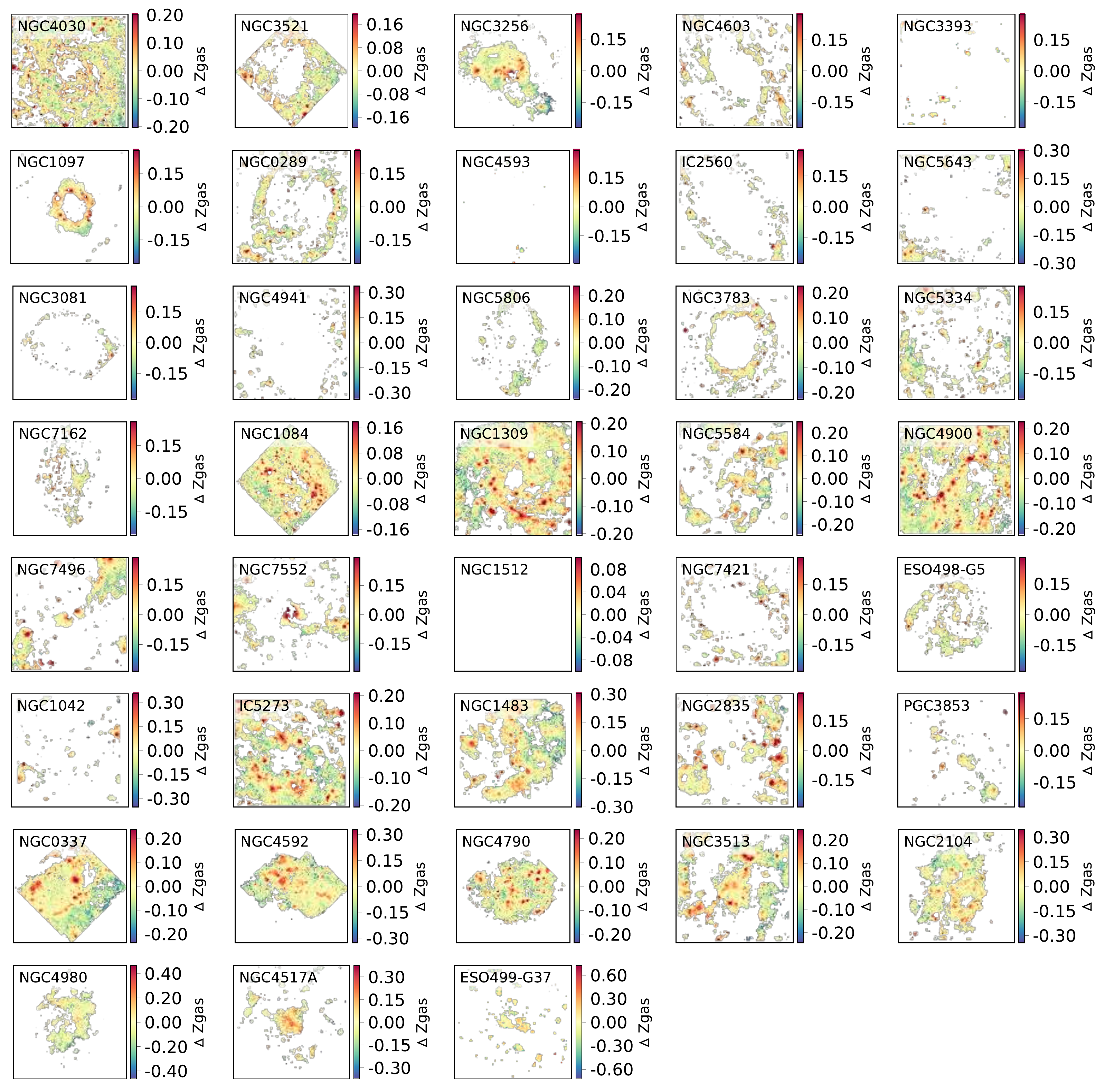}
\caption{Residual maps of the gas metallicity of the SF regions for the 38 MAD galaxies in the paper, ordered in decreasing total stellar mass. A model metallicity map for each galaxy has been generated from the measured metallicity gradients assuming azimuthally symmetric metallicity distributions. Then, the residual maps have been computed by subtracting these model maps to the metallicity map.}
\label{Zgas_residuals}
\end{center}
\end{figure*}

To further highlight the gas metallicity variations that are ubiquitously present in galactic disks  at any given radius, we present in Fig.~\ref{Zgas_residuals} ``residual'' maps generated by subtracting, from our 2D gas metallicity maps, the 2D maps generated from the measured metallicity gradients assuming azimuthally symmetric metallicity distributions. It is clear from those residual maps that the largest deviations from the linear fit (i.e., from the gradients) are usually found in the H{\sc ii} regions. 

\section{Complementary Figures to the Resolved mass-metallicity relation}
\label{App:AppendixD}

\begin{figure*}
\begin{center}
 \includegraphics[width=165mm]{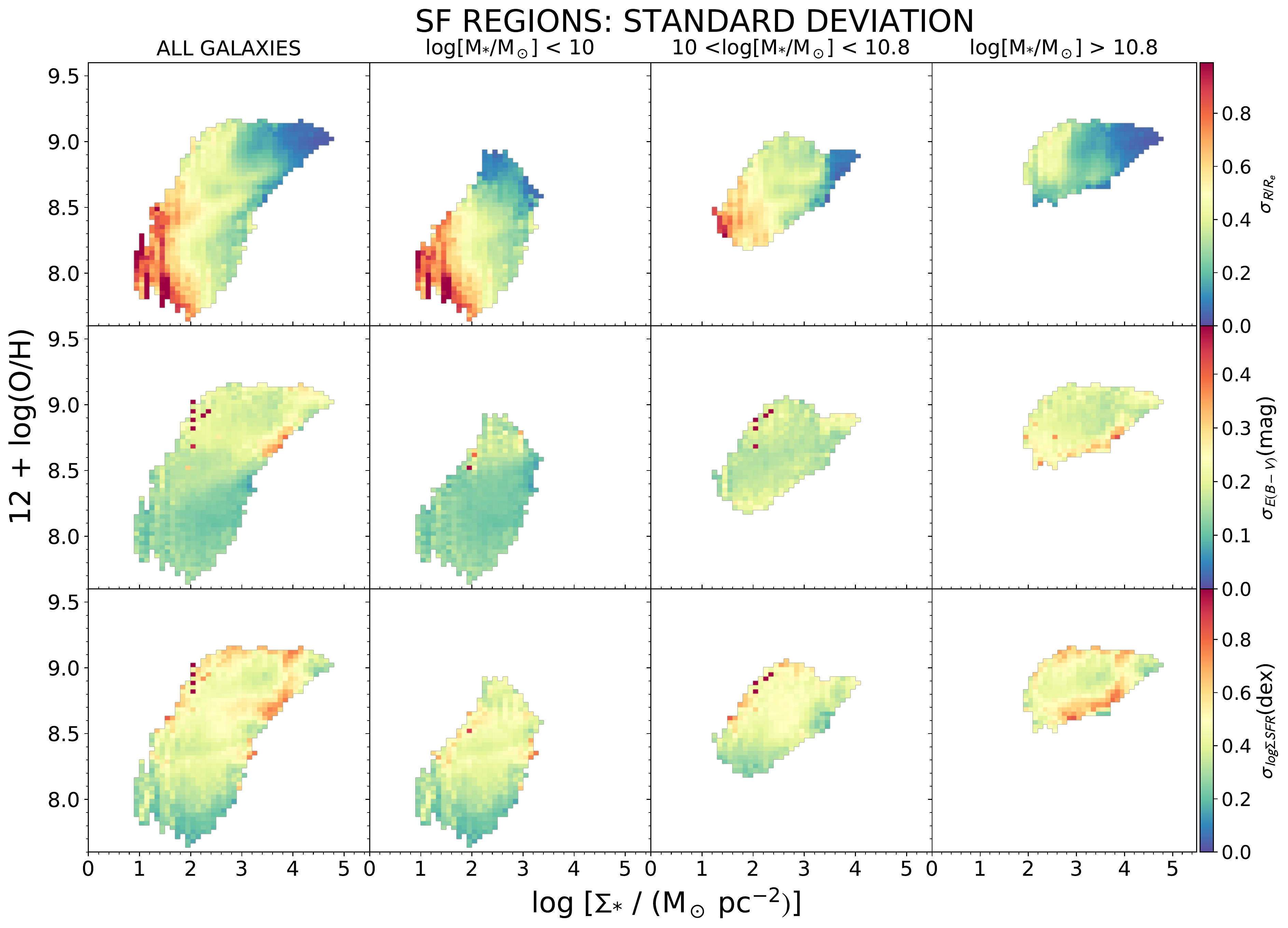}
\caption{As Fig.~\ref{supergalaxyA} but representing the standard deviation of each median distribution for each bin.}
\label{supergalaxyAsigma}
\end{center}
\end{figure*}

\begin{figure*}
\begin{center}
 \includegraphics[width=165mm]{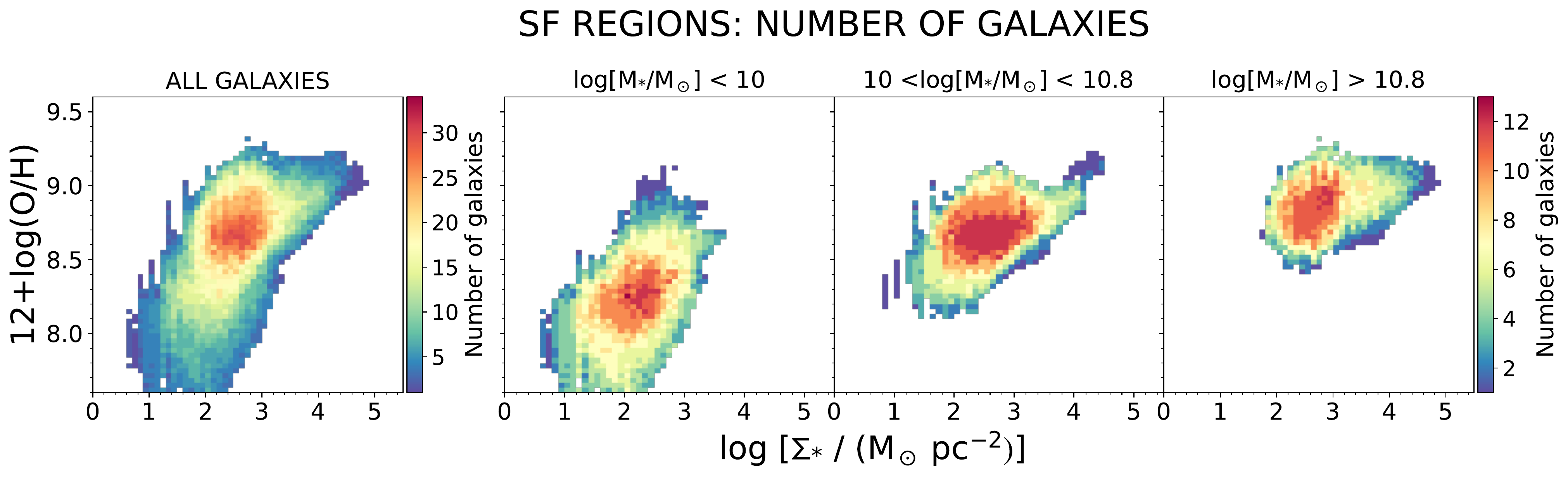}
\caption{Number of galaxies presented in each bin from Figs.~\ref{supergalaxyA} and \ref{supergalaxyAsigma}.}
\label{supergalaxyAnumber}
\end{center}
\end{figure*}

\begin{figure*}
\begin{center}
 \includegraphics[width=165mm]{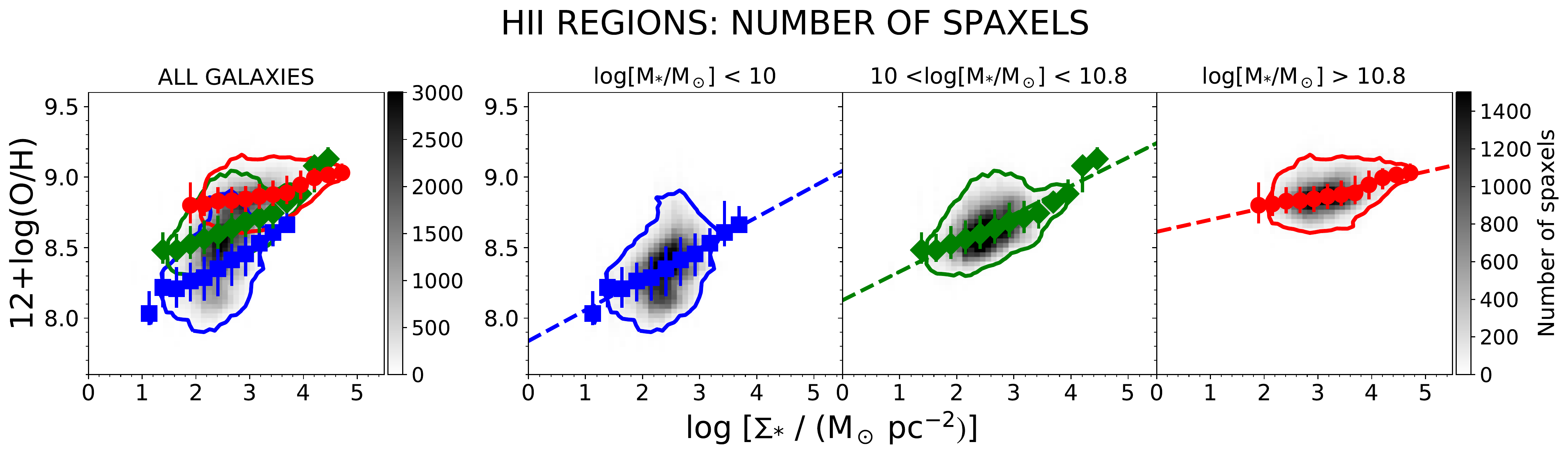}
 \includegraphics[width=165mm]{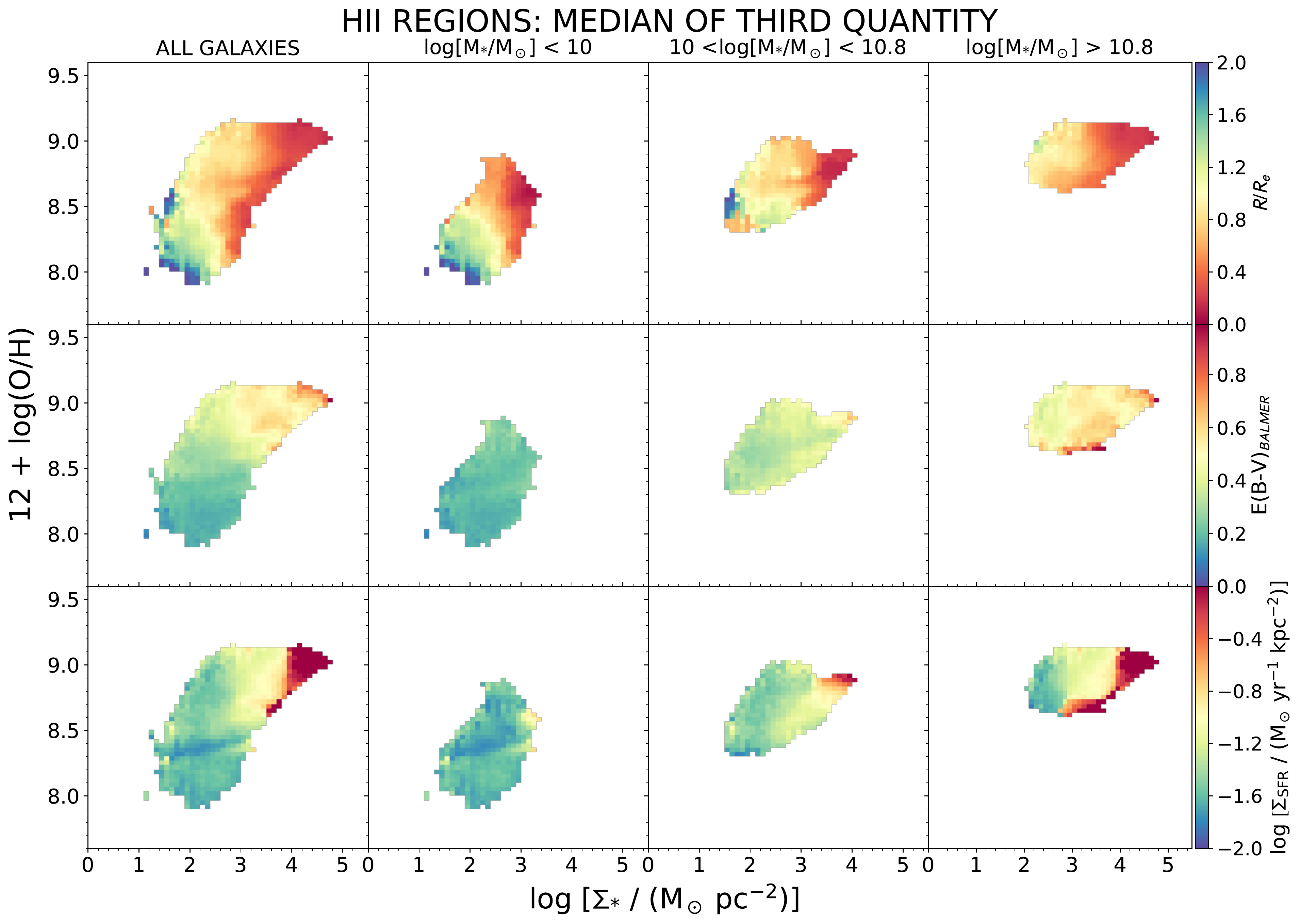}
\caption{As Fig.~\ref{supergalaxyA} but restricting to all the H{\sc ii} regions.}
\label{supergalaxyA_HII}
\end{center}
\end{figure*}

\begin{figure*}
\begin{center}
 \includegraphics[width=165mm]{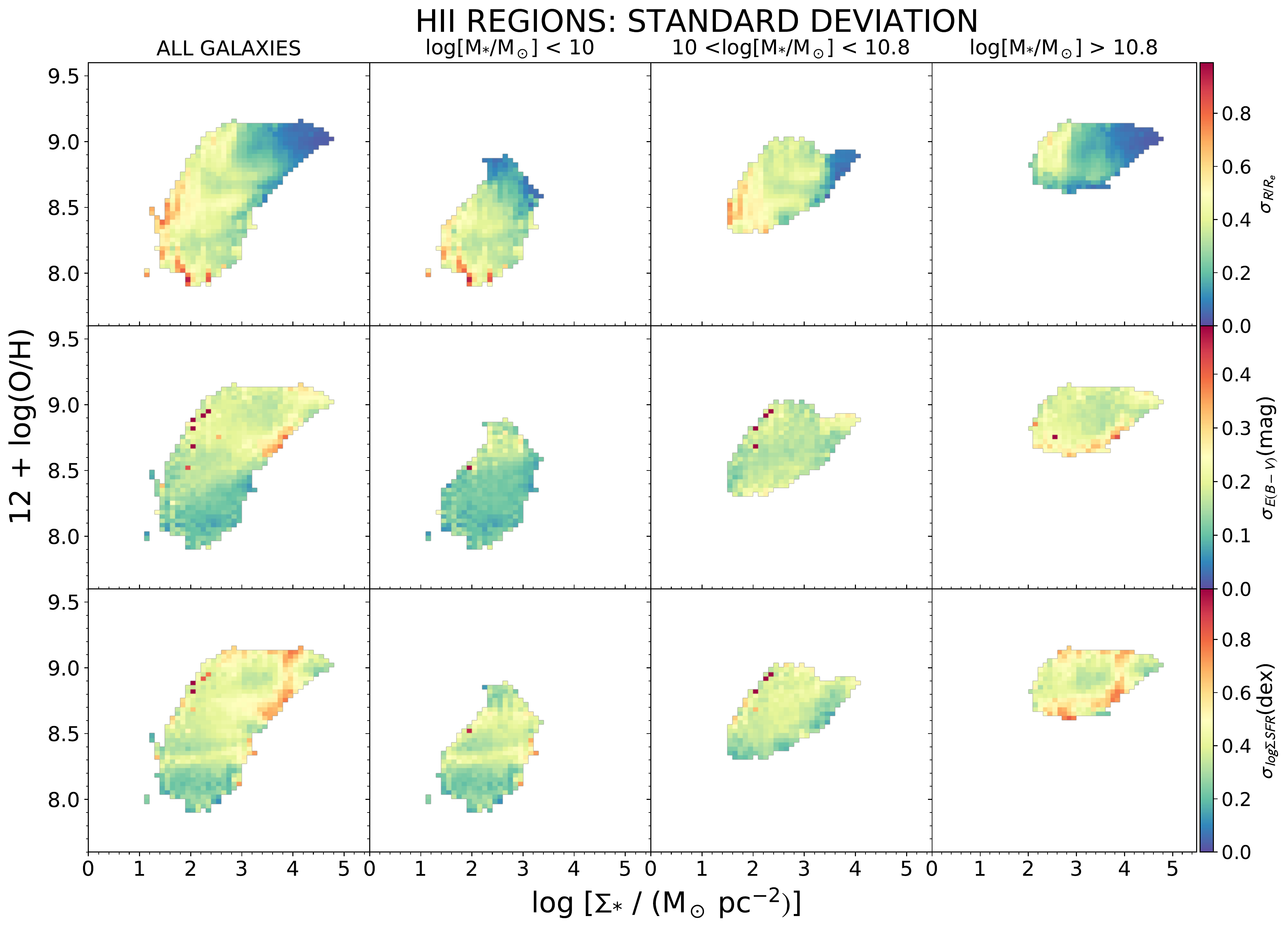}
\caption{As Fig.~\ref{supergalaxyA_HII} but representing the standard deviation of each median distribution for each bin.}
\label{supergalaxyAsigma_HII}
\end{center}
\end{figure*}

\begin{figure*}
\begin{center}
 \includegraphics[width=165mm]{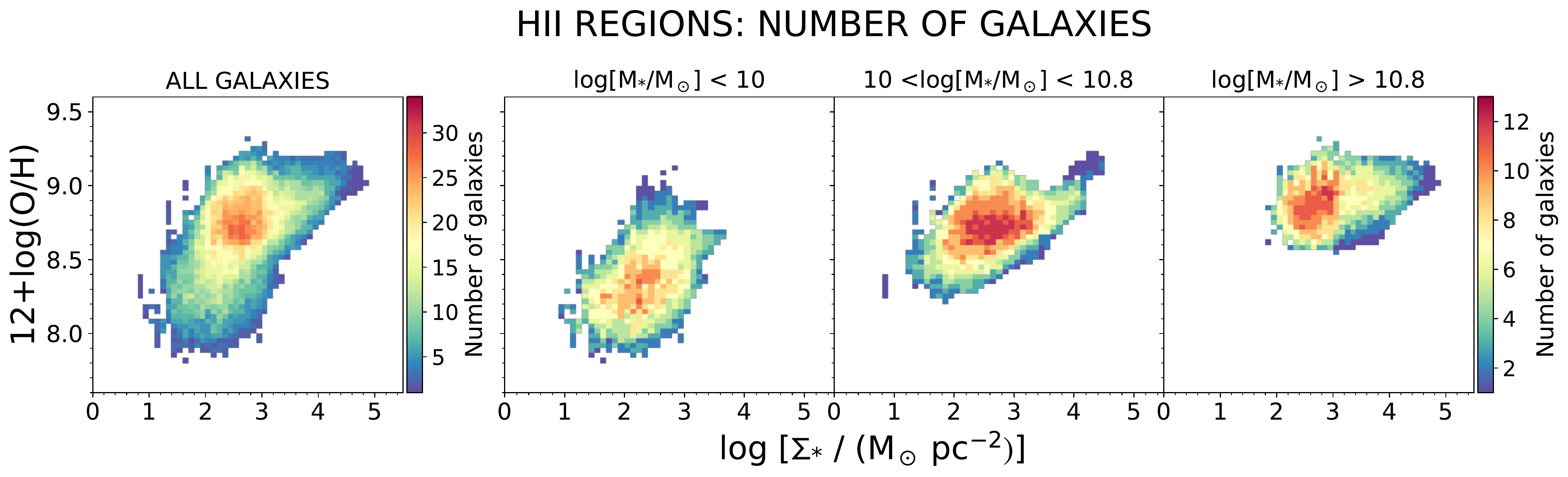}
\caption{Number of galaxies presented in each bin from Figs.~\ref{supergalaxyA_HII} and \ref{supergalaxyAsigma_HII}.}
\label{supergalaxyAnumber_HII}
\end{center}
\end{figure*}

\begin{figure*}
\begin{center}
 \includegraphics[width=165mm]{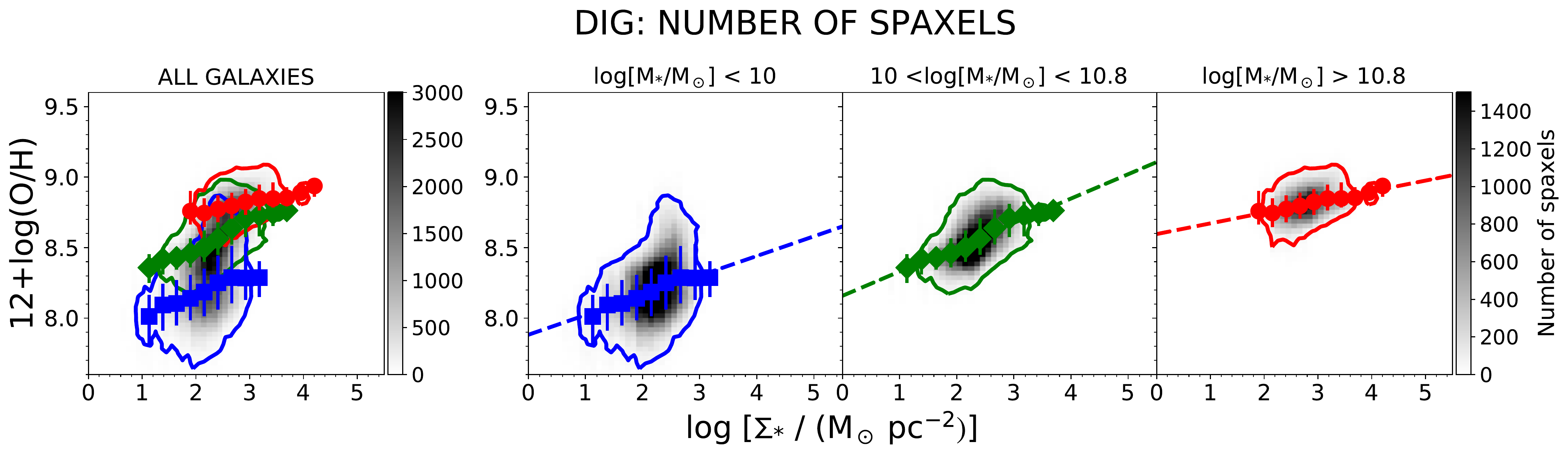}
 \includegraphics[width=165mm]{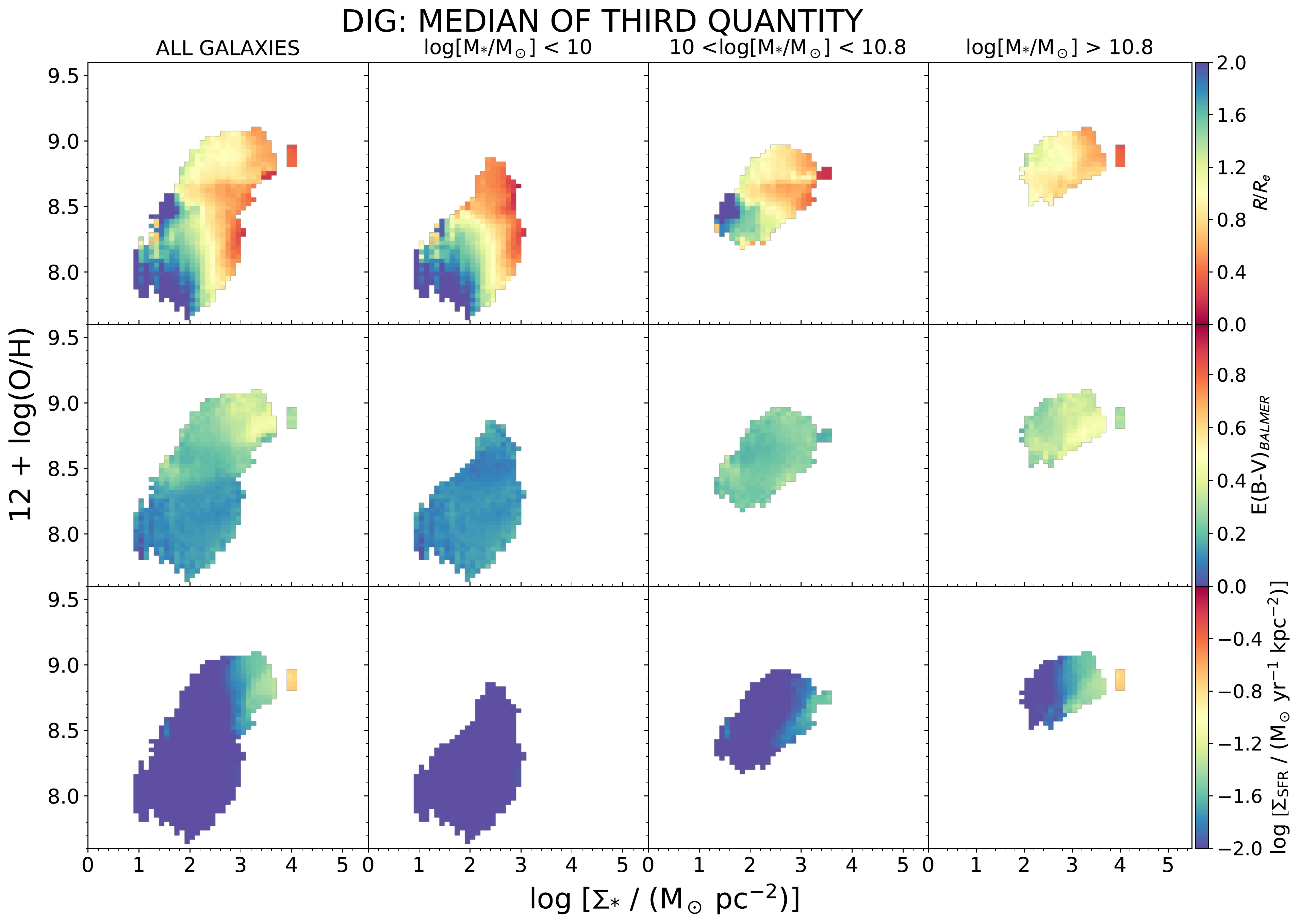}
\caption{As Fig.~\ref{supergalaxyA} but restricting to all the DIG.}
\label{supergalaxyA_DIG}
\end{center}
\end{figure*}

\begin{figure*}
\begin{center}
 \includegraphics[width=165mm]{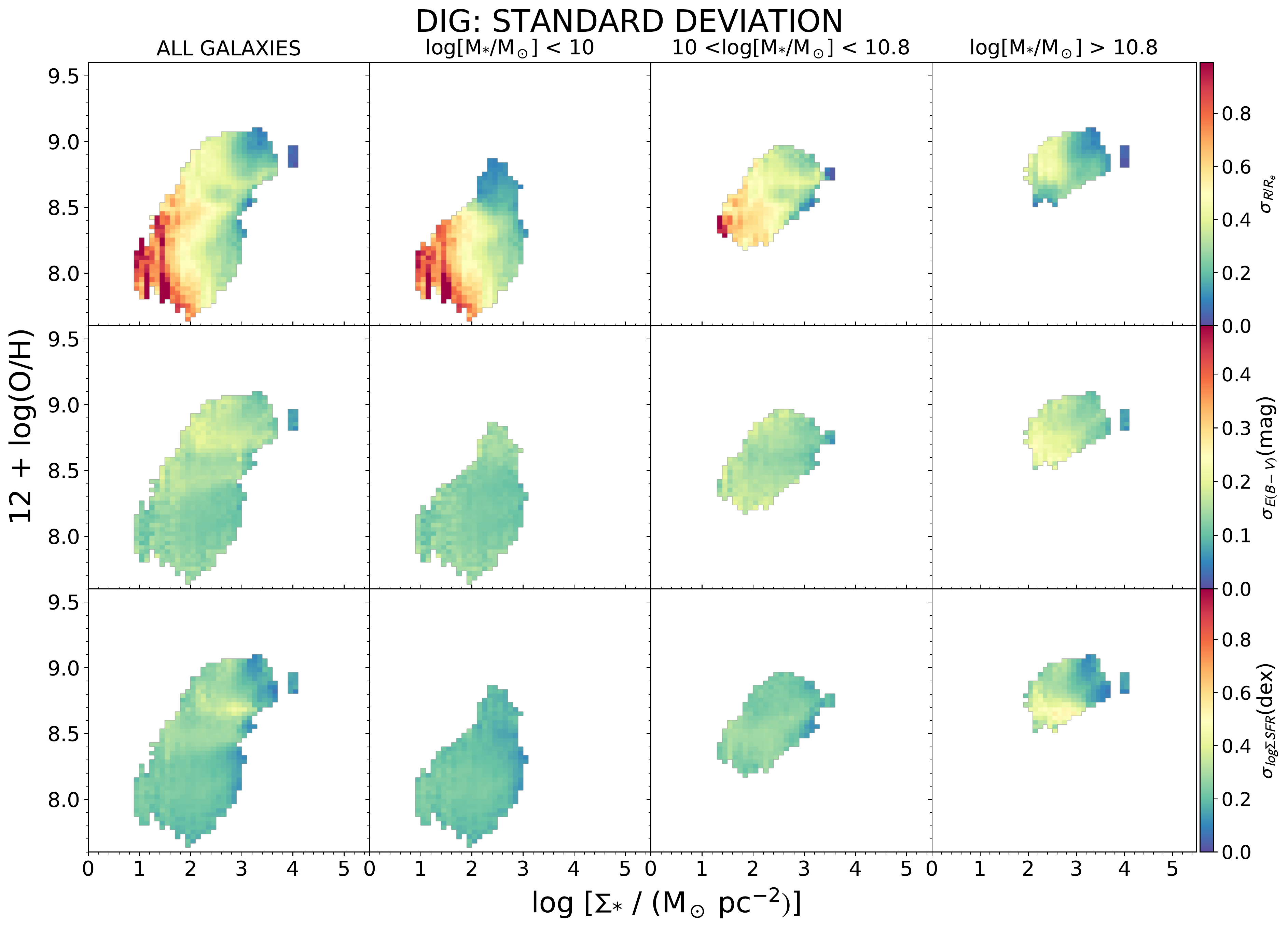}
\caption{As Fig.~\ref{supergalaxyA_DIG} but representing the standard deviation of each median distribution for each bin.}
\label{supergalaxyAsigma_DIG}
\end{center}
\end{figure*}

\begin{figure*}
\begin{center}
 \includegraphics[width=165mm]{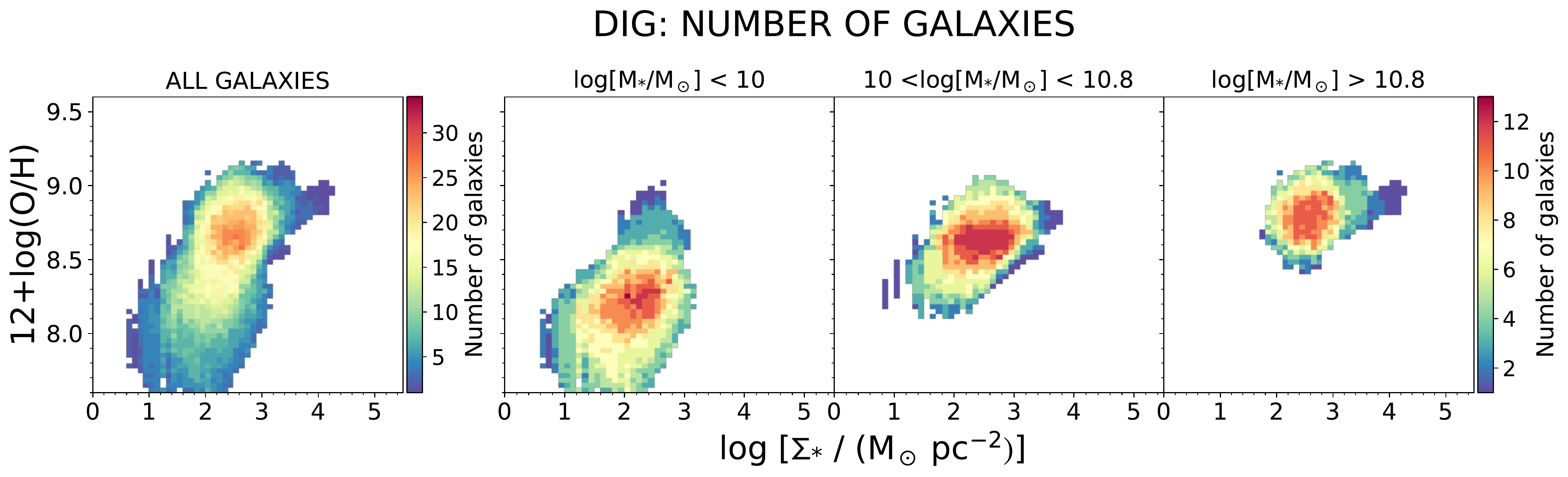}
\caption{Number of galaxies presented in each bin from Figs.~\ref{supergalaxyA_DIG} and \ref{supergalaxyAsigma_DIG}.}
\label{supergalaxyAnumber_DIG}
\end{center}
\end{figure*}

\section{Complementary Figures to the Resolved mass-SFR relation}
\label{App:AppendixE}

\begin{figure*}
\begin{center}
 \includegraphics[width=165mm]{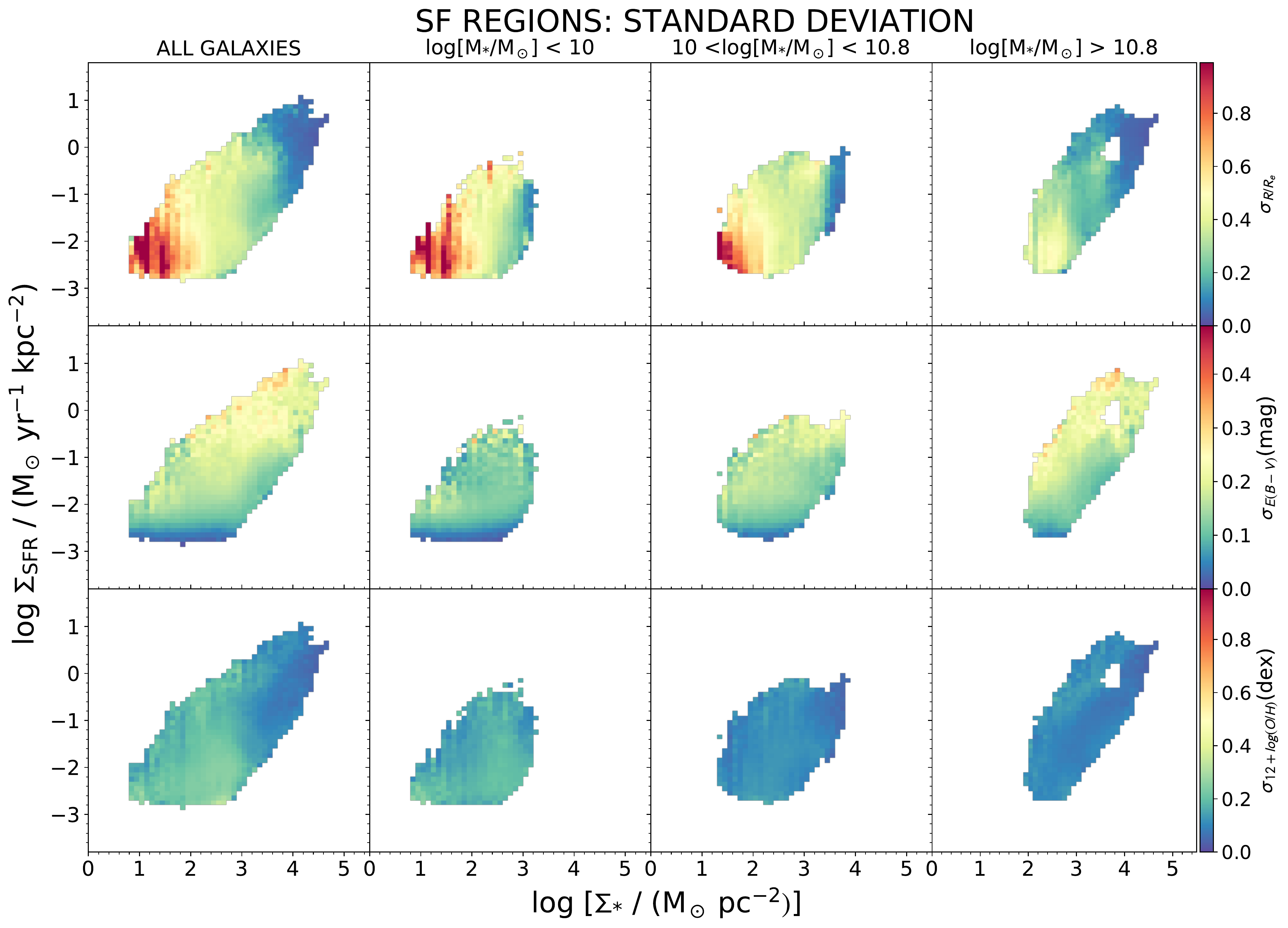}
\caption{As Fig.~\ref{supergalaxyB} but representing the standard deviation of each median distribution for each bin.}
\label{supergalaxyBsigma}
\end{center}
\end{figure*}

\begin{figure*}
\begin{center}
 \includegraphics[width=165mm]{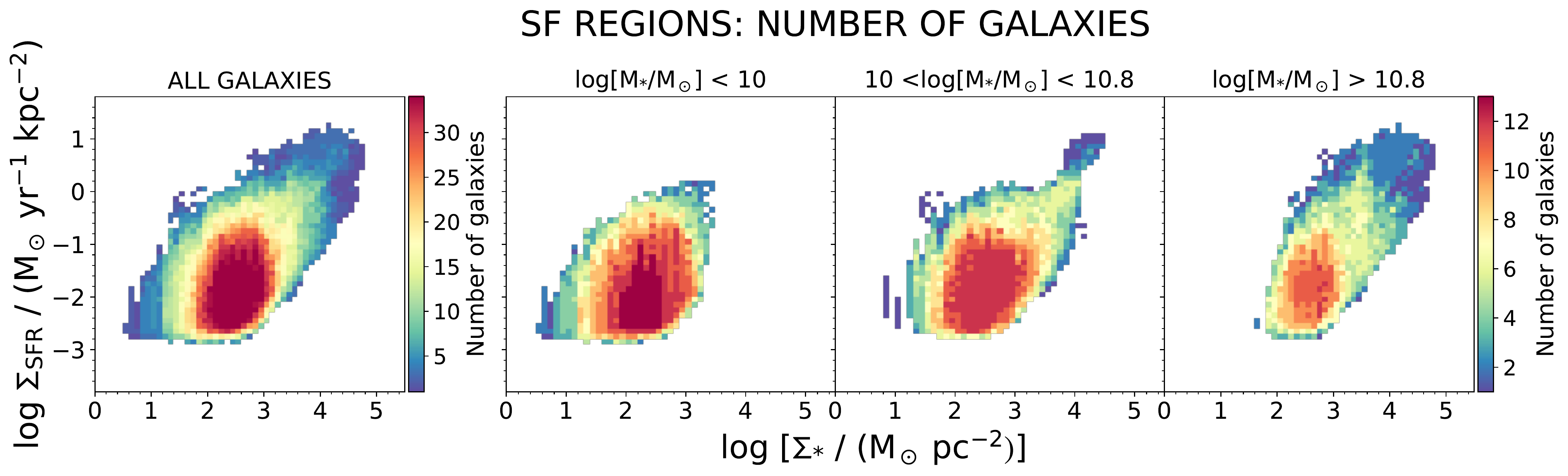}
\caption{Number of galaxies presented in each bin from Figs.~\ref{supergalaxyB} and \ref{supergalaxyBsigma}.}
\label{supergalaxyBnumber}
\end{center}
\end{figure*}

\begin{figure*}
\begin{center}
 \includegraphics[width=165mm]{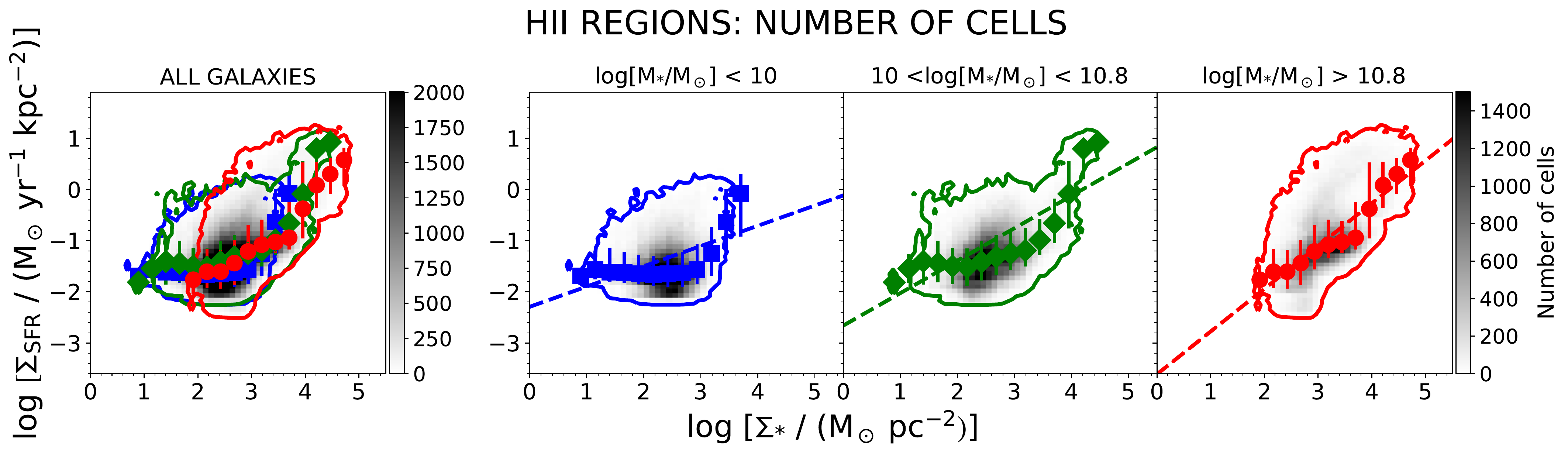}
 \includegraphics[width=165mm]{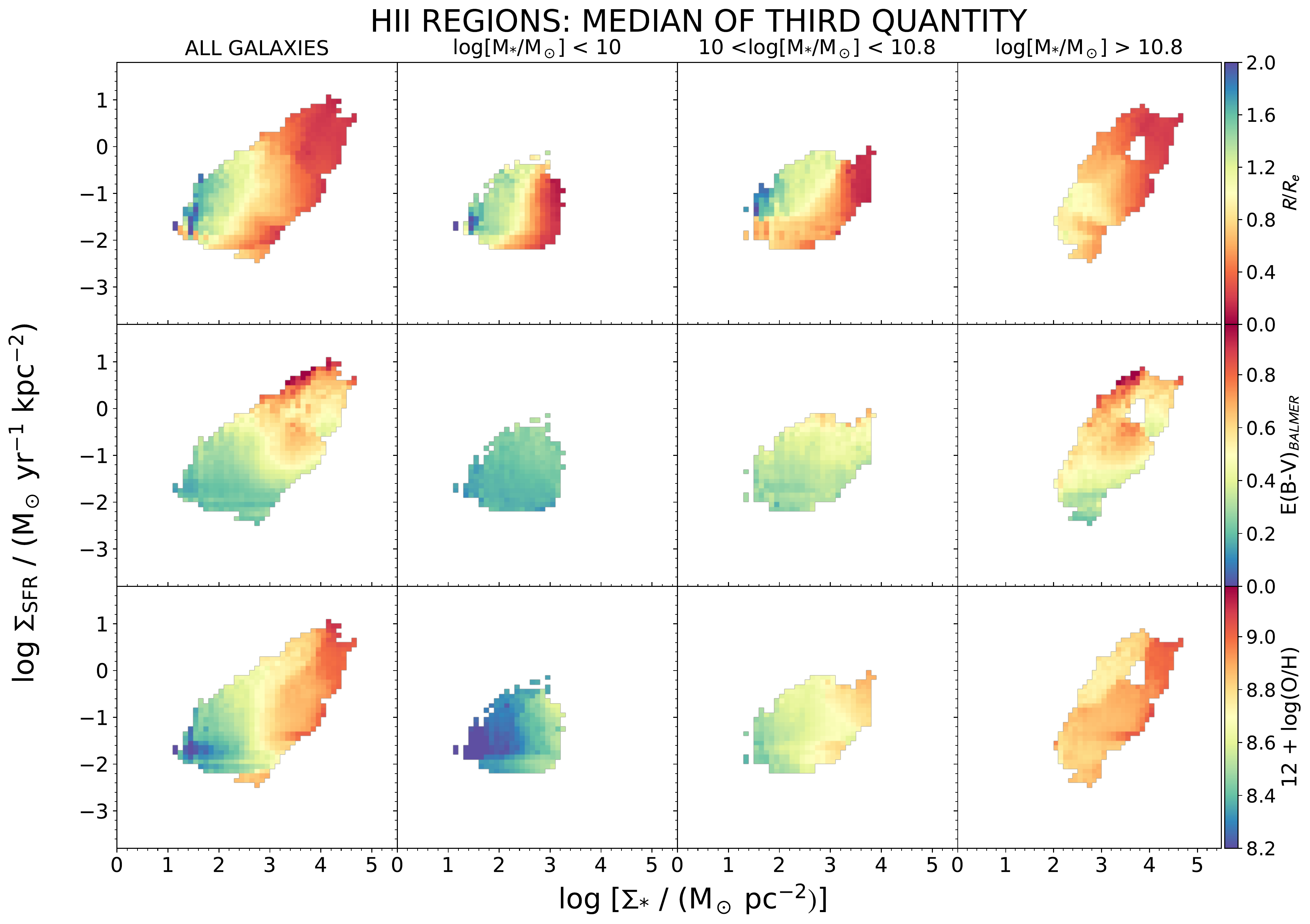}
\caption{As Fig.~\ref{supergalaxyB} but restricting to all the H{\sc ii} regions.}
\label{supergalaxyB_HII}
\end{center}
\end{figure*}

\begin{figure*}
\begin{center}
 \includegraphics[width=165mm]{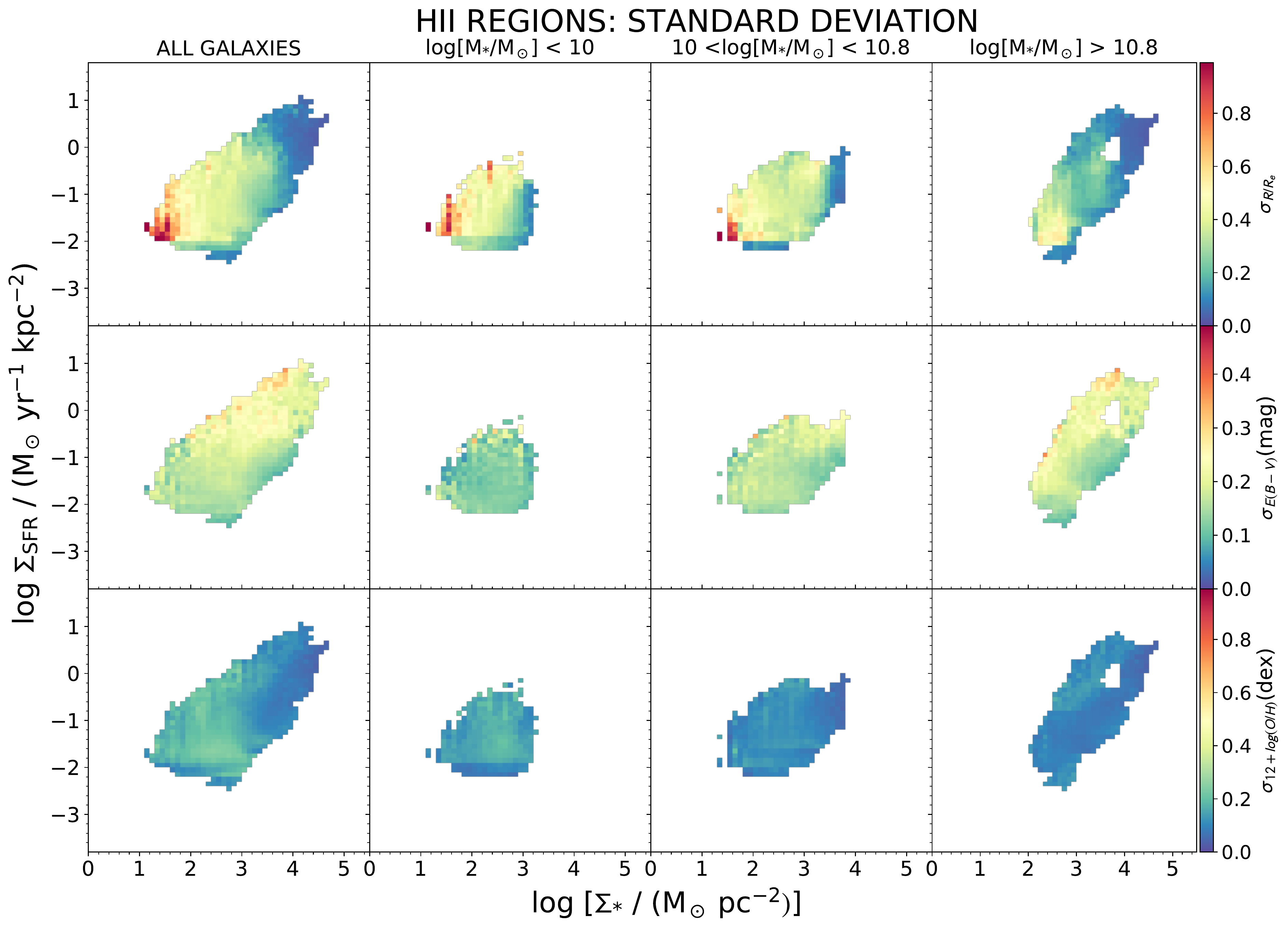}
\caption{As Fig.~\ref{supergalaxyB_HII} but representing the standard deviation of each median distribution for each bin.}
\label{supergalaxyBsigma_HII}
\end{center}
\end{figure*}

\begin{figure*}
\begin{center}
 \includegraphics[width=165mm]{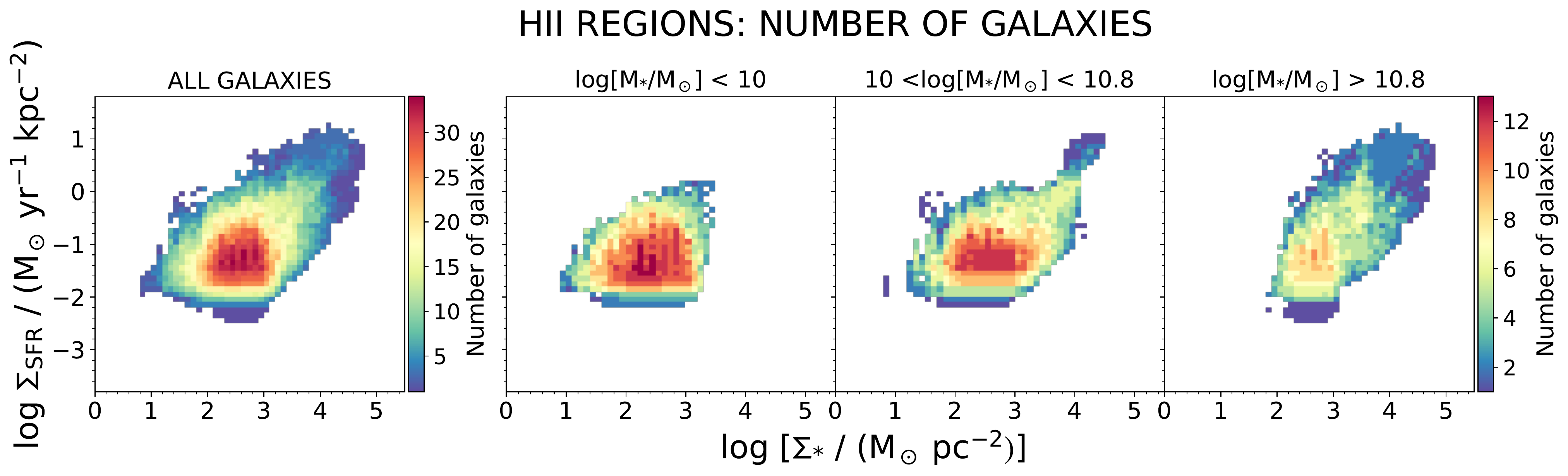}
\caption{Number of galaxies presented in each bin from Figs.~\ref{supergalaxyB_HII} and \ref{supergalaxyBsigma_HII}.}
\label{supergalaxyBnumber_HII}
\end{center}
\end{figure*}

\begin{figure*}
\begin{center}
 \includegraphics[width=165mm]{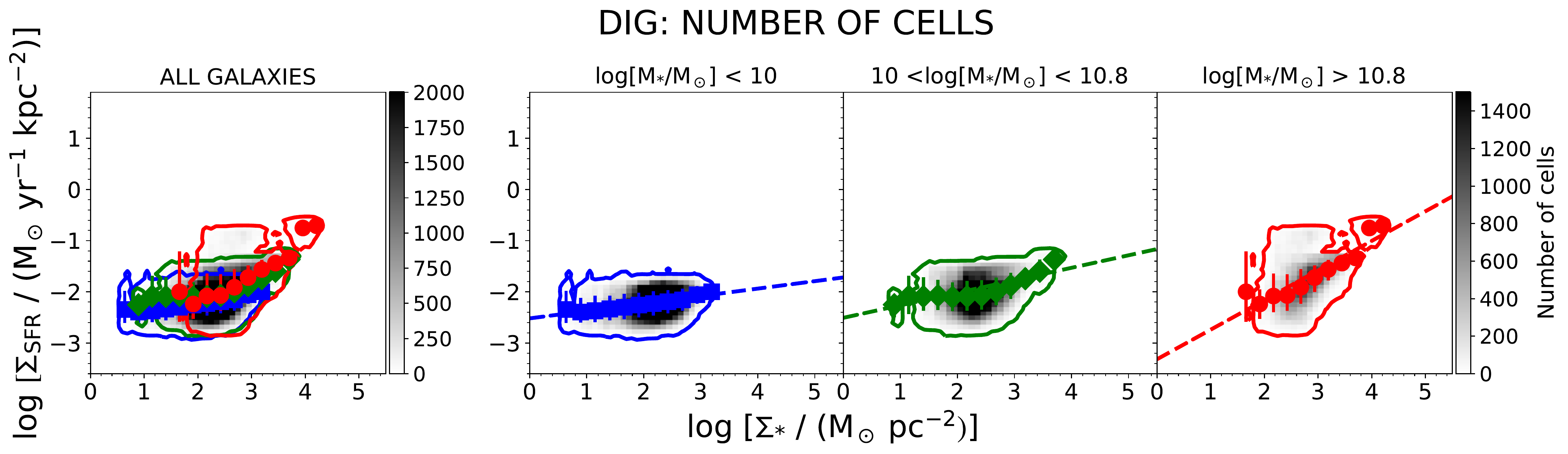}
 \includegraphics[width=165mm]{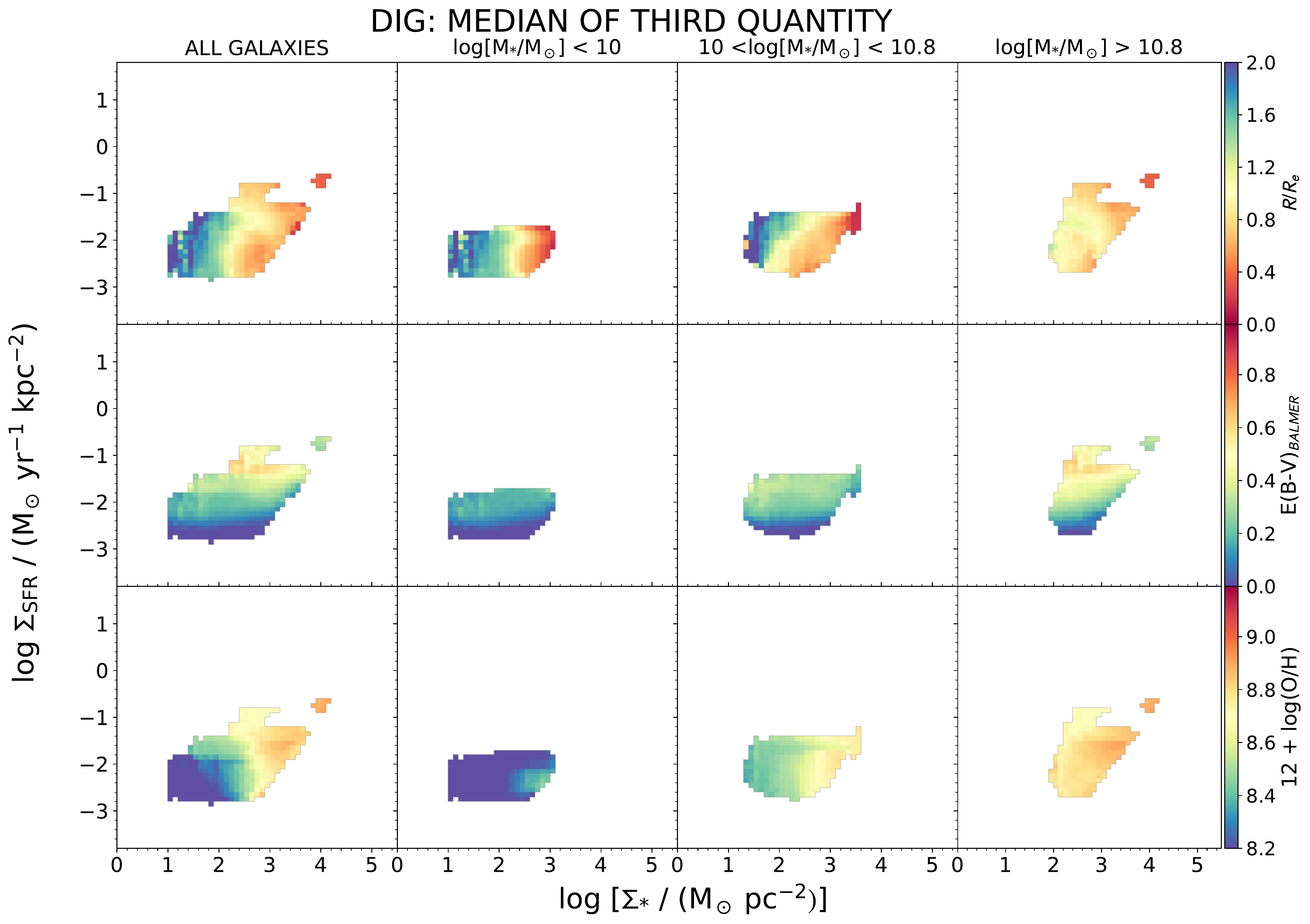}
\caption{As Fig.~\ref{supergalaxyB} but restricting to all the DIG.}
\label{supergalaxyB_DIG}
\end{center}
\end{figure*}

\begin{figure*}
\begin{center}
 \includegraphics[width=165mm]{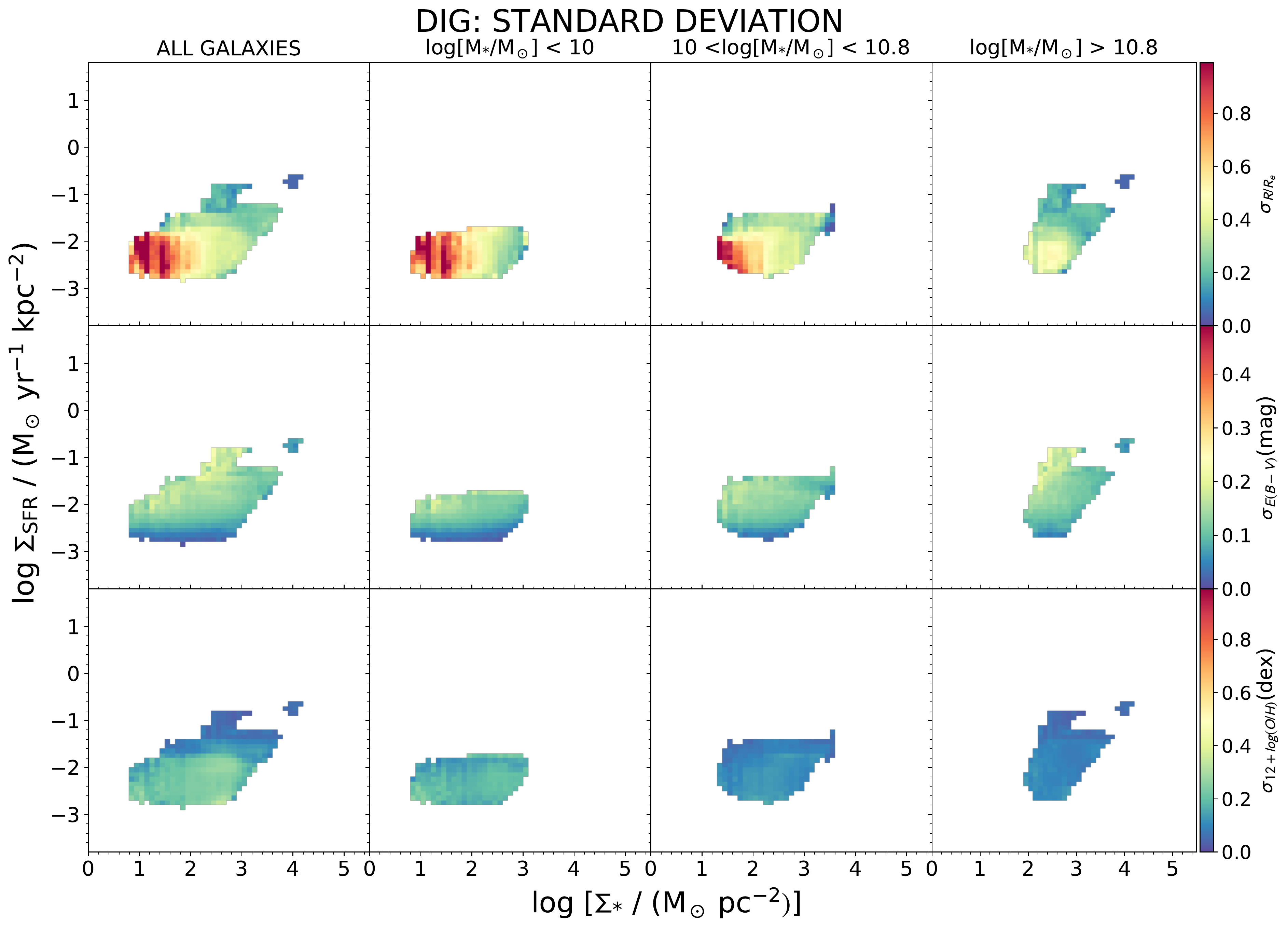}
\caption{As Fig.~\ref{supergalaxyB_DIG} but representing the standard deviation of each median distribution for each bin.}
\label{supergalaxyBsigma_DIG}
\end{center}
\end{figure*}

\begin{figure*}
\begin{center}
 \includegraphics[width=165mm]{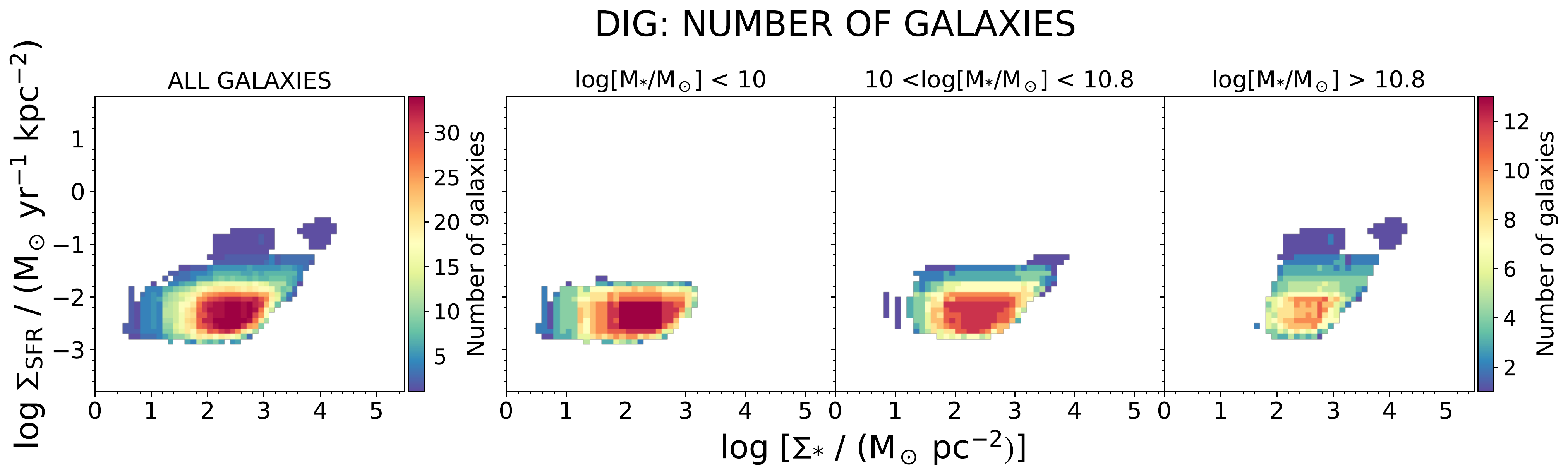}
\caption{Number of galaxies presented in each bin from Figs.~\ref{supergalaxyB_DIG} and \ref{supergalaxyBsigma_DIG}.}
\label{supergalaxyBnumber_DIG}
\end{center}
\end{figure*}

\end{document}